\documentclass[12pt]{article}

\pdfoutput=1

\usepackage{amsfonts}
\usepackage{amsmath}
\usepackage{amssymb}
\usepackage{array}
\usepackage{bigints}
\usepackage{booktabs}
\usepackage[nosort]{cite}
\usepackage{cancel}
\usepackage{color}
\usepackage{dsfont}
\usepackage{float}
\usepackage{framed}
\usepackage{graphicx}
\usepackage{indentfirst}
\usepackage{mathrsfs}
\usepackage{multirow}
\usepackage{setspace}
\usepackage{subdepth}
\usepackage{subfig}
\usepackage{titlesec}
\usepackage[dotinlabels]{titletoc}
\usepackage{wrapfig}
\usepackage[all]{xy}
\usepackage{young}
\usepackage[vcentermath]{youngtab}
\usepackage{slashed}

\usepackage{hyperref}
\hypersetup{colorlinks=true}
\hypersetup{linkcolor=black}
\hypersetup{citecolor=black}
\hypersetup{urlcolor=black}

\numberwithin{equation}{section}

\usepackage[left=2.5cm,right=2.5cm,top=2.5cm,bottom=3cm]{geometry}
\def\tilde{\widetilde}
\def\t{\tilde}
\def\hat{\widehat}

\def\bar{\overline}

\def\half{{1 \over 2}}

\def\ep{\varepsilon}
\def\1{{\mathds 1}}

\def\Re{\mathop{\rm Re}}
\def\Im{\mathop{\rm Im}}
\DeclareMathOperator{\tr}{tr}

\def\alphadot{{\dot\alpha}}

\newcommand{\Z}{{\mathbb Z}}
\newcommand{\C}{{\mathbb C}}
\newcommand{\R}{{\mathbb R}}
\def\P{\hbox{$\mathbb P$}}

\def\SL{{\mathscr L}}

\def\SV{{\mathscr V}}

\def\CF{{\mathcal F}}

\def\CH{{\mathcal H}}

\def\CM{{\mathcal M}}
\def\CN{{\mathcal N}}
\def\CO{{\mathcal O}}

\def\CQ{{\mathcal Q}}
\def\CR{{\mathcal R}}

\def\CT{{\mathcal T}}

\def\CV{{\mathcal V}}
\def\CW{{\mathcal W}}

\def\be{\begin{equation}}
\def\ee{\end{equation}}
\def\bea{\begin{eqnarray}}
\def\eea{\end{eqnarray}}

\DeclareFontShape{OT1}{cmr}{mx}{n}%
    {<->cmr10}{}
\newcommand{\mytitlefont}{\fontseries{mx}\selectfont}
\DeclareMathAlphabet{\titlemath}{OT1}{cmr}{mx}{n}

	\usepackage[usenames,dvipsnames]{xcolor} 
	\usepackage{tikz,tikz-3dplot, pgfplots,tkz-graph}
	\usetikzlibrary{arrows,shapes}
	\usetikzlibrary[positioning,patterns,decorations.markings] 
	\usepackage[rightcaption]{sidecap}
	\tikzstyle arrowstyle=[scale=1]
	\tikzstyle directed=[postaction={decorate,decoration={markings,
    mark=at position .65 with {\arrow[arrowstyle]{stealth}}}}]
	\tikzstyle reverse directed=[postaction={decorate,decoration={markings,
    mark=at position .65 with {\arrowreversed[arrowstyle]{stealth};}}}]
    
\def\a{{\alpha}}
\def\b{{\beta}}

\newcommand{\bma}{\begin{matrix}}
\newcommand{\ema}{\cr\end{matrix}}
\newcommand{\<}{\langle}
\renewcommand{\>}{\rangle}

	\def\cF{{\cal F}}
	
	\def\cH{{\cal H}}

	\def\cL{{\cal L}}
	
	\def\cN{{\cal N}}
	\def\cO{{\cal O}}
	
	\def\cQ{{\cal Q}}
	
	\def\cS{{\cal S}}
	\def\cT{{\cal T}}

	\def\mA{\mathfrak{A}}
	\def\mB{\mathfrak{B}}

	\def\mJ{\mathfrak{J}}

\def\ZZ{{\mathbb Z}}
\def\RR{{\mathbb R}}

\def\Re{{\rm Re \,}}
\def\Im{{\rm Im \,}}
\def\tr{{\rm tr}}

\def\det{{\rm det \,}}

\def\half{{1\over 2}}
\def\thalf{{\tfrac{1}{2}}}
\def\p{\partial}

\def\a{\alpha}
\def\b{\beta}

\def\ep{\varepsilon}

\def\no{\nonumber}
\def\sm{\smallskip}

\def\AA{{\mathsf C}}
\def\BB{{\mathsf H}}

\def\supp{{(+)}}
\def\supm{{(-)}}
\def\suppm{{(\pm)}}

\newcommand{\ba}[1]{\begin{align} #1 \end{align} }

\newcommand{\bs}[1]{\begin{split} #1 \end{split} }

\begin{document}


\begin{titlepage}

\begin{center}

~\\[0.9cm]

{\fontsize{24pt}{0pt} \mytitlefont Cascading from~$\CN=2$ Supersymmetric\\[4pt] Yang-Mills Theory to Confinement and\\[4pt] Chiral Symmetry Breaking in Adjoint QCD}

\vskip20pt

Eric D'Hoker,$^1$ Thomas T.~Dumitrescu,$^1$ Efrat Gerchkovitz,$^2$ and Emily Nardoni$\,^3$

\bigskip

$^1${\it  Mani L.\,Bhaumik Institute for Theoretical Physics, Department of Physics and Astronomy,}\\[-2pt]
       {\it University of California, Los Angeles, CA 90095, USA}\\[4pt] 

$^2${\it Department of Physics of Complex Systems, Weizmann Institute of Science,\\[-2pt] Rehovot, Israel}

$^3${\it Department of Physics and Astronomy, Vassar College, 124 Raymond Avenue,\\[-2pt] Poughkeepsie, New York 12604, USA}

\end{center}

\bigskip

\noindent  We argue that adjoint QCD in $3+1$ dimensions, with any $SU(N)$ gauge group and two Weyl fermion flavors (i.e. one adjoint Dirac fermion), confines and spontaneously breaks its chiral symmetries via the condensation of a fermion bilinear. We flow to this theory from pure $\cN=2$ SUSY Yang-Mills theory with the same gauge group, by giving a SUSY-breaking mass $M$ to the scalars in the $\cN = 2$ vector multiplet. This flow can be analyzed rigorously at small $M$, where it leads to a deconfined vacuum at the origin of the $\cN=2$ Coulomb branch. The analysis can be extended to all $M$ using an Abelian dual description that arises from the $N$ multi-monopole points of the $\cN = 2$ theory. At each such point, there are $N-1$ hypermultiplet Higgs fields $h_m^{i = 1, 2}$, which are $SU(2)_R$ doublets. We provide a detailed study of the phase diagram as a function of~$M$, by analyzing the semi-classical phases of the dual using a combination of analytic and numerical techniques. The result is a cascade of first-order phase transitions, along which the Higgs fields $h_m^i$ successively turn on, and which interpolates between the Coulomb branch at small $M$, where all $h_m^i = 0$, and a maximal Higgs branch, where all $h_m^i \neq 0$, at sufficiently large $M$. We show that this maximal Higgs branch precisely matches the confining and chiral symmetry breaking phase of two-flavor adjoint QCD, including its broken and unbroken symmetries, its massless spectrum, and the expected large-$N$ scaling of various observables. The spontaneous breaking pattern $SU(2)_R \to U(1)_R$, consistent with the Vafa-Witten theorem, is ensured by an intricate alignment mechanism for the $h_m^i$ in the dual, and leads to a $\C\P^1$ sigma model of increasing radius along the cascade.  
\end{titlepage}


\setcounter{tocdepth}{3} 
\tableofcontents

\newpage


\section{Introduction and main results}
\label{sec:intro}

In this paper we will argue that adjoint QCD in~$3$+$1$ dimensions, with any~$SU(N)$ gauge group and~$N_f = 2$ adjoint Weyl fermion flavors, confines and spontaneously breaks chiral symmetry via the condensation of a fermion bilinear. We will do so  by utilizing the relationship of this (non-supersymmetric) theory to~$SU(N)$ supersymmetric Yang-Mills (SYM) theory with~$\mathcal{N}=2$ extended supersymmetry (SUSY), and no hypermultiplet matter. Upon turning on a SUSY-breaking deformation, this theory motivates a detailed and powerful dual description for adjoint QCD that predicts confinement and chiral symmetry breaking. In the (entirely self contained) introduction below, we briefly recall some background material about adjoint QCD, before explaining our approach and summarizing our main results.

\subsection{Adjoint QCD in 3+1 dimensions}

In this paper we are concerned with adjoint QCD in 3+1 dimensions (spacetime is~$\R^{3,1}$). We take the gauge group to be~$G = SU(N)$,\footnote{~Much of our discussion can be generalized to other gauge groups~$G$.} and there are~$N_f$ flavors of massless,\footnote{~Turning on quark masses always breaks some chiral symmetries. It is therefore both meaningful and interesting to study the massless theory.} two-component Weyl fermions (or quarks)~$\lambda^i_\alpha$ in the adjoint representation of~$G$,  
\begin{equation}
G = SU(N)~, \qquad N_f = N_f^\text{Weyl} = \half N_f^\text{Dirac} \quad \text{ adjoint fermions } \quad\lambda^i_\alpha~.
\end{equation}
Here~$\alpha = 1, 2$ is a left-handed Weyl spinor index and~$i = 1, \ldots, N_f$ is a flavor index.  We also take~$\lambda^i_\alpha$ to be valued in the (Hermitian) $SU(N)$ generators.\footnote{~Starting in section \ref{sec:SYM} we spell out the conventions we use in detail; a summary appears in appendix \ref{app:conv}.} 

Let us summarize some facts about these theories with a very broad brush:
\begin{itemize}
\item Asymptotic freedom requires~$N_f \leq 5$. An examination of the two-loop $\beta$-function suggests that the theories with~$N_f = 5$, and perhaps also~$N_f = 4$, flow to conformal field theories (CFTs) of Banks-Zaks type~\cite{Banks:1981nn}. For these values of~$N_f$ the two-loop $\beta$-function has a zero at a value of the coupling that is numerically somewhat small~\cite{Shifman:2013yca},\footnote{~The fixed-point value of the~$SU(N)$ gauge coupling~$g$ as computed from the 2-loop~$\beta$-function is~\cite{Shifman:2013yca},
$$
{g^2 N \over (4 \pi)^2} = {1 \over 46} \quad (N_f = 5)~, \qquad {1 \over 10} \quad (N_f = 4)~, \qquad {5 \over 14} \quad (N_f = 3)~.
$$
Note that the fixed point is naturally set by the 't Hooft coupling~$g^2 N$. 
}
However, the coupling cannot be made parametrically small,\footnote{~This is unlike QCD with~$SU(N)$ gauge group and~$N_f$ fundamental quarks in the Veneziano limit~$N, N_f \to \infty$ with~$x = N_f/N$ fixed. Because~$x$ is quasi-continuous at large~$N$, the fixed point 't Hooft coupling~$g^2 N$ can be made parametrically small by dialing~$x$ parametrically close to the asymptotic freedom bound.} and thus these considerations are not rigorous.  For~$N_f = 3$ the two-loop $\beta$-function has a zero at strong coupling; the zero disappears (i.e.~becomes complex) when~$N_f \leq  2$. 

\item As is typical in the absence of supersymmetry, there is no analytic argument that determines the lower end of the conformal window, i.e.~the critical number of flavors~$N^\text{crit.}_f(N)$ below which the theory no longer flows to an interacting CFT in the IR. Attempts to determine~$N^\text{crit.}_f(N)$ using numerical lattice simulations have been reported in~\cite{ Athenodorou:2021wom,Bennett:2022bhc,Bergner:2022trm,Athenodorou:2024rba,Bennett:2024qik}; see~\cite{Bergner:2022snd} for a relatively recent review with references. These calculations are very challenging because the gauge coupling necessarily passes through a region of slow running near the edge of the conformal window, delaying the approach to the continuum limit. For this reason there are no definitive lattice results for~$N^\text{crit.}_f(N)$.

\item The theory with~$N_f = 1$ is the minimally, $\CN=1$ supersymmetric Yang-Mills (SYM) theory in 3+1 dimensions (with the single adjoint Weyl fermion~$\lambda_\alpha$ playing the role of the gaugino), which is relatively much better understood (see the review~\cite{Intriligator:1995au}, and references therein). In particular, it is known to be gapped and confining,\footnote{~Here confinement means that the~$\Z_N^{(1)}$ one-form symmetry~\cite{Gaiotto:2014kfa} associated with the center of the~$SU(N)$ gauge group is unbroken. A slightly stronger statement, also believed to be true, is that the theory has finite-tension confining strings (first studied in~\cite{Douglas:1995nw}), which are charged under this symmetry.} and to spontaneously break a~$\Z_{2N}$ discrete chiral symmetry acting on~$\lambda_\alpha$ (i.e.~an~$R$-symmetry) to its~$\Z_2^F$ fermion-parity subgroup via gaugino condensation,
\begin{equation}
\label{eq:n1LLcond}
 \langle \tr (\lambda^\alpha \lambda_\alpha)  \rangle \sim \Lambda^3 \, e^{2 \pi i k \over N}~, \qquad k = 1, \ldots, N~.
\end{equation}
Here~$\Lambda$ is the strong-coupling scale of the theory, in a suitable renormalization scheme.\footnote{~The proportionality constant omitted in~\eqref{eq:n1LLcond}, indicated by the~$\sim$ there, is thus scheme-dependent. It can be computed exactly once a suitable supersymmetric scheme has been specified, as reviewed in~\cite{Intriligator:1995au}.} This leads to~$N$ degenerate vacua (each of which is trivially gapped, i.e.~the low-energy theory in each vacuum is an invertible TQFT), in agreement with the Witten index of the theory~\cite{Witten:1982df}. Thus, the lower endpoint of the conformal window satisfies
\begin{equation}
N_f^\text{crit.}(N) \geq 2~.
\end{equation}

\end{itemize}

\noindent Since the adjoint QCD theories on~$\R^{3,1}$ with~$N_f \geq 2$ are not supersymmetric, there is no general analytic strategy for studying them. Their lattice versions have been studied numerically in~\cite{ Athenodorou:2021wom,Bennett:2022bhc,Bergner:2022trm,Athenodorou:2024rba,Bennett:2024qik} (as reviewed in~\cite{Bergner:2022snd}), but sharp conclusions about the massless point are not yet available (but may well be soon). In particular, the possibility that the conformal window may in principle extend all the way down to~$N_f^\text{crit.}(N) = 2$, corresponding to~one Dirac fermion, has not yet been definitely ruled out by lattice simulations. The fate of the adjoint theory with~$N_f=2$ flavors in 3+1 dimensions is therefore particularly interesting.

In this paper we will provide compelling evidence that the~$N_f = 2$ adjoint theories are not in the conformal window for any number of colors~$N$ (so that~$N_f^\text{crit.}(N) \geq 3$), but rather realize the confining and chiral-symmetry-breaking scenario summarized in section~\ref{sec:confchisbIntro} below. Our approach, following~\cite{Cordova:2018acb} and reviewed in section~\ref{intro:susy} below, is based on the close relationship between these adjoint theories and pure~$\CN=2$ SYM theories with gauge group~$G = SU(N)$. The two-color case~$N=2$ was analyzed in \cite{Cordova:2018acb}; here we are primarily interested in generalizing these results to all~$N \geq 3$. 

We pause to mention that non-supersymmetric adjoint QCD with~$N_f \geq 2$ does become analytically tractable when compactified on a sufficiently small spatial circle (with periodic boundary conditions for fermions), as first explored in~\cite{Unsal:2007vu,Unsal:2007jx} (see~\cite{Dunne:2016nmc, Poppitz:2021cxe}  for reviews with references to subsequent work). There it is argued that adjoint QCD confines and spontaneously breaks a discrete chiral symmetry, while leaving the continuous chiral symmetry unbroken;\footnote{~For the~$N_f =2$ case studied here there is therefore at least one phase transition, associated with continuous chiral symmetry breaking, as a function of radius.} this leads to~$N$ vacua, each of which harbors massless two-component Weyl fermions that weakly interact via irrelevant operators.

By contrast, the behavior of the~$N_f = 1$ theory on a spatial circle can be determined for any radius, thanks to the unbroken supersymmetry:\footnote{~There are four supercharges, corresponding to~$\CN=1$ in 3+1 dimensions and~$\CN=2$ in 2+1 dimensions.}  the~$N$ gapped, confining vacua of $\CN=1$ SYM theory in 3+1 dimensions smoothly evolve as a function of the radius, without encountering a phase transition, as follows from~\cite{Affleck:1982as,Seiberg:1996nz}.

\subsection{Some facts and lore about adjoint QCD with~$N_f = 2$ flavors}
\label{intro:nf2adjoint}

\subsubsection{Global symmetries of~$N_f = 2$ adjoint QCD}\label{sec:symintro}

In this paper we will focus on~$G = SU(N)$ adjoint QCD theories with~$N_f = 2$ Weyl flavors~$\lambda_\alpha^{i}~(i = 1,2)$,  i.e.~one full Dirac flavor, on~$\R^{3, 1}$. This theory has the following zero-form\footnote{~In the terminology of~\cite{Gaiotto:2014kfa}, an ordinary zero-form symmetry acts on gauge-invariant local operators supported at spacetime points, while one-form symmetries act on extended defects supported on lines.} and one-form global symmetries (see section~\ref{sec:symm} for more detail): 
\begin{itemize}
\item A continuous chiral~$SU(2)_R$ flavor symmetry under which the~$\lambda_\alpha^i$ transform as doublets. Calling this symmetry~$SU(2)_R$ is natural from the point of view of the~$\CN=2$ SYM theory associated with two-flavor adjoint QCD (see section~\ref{intro:susy} below). 
\item A discrete~$\Z_{4N} \subset U(1)_r$ chiral symmetry under which the~$\lambda_\alpha^{i}$ have charge~$1$. If we denote the~$\Z_{4N}$ generator by~$r$, then~$r^{2N}$ is identified with the central~$-\1_{2 \times 2} \in SU(2)_R$, and both of them are further identified with fermion parity~$(-1)^F$, which is necessarily unbroken in a Lorentz-invariant vacuum. The faithfully acting chiral flavor symmetry acting on the~$\lambda_\alpha^i$ is therefore
\begin{equation}\label{chisym}
{SU(2)_R \times \Z_{4N} \over \Z_2}~.
\end{equation}
\item A~$\ZZ_N^{(1)}$ one-form global symmetry associated with the center of the~$SU(N)$ gauge group~\cite{Gaiotto:2014kfa}, whose realization diagnoses confinement. 
\item An anti-unitary time-reversal symmetry~$T$ and (for~$N \geq 3$) a unitary charge-conjugation symmetry~$C$.\footnote{~By the~$CPT$ theorem, there is therefore also a unitary parity symmetry~$P$ whose realization is correlated with  that of~$C$ and~$T$.}
\end{itemize}

\subsubsection{Does~$N_f = 2$ adjoint QCD confine and break chiral symmetry?} \label{sec:confchisbIntro}

If the~$N_f = 2$  adjoint QCD theories are not in the conformal window, it is generally expected -- though by no means certain\footnote{~Exotic alternatives were considered for~$N=2$ colors in~\cite{Anber:2018iof, Cordova:2018acb,Bi:2018xvr}. The scenarios proposed in~\cite{Cordova:2018acb} are (by construction) compatible with 't Hooft anomaly matching; by contrast, the putative scenarios explored in~\cite{Anber:2018iof, Bi:2018xvr} do not match all anomalies, as explicitly shown in~\cite{Cordova:2018acb, Wan:2018djl,Cordova:2019bsd,Cordova:2019jqi}.}  --  that they confine and spontaneously break their chiral symmetries via the condensation of a quark (or~$\CN=2$ gaugino, see below) bilinear,
\begin{equation}\label{eq:introll}
\langle \tr (\lambda^{\alpha(i} \lambda_\alpha^{j)})\rangle \neq 0~.
\end{equation}
Let us elaborate on this scenario. 

This complex order parameter~\eqref{eq:introll} transforms as a triplet of~$SU(2)_R$ and has charge~$2$ under the~$\ZZ_{4N}$ discrete chiral symmetry. It is therefore convenient to introduce\footnote{~Note that~$\vec \CO$ defined here differs from that defined in~\cite{Cordova:2018acb} by a sign: $\vec \CO_\text{here} = - \vec \CO_\text{there}$.}
\begin{equation}\label{eq:vecOdef}
\vec \CO = i \tr\left( \lambda^{\alpha i} {\vec \sigma_i}^{\;\; j} \lambda_{\alpha j} \right)~, \qquad \langle \vec \CO \rangle \neq 0~.
\end{equation}
Here~$\vec \sigma$ (with the indicated index placement) are the standard Pauli matrices.

Since the Cartan~$U(1)_R \subset SU(2)_R$ acts on the quarks in a vector-like fashion, compatible with a standard Dirac mass, the Vafa-Witten theorem states that it cannot be spontaneously broken~\cite{Vafa:1983tf}. Similarly, a certain notion of time-reversal symmetry must also remain unbroken~\cite{Vafa:1984xg}. This implies that the real and imaginary parts of the complex~$SU(2)_R$ triplet order parameter~\eqref{eq:vecOdef} are suitably aligned, leading to the following symmetry-breaking pattern:\footnote{~See~\cite{Cordova:2018acb} for a detailed discussion in the~$N=2$ case.} 
\begin{itemize}
\item The chiral symmetry in~\eqref{chisym} spontaneously breaks as follows,
\begin{equation}\label{chiSB}
{SU(2)_R \times \Z_{4N} \over \Z_2} \qquad \longrightarrow \qquad O(2)_R = U(1)_R \rtimes \Z_2~.\quad 
\end{equation}
Here the unbroken~$\Z_2$ extending~$U(1)_R$ to~$O(2)_R$ is generated by the product of~$r^{N}$ and the~$SU(2)_R$ Weyl reflection associated with the~$U(1)_R$ Cartan. 

The symmetry breaking pattern~\eqref{chiSB} leads to~$N$ disconnected vacuum sectors, each of which contains one copy of a 
\begin{equation}
\C\P^1 = {SU(2)_R \over U(1)_R}
\end{equation}
non-linear sigma-model for two massless Nambu-Goldstone Bosons, which furnish the only IR degrees of freedom in each vacuum. The different~$\C\P^1$s are cyclically permuted by the broken~$\ZZ_{4N}$ symmetry, as shown schematically in figure~\ref{fig:CP1} below. The fact that there are precisely~$N$ distinct~$\C\P^1$s is due to the fact that~$r^N$, which generates a~$\ZZ_4 \subset \ZZ_{4N}$ subgroup, negates the quark bilinear~\eqref{eq:introll} and thus acts on a fixed~$\C\P^1$ as orientation reversal.\footnote{~This also explains why combining it with an~$SU(2)_R$ Weyl reflection leads to the unbroken~$\Z_2$ symmetry on the right side of~\eqref{chiSB}.} By contrast the~$N$ distinct~$\C\P^1$s are cyclically permuted  by the~$\Z_N = \ZZ_{4N} / \ZZ_4$ quotient group.

\bigskip

\begin{figure}[t!]
\begin{center}
\begin{tikzpicture}[scale=1]
\scope[xshift=0cm,yshift=0cm]
\draw[scale=1, domain=0:90, variable=\x, line width=1] plot ({4*cos(\x)}, {sin(\x)});
\draw[scale=1, domain=90:180, variable=\x, line width=1] plot ({4*cos(\x)}, {sin(\x)});
\draw[scale=1, domain=180:270, variable=\x, line width=1] plot ({4*cos(\x)}, {sin(\x)});
\draw[scale=1, domain=270:360, variable=\x, line width=1] plot ({4*cos(\x)}, {sin(\x)});
\draw[scale=1, domain=260:285, variable=\x, line width=1,->] plot ({2.8*cos(\x)}, {0.7*sin(\x)});
\draw[scale=1, domain=285:310, variable=\x, line width=1] plot ({2.8*cos(\x)}, {0.7*sin(\x)});
\draw[blue] (4,0) node{\large $\bullet$};
\draw[blue]  (2.4939,0.78183) node{\large $\bullet$};
\draw[blue] (-0.89, 0.97492) node{\large $\bullet$};
\draw[blue] (-3.6038,0.43388) node{\large $\bullet$};
\draw[blue] (2.4939,-0.78183) node{\large $\bullet$};
\draw[blue] (-0.89, -0.97492) node{\large $\bullet$};
\draw[blue] (-3.6038,-0.43388) node{\large $\bullet$};
\draw (4.6,0) node{$\C\P^1$};
\draw (2.8,-1.2) node{$\C\P^1$};
\draw (2.8,1.25) node{$\C\P^1$};
\draw (-0.9,1.45) node{$\C\P^1$};
\draw (-0.9,-1.4) node{$\C\P^1$};
\draw (-3.7,0.9) node{$\C\P^1$};
\draw (-3.7,-0.9) node{$\C\P^1$};
\draw (0.7,-0.3) node{$\ZZ_N = \ZZ_{4N} / \ZZ_4$};
\draw (0,0) node{\Large $\bullet$};
\endscope
\end{tikzpicture}
\caption{The~$N$ disconnected vacuum sectors, each of which contains a single $\C\P^1$ non-linear sigma model (represented by a blue dot), associated with the breaking pattern~\eqref{chiSB}. These sectors are cyclically permuted by the spontaneously broken $\ZZ_{4N}$ symmetry, whose~$\ZZ_4 \subset \ZZ_{4N}$ subgroup does not permute distinct~$\C\P^1$s. In the figure, we depict the case~$N = 7$. \label{fig:CP1} 
}\end{center}
\end{figure}
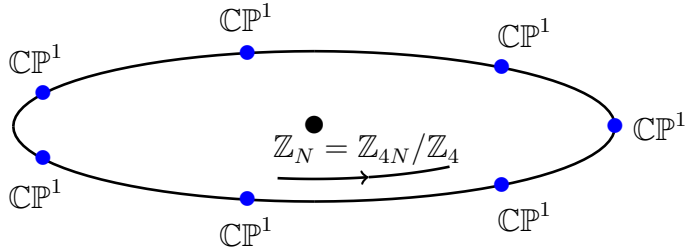

These observations amount to the statement that the complex~$SU(2)_R$ triplet order parameter defined in~\eqref{eq:vecOdef} has the following effective description in the deep IR,\footnote{~Here we are omitting subleading corrections to~$\vec \CO$, all of which involve derivatives of~$\vec n$. See~\cite{Dumitrescu:2024jko} for a related recent discussion.}
\begin{equation}\label{eq:vecOCn}
\vec \CO \quad \xrightarrow{\text{IR}} \quad |\langle \vec \CO \rangle | e^{i k \pi \over N} \vec n~, \qquad |\langle \vec \CO \rangle | > 0~, \qquad \vec n^2 = 1~.
\end{equation}
Here the phase of~$\langle \vec \cO \rangle$ is determined by the integer
\begin{equation}
k = 0, \ldots, N-1~,
\end{equation}
which labels the~$N$ disconnected vacuum sectors, while $\vec n$ is a unit vector parametrizing the~$\C\P^1$ in that sector. Since the different sectors are physically identical (being related by a broken symmetry), it suffices to focus on one at a time, which we take to be the~$k  = 0$ sector. In this sector, the unbroken time-reversal symmetry is given by
\begin{equation}\label{eq:untbrokenTtintro}
\t T = r^N T~, \qquad \t T : \vec \CO(t, \vec x) \to  \vec \CO(-t, \vec x)~.
\end{equation}

\item 't Hooft anomaly matching (see~\cite{Cordova:2018acb} for a detailed discussion of the~$SU(2)$ case, with related discussions and generalizations in~\cite{Wan:2018djl,Hsin:2019fhf,Cordova:2019bsd,Cordova:2019jqi,Brennan:2023vsa}) requires each~$\C\P^1$ model to be furnished with a discrete~$\theta$-angle. As originally discussed in~\cite{Witten:1983tw,Witten:1983tx}, this~$\theta$-angle is associated with~$\pi_4(\C\P^1) = \Z_2$, i.e.~it is a sign in the Euclidean path integral.\footnote{~This sign can be defined in a fully local fashion, see for instance~\cite{Freed:2006mx,Lee:2020ojw,Yonekura:2020upo}.} In our context the~$\theta$-angle is activated when~$N$ is even, and absent when~$N$ is odd.\footnote{~The~$\theta$-angle in the~$\C\P^1$ sigma model matches the~$\ZZ_2$-valued Witten anomaly~\cite{Witten:1982fp} associated with~$SU(2)_R$, which counts the~$N^2 - 1$ adjoint fermion~$SU(2)_R$ doublets of the UV~$SU(N)$ gauge theory modulo~$2$.}

\item The~$\Z_N^{(1)}$ symmetry is unbroken in every vacuum, so that the theory is confining. 

\item For~$N\geq 3$ the charge-conjugation symmetry~$C$ is unbroken. 
\end{itemize}

An appealing feature of the scenario above is that it reduces to the~$N$ gapped confining vacua of pure~$\CN=1$ SYM, upon giving an arbitrarily small mass to one of the two adjoint quarks, as can be checked using~\eqref{eq:vecOCn}. With the benefit of hindsight, this points to the common origin of confinement and chiral symmetry breaking in~$\CN=1$ SYM and~$N_f = 2$ adjoint QCD -- a common origin furnished by~$\CN=2$ SYM theory, as we will now explain.

\subsection{An approach via broken~$\CN=2$ supersymmetry}
\label{intro:susy}

Our strategy, following~\cite{Cordova:2018acb}, is to flow to~$N_f = 2$ adjoint QCD starting from pure~$\CN=2$ SYM (with the same~$SU(N)$ gauge group) via a non-holomorphic mass~$M$ for the complex adjoint scalar~$\phi$ in the~$\CN=2$ vector multiplet,\footnote{~Here~$g$ is the gauge coupling of the~$\CN=2$ SYM theory and the factors are chosen so that~$M$ is the tree-level pole mass of~$\phi$.} 
\begin{equation}\label{eq:vsbintro}
\SV_\text{\cancel{SUSY}} = {2 M^2 \over g^2} \tr \left( \bar \phi \phi\right)~,
\end{equation}
which completely breaks supersymmetry. If~$M \gg \Lambda$ is much larger than the strong-coupling scale~$\Lambda$ of the~$\CN=2$ SYM theory, we can safely integrate out the scalar~$\phi$ and flow to~$N_f = 2$ adjoint QCD,
\begin{equation}\label{eq:mrgintro}
\left(\CN=2 \text{ SYM} \right)+  \SV_{\text{\cancel{SUSY} }}  \quad \xrightarrow{M \gg \Lambda} \quad N_f = 2 \text{ adjoint QCD}~.
\end{equation}
Two comments are in order:
\begin{itemize}
\item We will study the RG flow triggered by the SUSY-breaking scalar mass~\eqref{eq:vsbintro} as a function of the ratio~$M/\Lambda$. We start in the controlled regime~$M \ll \Lambda$, where SUSY is only weakly broken, and ultimately extrapolate to large~$M$ to make contact with adjoint QCD, as in~\eqref{eq:mrgintro}.
\item The scalar mass~$M$ in~\eqref{eq:vsbintro} preserves all symmetries other than SUSY. Therefore all 't Hooft anomalies of the~$\CN=2$ SYM theory, including subtle global anomalies~\cite{Cordova:2018acb}, must be matched by the deformed theory, including (when~$M \gg \Lambda$) by adjoint QCD. Conversely, if we systematically analyze the fate of the~$\CN=2$ theory upon dialing~$M$ we are guaranteed to find an IR phase that matches all 't Hooft anomalies. 

\end{itemize}

\subsubsection{The small-SUSY-breaking regime: $M \ll \Lambda$}

\label{smallMintro}

Even though the SUSY-breaking mass term~$\tr (\bar \phi \phi)$ in~\eqref{eq:vsbintro} is not holomorphic, it turns out to nevertheless be protected by supersymmetry because it furnishes the primary~$\CT$ of the~$\CN=2$ stress-tensor supermultiplet~\cite{Luty:1999qc, Abel:2011wv, Cordova:2016xhm, Cordova:2016emh, Cordova:2018acb}, 
\begin{equation}\label{eq:CTdefintro}
\CT = {2 \over g^2} \, \tr (\bar \phi \phi)~.
\end{equation}

At leading order in the SUSY-breaking mass~$M$, it is therefore sufficient to track the operator~$\CT$ from the UV to the IR in the undeformed~$\CN=2$ SYM theory. As is well known from the work of Seiberg and Witten~\cite{Seiberg:1994rs,Seiberg:1994aj} and its generalizations~\cite{Klemm:1994qs, Argyres:1994xh, Klemm:1995wp}, the low-energy description of the~$\CN=2$ theory involves a Coulomb branch of vacua, parametrized by~$N-1$ holomorphic moduli~$u_I = \tr (\phi^I)$  with $I=2, \ldots, N$. At generic points, the low energy-theory is an Abelian~$U(1)^{N-1}$ gauge theory with~$\CN=2$ SUSY. The two-derivative effective Lagrangian of this theory -- including importantly its non-holomorphic K\"ahler potential~$K(u)$ -- is completely captured by the dependence of the holomorphic Seiberg-Witten periods~$(a_m(u), a_{Dm}(u))$ (with $m=1,\ldots , N-1$) on the Coulomb-branch moduli~$u_I$. Precisely this dependence was deduced in~\cite{Seiberg:1994rs,Seiberg:1994aj, Klemm:1994qs, Argyres:1994xh, Klemm:1995wp}, for all~$SU(N)$ gauge groups. 

As we will show in section \ref{sec:stmult}, the~$\CN=2$ stress-tensor primary~$\CT$ in~\eqref{eq:CTdefintro} that controls the non-holomorphic scalar mass flows to a certain (globally well-defined) choice of K\"ahler potential on the Coulomb branch,
\begin{equation}
\CT_\text{UV} = {2 \over g^2} \, \tr (\bar \phi \phi) 
\qquad \longrightarrow \qquad 
\CT_\text{IR} = K(u_I) = {1 \over 2 \pi} \sum_{m = 1}^{N-1} \Im \left(\bar a_m a_{Dm}\right)~.
\end{equation}
At leading order in small~$M \ll \Lambda$, we can therefore reliably analyze the effect of SUSY-breaking by approximating~\eqref{eq:vsbintro} as
\begin{equation}\label{eq:vsbkintro}
\SV_{\text{\cancel{SUSY} }} = M^2 K(u_I) + \CO(M^4)~,
\end{equation}
in the low-energy effective theory on the Coulomb branch. Note that this explicitly depends on the K\"ahler potential, which is calculable thanks to~$\CN=2$ supersymmetry. As indicated in~\eqref{eq:vsbkintro}, there are higher-order corrections in~$M^2$ that are not calculable,\footnote{~They receive contributions from full~$\CN=2$ $D$-terms, which (much like K\"ahler potentials in~$\CN=1$ theories) are not subject to any non-renormalization theorems.} so that we cannot explore the large-$M$ regime that governs adjoint QCD in a controlled way. We will circumvent this obstacle below by formulating a dual description whose utility extends beyond the small-$M$ regime accessible via~\eqref{eq:vsbkintro}.

\begin{figure}[t!]
\centering
\includegraphics[width=\textwidth]{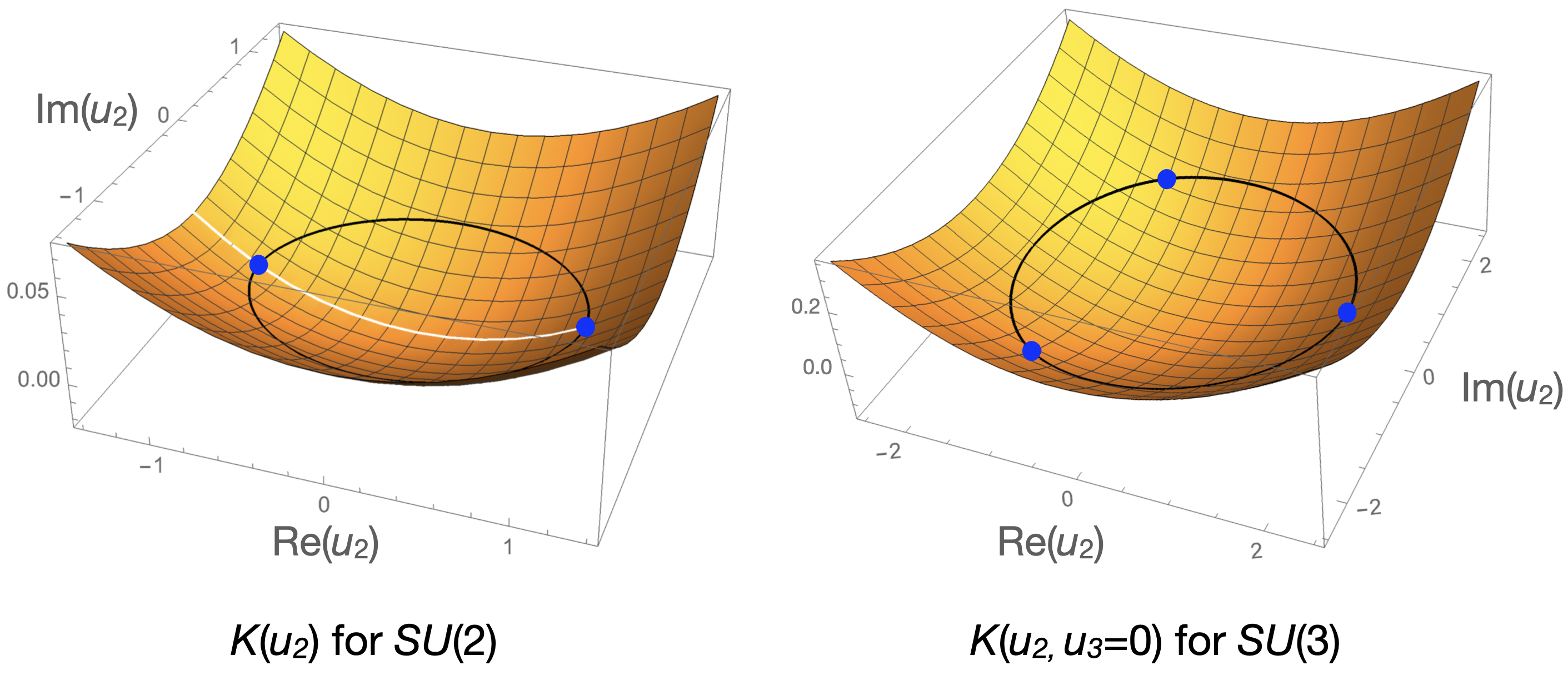}
\caption{  The K{\"a}hler potential~$K$ for  $SU(2)$ (left) and $SU(3)$ (right), plotted in the complex $u_2$ plane, has a unique minimum at the origin. The black curves (where~$K = 0$) define the strong-coupling region surrounding the origin. The multi-monopole points, indicated by blue dots, lie on these~$K = 0$ curves. 
\label{fig:kahlerintro}}
\end{figure}

For~$SU(N)$ gauge group, the Seiberg-Witten K\"ahler potential~$K(u_I)$ turns out to be a rather well-behaved function~\cite{DHoker:2022loi}: it is convex, with a unique minimum at the origin of the Coulomb branch, where all~$u_I =0$. (See figure~\ref{fig:kahlerintro}.) Thus, even though~$K$ is in principle an unwieldy function of~$N-1$ variables, its qualitative behavior is not substantially different from the K\"ahler potential of the~$SU(2)$ theory analyzed in~\cite{Luty:1999qc,Cordova:2018acb}.

The convexity of~$K(u_I)$ leads to the following predictions for the small SUSY-breaking regime~$M \ll \Lambda$:
\begin{itemize}
\item There is a single SUSY-breaking vacuum at the origin of the Coulomb branch, where all~$u_I=0$. Only the~$N-1$ Abelian vector-multiplet scalars get a mass~$\sim M$ from SUSY breaking, while their~$\CN=2$ superpartners -- the ~$U(1)^{N-1}$ gauge fields and gauginos --  remain massless. 
\item The vacuum is in a Coulomb phase, with spontaneously broken~$\Z_N^{(1)}$ symmetry.
\item All zero-form symmetries (i.e.~the~$\left(SU(2)_R \times \Z_{4N}\right)/ \Z_2$ symmetry, as well as $C$ and~$T$) are not spontaneously broken.
\end{itemize}

\subsubsection{A dual description for all values of~$M$ motivated by BPS states} 
\label{sec:dualintro}

In order to describe the behavior of the~$M$-deformed theory beyond the rigorously controlled~$M \ll \Lambda$ regime analyzed in the previous subsection, we will formulate a useful dual description of the physics.

This dual description can be motivated by recalling that~$\CN=2$ supersymmetry not only controls the massless degrees of freedom in the deep IR, but also determines the fate of BPS particles on the Coulomb branch. These particles reside in short multiplets of~$\CN=2$ supersymmetry and are generically massive, but they can become massless at certain singular loci on the Coulomb branch. An example of such loci in~$\CN=2$~$SU(N)$ SYM -- and the one that will be most relevant to our discussion below -- is furnished by the multi-monopole points, i.e.~the generalization of the~$SU(2)$ monopole and dyon points of Seiberg and Witten to~$SU(N)$ gauge group. At these points,  a maximal number (namely~$N-1$) of mutually local dyons become massless simultaneously \cite{Douglas:1995nw}. There are~$N$ such points, labeled by~$k = 0, \ldots, N-1$, which are cyclically permuted by the broken~$\ZZ_{N} = \ZZ_{4N}/\ZZ_4$ quotient symmetry (see figure~\ref{fig:CP1}).\footnote{~As we shall see, the integer~$k = 0, \ldots, N-1$ labeling the multi-monopole points is precisely the same as the one that dictates the phase of the fermion bilinear condensate~$\langle \vec \CO\rangle$ in~\eqref{eq:vecOCn}.} At the point labeled by~$k = 0$ -- referred to throughout as {\it the} multi-monopole point -- these dyons are magnetic monopoles from the perspective of the UV theory.

At the origin of the Coulomb branch (where all~$u_I = 0$) there are~$N (N - 1)$ massive BPS particles~\cite{Lerche:2000uy,Alim:2011kw},\footnote{~Here we are counting full BPS hypermultiplets; in particular, we are not separately counting the particles and anti-particles within a given hypermultiplet.} which we denote by
\begin{equation}
\text{BPS particles at } u_I = 0: \quad \mu_{km}~, \quad k = 0, \ldots, N-1~, \quad m = 1, \ldots, N-1~.
\end{equation}
These BPS particles are mutually non-local dyons, but they can be grouped into~$N$ towers (labeled by~$k$), such that the~$N-1$ dyons within a given tower (labeled by~$m$) are mutually local. In fact each tower comprises a maximal set of mutually local dyons.  The unbroken~$\Z_{4N}$ symmetry at the origin cyclically permutes the tower index~$k$, so that the BPS masses~$M_\text{BPS}(\mu_{km})$ only depend on~$m$. Omitting an~$N$-dependent~$\CO(1)$ pre-factor (indicated by~$\sim$ below), the BPS masses at the origin take the following form, 
\begin{equation}\label{eq:bpsMintro}
M_\text{BPS}(\mu_{km}) \sim \Lambda \sin{\pi m \over N}~.
\end{equation}
In addition to the degeneracy in~$k$, there is a further degeneracy due to the unbroken charge-conjugation symmetry~$C : m \leftrightarrow N-m$ at the origin.

All BPS states in~\eqref{eq:bpsMintro} are massive; the lightest ones have masses \begin{equation}
M_\text{BPS}(\mu_{k1}) = M_\text{BPS}(\mu_{k, N-1})\sim {\Lambda \over N}~.
\end{equation}
This is also the natural UV cutoff of the massless effective theory at the origin that we used in section~\ref{smallMintro} to analyze the effects of the SUSY-breaking mass~$M$ there, i.e.~we should not trust our soft SUSY-breaking analysis once~$M \gtrsim {\Lambda \over N}$. Note that this UV cutoff vanishes in the large-$N$ limit, as first noted by \cite{Douglas:1995nw}, raising possible concerns about the utility of the Seiberg-Witten IR effective theory in the large-$N$ regime. Pleasingly and reassuringly, we will find exactly the opposite: our results below agree rather nicely with the expected large-$N$ behavior of adjoint QCD. 

Once the SUSY-breaking mass~$M \sim {\Lambda \over N}$ becomes comparable to the mass of the lightest BPS states at the origin, one cannot integrate them out and instead must take into account their effect on the dynamics. The same goes for all other BPS states at the origin, once we dial~$M$ through their mass thresholds~\eqref{eq:bpsMintro}. We thus require an effective description that includes all massive BPS states at the origin (as well as the massless degrees of freedom already discussed above), but this is complicated by the fact that they are mutually non-local. 

\begin{figure}[t!]
\begin{center}
\begin{tikzpicture}[scale=1]
\scope[xshift=0cm,yshift=0cm]
\draw[very thick, ->] (0,0) -- (0,6.5);
\draw (0.9,6.8) node{$M_\text{BPS}(\mu_{km})$};
\draw[scale=1, domain=0:90, variable=\x, line width=1.5] plot ({4*cos(\x)}, {sin(\x)});
\draw[scale=1, domain=90:180, variable=\x, line width=1.5] plot ({4*cos(\x)}, {sin(\x)});
\draw[scale=1, domain=180:270, variable=\x, line width=1.5] plot ({4*cos(\x)}, {sin(\x)});
\draw[scale=1, domain=270:360, variable=\x, line width=1.5] plot ({4*cos(\x)}, {sin(\x)});

\draw[red, thick] (2.4939,0.78183) -- (0,2.6033);
\draw[red, thick] (2.4939,0.78183) -- (0,4.6909);
\draw[red, thick] (2.4939,0.78183) -- (0,5.8495);

\draw[blue, thick] (4,0) -- (0,2.6033);
\draw[blue, thick] (4,0) -- (0,4.6909);
\draw[blue, thick] (4,0) -- (0,5.8495);

\draw[yellow, thick] (2.4939,-0.78183) -- (0,2.6033);
\draw[yellow, thick] (2.4939,-0.78183) -- (0,4.6909);
\draw[yellow, thick] (2.4939,-0.78183) -- (0,5.8495);

\draw[black, thick] (-0.89, 0.97492) -- (0,2.6033);
\draw[black, thick] (-0.89, 0.97492) -- (0,4.6909);
\draw[black, thick] (-0.89, 0.97492) -- (0,5.8495);

\draw[cyan, thick] (-3.6038,0.43388) -- (0,2.6033);
\draw[cyan, thick] (-3.6038,0.43388) -- (0,4.6909);
\draw[cyan, thick] (-3.6038,0.43388) -- (0,5.8495);

\draw[purple, thick] (-3.6038,-0.43388) -- (0,2.6033);
\draw[purple, thick] (-3.6038,-0.43388) -- (0,4.6909);
\draw[purple, thick] (-3.6038,-0.43388) -- (0,5.8495);

\draw[green, thick] (-0.89, -0.97492) -- (0,2.6033);
\draw[green, thick] (-0.89, -0.97492) -- (0,4.6909);
\draw[green, thick] (-0.89, -0.97492) -- (0,5.8495);

\draw[blue] (4,0) node{\large $\bullet$};
\draw[blue] (2.4939,0.78183) node{\large $\bullet$};
\draw[blue] (-0.89, 0.97492) node{\large $\bullet$};
\draw[blue] (-3.6038,0.43388) node{\large $\bullet$};
\draw[blue] (2.4939,-0.78183) node{\large $\bullet$};
\draw[blue] (-0.89, -0.97492) node{\large $\bullet$};
\draw[blue] (-3.6038,-0.43388) node{\large $\bullet$};

\draw[red] (0,0) node{\Large $\bullet$};
\draw (0,2.6033) node{ \large $\bullet$};
\draw (0,4.6909) node{\large $\bullet$};
\draw (0,5.8495) node{\large $\bullet$};

\endscope
\end{tikzpicture}
\caption{The $N(N-1)$ massive BPS states (counting full hypermultiplets)  at the origin of the Coulomb branch (indicated by the red dot), where $u_I=0$ for all $I=2,\ldots, N$. The figure corresponds to the gauge group $SU(7)$. The 3 distinct BPS masses at the origin (given by~\eqref{eq:bpsMintro}) are indicated by the black dots on the vertical axis, each of which describes~$14$ degenerate hypermultiplets. There are 7 multi-monopole points (indicated by the blue dots); at each such point 2 BPS hypermultiplets from every black dot become massless, as indicated by the thin colored lines.  } \label{fig:BPSintro} 
\end{center}
\end{figure}
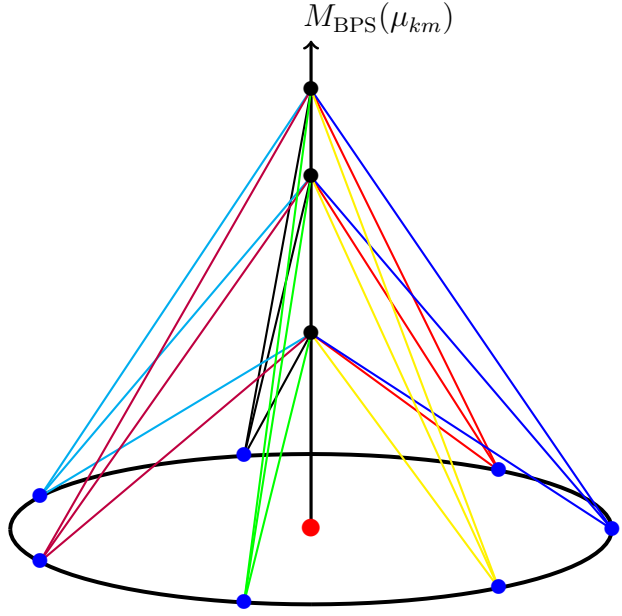

As already mentioned above, for fixed~$k$ the~$N-1$ BPS states~$\mu_{k m}$ labeled by~$m = 1, \ldots, N-1$ comprise a maximal set of mutually local dyons; moreover, precisely this set becomes massless at the multi-monopole point labeled by~$k$. (See figure~\ref{fig:BPSintro}.) Thus we adopt the strategy of using the~$N$ multi-monopole points as a dual description that includes all the BPS states at the origin of the Coulomb branch. Since the multi-monopole points are cyclically permuted by the~$\Z_{N} = \Z_{4N}/\Z_4$ symmetry, it suffices to focus on the dual description associated with any one of them -- which we take to be \textit{the} multi-monopole point labeled by~$k = 0$. The price to pay is that this description does not manifest the unbroken~$\Z_{4N}$ symmetry at the origin of the Coulomb branch, which is therefore an accidental symmetry of the dual description (as is common in many dualities).

Famously \cite{Seiberg:1994rs, Seiberg:1994aj,Douglas:1995nw}, the IR effective theory at the multi-monopole point is an~$\CN=2$ Abelian Higgs model (also referred to as SQED) with~$N-1$ dual magnetic Abelian vector multiplets, whose scalar bottom components are precisely the magnetic periods~$a_{Dm}$ that all vanish at the multi-monopole point, and~$N-1$ hypermultiplets~$h_m^i$ (plus their fermionic superpartners) which represent the massless BPS monopoles. Thus they carry electric charge~$+1$ under the dual magnetic~$U(1)_D^{N-1}$ gauge group.  Note that the hypermultiplet scalars~$h_m^i$ are doublets under the~$SU(2)_R$ symmetry. 

The formula~\eqref{eq:vsbkintro} for the leading SUSY-breaking potential on the Coulomb branch can be extended to the multi-monopole point by including the hypermultiplets~$h_m^i$ and performing a standard matching calculation. This leads to the following form for the scalar potential at the multi-monopole point,
\begin{equation}\label{eq:ahmvtotalintro}
\SV_\text{total} = \SV_\text{SUSY} + \SV_{\text{\cancel{SUSY} }}~,
\end{equation}
where the supersymmetric part of the potential takes the form
\begin{equation}\label{eq:susyvintro}
\SV_\text{SUSY} = \sum_{m = 1}^{N-1} 2 |a_{Dm}|^2 (\bar h^i h_i)_m + \sum_{m, n = 1}^{N-1} (t^{-1})_{m,n} \left((\bar h^i h_j)_m(\bar h^j h_i)_n - \half (\bar h^i h_i)_m (\bar h^j h_j)_n\right)~,
\end{equation}
and the SUSY-breaking potential is
\begin{equation}\label{eq:VSBintro}
\SV_{\text{\cancel{SUSY} }} = M^2 \left({N \Lambda \over \pi^2}  \sum_{m = 1}^{N-1} \Im (a_{Dm}) \sin {\pi m \over N} + \sum_{m,n = 1}^{N-1} t_{mn} a_{Dm} \bar a_{Dn} - \half \sum_{m =1}^{N-1} (\bar h^i h_i)_m + \CO(a_D^3) \right)~.
\end{equation}
We will now discuss the ingredients that go into these two formulas in some detail. 
\begin{itemize}
\item An essential ingredient in our story is the matrix~$t_{mn}(\mu)$ of effective~$U(1)_D^{N-1}$ gauge couplings, which is a feature of the~$\CN=2$ theory and unrelated to SUSY-breaking. Crucially, this matrix has off-diagonal elements, in addition to the well-known diagonal logarithmic running due to the massless charge-1 monopoles,
\begin{equation}\label{eq:tdefintro}
t_{mn}(\mu) = {1 \over (2 \pi)^2} \left(\delta_{mn} \log { \Lambda \over \mu } + \log \Lambda_{mn}\right)~.
\end{equation}
Here~$\mu$ is the renormalization (or RG) scale. The dimensionless threshold corrections~$\Lambda_{mn}$ (not to be confused with the dimensionful strong coupling scale~$\Lambda$) were computed in~\cite{DHoker:2020qlp} from the exact~$\CN=2$ Seiberg-Witten solution,\footnote{~See also~\cite{Bonelli:2017ptp} for an alternative approach using topological strings and matrix models.} 
\begin{equation}
\Lambda_{mm} = 16 N \sin^3{ \pi m \over N}~, \qquad \Lambda_{m \neq n}  =  { 1-\cos \frac{(m+n) \pi}{N} \over 1-\cos \frac{(m-n)\pi}{N}}~.
\end{equation} 
Intuitively, these threshold corrections reflect the properties of the massive BPS particles that we have integrated out at the multi-monopole point, but whose effect we nevertheless wish to capture. In other words, the matrix~$t_{mn}(\mu)$ represents detailed dynamical input, which remembers the underlying~$SU(N)$ gauge theory\footnote{~Indeed, one can equivalently think of the massive BPS particles at the multi-monopole point as the W-bosons of the~$SU(N)$ gauge theory, which are related to the dyons inside the strong-coupling region surrounding the origin by wall crossing.} and is not determined by general considerations such as symmetries or anomaly matching. 

\item Since the matrix~$t_{mn}(\mu)$ is positive definite, the~$a_{Dm}$ fields all acquire positive masses. 

\item By contrast, the monopoles~$h_m^i$ have tachyonic masses and want to condense (though they are prevented from doing so for small SUSY-breaking~$M$, see below). 

\item The quartic terms~$\sim (t^{-1})(\bar h h)^2 \subset \SV_\text{SUSY}$ in~\eqref{eq:susyvintro} are nothing but the~$\CN=2$ Abelian~$D$-term potential for the charged hypermultiplet scalars $h_{im}$ and their complex conjugates $\bar h_m^i$.\footnote{~We raise and lower~$SU(2)_R$ doublet indices from the left using the standard~$\ep^{ij}$ and~$\ep_{ij}$ symbols (see appendix~\ref{app:conv}). Thus~$h^i_m = \ep^{ij} h_{jm}$ etc. The complex conjugate fields are defined via~$\bar h^i_m = (h_{im})^\dagger$.} This also explains why those terms are completely determined by the matrix~$t_{mn}$ of effective~$U(1)_D^{N-1}$ gauge couplings. 

This~$D$-term potential has a dramatic effect on the orientation of the~$h_{m}^i$ in~$SU(2)_R$ space, manifest upon introducing the gauge-invariant, real~$SU(2)_R$ triplet vectors
\begin{equation}
\vec S_m = \bar h_m^i  {\vec \sigma_i}^{\; \; j}  h_{jm}~,
\end{equation}
where $\vec \sigma$ are the standard Pauli matrices (with the same index placement as in~\eqref{eq:vecOdef}). 
In terms of these variables, the parts of the~$D$-term potential that depend on the orientation of the~$h_m^i$, rather than just their~$SU(2)_R$-invariant magnitudes, take the form of a spin-chain with all-to-all interactions determined by~$(t^{-1})_{mn}$, 
\begin{equation}\label{eq:spinchainintro}
\left(\CN = 2 \text{  $D$-terms}\right) \supset \sum_{m < n} (t^{-1})_{mn} \, \vec S_m \cdot \vec S_n~.
\end{equation}
It is an important feature of the matrix~$t_{mn}(\mu)$ in~\eqref{eq:tdefintro}, to be discussed in detail below, that the off-diagonal elements of its matrix inverse are strictly negative,
\begin{equation}
(t^{-1})_{mn} < 0 \qquad \hbox{ for } m \not=n~.
\end{equation}
This property implies that the spin chain~\eqref{eq:spinchainintro} has purely ferromagnetic couplings, i.e.~in the ground state all spins~$\vec S_m$ are perfectly aligned in~$SU(2)_R$ space. In turn, this result  implies that, whenever some Higgs fields have a non-vanishing expectation value~$h_m^i \neq 0$, so that~$SU(2)_R$ is spontaneously broken, they must align in such a way that the symmetry-breaking pattern is~$SU(2)_R \to U(1)_R$, in precise agreement with the expectation for adjoint QCD from the Vafa-Witten theorem discussed below~\eqref{eq:vecOdef}.

\end{itemize}

In this paper we will explore the fate of the~$\CN=2$ SYM theory with gauge group~$SU(N)$ and SUSY-breaking mass~$M$ by analyzing the Abelian dual at the multi-monopole point. Moreover, we will do so semi-classically, by minimizing the scalar potential~\eqref{eq:ahmvtotalintro}.\footnote{~Note that the classical limit of the dual is neither classical nor weakly coupled from the point of view of the~$SU(N)$ theory in the UV.} As usual, this analysis involves (i) solving for the critical points of the potential, (ii) assessing their local stability by examining the Hessian around each critical point, and (iii) determining which locally stable solution has the lowest potential, making it the globally stable true vacuum. 

As we will summarize below, this leads to a compelling picture for all values of~$M$ and all~$SU(N)$ gauge groups that beautifully matches the confining and chiral symmetry-breaking phase of adjoint QCD reviewed in section~\ref{sec:confchisbIntro} in the appropriate large-$M$ regime, $M \gtrsim \Lambda$.

\subsection{Main result: a cascade of phase transitions}

\begin{figure}[t!]
\centering
\includegraphics[width=\textwidth]{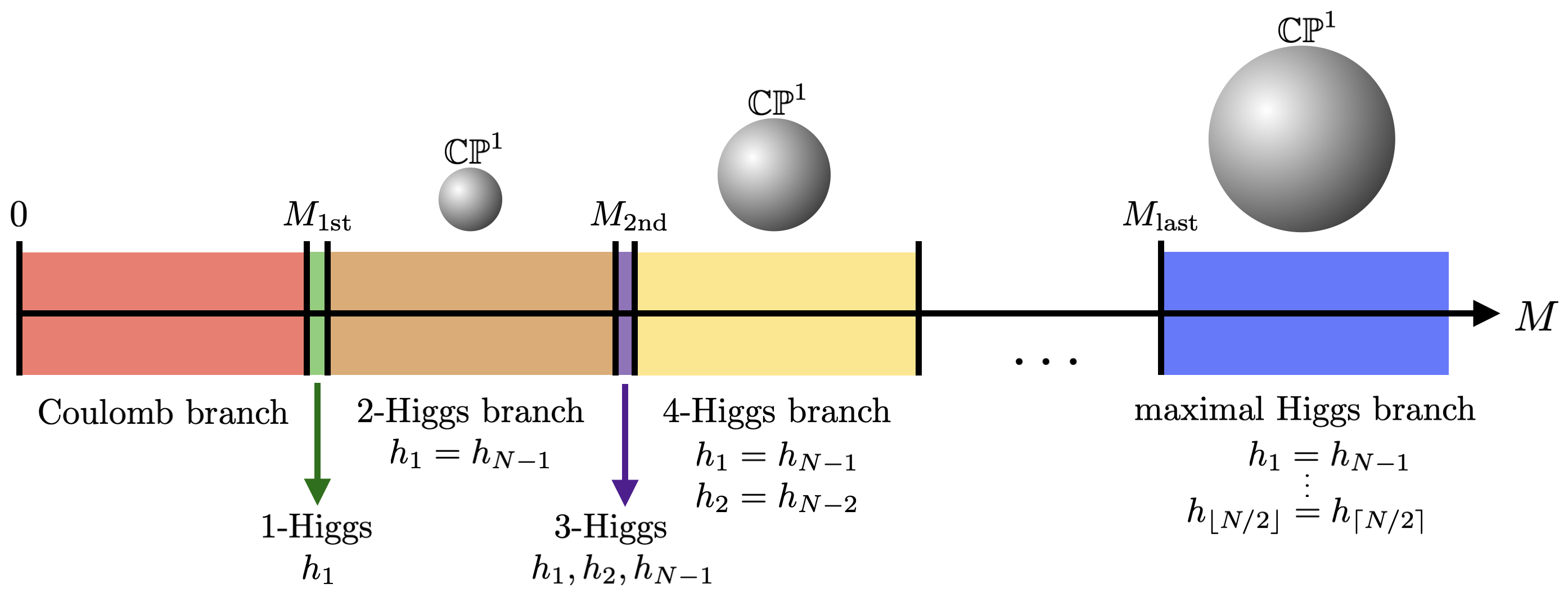}
\caption{\label{fig:phases} Cascade of first-order phase transitions interpolating between the Coulomb branch at small SUSY-breaking~$M$, where all Higgs fields vanish, $h_m^i = 0$, and the maximal Higgs branch (HB), where all Higgs fields are condensed and aligned, $h_m^i \neq 0$. (This alignment, discussed around~\eqref{eq:spinchainintro}, means that we can omit the~$SU(2)_R$ indices in the figure.) The coarse structure of the cascade consists of the~$C$-even large phases (e.g.~Coulomb in red, 2-Higgs in brown, etc.), while the fine structure generically leads to~$C$-breaking small interpolating phases (e.g.~1-Higgs in green, 3-Higgs in purple etc.). Note that the~$C$-odd 1-Higgs phase in green only exists for~$SU(3)$ gauge group. Once the first Higgs field turns on, the~$SU(2)_R \to U(1)_R$ breaking leads to a~$\C\P^1$ sigma-model of increasing radius  along the cascade. 
}
\end{figure}

We will now present the semi-classical phase structure of the Abelian dual at the multi-monopole point, by minimizing the scalar potential~\eqref{eq:ahmvtotalintro} as a function of the SUSY-breaking mass~$M$. This phase structure, which is essentially uniform for all~$SU(N)$ gauge groups,\footnote{~An exception concerning the first phase transition of the cascade is discussed in section~\ref{sec:cascadeIntro} below.} is obtained through a combination of exact analytic and numerical calculations (for~$N \leq 6$), and via an approximate, perturbative analytic scheme that is valid for all~$N$.\footnote{~All approaches agree within their overlapping regimes of validity.} The result is a cascade of phases and transitions, which are summarized in figure~\ref{fig:phases}. A detailed account of the cascade, synthesizing all results in the paper, appears in section~\ref{sec:summary}. What follows below is an abbreviated version. 

We will begin at small~$M \ll \Lambda$, where the dual correctly recovers the exact small-SUSY-breaking regime analyzed in section~\ref{smallMintro} above. We refer to this regime as the Coulomb branch (CB). As we increase the SUSY-breaking mass~$M$ we will encounter a cascade of first-order phase transitions,\footnote{~See~\cite{Armoni:2019lgb} for a similar cascade of first-order phase transitions and its dual description in large-$N$ QCD$_3$.} which occur roughly (but not exactly) when the SUSY-breaking mass passes through the BPS thresholds~\eqref{eq:bpsMintro} at the origin,  $M \sim M_\text{BPS}(\mu_{km}) \sim  \Lambda \sin (\pi m/N)$  (with~$m = 1, \ldots, N-1$). 

Finally, when~$M \gtrsim \Lambda$, we find a maximal Higgs branch (HB), where all monopole Higgs fields of the Abelian dual are non-vanishing, $h_{m}^i \neq 0$ for all~$m = 1, \ldots, N-1$.  As we will explain in section~\ref{sec:HBadjointIntro} below, this phase precisely matches the confining and chiral-symmetry breaking phase of adjoint QCD reviewed in section~\ref{sec:confchisbIntro}. In particular, the broken and unbroken symmetries, as well as the massless spectrum match exactly. Even more reassuringly, and contrary to the concerns raised in~\cite{Douglas:1995nw}, we find that the Abelian dual correctly captures the expected large-$N$ dependence of various observables.

\subsubsection{The Coulomb branch (CB) at small~$M$}

An important consistency check of the dual Abelian Higgs model introduced in~section~\ref{sec:dualintro} above is whether it correctly reproduces the small-SUSY-breaking regime~$M \ll \Lambda$ (the red phase in figure~\ref{fig:phases}), because this regime was analyzed exactly in section~\ref{smallMintro} above. There we found a unique vacuum at the origin~$u_I =0$ of the Coulomb branch, with~$N-1$ massless photons and gaugino~$SU(2)_R$ doublets, but no massless scalars. Thus the~$\Z_N^{(1)}$ one-form symmetry is spontaneously broken, while all zero-form symmetries, i.e. $(SU(2)_R \times \ZZ_{4N} ) / \Z_2$, as well as~$C$ and~$T$, are unbroken. All charged BPS particles are massive, with masses given by~\eqref{eq:bpsMintro}. In analogy with supersymmetric terminology, we will refer to this non-supersymmetric vacuum at~$u_I = 0$ and small~$M$ as the Coulomb branch (CB).

In section~\ref{sec:dualsmallM}, we will analyze the potential~\eqref{eq:ahmvtotalintro} of the Abelian dual in the small-$M$ regime, and we will give a detailed account of how the small-$M$ Coulomb branch summarized above is indeed reproduced in the dual description. Here we restrict our attention to two important comments: 
\begin{itemize}

\item None of the hypermultiplet scalars~$h_{m}^i$ condense on the CB; rather, they (and their fermionic superpartners) are massive, and their masses as predicted by the dual are
\be
\label{eq:dualmassintro}
(\text{mass of hypermultiplet } h_{m}^i) \sim  \Lambda N  \sum_{n=1}^{N-1} (t^{-1})_{mn} \sin\frac{n \pi}{N} \sim \Lambda \sin\frac{\pi m}{N}~.
\ee
This is in good agreement with the exact BPS mass formula~\eqref{eq:bpsMintro}, up to~$\CO(1)$ constants indicated by~$\sim$ in~\eqref{eq:dualmassintro}. 

The~$N-1$ hypermultiplets (labeled by~$m = 1, \ldots, N-1$) from the~$k$-th multi-monopole point (with~$k = 0, \ldots, N-1$), describe all~$N(N-1)$ BPS hypermultiplets~$\mu_{km}$ at the origin, which were discussed above~\eqref{eq:bpsMintro} (see also figure~\ref{fig:BPSintro}). Since the masses~\eqref{eq:dualmassintro} do not depend on~$k$, they display the unbroken~$\Z_{4N}$ symmetry rotating the BPS states at the origin, even though this symmetry is not manifest in the dual (see below). 

\item The discrete~$\Z_{4N}$ $R$-symmetry is unbroken on the CB (where~$u_I = 0$). This is partially obscured in the Abelian dual at the multi-monopole point because the~$\ZZ_{N} = {\ZZ_{4N} / \ZZ_{4}}$ quotient permutes the different multi-monopole points. It is therefore an emergent symmetry of the dual in the small-$M$ regime describing the CB. All other symmetries are manifest in the dual, and their realization exactly matches the CB at small~$M$. See section~\ref{sec:symcascade} for further details regarding the realization of global symmetries. 
\end{itemize}

\subsubsection{Cascade of phase transitions at intermediate~$M$: coarse structure}\label{sec:cascadeIntro}

As we increase the SUSY-breaking mass~$M$, the dual predicts a cascade of first-order phase transitions, depicted in figure~\ref{fig:phases}. As is plainly visible, the cascade has a two-tier structure, that we term its coarse and fine structure:
\begin{itemize}
    \item[(i)] The {\it coarse structure} of the cascade (further discussed below) consists of the phases that are drawn large in figure~\ref{fig:phases}, e.g.~the red Coulomb branch, or the brown 2-Higgs branch. As we will explain below, all of these phases preserve charge-conjugation symmetry~$C$. 
    \item[(ii)] The {\it fine structure} of the cascade (further discussed in section~\ref{FSintro} below) is indicated by the phases that are drawn small in figure~\ref{fig:phases}, e.g.~the green~$1$-Higgs phase, or the purple~$3$-Higgs phase. These phases only open up for a very short range of~$M$-values (if at all), and they all spontaneously break~$C$-symmetry. Collapsing them reduces the cascade to its coarse structure in point (i) above. 
\end{itemize}

\noindent We will now describe in more detail the coarse structure of the cascade, i.e.~those phases that are drawn large in figure~\ref{fig:phases}:
\begin{itemize}
\item At the coarse level, the cascade proceeds by turning on pairs of Higgs fields, leading to the following~$C$-symmetric sequence of first-order phase transitions interpolating between the Coulomb branch (CB, no Higgs fields turned on) and the maximal Higgs branch (HB, all Higgs fields turned on), as we dial from small to large~$M$,\footnote{~Note that for even~$N = 2\nu$, the last transition only involves turning on one~$C$-even Higgs field~$h_\nu$.}
\begin{equation}\label{eq:coarscascadeintro}
\text{coarse cascade}\; : \; \text{CB} \to \left\{h_1^i = h_{N-1}^i \neq 0 \right\} \to \left\{ \begin{matrix}
    h_1^i = h_{N-1}^i \neq 0 \\ h_2^i = h_{N-2}^i \neq 0
\end{matrix}
\right\} \to \cdots \to \text{HB}
\end{equation}

\item The transition that involves turning on~$h_m^i = h_{N-m}^i$ occurs approximately at
\begin{equation}\label{eq:ksdiagbisintro}
    M = M_{*m} \sim {N \Lambda} (t^{-1} s)_m~,
\end{equation}
which agrees with the masses of the BPS states at the origin of the Coulomb branch (see~\eqref{eq:dualmassintro}), up to an~$\CO(1)$ constant indicated by~$\sim$ in~\eqref{eq:ksdiagbisintro}. As anticipated above, the BPS masses at the origin roughly (though not exactly, see below) determine the thresholds in~$M$ at which a phase transition occurs. 

Two special cases of~\eqref{eq:ksdiagbisintro}, which will be important below, are the first transition out of the CB, and the last transition into the maximal HB, which occur at
\begin{equation}\label{eq:firstlastintro}
    M_\text{1st} \sim {\Lambda \over N}~, \qquad M_\text{last} \sim \Lambda~,
\end{equation}
Here~$M_\text{1st}$ and~$M_\text{last}$ are the transition points shown in figure~\ref{fig:phases}.

\end{itemize}

\begin{figure}[t!]
\centering
\includegraphics[width=0.49\textwidth]{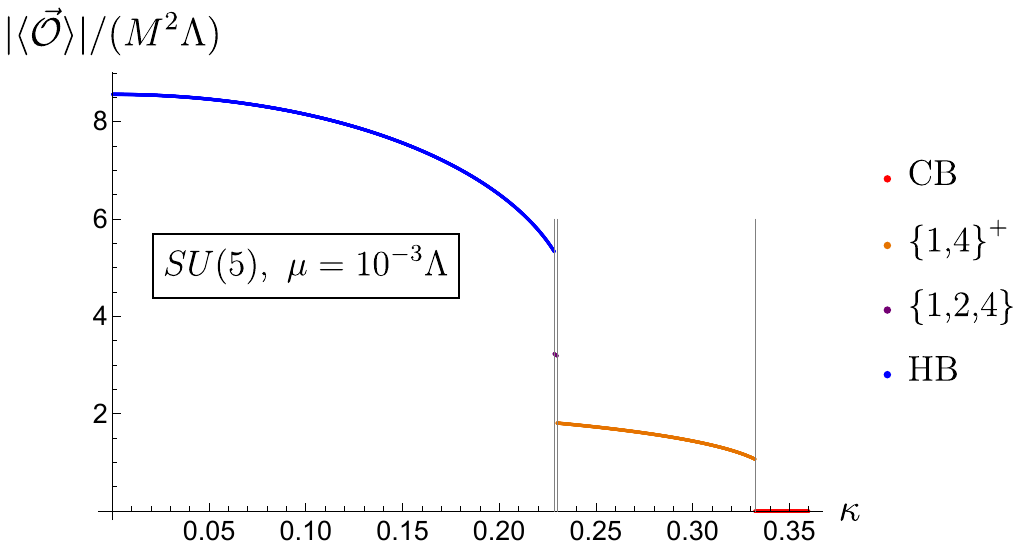}
\includegraphics[width=0.49\textwidth]{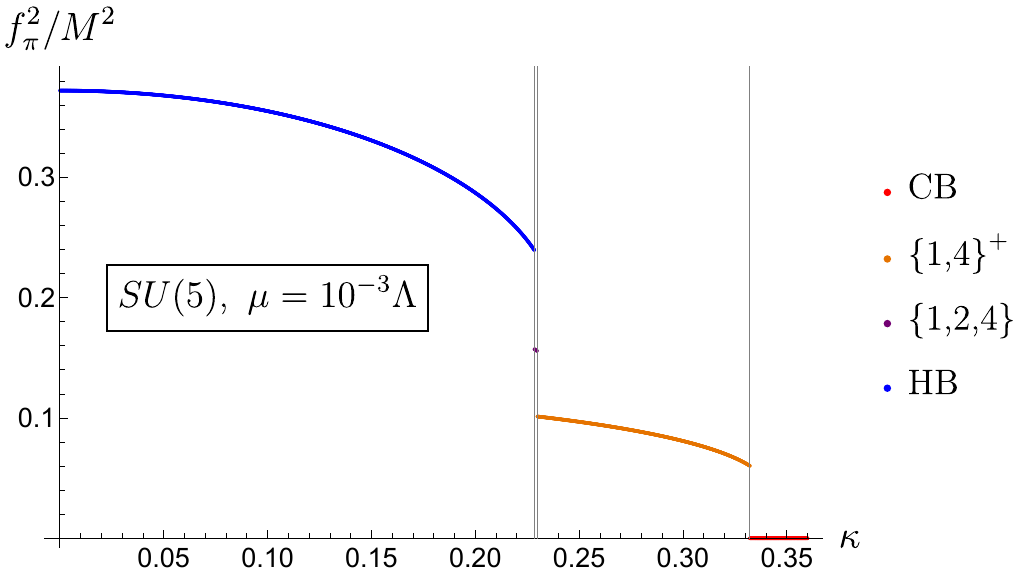}
	\caption{Rescaled vev of the gaugino bilinear~$\vec \CO$ in~\eqref{eq:vecOdef}, and rescaled radius-squared~$f_\pi^2$ of the~$\C\P^1$ sigma model in~\eqref{eq:cp1lagintro}, as functions of~$\kappa = {N \Lambda/ (2 \pi^2 M)}$ for~$SU(5)$ gauge group and RG scale~$\mu = 10^{-3}\Lambda$. Here we use a condensed notation for the branches introduced in section~\ref{sec:sun}: the $C$-even 2-Higgs branch with~$h_1 = h_4 > 0$ (shown in orange) is denoted by~$\{1, 4\}^+$, and the~$C$-odd 3-Higgs branch with~$h_1, h_2, h_4 > 0$ (shown in purple) by~$\{1, 2, 4\}$. As always, CB (red) and HB (blue) denote the Coulomb and the maximal Higgs branches, respectively. The fact that the~$C$-odd~3-Higgs branch is barely visible relative to the other~$C$-even branches reflects the coarse and fine structure of the cascade.}	\label{fig:condfpiIntro}
\end{figure}

Recall that the hypermultiplet Higgs fields~$h_{m}^i$ (with~$m = 1, \ldots, N-1$) are~$SU(2)_R$ doublets, with~$i=1,2$ the associated $SU(2)_R$ index. Thus, the moment the first Higgs field condenses, it spontaneously breaks~$SU(2)_R \to U(1)_R$, as in~\eqref{chiSB}. This leads to two massless Nambu-Goldstone bosons parametrizing a~$\C\P^1$ non-linear sigma model with radius (or decay constant)~$f_\pi$, described by the following Lagrangian,
\begin{equation}\label{eq:cp1lagintro}
    \SL_{\mathbb C \mathbb P^1}= -{f_\pi^2 \over 2} \p^\mu \vec n \cdot \p_\mu \vec n~, \qquad \vec n^2 = 1~.
\end{equation}
If more Higgs fields turn on along the cascade, the vacuum alignment mechanism discussed around~\eqref{eq:spinchainintro} ensures that the symmetry-breaking pattern remains~$SU(2)_R \to U(1)_R$,  with the~$U(1)_R$ Cartan always unbroken (as required by the Vafa-Witten theorem in adjoint QCD).  Note that the~$\C\P^1$ radius~$f_\pi$ in~\eqref{eq:cp1lagintro} depends on~$M$, and jumps discontinuously across the first order phase transitions along the cascade. This is depicted schematically in figure~\ref{fig:phases}, and quantitatively in the right panel of figure~\ref{fig:condfpiIntro}, for~$SU(5)$ gauge group. 

It is natural to ask whether the breaking of the~$SU(2)_R$ symmetry due to the monopole hypermultiplets~$h^i_m$ in the Abelian dual gives rise to a vev for the non-Abelian gaugino bilinear~$\vec \CO$ in~\eqref{eq:vecOdef}. This is indeed the case, as we show explicitly in section~\ref{sec:ordparlargeN}. The resulting vev~$\langle \vec \CO\rangle$ is plotted for gauge group~$SU(5)$ in the left panel of figure~\ref{fig:condfpiIntro}, where we show that it also grows along the cascade. 

We conclude our discussion of the coarse structure of the cascade with the following observations:
\begin{itemize}

\item Let~$\# > 0$ denote the number of Higgs fields that have condensed in a given phase along the cascade. Then the only massless fields in the IR are the two Nambu-Goldstone bosons parametrizing the~$\C\P^1$ sigma model in~\eqref{eq:cp1lagintro}, as well as~$N-1 - \#$ massless Abelian gauge bosons, and~$N-1-\#$ massless~$SU(2)_R$ gaugino doublets.\footnote{~This implies that the~$\C\P^1$ sigma model in~\eqref{eq:cp1lag} requires a discrete~$\theta$-angle to match the~$SU(2)_R$ Witten anomaly if and only if~$\#$ is odd.} Thus the number of massless fields decreases along the cascade. 

\item At the coarse level, the realization of the discrete zero-form symmetries is as follows:\footnote{~This discussion of the unbroken discrete zero-form symmetries must be amended once the fine-structure of the cascade (see section~\ref{FSintro}) is taken into account, most visibly because~$C$-symmetry can then be spontaneously broken. This is further discussed in section~\ref{sec:symcascade}.}
\begin{itemize}
    \item~$C$-symmetry is unbroken, as is clearly reflected in~\eqref{eq:coarscascadeintro}. 
    \item In most vacua along the cascade, the discrete~$\Z_{4N}$ $R$-symmetry (with generator~$r$) is spontaneously broken. (The CB at small~$M$ is the only exception.) By combining the broken~$r^N$ generator with an~$SU(2)_R$ Weyl reflection, one finds an unbroken~$\Z_2$ symmetry that extends~$U(1)_R$ to~$O(2)_R$, exactly as in~\eqref{chiSB}.
\item The time-reversal symmetry~$\t T = r^N T$ in~\eqref{eq:untbrokenTtintro} is unbroken.\footnote{~Note that this is the unbroken time-reversal symmetry at the multi-monopole point, corresponding to~$k = 0$ in the discussion above~\eqref{eq:untbrokenTtintro}.}
\end{itemize}

\item The~$\Z_N^{(1)}$ one-form symmetry is generically spontaneously broken completely. There are two exceptions: (i) on the maximal HB (see below), $\Z_N^{(1)}$ is completely unbroken, indicating confinement in adjoint QCD; (ii) when~$N = 2\nu$ is even, then the penultimate phase  of the cascade, where the only vanishing Higgs field is~$h_\nu = 0$, has an unbroken~$\Z_\nu^{(1)}$ one-form symmetry, i.e. the symmetry breaking pattern is $\Z_N^{(1)} \to \Z_\nu^{(1)}$. 
\end{itemize}

\subsubsection{Fine structure of the cascade and spontaneous~$C$-breaking}\label{FSintro}
We will now describe how the coarse structure described in section~\ref{sec:cascadeIntro} above is -- generically but briefly -- interrupted by the appearance of~$C$-odd phases that are drawn small in figure~\ref{fig:phases}. This is the fine structure of the cascade.

Consider the transition between the~$C$-even phases of the coarse cascade~\eqref{eq:coarscascadeintro} that involves turning on the following two Higgs fields in a $C$-symmetric fashion,\footnote{~Again, the case~$N = 2\nu$ and~$m = \nu = N-m$ is an exception; in that case only the~$C$-even Higgs field~$h_\nu$ turns on and there is no interesting fine structure associated with that transition.} 
\begin{equation} \label{eq:hmOnintro}
h_m^i = h_{N-m}^i \neq 0~,
\end{equation}
which occurs around~$M \simeq M_{*m}$ in~\eqref{eq:ksdiagbisintro}. The fine structure of the cascade manifests as the splitting of this transition into two closely spaced first-order phase transitions, each of which only involves turning on a single Higgs field. These transitions occur at~$M \simeq M_{*m} \pm \Delta M$, where the splitting~$\Delta M$ is much smaller than the size of the~$C$-even phases in~$M$-space. This is clearly visible in figures~\ref{fig:phases} and~\ref{fig:condfpiIntro}.

At~$M \simeq M_{*m} - \Delta M$, we transition from a~$C$-even phase with the following non-vanishing (and~$SU(2)_R$ aligned) Higgs fields,  already realized in the coarse cascade~\eqref{eq:coarscascadeintro},
\begin{equation}\label{eq:cevenhintro}
C\text{-even} : \quad  h_1^i = h_{N-1}^i \neq 0, \quad h_2^i = h_{N-2}^i \neq 0~, \quad \cdots \quad h_{m-1}^i = h_{N-1-m}^i \neq 0~, 
\end{equation}
to a branch that spontaneously breaks~$C$-symmetry because we only turn on one of the two Higgs fields in~\eqref{eq:hmOnintro}, 
\begin{equation}\label{eq:coddphaseintro}
C\text{-odd} : \quad h_m^i \neq 0~, \; h_{N-m}^i = 0 \qquad \text{or} \qquad h_m^i = 0~, \; h_{N-m}^i \neq 0~.
\end{equation}
These are precisely the short phases indicated for~$m = 1$ and~$m = 2$ in figure~\ref{fig:phases}. 

The~$C$-odd interpolating phase~\eqref{eq:coddphaseintro} only appears very briefly: at~$M \simeq M_{*M} + \Delta M$ we immediately transition back to the next~$C$-even phase already present in the coarse cascade~\eqref{eq:coarscascadeintro}, where all the Higgs fields in~\eqref{eq:cevenhintro} and~\eqref{eq:hmOnintro} are turned on in a charge-conjugation preserving fashion.

Several comments are in order:
\begin{itemize}
    \item Clearly, the coarse and fine structure of the cascade discussed above, which is an exact feature of the classical potential~\eqref{eq:ahmvtotalintro} that we are minimizing, calls for an explanation in some perturbative scheme in which the coarse structure arises at leading order, while the fine-structure arises at higher orders in perturbation theory. Precisely such a perturbative scheme is devised in section~\ref{sec:cascade}.\footnote{~The perturbative approach of section~\ref{sec:cascade} involves expanding the matrix~$t_{mn}(\mu)$ of effective gauge couplings in~\eqref{eq:tdefintro} in powers of its off-diagonal entries (modulo fine print that is explained there).}

    There we show that the three branches discussed above -- the first with only the Higgs fields~\eqref{eq:cevenhintro} turned on; the second obtained by adding the single Higgs fields in~\eqref{eq:coddphaseintro} to the first; and the third by adding both Higgs fields in~\eqref{eq:hmOnintro} -- are exactly degenerate at leading order in the perturbative scheme of section~\ref{sec:cascade}. As is familiar from elementary examples of perturbation theory, this degeneracy is lifted at higher orders, but can naturally give rise to the~$C$-odd phase~\eqref{eq:coddphaseintro} and the small splittings~$\Delta M$ above.

\item The first transition out of the Coulomb branch (CB), at~$M \simeq M_\text{1st}$ in figure~\ref{fig:phases}, requires a separate discussion: as we show in section~\ref{sec:aD3}, the potential~\eqref{eq:ahmvtotalintro} never gives rise to the green 1-Higgs phase in figure~\ref{fig:phases}, because the three phases CB, 1-Higgs with~$h_1^i \neq 0$, and $2$-Higgs with~$h_1^i = h_{N-1}^i \neq 0$, are  exactly degenerate. This accidental degeneracy can be broken by considering effects that we have so far neglected, e.g.~quantum corrections in the dual. We will not consider those, and instead focus on the effects of the~$\CO(a_D^3)$ terms in the effective~$\CN=2$ K\"ahler potential, that we have so far neglected in~\eqref{eq:susyvintro} and~\eqref{eq:VSBintro}. The upshot (see section~\ref{sec:aD3}) is that the $C$-odd 1-Higgs branch comes down in energy for~$SU(3)$, but is lifted for higher~$SU(N)$, leading to the following picture for the first phase transition(s) out of the Coulomb branch,
\begin{align}
SU(2) \; & : \; \text{CB} \quad \to \quad \text{HB} =  \{h_1^i \neq 0 \}  \\
          SU(3) \; & : \; \text{CB} \quad \to \quad \{h_1^i \neq 0 \text{ or } h_2^i \neq 0\} \quad \to \quad \text{HB} = \{h_1^i = h_2^i \neq 0\}   \\
      SU(N \geq 4) \; & : \; \text{CB} \quad \to  \quad \{h_1^i = h_{N-1}^i \neq 0\}
\end{align}
\end{itemize}

\subsubsection{The maximal Higgs branch (HB) at large~$M$ and adjoint QCD} \label{sec:HBadjointIntro}

Finally, at the last step of the cascade at~$M \sim M_\text{last} \sim \Lambda$ (see~\eqref{eq:firstlastintro} and figure~\ref{fig:phases}), we transition to the maximal Higgs branch (HB), where all~$h_m^i \neq 0$ (with~$m = 1, \ldots, N-1$) are non-zero and aligned. In particular, all gauge fields in the Abelian dual at the multi-monopole point are Higgsed, so that the~$\Z_N^{(1)}$ symmetry is unbroken. This is the dual Higgs description of confinement already familiar from~\cite{Seiberg:1994rs, Douglas:1995nw}, where it was used to demonstrate confinement in pure~$\cN=1$ SUSY Yang-Mills theory.

By contrast, the fact that the continuous~$SU(2)_R$ symmetry under which the~$h_m^i$ are doublets is broken to its Cartan,
\begin{equation}\label{eq:su2rbreakintro}
    SU(2)_R \to U(1)_R~,
\end{equation}
consistent with the chiral symmetry breaking pattern (and the Vafa-Witten theorem) for adjoint QCD in~\eqref{chiSB}, requires the novel vacuum alignment mechanism explained  around~\eqref{eq:spinchainintro}.\footnote{~For~$SU(2)$ gauge group, the fact that $SU(2)_R \to U(1)_R$ was found in \cite{Cordova:2018acb}, but because there is only a single~$SU(2)_R$ doublet Higgs field~$h^i$ in that case, no vacuum alignment was needed.} In fact, the discrete zero-form symmetries on the maximal Higgs branch are also realized exactly as in the confining and chiral symmetry breaking scenario for adjoint QCD summarized in section~\ref{sec:confchisbIntro}: charge-conjugation symmetry~$C$ is unbroken, the~$\Z_{4N}$ chiral symmetry is spontaneously broken as in~\eqref{chiSB}, and the time-reversal symmetry~$\t T = r^N T$ is unbroken.\footnote{~This is true at the multi-monopole point; the unbroken time-reversal symmetries in the other~$N-1$ disjoint vacuum sectors are obtained by conjugating with the spontaneously broken~$\Z_{4N}$ symmetry, which cyclically permutes the~$N$ multi-monopole points.}

The only massless fields on the maximal Higgs branch are the two Nambu-Goldstone bosons associated with the continuous chiral symmetry breaking~\eqref{eq:su2rbreakintro}, which are described by the non-linear sigma model~\eqref{eq:cp1lagintro} with target space~$\C\P^1 = SU(2)_R/U(1)_R$ and radius~$f_\pi$. We compute this radius in~\eqref{eq:fpiHB}, and determine its large-$N$ limit in~\eqref{eq:finalfpiHB}, both of which we repeat here,\footnote{~See also the right panel of figure~\ref{fig:condfpiIntro}, where~$f_\pi^2$ is plotted for~$SU(5)$ gauge group. The maximal HB corresponds to~$\kappa \to 0$.} 
\begin{equation}\label{eq:fpiHBIntro}
        f_\pi^2 = {1 \over 4} M^2 \sum_{m,n = 1}^{N-1} t_{mn} \quad \to \quad  {7 \zeta(3) \over 8\pi^4} N^2 M^2 \quad \text{as} \quad N \to \infty~.
\end{equation}
This is precisely the expected large-$N$ scaling for~$f_\pi^2$, since the Lagrangian~\eqref{eq:cp1lagintro} should be~$\CO(N^2)$ in a theory with only adjoint fields. Note that the expected scaling in adjoint QCD is actually~$\CO(N^2 \Lambda^2)$, rather than~$\CO(N^2 M^2)$.\footnote{~This glosses over a slight mismatch between the strong-coupling scales of adjoint QCD and~$\CN=2$ SYM, a quantum effect that is due to the fact that these theories have different UV~$\beta$-functions.} However, since the transition to the maximal Higgs branch occurs at~$M \sim \Lambda$, which is also the scale at which the vector-multiplet scalar~$\phi$ with SUSY-breaking mass~$M$ in~\eqref{eq:vsbintro} decouples, it is reasonable to hope that~\eqref{eq:fpiHBIntro} saturates at that scale, giving the expected adjoint QCD scaling. 

Finally, the vev of the gaugino bilinear~$\vec \CO = i \tr(\lambda \vec \sigma \lambda)$ in~\eqref{eq:vecOdef}, which serves as the order parameter for the chiral symmetry breaking pattern~\eqref{chiSB} in adjoint QCD, is computed on the maximal HB in~\eqref{eq:smallkOvev}. (See for instance the left panel of figure~\ref{fig:condfpiIntro}, where~$|\langle \vec O \rangle|$ is plotted for~$SU(5)$ gauge group; the maximal HB corresponds to~$\kappa \to 0$.) Here we limit ourselves to the large-$N$ limit of that expression, which is determined in~\eqref{eq:largeNllfinal},
\begin{equation}\label{eq:finlargNvecOintro}
      \langle \vec \CO\rangle  \quad \to \quad  {4 \sqrt 2 N \over \pi} M^2 \Lambda \, \vec e_3 \quad \text{as} \quad N \to \infty~.
\end{equation}
Here~$\vec e_3$ is the unit vector along the~$3$-axis in~$SU(2)_R$ triplet space, corresponding to the vacuum~$\vec n = \vec e_3$ in~\eqref{eq:vecOCn}, i.e.~the north pole of the~$\C\P^1$. Note that the coefficient of~$\vec n = \vec e_3$ in~\eqref{eq:finlargNvecOintro} is real and positive, corresponding to the complex phase~$k = 0$ in~\eqref{eq:vecOCn}. This is because we are working in the dual at the multi-monopole point; the other values~$k = 1, \ldots, N-1$ in~\eqref{eq:vecOCn} arise from the other~$N-1$ multi-monopole points. Again, we find that~\eqref{eq:finlargNvecOintro} has the correct~large-$N$ scaling for the gaugino bilinear~$\vec \CO$, if we assume that the expression saturates at~$M \sim \Lambda$, roughly at the transition to the maximal Higgs branch. 

In summary, the maximal Higgs branch of our dual description matches the confining and chiral symmetry breaking phase of adjoint QCD spelled out in section~\ref{sec:confchisbIntro} in great detail. We view this as strong evidence that this phase is actually realized in adjoint QCD.

\subsection{Reading guide}

\noindent The remainder of the paper is organized as follows:

In section~\ref{sec:SYM} we review pure $\cN=2$ supersymmetric Yang-Mills theory with gauge group $SU(N)$; the corresponding Seiberg-Witten solution; the effective Abelian Higgs model valid near one of the multi-monopole points; and the action  of relevant symmetries.

In section~\ref{sec:general} we introduce the SUSY-breaking scalar mass~\eqref{eq:vsbintro}, and track it onto the Coulomb branch, and to the multi-monopole points of the~$\cN = 2$ theory. We motivate the proposed dual Abelian Higgs model with supersymmetry breaking; exhibit relevant properties of the effective matrix $t$ of couplings and mixings of the $U(1)^{N-1}$ gauge fields; prove vacuum alignment; and formulate the semi-classical analysis problem for arbitrary $SU(N)$ in terms of existence, local stability, and global stability of solutions.   We also show that the dual correctly captures the small-$M$ Coulomb branch.

In sections~\ref{sec:su2} and~\ref{sec:su3} we present the analytical semi-classical phase diagram of the Abelian dual for gauge groups $SU(2)$ and $SU(3)$, respectively.

In section~\ref{sec:sun} we develop a general classification and taxonomy of the different branches of solutions of the dual, reduce the field equations, the effective potential, and the local stability conditions  in each branch, and introduce  analytical tools for the comparison of the effective potential in the different branches. We also discuss the fate of charge conjugation symmetry in each branch, and derive general properties of the Coulomb and maximal Higgs branches.

In section~\ref{sec:num} we obtain the phase diagrams for the gauge groups $SU(4)$, $SU(5)$, and $SU(6)$ using numerical analysis of the various branches of solutions to the Abelian dual. Drawing on the results from these low rank cases we present systematic evidence for the existence of a cascade of phase transitions from the Coulomb branch at small $M$, to the maximal Higgs branch at large $M$, passing through a sequence of mixed Coulomb/Higgs branches. These numerical cascades display the~$C$-even coarse structure and~$C$-breaking fine structure already discussed above.

In section \ref{sec:cascade}, we develop a perturbative approach to solving the field equations of the Abelian dual by expanding the matrix $t$ in powers of its off-diagonal entries, which is valid for sufficiently small $\mu / \Lambda$. To leading order, the dual decouples into $N-1$ analytically solvable models akin to the $SU(2)$ case; their solution confirms the coarse cascade structure of phases already identified in section~\ref{sec:num} using exact numerics. Higher order corrections are evaluated as well, shown to lift various accidental degeneracies present to leading order, and lead to the fine structure of the cascade, including the brief existence of phases with spontaneously broken charge-conjugation symmetry~$C$.

In section~\ref{sec:11} the mass spectra are discussed with emphasis on light and massless states.

Finally, section~\ref{sec:summary} provides a comprehensive summary of all the evidence for the cascading phase structure amassed in the paper, how the symmetries are realized in each phase, and how this picture is consistent with the confining and chiral symmetry breaking phase for~$SU(N)$ adjoint QCD with~$N_f = 2$ Weyl fermion  flavors. In particular, we comment on the large-$N$ scaling of various quantities of physical interest.

Appendix~\ref{sec:conventions} summarizes our conventions, focusing on spinors, the $SU(2)_R$ symmetry, and supersymmetry; appendix~\ref{app:B} explains the numerical methods used to evaluate the effective potential on the various branches; appendix~\ref{app:appt} establishes numerous properties of the matrix $t$ needed throughout the paper; and appendix~\ref{sec:maxHglob} presents a proof that the maximal Higgs branch is always globally stable as $M \to \infty$.

\subsection{Acknowledgements}

\noindent We thank G.~Bergner, C.~C\'ordova, T.~DeGrand, K.~Intriligator, N.~Seiberg, and M.~\"Unsal for useful comments and discussions.  The research of ED is supported in part by NSF grant PHY-22-09700. The work of TD is supported in part by DOE awards DE-SC0020421 and DE-SC0025534, as well as by the Simons Collaboration on Global Categorical Symmetries. The work of EG is in part supported by the Israel National Postdoctoral Award Program for
Advancing Women in Science.

\newpage

\section{Pure~$\CN=2$ SYM with $SU(N)$ gauge group}
\label{sec:SYM}

In this section we review the salient features of pure $\cN=2$ super Yang-Mills (SYM) theory with gauge group $SU(N)$  (and no matter hypermultiplets) that are required for the analysis in this paper, starting with the UV Lagrangian and its symmetries. We then recall the Seiberg-Witten description of the IR effective action at generic points on the Coulomb branch of supersymmetric vacua in terms of an $\CN=2$ gauge theory with gauge group $U(1)^{N-1}$, and its extension to the~$N$ multi-monopole points on the Coulomb branch. At each of these points, $N-1$ mutually local monopoles described by~$\CN=2$ hypermultiplets become massless. We also explain how to track the~$\CN=2$ stress tensor supermultiplet from the UV to the IR and show that its scalar primary operator~$\CT$ flows to a suitably well-defined choice of effective K\"ahler potential on the Coulomb branch, with an important modification at the multi-monopole points. 

We will also recall relevant results from two earlier companion papers:

\begin{itemize}
\item[1.)] In~\cite{DHoker:2022loi} we analyzed the behavior of the Seiberg-Witten periods and the effective K\"ahler potential in the strong-coupling region surrounding the origin of the Coulomb branch. 

\item[2.)] In~\cite{DHoker:2020qlp} we determined the matrix of effective~$U(1)^{N-1}$ gauge couplings, which is non-diagonal and mixes the different~$U(1)$ factors, near the multi-monopole points. 
\end{itemize}

\noindent We will explicitly spell out our conventions below.\footnote{~They largely agree with those of~\cite{DHoker:1997mlo,DHoker:2020qlp}.}

\subsection{UV Lagrangian and Coulomb branch of vacua}

Throughout, we shall adopt Einstein conventions for summation over repeated indices,  Wess and Bagger conventions  for spinors, including signature $(-+++)$ for Minkowski space.  Any field in the adjoint representation of $SU(N)$  is denoted by $\chi= \chi ^a T^a$ where $a=1,\dots , N^2-1$ and $T^a$ are the Hermitian $SU(N)$ generators in the defining (fundamental) representation, normalized so that $\tr(T^a T^b) = \half \delta^{ab}$. A more detailed summary of our conventions can be found in appendix~\ref{app:conv}.

Pure $\cN=2$ SYM in four spacetime dimensions consists of an $\cN=2$ vector multiplet in the adjoint representation of the gauge group, which we here take to be~$SU(N)$. Under an $\cN=1$ subalgebra the $\cN=2$ vector multiplet decomposes into an $\cN=1$ vector multiplet $\CV$ and an $\cN=1$ chiral multiplet $\Phi$, both in the adjoint of $SU(N)$. The Lagrangian can then be written in~$\CN=1$ superspace,
\bea
\label{2.Lsusy}
{\mathscr L}_{\text{SU(N)}} = { 2 \over g^2} \int d^4 \theta \, \tr \left ( \bar \Phi e^{-2 \CV} \Phi \right ) 
+ { 1 \over g^2} \, \Re \! \int d^2 \theta \, \tr \big ( W^\a W_\a \big )~,
\eea
where $W_\a= - \tfrac{1}{4} \bar D^2 \left ( e^{-\CV} D_\a e^{\CV} \right )$. We do not include a possible theta angle in the Lagrangian, since it can be removed in the quantum theory thanks to an ABJ anomaly (see below). 

In terms of components, we have a gauge field~$v_\mu$ with field strength~$v_{\mu\nu}$, two Weyl gauginos~$\lambda_\alpha^i$ (with~$i = 1,2$ an~$SU(2)_R$ doublet index, see below), a complex scalar field~$\phi$, and a real~$SU(2)_R$ triplet of auxiliary fields~$D^{(ij)} = (D_{(ij)})^\dagger$. All fields are in the adjoint representation of the~$SU(N)$ gauge group. The component form of the Lagrangian~\eqref{2.Lsusy} is\footnote{~The covariant derivative acting on fields in the adjoint representation is given by $D_\mu \phi = \p_\mu \phi - i [v_\mu, \phi]$ and the field strength is given by $v_{\mu \nu } = \p_\mu v_\nu - \p_\nu v_\mu - i [v_\mu , v_\nu]$. Finally, $SU(2)_R $ indices are raised and lowered using $\lambda_i = \ep _{ij} \lambda ^j$, $\lambda^i = \ep^{ij} \lambda_j$ with $\ep^{12} = - \ep_{12} = 1$, and Hermitian conjugate fields are denoted by bars, e.g.~$\bar \lambda _{\dot \a i} = (\lambda ^i_\a)^\dagger$ and~$\bar \phi = \phi^\dagger$. See appendix~\ref{app:conv} for more detail.}
\bea
\label{2.Lcomp}
{\mathscr L} _{\text{SU(N)}} & = & 
{ 2 \over g^2} \, \tr \bigg ( -\frac{1}{4}  v^{\mu \nu} v_{\mu \nu} 
- D^\mu \bar \phi D_\mu \phi -  i \ \bar \lambda _i \bar \sigma ^\mu D_\mu \lambda ^i  
\no \\ && \hskip36pt
- \half [\bar \phi, \phi]^2  +  {1 \over \sqrt{2}}   [\bar \phi, \lambda^i] \lambda_i 
+ {1 \over  \sqrt{2}}  [\phi, \bar \lambda _i] \bar \lambda ^i + \frac{1}{4} D^{ij} D_{ij} \bigg )
\eea  
The~$\CN=2$ supersymmetry transformations of the component fields under which this Lagrangian is invariant are summarized in appendix~\ref{app:conv}. The gauge coupling~$g$ is asymptotically free and the theory dynamically generates a strong-coupling scale that we denote by~$\Lambda$.\footnote{~A choice of scheme for~$\Lambda$ will be implicit in our conventions for the Seiberg-Witten solution of the~$SU(N)$ theory reviewed in section~\ref{sec:swsol} below.}

The scalar potential $  \tr( [\bar \phi, \phi]^2)$ is non-negative, and vanishes along flat directions where $\phi$ satisfies $[\bar \phi, \phi]=0$. This condition is solved by restricting the expectation value of~$\phi$ to lie in a Cartan subalgebra of~$SU(N)$. A generic such expectation value corresponds to a supersymmetric vacuum in which the gauge symmetry is spontaneously broken to
\bea
U(1)^{N-1} = \prod_{m = 1}^{N-1} U(1)_m
\eea
In the quantum theory, the family of vacua described by such an Abelian gauge theory at low energies is known as the Coulomb branch. It is locally parametrized by~$N-1$ complex moduli, or Coulomb-branch coordinates. A gauge-invariant choice consists of the following traces of powers of $\phi$ (recall that~$\tr (\phi) =0$),
\bea
\label{2.moduli}
u_I = \tr ( \phi^{I}  )~, \qquad I = 2, \ldots, N
\eea
Note that we reserve capital letters~$I, J, \text{etc.} = 2, \ldots, N$ for the~$u$'s and  use lowercase letters such as~$m, n, \text{etc.}= 1, \ldots, N-1$ for~$U(1)^{N-1}$ gauge-group indices on the Coulomb branch.\footnote{~We apologize that these conventions are reversed relative to the companion paper~\cite{DHoker:2022loi}.}

\subsection{Symmetries}

\label{sec:symm}

In addition to supersymmetry, the~$\CN=2$ SYM theory reviewed above has a wealth of other global symmetries that will feature in our analysis:

\begin{itemize}
\item{\it $R$-Symmetries:} The supercharges~$Q_\alpha^i$ are acted on by an~$R$-automorphism
\bea
{SU(2)_R \times U(1)_r \over \ZZ_2} 
\eea
Here~$Q^i_\alpha$ is an~$SU(2)_R$ doublet and has charge~$-1$ under~$U(1)_r$. As already stated above~\eqref{2.Lcomp}, the gauginos~$\lambda_\alpha^i$ are~$SU(2)_R$ doublets and the auxiliary field~$D^{(ij)}$ is an~$SU(2)_R$ triplet. The~$U(1)_r$ charges of the component fields are~$+2$ for the scalar~$\phi$, $+1$ for the gauginos~$\lambda_\alpha^i$, and~$0$ for the gauge and auxiliary fields~$v_\mu, D^{ij}$, as required by consistency with their supersymmetry transformations (see appendix~\ref{app:conv}). 
\sm

While the~$SU(2)_R$ symmetry is quantum mechanically exact, $U(1)_r$ is explicitly broken to its~$\ZZ_{4N}$ cyclic subgroup by an Adler-Bell-Jackiw (ABJ) anomaly,
\bea
\text{ABJ}: \quad U(1)_r \quad \longrightarrow \quad \ZZ_{4N}
\eea
We will denote the generator of~$\ZZ_{4N}$ by~$r$, so that~$r^{4N} = 1$ on all gauge-invariant fields.\footnote{~Note that the order-2 element~$r^{2N} \in \ZZ_{4N}$ is identified with the order-2 central element of~$SU(2)_R$ as well as with~$(-1)^F$ fermion parity.} Since~$\phi$ has~$U(1)_r$ charge~$2$, we conclude from~\eqref{2.moduli} that the gauge-invariant moduli~$u_I$ transform as follows,
\bea\label{2.ronu}
r : u_I \; \to \; e^{{2 \pi i \over 2N  } \, I } \, u_I 
\eea
Generically, for~$N \geq 3$, a~$\Z_{2N}$ quotient of~$\Z_{4N}$ acts faithfully on the~$u_I$.\footnote{~Equivalently, only~$(-1)^F = r^{2N}$ does not act.} In particular, note that~$r^{N} : u_I \to (-1)^{I} u_I$, which thus acts non-trivially as long as~$I$ can be odd, i.e.~when~$N \geq 3$. An exception occurs for~$N = 2$, where only a~$\Z_2$ quotient of~$\Z_{4N} = \Z_8$ acts on the Coulomb branch, because~$u_2$ is the only modulus and~$ r : u_2 \to - u_2$. 

\item{\it 1-Form Symmetry:} All fields transform in the adjoint representation of the~$SU(N)$ gauge group. The theory therefore has a~$\ZZ_N^{(1)}$ 1-form symmetry associated with the center of~$SU(N)$, commonly referred to as center symmetry. The~$\ZZ_N^{(1)}$ charge of a Wilson loop~$\CW_\CR$ in an~$SU(N)$ representation~$\CR$ is given by the~$N$-ality of~$\CR$. 

The~$\ZZ_N^{(1)}$ center symmetry is unbroken  if the expectation values of all large Wilson loops charged under it decay faster than perimeter-law scaling.  Unbroken center symmetry is a sharp way to characterize a confining phase~\cite{Gaiotto:2014kfa}. Standard linear confinement with finite-tension strings requires the stronger assumption that large loops decay according to the usual area law.

\item{\it Charge Conjugation:} For~$N \geq 3$, the~$SU(N)$ theory has a~$\ZZ_2$ charge-conjugation symmetry~$C$, which commutes with the supercharges. Any field~$\chi = \chi^a T^a$ in the adjoint representation of~$SU(N)$ then transforms under~$C$ as follows,
\bea\label{eq:cdef}
C : \chi^a T^a \; \to \; - \chi^a (T^a)^*
\eea
Thus~$C$ effectively maps the generators~$T^a$ in the fundamental representation of~$SU(N)$ to the generators~$-(T^a)^*$ of the complex-conjugate anti-fundamental representation (which are gauge-inequivalent for~$N \geq  3$). Applying this to~\eqref{2.moduli}, we find that the Coulomb branch moduli~$u_I$ transform as follows under charge conjugation,
\be\label{2.conu}
C : u_I \rightarrow (-1)^I u_I  
\ee

\item{\it Time Reversal and Parity:} It can be checked that the~$\CN=2$ SYM theory is invariant under (anti-unitary) time reversal~$T$ and (unitary) parity~$P$ symmetries. The CPT theorem guarantees that these are not independent, and we will therefore focus on the time-reversal symmetry~$T$. Following~\cite{Cordova:2018acb}, we define~$T$ as follows,
\bea \label{2.Tdef}
T : \phi^a \to \phi^a~, \qquad \lambda_\a^{i a} \to i \lambda_i^{\a a}~, \qquad v_\mu^a \to {T_\mu}^\nu  v_\nu^a~,  \qquad D^{ij a} \to D_{ij}^a
\eea
Here~${T_\mu}^\nu = \text{diag}(-1, 1,1,1)$ is the standard geometric time-reversal element of the Lorentz group. In order to avoid clutter, we have given the action of~$T$ on adjoint-valued fields~$\chi^a$ rather than~$\chi = \chi^a T^a$ since time reversal is anti-unitary and complex conjugates the generators~$T^a$. The action of~$T$ in~\eqref{2.Tdef} also acts on the argument of every field as follows, 
\be
T: x^\mu = (x^0, \vec x) \to {T^\mu}_\nu x^\nu = (-x^0, \vec x)~.
\ee

Our choice of~$T$ is consistent with the Lorentz transformation properties of spinors (which requires raising the spinor index~$\a$ on~$\lambda_\alpha^{ia}$) and has the additional property of commuting with the unitary operators that implement~$SU(2)_R$ transformations (which requires lowering the~$SU(2)_R$ indices on~$\lambda_\alpha^{ia}$ and~$D^{ija}$). It can be verified that
\bea\label{eq:tonq}
T : Q_\alpha^i \to -i Q^\alpha_i~,
\eea
and that
\be
T^2 = 1 
\ee
on gauge-invariant local operators.  These relations are consistent with the supersymmetry algebra and imply that~$T$ generates an anti-unitary~$\ZZ_2$ symmetry. Finally, we note that the Coulomb branch coordinates~$u_I$ in~\eqref{2.moduli} are~$T$-invariant operators,
\bea\label{2.Tonu}
T : u_I \to u_I
\eea
By contrast, expectation values~$\langle u_I\rangle$ of~$u_I$ are~c-numbers that are complex conjugated by~$T$. Thus unbroken~$T$-symmetry requires all~$\< u_I\>$ to be real.

\end{itemize}

Below it will be useful to define two symmetries that arise from mixing charge conjugation and time reversal with the order-four element~$r^N \in \ZZ_{4N}$:
\begin{itemize}
\item Mixing~$C$ with~$r^N$, defines a unitary~$\ZZ_4$ symmetry with generator
\bea
\t C = r^N C
\eea
Note that~$\t C : Q_\alpha^i \to -i Q_\alpha^i$ is an~$r$-symmetry that squares to fermion parity, $\t C^2 = r^{2N} = (-1)^F$. It follows from~\eqref{2.ronu} and~\eqref{2.conu} that
\bea\label{2.ctdef}
\t C : u_I \rightarrow u_I
\eea
Thus the~$\ZZ_4$ symmetry generated by~$\t C$ is unbroken at every point on the Coulomb branch. Note that this remains true for~$N = 2$, where~$C$ does not exist and~$\t C = r^2$ acts trivially on the Coulomb branch. 

\item  Mixing~$T$ with~$r^N$ defines an anti-unitary~$\ZZ_2$ symmetry generated by
\bea\label{2.ttdef}
\t T = r^N T
\eea
Note that~$\t T^2 = 1$. Using~\eqref{2.ronu} and~\eqref{2.Tonu} we conclude that
\bea
\t T : u_I \rightarrow (-1)^I u_I
\eea

\end{itemize}

\subsection{Seiberg-Witten description of the IR effective theory}
\label{sec:swsol}

Seiberg-Witten theory~\cite{Seiberg:1994rs,Seiberg:1994aj, Klemm:1994qs, Argyres:1994xh, Klemm:1995wp} describes the exact low-energy Lagrangian on the Coulomb branch, valid in the deep IR (much below the scale~$\Lambda$), and a way to compute the masses of BPS states given their charge spectrum.\footnote{~Determining the spectrum of BPS states is in general a hard problem, and except in rare cases explicit answers are typically only available at special loci on the Coulomb branch.}  For generic values of the~$N-1$ Coulomb branch moduli~$u_I~(I = 2, \ldots, N)$, the low-energy theory is an~$\CN=2$ Abelian gauge theory of rank~$N-1$, with gauge group~$U(1)^{N-1} = \prod_{m=1}^{N-1} U(1)_m$. Each~$\CN=2$ vector multiplet can be decomposed into an~$\CN=1$ vector superfield~$\CV_m$ and an~$\CN=1$ chiral superfield~$A_m$, with~$m = 1, \ldots, N-1$. Here the~$A_m$ are uncharged under any of the~$\CV_m$. Together the components of~$A_m$ and~$\CV_m$ consist of a complex scalar~$a_m$ (the bottom component of~$A_m$), an~$SU(2)_R$ doublet of gauginos, and the~$U(1)_m$ gauge field (together with the auxiliary fields of both superfields).\footnote{~We do not give explicit names to the~$U(1)^{N-1}$ gauginos and gauge fields, to avoid confusion with their magnetic duals, which are introduced in section~\ref{sec:MMLag} below and used throughout the paper.}

The two-derivative Seiberg-Witten IR Lagrangian, constrained by~$\CN=2$ supersymmetry, is encoded in a locally holomorphic pre-potential~$\cF(A)$ that depends on the fields $A_k$ but does not involve any derivatives of $A_k$. In~$\CN=1$ superspace, it takes the form\footnote{~In~$\CN=2$ superspace the Lagrangian~\eqref{2.LIR} can be written as a chiral superspace integral $\int d^4 \theta_{\CN=2} \CF(A_{\CN=2, m})$, with~$A_{\CN=2, m}$ the~$\CN=2$ vector multiplet containing~$A_m, \CV_m$. This form makes manifest the unbroken~$\ZZ_4$ $r$-symmetry generated by~$\t C$ discussed around~\eqref{2.ctdef}, under which both~$A_{\CN=2, m}$ and~$d^4\theta_{\CN=2}$ are invariant. We will not need the detailed component form of~\eqref{2.LIR}, except at special points on the Coulomb branch (see section~\ref{sec:AHM} below). }
\bea
\label{2.LIR}
{\mathscr L}= { 1 \over 2\pi} \sum _{m=1}^{N-1}  \Im \int d^4 \theta \, A_{Dm} \bar A_m 
+ { 1 \over 4 \pi } \sum _{m,n=1}^{N-1} \Im \int d^2 \theta \, \tau_{mn}  W^\a{} _m \, W_{\a  n}
\eea
Here the Abelian field strength superfields are given by $W_{\a  n} = - \frac{1}{4} \bar D^2 D_\a \CV_n$, while the magnetic dual chiral superfields $A_{Dm}$ and the~$(N-1) \times (N-1)$ matrix~$\tau_{mn}$ of complexified $U(1)^{N-1}$ gauge couplings are defined in terms of the pre-potential as follows, 
\bea
\label{ADtau}
A_{Dm} = { \p \cF \over \p A_m}
\hskip 1in
\tau_{mn} = { \p^2 \cF \over \p A_m \p A_n} = {\p A_{Dm} \over \p A_n}
\eea
The~$\CN=1$ K\"ahler potential of the low-energy sigma model, which can be read off from~\eqref{2.LIR}, is given by
\be
\label{2.Kdef}
K = {1 \over 2 \pi} \sum_{m = 1}^{N-1} \Im A_{Dm} \bar A_m~.
\ee
This formula will play an important role throughout our analysis. The K\"ahler metric~$g_{m \bar n}$ derived from~$K$ is given by
\be
g_{m \bar n} = {1 \over 2\pi} \Im \tau_{mn}
\ee

Note that the Lagrangian~\eqref{2.LIR} depends on the pre-potential $\cF$ only through~$\tau_{mn}$ and the K\"ahler metric, and hence it is invariant under the following shift of the pre-potential,
\be\label{eq:fshift}
\CF \to \CF + \sum_{m=1}^{N-1} C_m A_m + D~, \qquad C_m, D \in \C
\ee
Here~$C_m, D$ are constants.  This in turn leads to the shifts
\be
A_{Dm} \to A_{Dm} + C_m~, \qquad K \to K + {1 \over 2\pi} \sum_{m=1}^{N-1} \Im C_m \bar A_m 
\ee
We recognize this as the freedom to perform K\"ahler transformations (constrained by~$\CN=2$ supersymmetry to be linear in the~$A_m$). However, it was already emphasized in~\cite{Seiberg:1994rs} that this freedom is not realized, because the central charge in the supersymmetry algebra (see below) for magnetically charged particles explicitly depends on~$A_{Dm}$, which is thus physically meaningful. It follows that the constants~$C_m$ must vanish, so that~$A_{Dm}$, and hence~$K$ itself, are not subject to any ambiguities associated with K\"ahler transformations. In section~\ref{sec:stmult} we will present another argument that~$K$ is physical, and hence single-valued, by relating it to the stress-tensor supermultiplet of the~$\CN=2$ gauge theory. 

Since~$C_m = 0$ it follows from~\eqref{eq:fshift} that the pre-potential~$\CF$ is at most ambiguous by constant shifts. In fact this ambiguity can also be fixed by arguing that~$\CF$ appears directly in the following formula for the single-valued Coulomb branch modulus~$u_2 = \tr (\phi^2)$, 
\be\label{rgeq}
{N \, u_2 \over 2 \pi i} =  2 \CF -  \sum_{m=1}^{N-1} a_{Dm} a_m~.
\ee
Here~$a_m, a_{Dm}$ are the bottom components of~$A_m, A_{Dm}$. This equation (originally found in~\cite{Matone:1995rx} for~$SU(2)$ and extended to other~$SU(N)$ in \cite{DHoker:1996yyu}) can be derived by promoting the strong-coupling scale~$\Lambda$ to an~$\CN=2$ chiral superfield that couples to~$u_2$ and tracking this coupling from UV to IR (see for instance~\cite{Luty:1999qc}). The coefficient~$N$ arises from the 1-loop beta function~$\beta \sim N$ that governs this coupling in the UV. For this reason~\eqref{rgeq} is sometimes referred to as a renormalization group equation. 

The Seiberg-Witten solution gives formulas for the scalar bottom components~$a_m$ and~$a_{Dm}$ (also known as the Seiberg-Witten periods) of the chiral superfields $A_m$ and $A_{Dm}$ in terms of the gauge-invariant Coulomb branch moduli $u_I$ defined in (\ref{2.moduli}). Because the pre-potential $\cF$ depends only on the fields $A_m$ and not on their derivatives, $\cF$ may be obtained by evaluating $a_{Dm}=\p \cF / \p a_m$ on the vacuum expectation values of $a_m$ and $a_{Dm}$.\footnote{~This is true up to an integration constant in~$\cF$, which can be fixed using~\eqref{rgeq}.} The basis for this construction is the Seiberg-Witten curve $\Sigma=\Sigma(u)$,  defined for pure $SU(N)$ gauge theory (without hypermultiplets) by
\bea\label{2.SWcurve}
 y^2 = C(x)^2 -1 
\hskip 1in
C(x) = 2^{N-1} \left(x^N - \sum_{I = 2}^N {u_I \over I (2 \Lambda)^I } x^{N-I}\right)
\eea
Here~$u_I  = \tr (\phi^I)$, as defined in~\eqref{2.moduli}. The curve $\Sigma$ is hyperelliptic and has genus $N-1$. A canonical basis for  its homology group $H_1(\Sigma, \ZZ) \approx \ZZ^{2N-2}$ consists of cycles $\mA_m$ and $\mB_m$ with $m=1, \ldots, N-1$. Their canonical intersection pairing $\mJ$ is given by $\mJ(\mA_m, \mA_n) = \mJ(\mB_m, \mB_n)=0$ and $\mJ(\mA_m, \mB_n)=\delta_{mn}$. Then $a_m$ and $a_{Dm}$ are given as period integrals of the Seiberg-Witten differential $\lambda_\text{SW}$,
\bea
\label{2.periods}
2 \pi i \, a_m = \oint _{\mA_m} \lambda_\text{SW} \quad\qquad 2 \pi i \, a_{Dm} = \oint _{\mB_m} \lambda_\text{SW}
\qquad\quad
\lambda_\text{SW} = (2 \Lambda) { x C'(x) dx \over y}
\eea
The pre-potential $\cF$ can then be determined by integrating~$a_{Dm} = \p \cF/ \p a_m$. This is possible, thanks to the fact that the variations $\p \lambda _\text{SW} / \p u_I$ are holomorphic Abelian differentials.

Seiberg-Witten theory also gives an exact formula for the central charge~$Z$ in the~$\CN=2$ supersymmetry algebra, and hence for the masses of BPS states. The central charge $Z$ and mass $M_\text{BPS}$ of a BPS state with $U(1)^{N-1}$ electric-magnetic  charge vector $(q^m, q_D^m) \in \ZZ^{2N-2}$ are given as follows,
\be
\label{2.BPS}
M_{\text{BPS}} = |Z|~,  \qquad 
Z =  \sqrt{2}  \sum_{m=1}^{N-1} \Big ( q^m a_m + q_D^m a_{Dm} \Big )
\ee
The modular group $Sp(2N-2,\ZZ)$ acts on the cycles $\mA_m$ and $\mB_m$, hence also on the periods~$a_m, a_{Dm}$ and the pre-potential~$\CF$,  while leaving the intersection pairing $\mJ$ invariant. Its action on the low-energy Abelian gauge theory is via electric-magnetic duality, which leaves the Dirac pairing between electric and magnetic charges invariant. However duality does act on the charges~$(q^m, q_D^m)$ themselves, which therefore depend on the duality frame under consideration. The actions on periods and charges are conjugate to each other, so that the central charge in~\eqref{2.BPS} is duality invariant.

Note that the expression~\eqref{2.Kdef} for the K\"ahler potential~$K$ is invariant under the action of~$Sp(2N-2,\ZZ)$ on the periods. As was emphasized in~\cite{Seiberg:1994rs}, the fact that duality acts on the periods homogenously, via the standard vector representation of~$Sp(2N-2,\ZZ)$, is a special feature of the pure~$SU(N)$ gauge theory. (In gauge theories with matter hypermultiplets, the periods may receive inhomogenous shifts under duality.) The invariance of~$K$ under duality transformations will play an important role below. In section~\ref{sec:stmult} we will interpret the single-valuedness of~$K$ on the entire Coulomb branch, and in every duality frame, in terms of the stress-tensor supermultiplet of the~$\CN=2$ gauge theory.

\subsection{The origin of the Coulomb branch}

The origin of the Coulomb branch is the point where all~$u_I = 0$. At this point the entire~$\ZZ_{4N}$ $r$-symmetry is unbroken. 

\subsubsection{Curve and K\"ahler potential} 

\label{ssec:KPorigin}

\begin{figure}[t!]
\centering
\includegraphics[width=\textwidth]{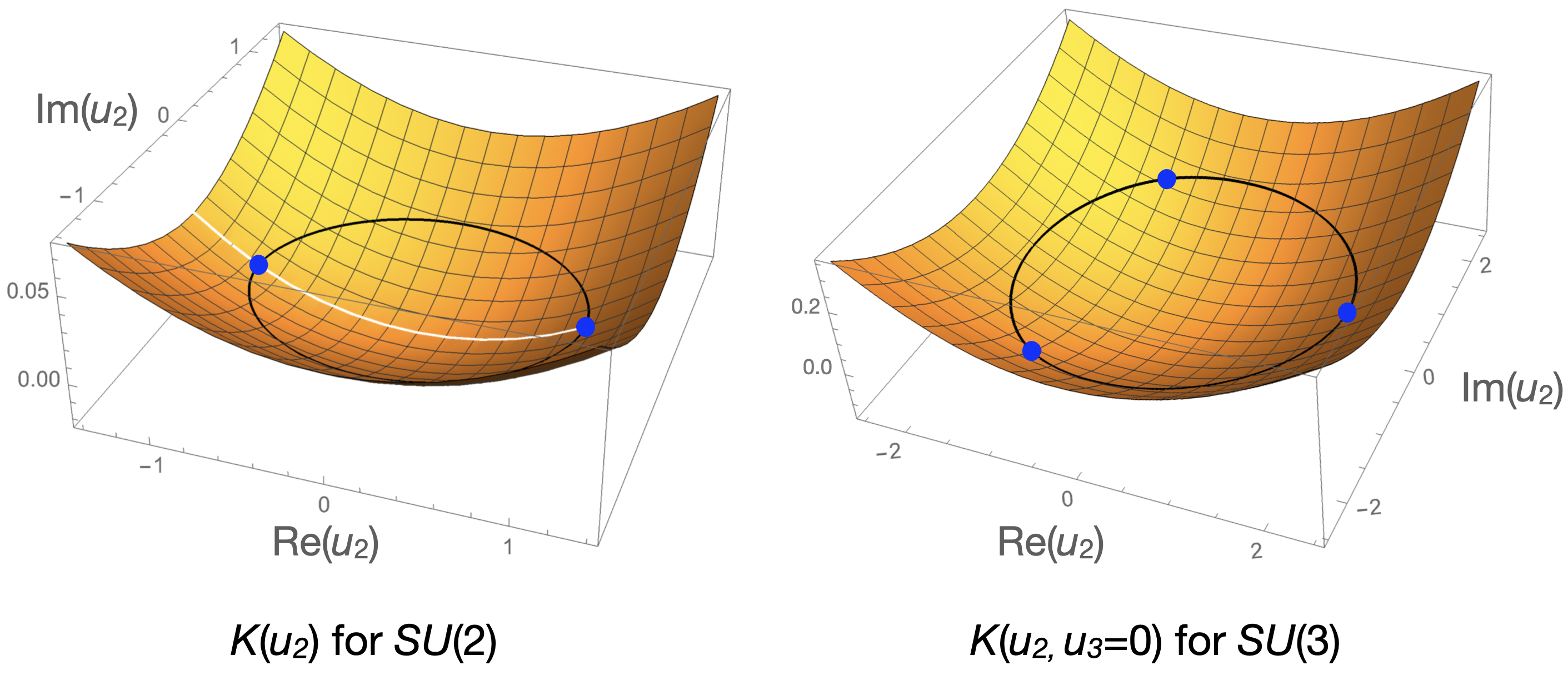}
\caption{  The K{\"a}hler potential~$K$ for  $SU(2)$ (left) and $SU(3)$ (right), plotted in the complex $u_2$ plane, has a unique minimum at the origin. The black curves indicate where~$K = 0$; they bound the strong-coupling region (defined by~$K< 0$) surrounding the origin. The multi-monopole points, indicated by blue dots, lie on these~$K = 0$ curves. 
\label{fig:kahler}}
\end{figure}

The Seiberg-Witten curve is non-singular (all BPS states are massive) and takes the following~$\ZZ_{2N}$ symmetric form,
\bea\label{2.z2ncurve}
y^2 = x^{2N} - 1
\eea
A detailed analysis of the Seiberg-Witten periods (and related quantities) in the vicinity of the origin was carried out in~\cite{DHoker:2022loi}, by systematically expanding around the~$\ZZ_{2N}$-symmetric curve (\ref{2.z2ncurve}) in the moduli~$u_I$.\footnote{~\label{fn:Lambdamatch}The analysis in~\cite{DHoker:2022loi} relied heavily on the~$\ZZ_{2N}$ symmetry of the curve at the origin. The conventions used there differ in several respects from those used here, as reflected for instance in the different choice of~$\mA$- and~$\mB$-cycles discussed in the main text. Additionally, the strong coupling scales are also normalized differently: $\Lambda_\text{here} = 2^{-{1 \over N}} \Lambda_\text{there}$.} Using the resulting formulas, the K\"ahler potential~$K$ defined in~\eqref{2.Kdef} was investigated and evidence (both analytical and numerical) was amassed for the conjecture that~$K$ is a convex function with a unique minimum at the origin of the Coulomb branch, where all~$u_I = 0$. Much of the evidence in~\cite{DHoker:2022loi}  for the convexity of~$K$ was in the strong-coupling region surrounding the origin, where~$K < 0$ (see below). As argued there, this is sufficient to ensure that the origin is the unique minimum of~$K$, and it is what we will assume here. The convexity of~$K$ is illustrated in figure~\ref{fig:kahler} (taken from~\cite{DHoker:2022loi}) for the examples of~$SU(2)$ and~$SU(3)$ gauge group.

The K\"ahler potential at the origin was computed in section 3.1 of~\cite{DHoker:2022loi},\footnote{~The formula for~$K$ in~\cite{DHoker:2022loi} is given in units where~$\Lambda_\text{there} = 2^{1 \over N} \Lambda_\text{here} = 1$. This explains the extra prefactor.}  
\bea
K(u_I = 0) = - \left(2^{1 \over N} \Lambda\right)^2 \,  {N \over 8 \pi^2} {\Gamma\left(\half + {1 \over 2N} \right)^2 \over \Gamma\left(1 + {1 \over 2N}\right)^2} \cot\left({\pi \over 2N}\right)
\eea
Note that it is negative, as required on general grounds for any critical point of~$K$ (see \cite{DHoker:2022loi}). It is interesting to expand this quantity at large~$N$, 
\bea\label{largeNK}
K(u_I = 0) \to -{N^2 \Lambda^2 \over 4 \pi^2} \qquad \text{as} \qquad N \to \infty
\eea

\subsubsection{Massive BPS states} \label{sec:BPSorigin}

We now turn to a description of the massive BPS particle states at the origin of the Coulomb branch, first investigated in~\cite{Lerche:2000uy}, where it was shown that there are~$2N(N-1)$ such states if we count both particles and their anti-particles.\footnote{~Together, a particle-antiparticle pair (plus its superpartners) comprise a full hypermultiplet, see for instance section~\ref{sec:symmahm} below for more detail.}  A more detailed description of these states was given in~\cite{Alim:2011kw}. As was shown there, the~$2N(N-1)$ states are comprised of~$N-1$ distinct, irreducible orbits of the unbroken~$\ZZ_{2N}$ symmetry acting on the Seiberg-Witten curve at the origin. If we label these orbits by~$m = 1, \ldots, N-1$ then the 1-cycle on the Seiberg-Witten curve corresponding to the~$k$-th BPS state in the~$m$-th orbit is\footnote{~Note that this notation differs slightly from the introduction, where we did not distinguish particles and antiparticles, so that~$k = 0, \ldots, N-1$ was restricted to half of its range here.}
$$
\mu_{km} \qquad k = 0, \ldots, 2N-1 \qquad m = 1, \ldots, N-1
$$
A detailed description of these cycles was given in~\cite{Alim:2011kw}.\footnote{~The conventions used in~\cite{Alim:2011kw} differ from those in~\cite{DHoker:2022loi} in a somewhat involved way, see appendix E of~\cite{DHoker:2022loi} for a detailed comparison.} For our purposes it is sufficient to know that the central charge of the BPS state~$\mu_{km}$ takes the form
\bea\label{2.ZBPS}
Z(\mu_{km}) = { \sqrt{2} \over 2 \pi i} \int_{\mu_{km}} \lambda_\text{SW} = \sqrt {2} \, \ep^{-k} f(N) s_m 
\eea
Here~$f(N)$ is a function of~$N$ (to be determined below) that does not depend on~$k$ or~$m$, while~$\ep$ and~$s_m$ are defined as follows
\bea\label{2.epsmdef}
\ep = e^{2 \pi i \over 2N} \qquad\qquad s_m = \sin {m \pi \over N}
\eea
Note that the~$\ep$-dependence of~\eqref{2.ZBPS} is dictated by the action of the unbroken~$\ZZ_{2N}$ symmetry on each orbit. Note also that~$Z(\mu_{k + N, m}) = -Z(\mu_{k m})$, because these pairs of cycles describe particles and antiparticles. So we can equivalently describe the~$2N$ BPS particles as~$N$ particle-antiparticle pairs. This description will be useful below, when we track the BPS states away from the origin. There the~$\Z_{2N}$ symmetry, and hence the degeneracy it implies among the~$N$ distinct pairs, is broken. However, particles and antiparticles necessarily remain degenerate. 

Let us also comment on the~$m$-dependence of~$Z(\mu_{km})$, which only involves the factor~$s_m$ on the right-hand side of~\eqref{2.ZBPS}.\footnote{~This was explicitly shown in~\cite{Alim:2011kw} for the case~$k = 0$.} As we will show explicitly in section~\ref{sec:symmahm} below, the charge-conjugation symmetry~$C$ defined in~\eqref{eq:cdef}, which is unbroken at the origin thanks to~\eqref{2.conu}, acts on the Seiberg-Witten periods as follows,
\be
C(a_\ell) = a_{N-\ell}~, \qquad C(a_{D\ell}) = a_{D, N-\ell}~, \qquad C(\mu_{km}) = \mu_{k, N-m}~.
\ee
This implies that the BPS states~$\mu_{km}$ and~$\mu_{k,N-m}$ must be degenerate, which indeed follows from~$s_m = s_{N-m}$.  Note that these orbits are distinct, except when~$N$ is even and~$m = N/2$. In that case~$\mu_{k, N/2}$ constitutes a single~$C$-invariant orbit. 

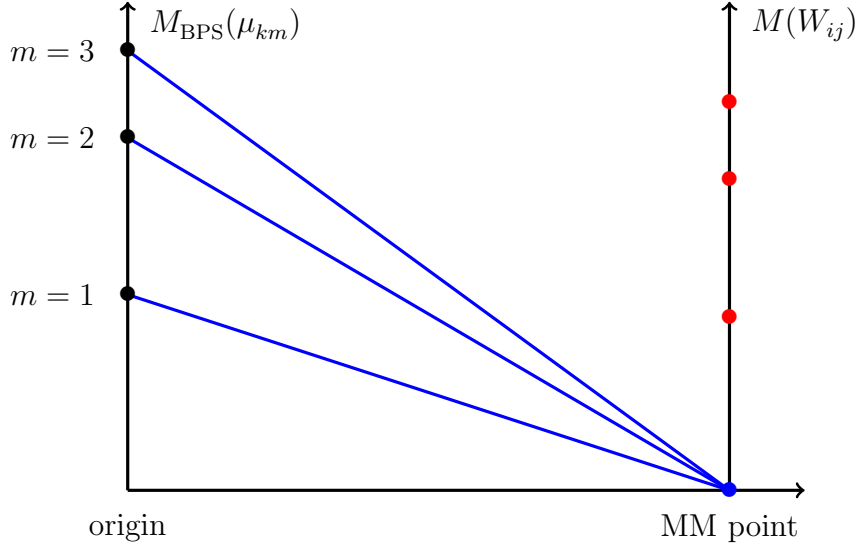
\begin{figure}[t!]
\begin{center}
\begin{tikzpicture}[scale=1]
\scope[xshift=0cm,yshift=0cm]
\draw[very thick, ->] (0,0) -- (0,6.5);
\draw[very thick, ->] (0,0) -- (9,0);
\draw[very thick, ->] (8,0) -- (8,6.5);

\draw[blue, very thick] (8,0) -- (0,2.6033);
\draw[blue, very thick] (8,0) -- (0,4.6909);
\draw[blue, very thick] (8,0) -- (0,5.8495);

\draw (1.3,6.2) node{$M_\text{BPS}(\mu_{km})$};
\draw (9,6.2) node{$M(W_{ij})$};

\draw (0,2.6033) node{ \large $\bullet$};
\draw (0,4.6909) node{\large $\bullet$};
\draw (0,5.8495) node{\large $\bullet$};
\draw[blue]  (8,0) node{ \large $\bullet$};

\draw[red] (8,2.29776) node{ \large $\bullet$};
\draw[red] (8,4.14042) node{ \large $\bullet$};
\draw[red] (8,5.16301) node{ \large $\bullet$};

\draw (-1,2.6) node{$m=1$};
\draw (-1,4.69) node{$m=2$};
\draw (-1, 5.85) node{$m=3$};

\draw (0,-0.5) node{origin};
\draw (8,-0.5) node{MM point};

\endscope
\end{tikzpicture}
\caption{Masses of BPS states at the origin of the Coulomb branch (black dots on the left vertical axis) and at the multi-monopole (MM) point (red and blue dots on the right vertical axis), plotted here for the case~$N = 7$. At the origin~$M_\text{BPS}(\mu_{km}) \sim \sqrt 2 \Lambda s_m$ (see~\eqref{2.BPSbigN}). Every level $m = 1, \ldots, 6$ contains $14$ degenerate BPS states in 7 full hypermultiplets. Since~$M_\text{BPS}(\mu_{km}) = M_\text{BPS}(\mu_{k, N-m})$ at the origin, there are only three distinct mass levels, the heaviest one being~$m = [N/2] = 3$.  Exactly one full hypermultiplet from each level~$m = 1, \ldots, 6$ becomes massless at the MM point, indicated by the blue lines (each of which denotes two full hypermultiplets); all other~$\mu_{km}$ remain massive. Crossing the MM point also involves crossing a wall of marginal stability, so that the massive BPS spectrum jumps discontinuously. To the right of the MM point, the massive BPS states include the $W$-bosons, with masses given by~\eqref{eq:wmass}. The lowest three $W$-boson masses are plotted in red on the right vertical axis.   \label{fig:BPSmassesbetter} 
}\end{center}
\end{figure}

In order to determine the function~$f(N)$ in~\eqref{2.ZBPS}, it suffices to look at the simplest cycle~$\mu_{01}$, which is related to the~$\mA$- and~$\mB$-cycles used in~\cite{DHoker:2022loi} (see in particular appendix E of that paper) as follows
\bea
\mu_{01} =  - \mA_1- \mB_1
\eea
The corresponding periods at the origin~$u_I = 0$ are computed in equation (3.4) of~\cite{DHoker:2022loi},\footnote{~Recall that formulas in~\cite{DHoker:2022loi} are given in units where~$1 = \Lambda_\text{there} = 2^{1 \over N} \Lambda_\text{here}$.}
\bea
- a_1 - a_{D1} = - 2 i \ep \left(2^{1 \over N} \Lambda\right)  Q_{N+1} s_1
\eea
Here the function~$Q_{N+1}$ at the  origin is given in equation (2.18) of \cite{DHoker:2022loi},
\bea
Q_{N+1}(u_I = 0) = {\Gamma\left(\half + {1 \over 2N}\right) \over 2 \sqrt{\pi} \, \Gamma\left(1 + {1 \over 2N}\right)}
\eea
Comparing with~\eqref{2.ZBPS} we see that~$f(N) = - 2 i (2^{1 \over N} \Lambda) \ep Q_{N+1}$. Substituting into~\eqref{2.ZBPS}, we find that the masses of the BPS states at the origin are given by the following formula
\bea\label{2.MBPS}
M_\text{BPS}(\mu_{km}) = \left|Z(\mu_{km})\right| = \sqrt {2\over \pi} \left(2^{1 \over N} \Lambda\right) {\Gamma\left(\half + {1 \over 2N}\right) \over \Gamma\left(1 + {1 \over 2N}\right)} \, s_m
\eea
Note that these masses do not depend on the label~$k$, leading to a~$2N$-fold degeneracy for each~$m = 1, \ldots, N-1$. As we did for the K\"ahler potential in~\eqref{largeNK} above, it is instructive to expand the prefactor of~$s_m$ in~\eqref{2.MBPS} in the large-$N$ limit, leading to
\be\label{2.BPSbigN}
M_\text{BPS}(\mu_{km}) \; \to \; \sqrt{2} \Lambda s_m \qquad \text{as} \qquad N \to \infty
\ee
Note that for~$N = 2$, the prefactor of~$\Lambda s_m$ in~\eqref{2.MBPS} is~$\simeq 1.53$, from which it monotonically drops to~$\sqrt{2} \simeq 1.41$ in the large-$N$ limit. The approximation~\eqref{2.BPSbigN} is thus excellent for all values of~$N$. In the large-$N$ limit, the sine function~$s_m = \sin {m \pi \over N}$  implies a BPS spectrum  ranging from equally spaced masses~$M_\text{BPS} \sim {m  \over N} \, \Lambda$ with~$m = \CO(1)$ to a dense spectrum of masses~$M_\text{BPS} \sim \Lambda$ for~$m = \CO(N/2)$. A plot of the BPS spectrum at the origin of the Coulomb branch for the case~$N = 7$ appears in the left half of figure~\ref{fig:BPSmassesbetter}.

\subsection{The multi-monopole points}\label{ssec:mmpts}

Multi-monopole points are defined as those points on the Coulomb branch where a maximal number~$N-1$ of mutually local dyons become massless. As a result of this definition, at each multi-monopole point, there exists an $Sp(2N-2,\ZZ)$ modular transformation to a duality frame where all the electric charges vanish, whence the terminology.  The multi-monopole points generalize the monopole and dyon points of the pure~$SU(2)$ theory~\cite{Seiberg:1994rs} to~$SU(N)$ and have been intensively studied starting with~\cite{Douglas:1995nw}, which we follow along with~\cite{DHoker:1997mlo,DHoker:2020qlp}.  There are precisely~$N$ such points on the Coulomb branch, which are mapped into each other by a spontaneously broken~$\ZZ_N$ quotient of the~$\ZZ_{4N}$ $r$-symmetry that acts on the moduli~$u_I$ by phase rotations. The dynamics at these~$N$ points is identical,\footnote{~This statement must be refined in the presence of background fields: due to a mixed 't Hooft anomaly between the~$\Z_{4N}$ $r$-symmetry and the~$\Z_N^{(1)}$ center symmetry, the~$N$ multi-monopole points constitute different SPT phases for the 1-form symmetry, see for instance~\cite{Gaiotto:2014kfa} for a discussion of this fact. See~\cite{Dumitrescu:2023hbe} for a recent discussion of SPT phases of gauge theories without 1-form symmetry.} and it suffices to study one of them -- referred to as \textit{the} multi-monopole point (and occasionally indicated by MM) -- which we now describe.

\subsubsection{Curve, pre-potential, and related quantities}
\label{sssec:periodsmmpt}

The Seiberg-Witten curve describing the multi-monpole point is given by
\be\label{tcheby}
y^2 = C(x)_{\text{MM}}^2 -1 
\hskip 1in 
C(x)_{\text{MM}} = \cos \left(N \arccos x\right)
\ee
Here~$C(x)_{\text{MM}}$ is the~$N$-th Chebyshev polynomial of the first kind, which has degree~$N$, real coefficients, and is an even or odd function of~$x$ according to whether~$N$ is even or odd. Comparing with~\eqref{2.SWcurve}, we conclude that at the multi-monopole point 
\be\label{eq:unmmpt}
u_{I    \text{ even}}\big|_\text{MM} \in \mathbb{R} \qquad u_{I   \text{ odd}}\big|_\text{MM}  = 0
\ee
It follows from~\eqref{2.ronu} that the~$\ZZ_{4N}$ $r$-symmetry is spontaneously broken to its~$\ZZ_4$ subgroup, which is generated by~$r^N$. (As mentioned above, the broken symmetry cyclically permuting the multi-monopole points is~$\ZZ_{4N} / \ZZ_4 = \ZZ_{N}$.) Since the only non-zero~$u_I$ have even~$I$, it follows from~\eqref{2.conu} that charge-conjugation symmetry~$C$ is unbroken at the multi-monopole point. And finally, it follows from~\eqref{2.Tonu} and the fact that the~$u_I$ are real at the multi-monopole point that time-reversal symmetry~$T$ is unbroken there. This also implies that the symmetries~$\t C = r^N C$ and~$\t T = r^N T$ defined in~\eqref{2.ctdef} and~\eqref{2.ttdef} are unbroken. 

At the multi-monopole point the~$N-1$ mutually local massless dyons are magnetic monopoles. In order to describe these light monopoles we work in a duality frame in which they are the fundamental electric charges, i.e.~we take the magnetic periods~$a_{Dm}$ to be fundamental. All of them vanish at the multi-monopole point,
\be
a_{Dm}\big|_\text{MM}  = 0~.
\ee

In the vicinity of the multi-monopole point, the Seiberg-Witten curve is a small deformation of~\eqref{tcheby}, which can be explicitly Taylor-expanded in~$a_{Dm}$~\cite{DHoker:1997mlo,DHoker:2020qlp}.\footnote{~The formulas in these papers are valid in units where the strong-coupling scale~$\Lambda = \half$.} Working to first order in small~$a_{Dm}$, this expansion takes the following form
\be\label{eq:mmcurvedef}
C(x) = C(x)_{\text{MM}} + {i \over 2 \Lambda N} C(x)'_{\text{MM}} \sum_{m = 1}^{N-1} {s_m a_{Dm} \over x - c_m} + \CO(a_D^2)~,
\ee
where we shall use the following notation here and throughout
\bea
\label{2.cksk}
c_m = \cos { m \pi  \over N}
\hskip 1in
s_m = \sin { m \pi  \over N}
\eea
The derivative $C(x)'_{\text{MM}}$, 
\bea
C(x)'_{\text{MM}} = 2^{N-1} N \prod_{m = 1}^{N-1} (x-c_m)
\eea 
has simple zeros at the~$N-1$ distinct values~$x = c_m$. The coefficients of the polynomial~$C(x)$ are in one-to-one correspondence with~$a_{Dm}$, e.g.~we can project onto~$a_{Dm}$ by computing a suitable residue integral of~$C(x)$ around~$x = c_m$. Comparing with the general form of the Seiberg-Witten curve in~\eqref{2.SWcurve}, this establishes the mapping between the periods~$a_{Dm}$ and the gauge-invariant moduli~$u_I$.  We will not spell out this mapping explicitly, except for the special case of~$u_2$ (see below), but we will use it in section~\ref{sec:symmahm} to infer the action of the global symmetries on the periods. 

The Seiberg-Witten effective Lagrangian at the multi-monopole point is encoded in a dual magnetic pre-potential~$\CF_D(a_D)$, which is related to the electric pre-potential~$\CF$ introduced around~\eqref{2.LIR} and~\eqref{ADtau} by the following Legendre transform,
\bea
\sum_{m=1}^{N-1} \left ( a_m {\p \cF (a) \over \p a_m}  + a_{Dm} { \p \cF_D (a_D) \over \p a_{Dm}} \right )=0
\eea
In particular this implies that
\be
\label{2.afd}
a_m =- {\p \CF_D \over \p a_{D m}} \qquad \qquad  \tau_{Dmn} = {\p^2 \CF_D \over \p a_{D m} \p a_{D n }} = - (\tau^{-1})_{mn}  
\ee
As in the discussion around~\eqref{rgeq}, the integration constants in~$\CF_D$ can be unambiguously fixed by using the Legendre-transformed version of the renormalization group equation,
\be\label{dualrg}
{N \, u_2 \over 2 \pi i} = 2 \CF_D + \sum_{m=1}^{N-1} a_{Dm} a_m
\ee
In particular, we will need the fact (which follows from comparing the Chebyshev polynomials in~\eqref{tcheby} with~\eqref{2.SWcurve}) that at the multi-monopole point, where all~$a_{Dm}$ vanish,
\be\label{u2atmm}
u_2(a_{D} = 0) = 2 N \Lambda^2 
\ee  
Substituting into~\eqref{dualrg}, we find that
\be\label{fdmm}
\CF_D(a_{D} = 0) = {N^2 \Lambda^2 \over 2\pi i}
\ee

A Taylor expansion for the dual pre-potential~$\CF_D$ around the multi-monopole point, based on direct evaluation of the Seiberg-Witten periods, was developed in~\cite{DHoker:1997mlo,DHoker:2020qlp}, where the terms up to and including~$\CO(a_D^3)$ were evaluated explicitly. A complementary approach based on matrix models for topological strings was pursued in~\cite{Bonelli:2017ptp}. The answers are in full agreement up to and including the~$\CO(a_D^3)$ terms that have been explicitly evaluated in both approaches (though both approaches in principle can be evaluated to higher orders). In our conventions, the dual pre-potential takes the following form,\footnote{~The constant term in~$\CF_D$ from~\eqref{fdmm} and the non-logarithmic~$\CO(a_D^2)$ terms are missing in~\cite{DHoker:1997mlo}. The latter were computed in~\cite{DHoker:2020qlp}, which also summarizes previous results in the literature.} 
\bea
\label{fdfull} 
\cF_D(a_D) &= &  {N^2 \Lambda^2 \over 2\pi i}
 - \frac{2 N \Lambda }{\pi} \sum_{m=1}^{N-1} s_m a_{Dm} 
 \no \\
 && 
 - \frac{i}{4\pi} \sum_{m,n = 1}^{N-1}  a_{Dm} a_{Dn}  \left( \delta_{m,n} \log \frac{-i a_{Dm}}{ \Lambda} - \log \Lambda_{mn}  - \frac{3}{2}\delta_{m,n} \right) 
 \no \\ &&
+  \frac{1}{32\pi N \Lambda} \sum_{m=1}^{N-1}  \left( \frac{a_{Dm}^3}{s_m^3} - 4 \sum_{n \neq m} \frac{a_{Dm}^2 a_{Dn} s_n}{(c_m -c_n)^2}\right)   + \CO(a_D^4)
\eea
Here we continue to use the abbreviations $c_m$ and $s_m$ defined in (\ref{2.cksk}) and we have introduced 
following the dimensionless, symmetric~$(N-1) \times (N-1)$ matrix~$\Lambda_{mn}$ (not to be confused with the dimensionful strong-coupling scale $\Lambda$), whose components are given by, 
\be\label{Lkldef}
\Lambda _{mm} = 16 N s_m^3
\hskip 1in 
\Lambda _{m \neq n} =  { 1-c_{m+n} \over 1-c_{m-n}} 
\ee
We will need explicit formulas for the $a_m$ as functions of $a_{Dn}$, which are given as follows,
\bea\label{aMMfirst}
a_m(a_D) &=& 
{2 N \Lambda \over \pi} s_m 
+{i \over 2 \pi} a_{D m} \left(\log {- i a_{Dm} \over \Lambda} - 1 \right) 
- {i \over 2\pi}  \sum_{n=1}^{N-1} a_{Dn} \log \Lambda_{mn} 
\nonumber \\
& & 
+ {1 \over 32 \pi N \Lambda} \left( - {3 a_{D m}^2 \over s_m^3} + 4 \sum_{n \neq m} {a_{Dn}^2 s_m + 2 a_{Dn} a_{Dm} s_n \over (c_m - c_n)^2} \right) + \CO(a_D^3)
\eea
and for $\tau_{Dmn}$ as a function of $a_{Dm}$ given by,
\bea
\tau_{D mn} (a_D) & = & 
-{i \over 2 \pi} \delta_{m,n} \log {- i a_{Dm} \over \Lambda} + {i \over 2 \pi} \log \Lambda_{mn} 
\no \\ &&
-{ \delta_{m,n}  \over 4 \pi N \Lambda}  \left [ { - 3 a_{Dm} \over 4 s_m^3} + \sum_{p\not= m} { s_p a_{Dp} \over (c_p-c_m)^2} \right ]
\no \\ &&
- { 1- \delta_{m,n} \over 4 \pi N \Lambda} \, { s_m a_{Dn} + s_n a_{Dm} \over (c_m-c_n)^2}
+ \cO(a_D^2)
\eea
These are obtained by substituting~\eqref{fdfull} into~\eqref{2.afd}. Note that~$\Im{\tau_{D mn}}$ is diagonal and positive definite sufficiently close to the monopole point, where all~$a_{Dm}$ vanish. The subleading terms~$\sim \log \Lambda_{mn}$, which can be thought of as threshold contributions due to massive states that have been integrated out, possess a rich non-diagonal structure that will play a crucial role throughout our analysis below. 

Finally, for future use, we substitute~\eqref{fdfull} and~\eqref{aMMfirst} into~\eqref{dualrg} to obtain
\be\label{eq:u2ofad}
u_2(a_D) = 2 N \Lambda^2 + \sum _{m=1}^{N-1} \Big ( - 4 i \Lambda s_m a_{Dm} - {1 \over 2N} a_{Dm}^2 \Big )  + \CO(a_D^3)~.
\ee

\subsubsection{Massive BPS states} 
\label{ssec:mmbps}

So far we have just discussed the massless BPS monopoles at the multi-monopole point. By contrast, the massive BPS spectrum is not strictly well-defined there, because the multi-monopole points lie on a wall of marginal stability, across which the massive BPS spectrum jumps discontinuously~\cite{Seiberg:1994rs,Douglas:1995nw}. If one approaches this wall from within the strong-coupling region surrounding the origin, the stable BPS states are those that are also present at the origin (see section~\ref{sec:BPSorigin} above). By contrast, in suitable weak-coupling regions of the Coulomb branch, the $W$-bosons one expects semi-classically are stable BPS particles,\footnote{~Semi-classically there is also an infinite number of stable dyons.} and they can remain stable up to the wall of marginal stability containing the multi-monopole points. An explicit example of a ray on the Coulomb branch that extends from the multi-monopole point to infinity and along which all $W$-bosons are stable was described in~\cite{Douglas:1995nw}. As was done there, we present the 1-cycles~$W_{ij}$ on the Seiberg-Witten curve that describe the $W$-bosons by formally extending our periods~$a_{m}$ with~$m = 1, \ldots, N-1$ to include~$a_0 = a_N = 0$. Then
\be \label{eq:Wcyc}
W_{ij} = - W_{ji} = a_i - a_{i-1} - a_j + a_{j-1}~, \qquad i, j = 1, \ldots, N~, \qquad a_0 = a_N = 0
\ee
This means that~$W_{ji}$ is the anti-particle of~$W_{ij}$. Counting both particles and anti-particles we therefore have~$N^2 - N$ non-vanishing~$W_{ij}$, which is exactly the number of $W$-bosons when~$SU(N)$ is Higgsed to~$U(1)^{N-1}$. Substituting into the BPS mass formula~\eqref{2.BPS}, and using the fact that~$a_m = {2 N \Lambda \over \pi} s_m$ at the multi-monopole point (see~\eqref{aMMfirst}), we find 
\be\label{eq:wmass}
M_\text{BPS}(W_{ij}) = {2 \sqrt 2 N \Lambda \over \pi} \left|s_i - s_{i-1} - s_j + s_{j-1}\right|~, \qquad i, j = 1, \ldots, N~.
\ee
Note that this formula is uniformly valid for all~$i, j$ because~$s_0 = s_N = 0$.  The heaviest $W$-bosons have mass~$\sim \Lambda$, while the lightest one is
\be\label{eq:wlight}
M_\text{BPS}(W_{12}) \to  {2 \sqrt 2 \pi^2 \Lambda \over N^2} \qquad \text{as} \qquad N \to \infty
\ee
As emphasized in~\cite{Douglas:1995nw}, this leads to a parametrically low UV cutoff for the Seiberg-Witten IR effective theory in the large-$N$ limit. We will subsequently explain how this fact impacts our analysis. 

\begin{figure}[t!]
\begin{center}
\begin{tikzpicture}[scale=1]
\scope[xshift=0cm,yshift=0cm]
\draw[very thick, ->] (0,0) -- (0,6.5);
\draw (0.9,6.8) node{$M_\text{BPS}(\mu_{km})$};
\draw[scale=1, domain=0:90, variable=\x, line width=1.5] plot ({4*cos(\x)}, {sin(\x)});
\draw[scale=1, domain=90:180, variable=\x, line width=1.5] plot ({4*cos(\x)}, {sin(\x)});
\draw[scale=1, domain=180:270, variable=\x, line width=1.5] plot ({4*cos(\x)}, {sin(\x)});
\draw[scale=1, domain=270:360, variable=\x, line width=1.5] plot ({4*cos(\x)}, {sin(\x)});

\draw[red, thick] (2.4939,0.78183) -- (0,2.6033);
\draw[red, thick] (2.4939,0.78183) -- (0,4.6909);
\draw[red, thick] (2.4939,0.78183) -- (0,5.8495);

\draw[blue, thick] (4,0) -- (0,2.6033);
\draw[blue, thick] (4,0) -- (0,4.6909);
\draw[blue, thick] (4,0) -- (0,5.8495);

\draw[yellow, thick] (2.4939,-0.78183) -- (0,2.6033);
\draw[yellow, thick] (2.4939,-0.78183) -- (0,4.6909);
\draw[yellow, thick] (2.4939,-0.78183) -- (0,5.8495);

\draw[black, thick] (-0.89, 0.97492) -- (0,2.6033);
\draw[black, thick] (-0.89, 0.97492) -- (0,4.6909);
\draw[black, thick] (-0.89, 0.97492) -- (0,5.8495);

\draw[cyan, thick] (-3.6038,0.43388) -- (0,2.6033);
\draw[cyan, thick] (-3.6038,0.43388) -- (0,4.6909);
\draw[cyan, thick] (-3.6038,0.43388) -- (0,5.8495);

\draw[purple, thick] (-3.6038,-0.43388) -- (0,2.6033);
\draw[purple, thick] (-3.6038,-0.43388) -- (0,4.6909);
\draw[purple, thick] (-3.6038,-0.43388) -- (0,5.8495);

\draw[green, thick] (-0.89, -0.97492) -- (0,2.6033);
\draw[green, thick] (-0.89, -0.97492) -- (0,4.6909);
\draw[green, thick] (-0.89, -0.97492) -- (0,5.8495);

\draw[blue] (4,0) node{\large $\bullet$};
\draw[blue] (2.4939,0.78183) node{\large $\bullet$};
\draw[blue] (-0.89, 0.97492) node{\large $\bullet$};
\draw[blue] (-3.6038,0.43388) node{\large $\bullet$};
\draw[blue] (2.4939,-0.78183) node{\large $\bullet$};
\draw[blue] (-0.89, -0.97492) node{\large $\bullet$};
\draw[blue] (-3.6038,-0.43388) node{\large $\bullet$};

\draw[red] (0,0) node{\Large $\bullet$};
\draw (0,2.6033) node{ \large $\bullet$};
\draw (0,4.6909) node{\large $\bullet$};
\draw (0,5.8495) node{\large $\bullet$};

\endscope
\end{tikzpicture}
\caption{The $2N(N-1)$ massive BPS states (counting both particles and anti-particles) at the origin of the Coulomb branch (indicated by the red dot), where $u_I=0$ for all $I=2,\ldots, N$. The figure corresponds to the gauge group $SU(7)$. The 3 distinct BPS masses at the origin (given by~\eqref{eq:bpsMintro}) are indicated by the black dots on the vertical axis, each of which describes~$28$ degenerate hypermultiplets. There are 7 multi-monopole points (indicated by the blue dots); at each such point 4 BPS states (in 2 full hypermultiplets) from every black dot become massless, as indicated by the thin colored lines.
  \label{fig:BPS} 
}\end{center}
\end{figure}
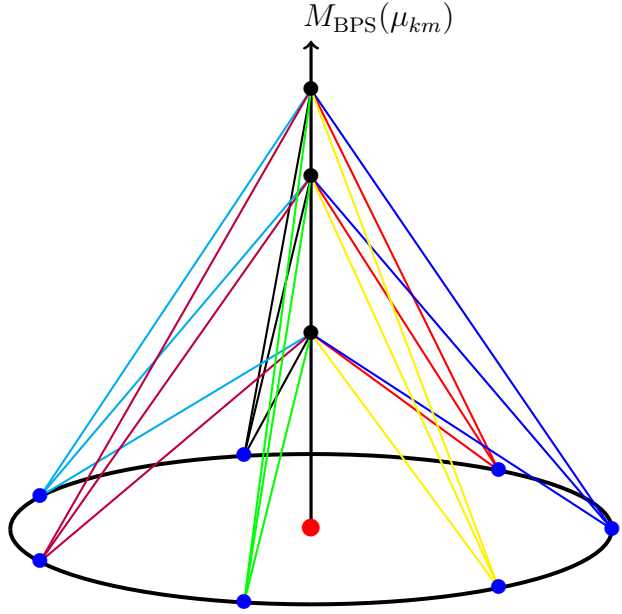

Upon crossing the wall of marginal stability intersecting the multi-monopole points, the $W$-bosons become unstable and decay into those~$2N(N-1)$ monopoles and dyons that are stable in the strong-coupling region surrounding the origin (see section~\ref{sec:BPSorigin}). We will not need a detailed description of these decays here. Instead we will describe qualitatively how the massive BPS states at the origin evolve as we move away from the origin, and toward one of the multi-monopole points. This is depicted in figures~\ref{fig:BPSmassesbetter} and~\ref{fig:BPS}.

The states at the origin are labeled by~$\mu_{km}$, with each~$m = 1, \ldots, N-1$ labeling a distinct orbit of the unbroken~$\Z_{2N}$ symmetry and~$k = 0, \ldots, 2N-1$ labeling the degenerate states in every such orbit. Recall from~\eqref{2.BPSbigN} that the masses scale as~$M(\mu_{km}) \sim \Lambda s_m$ in the large-$N$ limit. The BPS mass spectrum at the origin is indicated by black dots on the left vertical axis in figure~\ref{fig:BPSmassesbetter}, for~$N = 7$.  

Let us describe the fate of these states as we move away from the origin and toward one of the~$N$ multi-monopole points. (See figure~\ref{fig:BPS} where this is depicted for~$N = 7$.) As we approach the~$k$-th multi-monopole point, with~$k = 0, \ldots, N-1$, exactly one particle~$\mu_{km}$ and its anti-particle~$-\mu_{km} = \mu_{k + N, m}$ from each of the~$N-1$ distinct~$\Z_{2N}$ orbits labeled by~$m = 1, \ldots, N-1$ come down and become massless. This is indicated by the solid blue lines in figure~\ref{fig:BPSmassesbetter}. All other BPS states remain massive. Their masses are related to those of the massive $W$-bosons, which are stable outside the strong-coupling region, by wall crossing. The $W$-boson masses are indicated by red dots on the right vertical axis in figure~\ref{fig:BPSmassesbetter}, for~$N = 7$. 

The upshot is that all~$2N(N-1)$ BPS states at the origin of the Coulomb branch become massless at one of the~$N$ multi-monopole points, in the~$\Z_N$-symmetric fashion described above: the~$2(N-1)$ states~$\pm \mu_{km}~(m = 1, \ldots, N-1)$, described by~$N-1$ full hypermultiplets, become massless at the~$k$-th multi-monopole point ($k = 0, \ldots, N-1$). See figure~\ref{fig:BPS} for a three-dimensional representation.

\subsection{Effective Abelian Higgs model at the multi-monopole points}\label{sec:AHM}

At each of the~$N$ multi-monopole points, described in section~\ref{ssec:mmpts} above, $N-1$ mutually local charged BPS states become light, in addition to the massless~$U(1)^{N-1}$ vector multiplets. This in turn leads to singularities in the Seiberg-Witten periods (and related quantities) at those points, since the periods are computed in a description in which the charged BPS states have been integrated out. As in~\cite{Seiberg:1994rs, Douglas:1995nw}, this is remedied by explicitly including the additional massless states, leading to a well-defined Wilsonian effective description that we now review.  

\subsubsection{Wilsonian effective Lagrangian} 
\label{sec:MMLag}

As in section~\ref{sssec:periodsmmpt}, we use the broken~$\Z_N$ symmetry relating the~$N$ multi-monopole points to focus on the specific multi-monopole point whose Seiberg-Witten curve is given by~\eqref{tcheby}. At that point all~$a_{Dm} = 0$ and there are~$N-1$ massless, mutually local magnetic BPS monopoles, with electric and magnetic charges
\be\label{qmmm}
q^m = 0~, \qquad q_{D}^{m}  = \delta_{mn}~, \qquad m, n = 1, \ldots, N-1
\ee
In other words there is precisely one monopole of unit magnetic charge for every~$a_{Dm}$. As in section~\ref{sssec:periodsmmpt}, we work in a duality frame in which the~$a_{Dm}$ are fundamental, so that the theory is a magnetic dual gauge theory with gauge group
\be\label{eq:dualgg}
U(1)_D^{N-1} = \prod_{m =1}^{N-1} U(1)_{Dm}
\ee
In this description, the monopoles play the role of fundamental (i.e.~dual electric) charges. 
 
The massless BPS monopoles reside in hypermultiplets of~$\CN=2$ supersymmetry, whose coupling to the~$\CN=2$ magnetic vector multiplets with gauge group~\eqref{eq:dualgg} we explicitly spell out below. Once the monopole hypermultiplets have been included, the Wilsonian effective action -- furnished with a suitable UV cutoff~$\mu$ -- is local and non-singular. The effective dual pre-potential describing this action is then given by (see~\cite{Luty:1999qc} for a closely related discussion)\footnote{~Here we choose a scheme for~$\mu$ that eliminates the factor~$3 \over 2$ on the second line of~\eqref{fdfull}. In order to connect this scheme with standard supersymmetric perturbative schemes, such as~$\bar {\text{DR}}'$, one would have to rescale~$\mu$ by an~$\CO(1)$ constant that does not depend on~$N$. Since we will only analyze our effective field theory semi-classically (i.e.~at tree level) this will not cause problems. } 
\bea\label{eq:fdeff}
\cF_D^{\text{eff}} (a_D) & = & \cF_D(a_D) + { i \over 4 \pi} \sum _{m=1}^{N-1} a_{Dm}^2  \left ( \log { - i a_{Dm} \over \mu} - { 3 \over 2} \right )~,
\eea
with~$\CF_D(a_D)$ given by~\eqref{fdfull}. The~$\mu$-dependence of~$\CF_D^\text{eff}$ arises from the logarithmic, 1-loop exact running of the IR free magnetic gauge couplings that is due to the massless charge-1 monopole in every~$U(1)_{Dm}$ gauge group factor (see also~\eqref{eq:taueff} below).  

For future use, we explicitly spell out~\eqref{eq:fdeff}, 
\bea
\label{fdefffull}
\cF^{\text{eff}}_D(a_D) &= &  {N^2 \Lambda^2 \over 2\pi i}
 - \frac{2 N \Lambda }{\pi} \sum_{m=1}^{N-1} s_m a_{Dm} 
 + i \pi  \sum_{m,n =1}^{N-1}  t_{mn}(\mu) \, a_{Dm} a_{Dn}    
 \no \\ &&
+  \frac{1}{32\pi N \Lambda} \sum_{m = 1}^{N-1} \left(  \frac{a_{Dm}^3}{s_m^3} - 4 \sum_{n \neq m} \frac{a_{Dm}^2 a_{Dn} s_n}{(c_n-c_m)^2}\right)   + \CO(a_D^4)
\eea
Here we have defined the following~$\mu$-dependent~$(N-1) \times (N-1)$ symmetric matrix, which will play a starring role throughout our analysis below, 
\be
\label{eq:tdef}
t_{mn}(\mu) =     \frac{1}{(2\pi)^2} \left( \delta_{mn}   \log \frac{\Lambda}{\mu}   + \log \Lambda_{mn} \right)~.
\ee
Comparing with~\eqref{eq:taueff} below, we see that~$\tau_{Dmn}^\text{eff} = 2 \pi i t_{mn} + \CO(a_D)$, so that~$t_{mn}(\mu)$ is the matrix of~$U(1)_D^{N-1}$ coupling constants -- including, crucially, kinetic mixing between the different~$U(1)_D$ factors -- in the effective theory with cutoff~$\mu$. 
We also record the corresponding $a^{\text{eff}}_m$ and $\tau^{\text{eff}} _{D mn}$, obtained by using~$\CF_D^\text{eff}$ in~\eqref{2.afd},
\bea
\label{aMM}
a_m^{\text{eff}}(a_D) &=& {2 N \Lambda \over \pi} s_m  - 2 \pi i  \sum_{n=1}^{N-1} t_{mn}(\mu) a_{Dn} 
 \nonumber \\
& & + {1 \over 32 \pi N \Lambda} \left( - {3 a_{D m}^2 \over s_m^3} + 4 \sum_{n \neq m} {a_{Dn}^2 s_m + 2 a_{Dn} a_{Dm} s_n \over (c_n - c_m)^2} \right) + \CO(a_D^3)
\eea
and
\bea
\label{eq:taueff}
\tau_{Dmn}^{\text{eff}}(a_D)    & = &    
2 \pi i t_{mn}(\mu) -{ \delta_{m,n}  \over 4 \pi N \Lambda}  \left [ { - 3 a_{Dm} \over 4 s_m^3} + \sum_{p\not= m} { s_p a_{Dp} \over (c_p-c_m)^2} \right ]
\no \\ &&  \qquad
- { 1- \delta_{m,n} \over 4 \pi N \Lambda} \, { s_m a_{Dn} + s_n a_{Dm} \over (c_m-c_n)^2}
+ \cO(a_D^2)
\eea
Finally, we can substitute~$a_m^\text{eff}$ in~\eqref{aMM} above into~\eqref{2.Kdef} to obtain the effective~K\"ahler potential,
\bea
\label{eq:keffah}
 K^{\text{eff}}(a_D) & = & { N \Lambda \over \pi^2} \sum_{m=1}^{N-1}s_m \Im {a}_{Dm} 
	+\sum_{m, n=1}^{N-1} t_{mn} \, a_{Dm} \bar{a}_{Dn} 
 \\ &&  
 + {1 \over 64 \pi^2 N \Lambda} \sum_{m=1}^{N-1} \left({3 |a_{Dm}|^2 \Im a_{Dm} \over s_m^3} \right .
	\no \\ && \hskip 1in \left . 
	+ 4 \sum_{n \neq m} {\Im(a_{Dm} \bar a_{Dn}^2 ) s_m - 2 |a_{Dm}|^2 \Im a_{Dn} s_n \over (c_n-c_m)^2}\right) + \CO(a_D^4)~.
	\no
\eea

As expected, all dependence on~$a_{Dm}$ in the effective Wilsonian quantities above is analytic at the multi-monopole point~$a_{Dm} = 0$. However, positivity of the effective K\"ahler metric~$\Im \tau_{Dmn}^\text{eff} \sim t_{mn}(\mu)$ only holds if the cutoff~$\mu$ of the effective theory is sufficiently small. See section~\ref{sec:propt} below and appendix~\ref{app:appt}  for a detailed discussion of the restrictions on~$\mu$ that we will impose.  

Using formulas reviewed in appendix~\ref{app:ahmlag}, we can now write the effective~$\CN=2$  Lagrangian at the multi-monopole point in~$\CN=1$ superspace,
\be\label{eq:leffmm}
\begin{split}
\SL^\text{eff} = & \int d^4 \theta \, \left( K^\text{eff}(A_D) + \sum_\pm \sum_{m =1}^{N-1}  
\bar \CM_{m}^\suppm e^{\mp 2 \CV_{Dm}} \CM_{m}^\suppm \right)  \\
& +  \sqrt 2 \sum_{m =1}^{N-1} \int d^2 \theta \,  A_{Dm} \CM_{m}^\supp \CM_{m}^\supm + \left(\text{h.c.}\right) \\
& + {1 \over 4 \pi} \sum_{m,n=1}^{N-1} \Im \int d^2 \theta \, \tau_{Dmn}^\text{eff}(A_D) W^\alpha{}_m W_{\alpha n} 
\end{split}
\ee
Here~$A_{Dm}, \CV_{Dm}$ are~$\CN=1$ chiral and vector superfields that make up the~$U(1)_{Dm}$~$\CN=2$ vector multiplet. Similarly, $\CM^\suppm_m$ are~$\CN=1$ chiral multiplets which make up the~$\CN=2$ hypermultiplets describing the massless magnetic monopoles. The Lagrangian~\eqref{eq:leffmm} correctly captures all 2-derivative terms in the low-energy effective theory.

For most of this paper, it will be sufficient to concentrate on the (classically) marginal and relevant terms in the effective Lagrangian~\eqref{eq:leffmm}. For the purpose of constructing this renormalizable Lagrangian, we need only retain the quadratic~$\CO(a_D^2)$ terms in the effective K\"ahler potential~$K^\text{eff}$ on the first line of~\eqref{eq:keffah}. 
(Note however that the~$\CO(a_D)$ linear terms and, to a lesser extent,  the cubic~$\CO(a_D^3)$ terms will play an important role once we break supersymmetry; see section~\ref{sec:general} below.) We also only retain the constant term~$\tau_{Dmn}^\text{eff} \to 2 \pi i t_{m n}$. The resulting Lagrangian is given by equation~\eqref{eq:ahmssapp} in appendix~\ref{app:ahmlag}, which we repeat here, 
\be
\begin{split}
\SL = & \int d^4 \theta \, \left(\sum_{m, n = 1}^{N-1} t_{mn} \bar A_{Dm} A_{Dn} 
+ \sum_\pm \sum_{m =1}^{N-1}  \bar \CM_{m}^\suppm e^{\mp 2 \CV_{Dm}} \CM_{m}^\suppm \right)  \\
& + \int d^2 \theta \, \left({1 \over 4} \sum_{m, n = 1}^{N-1} t_{mn} W^\alpha{}_m W_{\alpha n} 
+ \sqrt 2 \sum_{m =1}^{N-1} A_{Dm} \CM_{m}^\supp \CM_{m}^\supm \right) + \left(\text{h.c.}\right) 
\end{split}
\ee
where the Hermitian conjugation applies to the second line only.

In appendix~\ref{app:ahmlag} we also review how to expand this Lagrangian in component fields:
\begin{itemize}
\item The~$U(1)_{Dm}$~$\CN=2$ vector multiplet is described by the $\CN=1$ superfields~$A_{Dm}$ and~$W_{\alpha m} = -{1 \over 4} \bar D^2 D_\alpha \CV_{Dm}$. After integrating out the auxiliary fields we are left with the following component fields,\footnote{~In order to avoid heavy notation, we do not explicitly add a subscript~$D$ to indicate that~$\rho_{\alpha, m}^i$ and~$f_{\mu\nu, m}$ are the superpartners of~$a_D$. Since we will never explicitly work in an electric duality frame, where~$a_m$ and its superpartners are the appropriate degrees of freedom, this will not cause any problems.}
\be
a_{Dm}~, \qquad \rho^i_{\alpha m}~, \qquad f_{\mu\nu  m} = \p_\mu b_{\nu m} - \p_\nu b_{\mu m}~.
\ee
Here~$a_{Dm}$ is the complex scalar that has already appeared copiously above, $\rho_{\alpha m}^i$ is the~$\CN=2$ gaugino (with~$SU(2)_R$ doublet index~$i = 1,2$), and~$b_{\mu m}$ is the~$U(1)_{Dm}$ gauge field, with field strength~$f_{\mu\nu m}$. 

\item The~$\CN=2$ monopole hypermultiplet that is charged under~$U(1)_{Dm}$ is described by the~$\CN=1$ superfields~$\CM^\suppm_m$ whose components (after integrating out the auxiliary fields) are as follows,
\be
h_{i m}~, \qquad \psi_{\alpha m}^\suppm ~.
\ee
Here~$h_{i m}$ has unit~$U(1)_{Dm}$ charge and is a doublet under the~$SU(2)_R$ symmetry. We denote its complex conjugate by~$\bar h_m^i = (h_{i m})^\dagger$. The fermions~$\psi_{\alpha m}^\suppm$ have~$U(1)_{Dm}$ charges~$\pm1$. 
\end{itemize}
The full component form of~$\SL$ is given in appendix~\ref{app:ahmlag}, starting with~\eqref{eq:applsum}. The purely bosonic terms take the form
\be\label{eq:Lbos}
\SL_\text{bosonic} = - \sum_{m, n  = 1}^{N-1} t_{m n} \left(
\p^\mu \bar a_{Dm} \p_\mu a_{Dn} 
+{1 \over 4} f^{\mu\nu}_m f_{\mu\nu n}   \right)   
- \sum_{m = 1}^{N-1} \left( D^\mu \bar h^i_m D_\mu h_{i m}  \right) - \SV_\text{SUSY}
\ee
where the supersymmetric scalar potential $\SV_\text{SUSY}$ is given by~\eqref{eq:appscalpot},  
\be 
\begin{split}\label{eq:vsimple}
\SV_\text{SUSY} = &  
\sum_{m = 1}^{N-1} 2 \left|a_{Dm}\right|^2 \bar h^i_m h_{i m} 
\\
& + \sum_{m, n = 1}^{N-1} \left(t^{-1}\right)_{mn} \left( 
\left(\bar h^i_m h_{i n}\right) \left(\bar h^j_n h_{j m}\right) 
- \half \left(\bar h^i_m h_{i m}\right) \left(\bar h^j_n h_{j n}\right)
\right)~.
\end{split}
\ee
This potential, together with a SUSY-breaking contribution described in section~\ref{sec:general} below, will play a starring role in our analysis.  Note that the second line of~\eqref{eq:vsimple} is a $D$-term potential, which results from integrating out the auxiliary fields~$D_m^{ij}$ in the off-shell~$\CN=2$ vector multiplets (see appendix~\ref{app:ahmlag}). 

Although they will feature less heavily in our analysis, we also list the (renormalizable) fermionic terms in the effective Lagrangian, 
\be
\label{eq:Lferm}
\SL_\text{fermionic} = - i \sum_{m,n=1}^{N-1} t_{mn} \bar \rho_{i m} \bar \sigma^\mu \p_\mu \rho_n^i 
- i \sum_{m=1}^{N-1} \sum_\pm \bar \psi_{m}^\suppm \bar \sigma^\mu D_\mu \psi_{m}^\suppm
+ \SL_\text{Yukawa}~, 
\ee
where the Yukawa couplings are given by 
\be\label{eq:lyukbis} 
\begin{split}
\SL_\text{Yukawa} =&  
\sqrt 2 \sum_{m = 1}^{N-1} \left( 
\bar h_{i m} \rho_m^i \psi_{m}^\supp - h_{i m} \rho^i_m \psi_{m}^\supm 
- h^i_m \bar \rho_{i m} \bar \psi_{m}^\supp 
- \bar h^i_m \bar \rho_{i m} \bar \psi_{m}^\supm \right) 
\\
& - \sqrt 2 \sum_{m = 1}^{N-1} \left(
a_{Dm} \psi_{m}^\supp \psi_{m}^\supm + \bar a_{Dm} \bar \psi_{m}^\supp \bar \psi_{m}^\supm \right)~.
\end{split}
\ee

\subsubsection{The matrix~$t_{mn}(\mu)$ of effective gauge couplings}
\label{sec:propt}

The matrix~$t_{mn}(\mu)$ was defined in~\eqref{eq:tdef} (see also \eqref{2.cksk}, \eqref{Lkldef}), and we repeat it here, 
\be
\label{eq:tdef1}
t_{mn}(\mu) =   \frac{1}{(2\pi)^2} \left( \delta_{mn}   \log \frac{\Lambda}{\mu}   + \log \Lambda_{mn} \right)
\ee
where $\Lambda_{mn} $ are given as follows,
\bea\label{eq:tdef1ii}
\Lambda _{mm} = 16 N s_m^3~, \hskip 1in \Lambda _{m \neq n} =  { 1-c_{m+n} \over 1-c_{m-n}} 
\eea
This matrix appears prominently in the~$\CN=2$ supersymmetric Lagrangian at the multi-monopole point, e.g.~it is the matrix of kinetic terms for~$a_{Dm}$ and~$f_{\mu\nu, n}$ in~\eqref{eq:Lbos}, and its inverse appears in the supersymmetric~$D$-term potential on the second line of~\eqref{eq:vsimple}. 

Here we summarize several properties of~$t_{mn}(\mu)$ and discuss their implications for the allowed range of UV cutoff~$\mu$ in our effective theory:
\begin{itemize}
\item The matrix~$t_{mn}(\mu)$ is bisymmetric, i.e.~it is symmetric about both of its diagonals, 
\be\label{eq:tccprop}
t_{mn}(\mu) = t_{nm}(\mu)~, \qquad  t_{N-m, N-n}(\mu) = t_{mn}(\mu)~.
\ee 
The second equation is required by charge-conjugation symmetry~$C$, as discussed in section~\ref{sec:symmahm} below. 

\item Let us decompose~$t_{mn}(\mu)$ into its diagonal~$\Delta_{mn}(\mu)$ and its~$\mu$-independent off-diagonal part~$\Omega_{mn}$,
\be\label{eq:tdo}
t_{mn}(\mu) = \Delta_{mn}(\mu) + \Omega_{mn}~, \qquad \Delta_{m\neq n} = 0~, \qquad \Omega_{mm} = 0~.
\ee
Then the off-diagonal entries~$\Omega_{mn}$ are positive,
\be\label{eq:omegaodp}
\Omega_{m \neq n} > 0~,
\ee
in light of the following trigonometric inequality,
\be
c_{m-n} - c_{m+n} = 2 s_m s_n >0~, \qquad m, n = 1, \ldots, N-1~.
\ee

\item  Since~$t_{mn}(\mu)$ appears as the matrix of kinetic terms for the vector-multiplet fields, it must be positive definite, i.e.~all of its eigenvalues~$\lambda_m(\mu) \sim g_{Dm}(\mu)^{-2}$ must be positive. Here~$g_{Dm}(\mu)$ are the magnetic gauge couplings at the scale~$\mu$, in the basis where the~$U(1)_{D}^{N-1}$ gauge-field kinetic terms have been diagonalized. Note that this is \textit{not} the basis which we will use in our analysis. 

The requirement that~$\lambda_m(\mu) >0$ for all~$m = 1, \ldots, N-1$ restricts the UV cutoff~$\mu$ of our effective theory.\footnote{~By contrast, the exact~$\tau_{Dmn}(a_D)$ appearing in the Seiberg-Witten solution automatically has positive-definite imaginary part, thanks to its relation to the period matrix of the Seiberg-Witten curve.} Numerical investigations carried out in appendix~\ref{app:appt} show that for~$N \leq 10$, positivity of~$t_{mn}(\mu)$ holds for all~$\mu \leq \Lambda$, while for larger values of~$N$ we find that
\be\label{eq:tposmu}
t_{mn}(\mu) \;\; \text{ positive definite if } \;\; \mu <  \mu_\text{pos} \simeq {107 \Lambda \over N^2}~.
\ee
For comparison, the mass~\eqref{eq:wlight} of the lightest~$W$-boson at the multi-monopole point, which furnishes a natural cutoff, is~$M_{W, \text{min}} \simeq {28 \Lambda \over N^2} \simeq {1 \over 4} \mu_\text{pos}$. 

Note that near the scale~$\mu \sim \mu_\text{pos}$ the smallest eigenvalue of~$t_{mn}(\mu_\text{pos})$ almost vanishes, and hence the corresponding magnetic gauge coupling is very strong. By contrast, the largest eigenvalue of~$t_{mn}(\mu_\text{pos})$ is numerically found to scale like
\be
\lambda_\text{max}(\mu_\text{pos}) \simeq {2N \over (2\pi)^2} + \CO(\log N)~.
\ee
This answer can also be established analytically at large~$N$, see appendix~\ref{app:appt}. Thus the weakest magnetic gauge coupling scales as~$g_{D, \text{min}}^2 \sim {1 \over N}$ at large~$N$.

\item  As $\mu \to 0$, the diagonal elements~$\Delta_{mn}(\mu) \sim \delta_{mn} \log{\Lambda \over \mu}$ of~$t_{mn}(\mu)$ become uniformly large and positive, and they grow to dominate the off-diagonal elements~$\Omega_{mn}$. In this regime, the effective $U(1)^{N-1}_D$ gauge couplings~$g_{D, m}^2 \sim 1/ \log {\Lambda \over \mu}$ become weak, and the mixing between different~$U(1)_{Dm}$ gauge group factors gradually decouples.  

The inverse matrix $t^{-1}(\mu)$ appearing in the effective potential~\eqref{eq:vsimple} is then also dominated by its diagonal entries and, for sufficiently small $\mu$, its off-diagonal elements are all negative. To show this, we use the decomposition~$t(\mu) = \Delta(\mu) + \Omega$ in~\eqref{eq:tdo} and the fact that all entries of~$\Delta(\mu)$ and all off-diagonal entries of~$\Omega$ are positive (see~\eqref{eq:omegaodp}). The inverse matrix~$t^{-1}(\mu)$ is then given by
\bea
t^{-1}(\mu) = (\Delta(\mu)+\Omega)^{-1} = \Delta^{-1}(\mu) - \Delta^{-1}(\mu) \, \Omega \, \Delta^{-1}(\mu)  + \CO\left(\Omega^2 \over \Delta^3\right) 
\eea 
For sufficiently small~$\mu$ we have large, positive~$\Delta(\mu)$. As a result the leading off-diagonal part of~$t^{-1}(\mu)$ is given by the off-diagonal part of~$- \Delta^{-1}(\mu) \, \Omega \, \Delta^{-1}(\mu)$, every entry of which is negative. 

The range~$\mu < \mu_\text{neg}$ for which the off-diagonal entries of $t^{-1}(\mu)$ are all negative is determined numerically in appendix~\ref{app:appt}. For~$N \leq 10$ the cutoff~$\mu_\text{neg}$ is given by~\eqref{tnegvalues}, which we repeat here,
\ba{
\begin{array}{c|cccccccccc}
N & 2 & 3 & 4 & 5 & 6 & 7 & 8 & 9 & 10 \\ \hline
\mu_\text{neg} /\Lambda  &1 & 1&  0.723& 0.577 & 0.451& 0.353 & 0.281& 0.228 & 0.188
\end{array}
\label{tnegvaluesbis}
}
For larger values of~$N$ we numerically obtain the approximate bound
\be\label{eq:tinvnegmu}
\left(t^{-1}\right)_{m \neq n}(\mu) < 0 \;\; \text{ if } \;\; \mu < \mu_\text{neg} \simeq {20 \Lambda \over N^2}~.
\ee
This restriction on~$\mu$ is somewhat more stringent that the condition~\eqref{eq:tposmu} required for positive definiteness of~$t_{mn}(\mu)$, since~$\mu_\text{pos} \simeq 5 \mu_\text{neg}$. Note that~$\mu_\text{neg} \simeq 0.7 M_{W, \text{min}}$ closely tracks the lightest~$W$-boson mass. Throughout our analysis in this paper, we will make the following

\noindent
{\bf Assumption}: The cutoff scale~$\mu$ of the effective field theory is restricted to a range $\mu < \mu_\text{neg}$ where the off-diagonal elements of $t^{-1}$ are all negative:
\be
\label{2.tinvneg}
(t^{-1})_{m n}(\mu) < 0 \qquad \hbox{ for all } \qquad m \not=n \quad \hbox{ and } \quad \mu < \mu_\text{neg},
\ee
with~$\mu_\text{neg}$ determined by~\eqref{tnegvaluesbis} or~\eqref{eq:tinvnegmu}, depending on~$N$. This in turn implies the positive-definiteness of~$t_{mn}(\mu)$ that is required for unitarity. The reason we make this slightly stronger restriction on~$\mu$ is that it streamlines several parts of our supersymmetry-breaking analysis below, allowing us to make some arguments analytically and uniformly in~$N$, rather than having to establish them numerically on a case-by-case basis. 
 
\end{itemize}

\subsubsection{Symmetries}\label{sec:symmahm}

In this section we discuss the global symmetries of the effective Lagrangian at the multi-monopole point, i.e.~those symmetries of the non-Abelian UV theory that are not spontaneously broken by the expectation values of the moduli~$u_n$ at the multi-monopole point. 

We start by examining the action of the unbroken symmetries on the magnetic periods~$a_{Dm}$. Comparing~\eqref{2.SWcurve} and~\eqref{eq:mmcurvedef}, we see that the degree-$N$ polynomial defining the Seiberg-Witten curve in the vicinity of the multi-monopole point can be written in the following two equivalent ways, 
\bea
\label{eq:mmcurvebis}
C(x) & = & 2^{N-1} \left(x^N - \sum_{I = 2}^N {u_I \over I (2 \Lambda)^I } x^{N-I}\right) 
\no \\ & = &
C(x)_{\text{MM}} + {i \over 2 \Lambda N} \, C(x)'_{\text{MM}} \sum_{m = 1}^{N-1} {s_m a_{Dm} \over x - c_m} + \CO(a_D^2)~,
\eea
where~$C_\text{MM}(x) = \cos ( N \arccos x)$ describes the curve~\eqref{tcheby} at the multi-monopole point. 

As discussed around~\eqref{eq:unmmpt}, the expectation values of the Coulomb-branch moduli~$u_I$ at the multi-monopole point spontaneously break the~$\Z_{4N}$ symmetry generated by~$r$ to the~$\Z_4$ subgroup generated by~$r^N$, which (according to~\eqref{2.ronu}) acts on the moduli as follows, 
\be\label{eq:rNonun}
r^N : u_I \to  (-1)^I u_I~.
\ee 
Comparing with~\eqref{eq:mmcurvebis}, we see that~$r^N : C(x) \to (-1)^N C(-x)$. Since we have~$C'_\text{MM}(-x) = (-1)^{N-1} C'_\text{MM}(x)$, we find that the following symmetry action on the periods,
\be\label{eq:rnonad}
r^N : a_{Dm} \to  a_{D( N-m)}~,
\ee
leads to the same transformation for~$C(x)$. 

We can similarly deduce the transformation rule of the periods under the unbroken charge-conjugation and time-reversal symmetries~$C$ and~$T$ at the multi-monopole point. Their action on the moduli was determined in~\eqref{2.conu} and~\eqref{2.Tonu}, which we repeat here, 
\be\label{eq:ctonun}
C : u_I \rightarrow (-1)^I u_I ~, \qquad T : u_I \to u_I~. 
\ee
We see that the action of~$C$ on the~$u_I$ is identical to that of~$r^N$ above,\footnote{~The distinction between them is that~$r^N$ is an~$r$-symmetry that does not commute with the supercharges, while~$C$ does commute with them (see below).} and therefore the same is true for their action on the periods,
\be\label{eq:conad}
C : a_{Dm} \to  a_{D(N-m)}~.
\ee
When discussing time-reversal~$T$, we must decide whether we treat~$u_n$ as operators which transform as in~\eqref{eq:ctonun} or as~c-number vevs, which are complex-conjugated by~$T$. We choose the former, so that~$T : C(x) \to C(\bar x)$. Comparing with~\eqref{eq:mmcurvebis}, we find that the periods are negated by time-reversal,
\be\label{eq:tonad}
T: a_{Dm} = -a_{Dm}~. 
\ee

The dual pre-potential~$\CF_D(a_D)$ at the multi-monopole point, given in~\eqref{fdfull}, is invariant under the symmetry actions of~$r^N$ and~$C$ in~\eqref{eq:rnonad} and \eqref{eq:conad}, and it transforms as~$T: \CF_D \to - \CF_D$ under the~$T$-symmetry in~\eqref{eq:tonad}. Via~\eqref{2.afd}, this in turn determines the action on the~$a_m$-periods,
\be\label{eq:symad}
r^N : a_m \to a_{N-m}~, \qquad C : a_m \to a_{N-m}~, \qquad T: a_m \to a_m~.
\ee
Note that the vev~$a_m = {2 N \Lambda \over \pi} s_m$ at the multi-monopole point is indeed invariant under these unbroken symmetries. 

\smallskip

Having determined the action of the unbroken symmetries on the periods, we now spell out all symmetries of the effective Lagrangian at the multi-monopole point. It is straightforward to check that these symmetries do in fact leave the renormalizable part~\eqref{eq:Lbos} and~\eqref{eq:Lferm} of the effective Lagrangian invariant, but in fact they are exact symmetries of~\eqref{eq:leffmm}, as well as all higher-derivative terms in the effective Lagrangian that we do not discuss explicitly. For future reference, the action of the discrete 0-form symmetries on the scalar fields of the model are summarized in table~\ref{tab:symahm}.

\renewcommand{\arraystretch}{1.6}
\renewcommand\tabcolsep{6pt}
\begin{table}[H]
  \centering
  \begin{tabular}{ |c|lc|lc|lc|}
\hline
{\bf Symmetry} & $a_m$   &  \multicolumn{1}{|c|}{  $a_{Dm}$ } &   \multicolumn{1}{c|}{$h_{i m}$} \\
\hline
\hline
$r^N$ & $a_{N-m}$ &  \multicolumn{1}{|c|}{$a_{D(N-m)}$} &  \multicolumn{1}{c|}{$\bar h_{i(N-n)}$} \\
\hline
$C$ & $a_{N-m}$ &  \multicolumn{1}{|c|}{$a_{D(N-m)}$} & \multicolumn{1}{c|}{$h_{i(N-m)}$} \\
\hline
$\t C$ & $a_{m}$ &  \multicolumn{1}{|c|}{$a_{D m}$} & \multicolumn{1}{c|}{$\bar h_{i m}$} \\
\hline
$T$ & $a_m$ &  \multicolumn{1}{|c|}{$-a_{Dm}$} & \multicolumn{1}{c|}{$h^i_{m}$} \\
\hline 
$C\t T$ & $a_{m}$ &  \multicolumn{1}{|c|}{$-a_{D m}$} & \multicolumn{1}{c|}{$\bar h^i_m$} \\
\hline
\end{tabular}
  \caption{The action of the symmetries~$r^N$, $C$, $T$, as well as their combinations~$\t C = r^N C$ and~$C\t T = C r^N T$ on the scalar fields of the effective Lagrangian at the multi-monopole point. The transformations in this table are for operator-valued fields, not their~c-number vevs; the latter are additionally complex conjugated by the anti-unitary~$T$ and~$C\t T$ symmetries.} 
\label{tab:symahm}
\end{table}

\begin{itemize}
\item {\it $SU(2)_R$ Symmetry:}  The fields transforming under this symmetry are the Abelian gauginos and the hypermultiplet scalars, which carry explicit~$SU(2)_R$ doublet indices, 
\be
\rho_{\alpha m}^i~, \qquad h_{i m}~.
\ee

\item {\it $\Z_4$ $r$-Symmetry:} This symmetry is generated by~$r^N$ and its action on the periods~$a_{Dm}$ was determined in~\eqref{eq:rnonad}. Since~$r^N$ represents a rotation by~$ \pi \over 2$ in the classical~$U(1)_r$ symmetry (broken to~$\Z_{4N}$ by the ABJ anomaly) under which the supercharges have charge~$-1$, it follows that
\be\label{eq:rnonq}
r^N(Q_\alpha^i) = - i Q_\alpha^i~.
\ee
Together with the supersymmetry transformations of the~$\CN=2$ vector multiplet in~\eqref{eq:qonvect}, this allows us to extend the action of~$r^N$ on~$a_{Dm}$ to its superpartners,
\be\label{eq:rnonadmult}
r^N: a_{D m} \to a_{D(N-m)}~, \quad  
\rho^i_{\alpha m} \to - i \rho^i_{\alpha(N-m)}~, \quad 
f_{\mu\nu m} \to -f_{\mu\nu(N-m)}~.
\ee
Note that~$r^{2N} = (-1)^F$, as required on gauge-invariant operators.  

The action of~$r^N$ on the hypermultiplet scalars~$h_{i m}$ is almost completely determined by three facts: $r^N$ commutes with the~$SU(2)_R$ symmetry;  it maps~$U(1)_{Dm} \to U(1)_{D, N-m}$; and it negates all~$U(1)_{Dm}$ gauge fields. The remaining freedom is a phase, which can be absorbed by a gauge transformation, leading to
\be\label{eq:rnonhmult}
r^N: h_{i m} \to \bar h_{i(N-n)}~, \qquad \psi_{ \alpha m}^\suppm \to \pm i \psi_{\alpha(N-m)}^\suppm ~.
\ee
Here we have used the hypermultiplet supersymmetry transformations~\eqref{eq:hypers2}, as well as~\eqref{eq:rnonq}, to deduce the action of~$r^N$ on the fermions. Note that~$r^{2N} \neq (-1)^F$ on the hypermultiplet fields; this is possible because they are not gauge invariant and carry fractionalized global symmetry quantum numbers, see for instance \cite{Cordova:2018acb, Brennan:2022tyl} for more detail.

\item{\it Charge Conjugation:} The action of charge-conjugation~$C$ on the~$a_{Dm}$ periods was determined in~\eqref{eq:conad}. Since~$C$ is unitary and commutes with the supercharges, its action is easily extended to the entire~$\CN=2$ vector multiplet containing~$a_{Dm}$,
\be
C : a_{Dm} \to a_{D(N-m)}~, \qquad 
\rho^i_{\alpha m} \to \rho^i_{\alpha( N-m)} ~, \qquad 
f_{\mu\nu m} \to f_{\mu\nu( N-m)}~. 
\ee
Thus~$C$ simply exchanges the~$U(1)_{Dm}$ and~$U(1)_{D(N-m)}$ gauge groups, without any further action on the gauge charges. Since~$C$ also commutes with the~$SU(2)_R$ symmetry and the supercharges, its action on the hypermultiplet fields is (up to a gauge transformation)
\be
C: h_{i m} \to h_{i(N-m)}~, \qquad \psi_{\alpha  m}^\suppm \to \psi_{\alpha(N-m)}^\suppm ~.
\ee
Note that the invariance of the renormalizable effective Lagrangian~\eqref{eq:Lbos} and~\eqref{eq:Lferm} under these~$C$-transformations follows from the fact that the matrix~$t_{mn}$ of effective~$U(1)_{D}^{N-1}$ gauge couplings satisfies~$t_{mn} = t_{(N-m)( N-n)}$, as discussed around~\eqref{eq:tccprop}. 

When~$N$ is even, $C$-symmetry groups the fields into~${N \over 2} -1$ charge-conjugate pairs, as well as the fields with gauge group index~$m = {N \over 2}$, which are invariant under~$C$.\footnote{~When~$N =2$ the~$C$-invariant fields with~$m = {N \over 2} = 1$ are the only fields in the theory, because~$SU(2)$ gauge theory does not admit a global charge-conjugation symmetry.} By contrast, when~$N$ is odd, there are no~$C$-invariant fields, and charge conjugation groups all fields into~${N-1} \over 2$ pairs. 

\item{\it Time Reversal:} The action of the anti-unitary time-reversal symmetry~$T$ on the periods was found in~\eqref{eq:tonad}. Given the action of~$T$ on the supercharges in~\eqref{eq:tonq},
\bea\label{eq:tonqbis}
T : Q_\alpha^i \to -i Q^\alpha_i~,
\eea
and the supersymmetry transformations of the~$\CN=2$ vector multiplet in~\eqref{eq:qonvect}, we determine its action the components fields to be  
\be
T : a_{Dm} \to - a_{Dm}~, \qquad 
\rho^i_{\alpha m} \to - i \rho^\alpha_{i m}~, \qquad 
f_{\mu\nu m} \to -{T_\mu}^\lambda {T_\nu}^\rho f_{\lambda \rho m}~, 
\ee
with~${T_\mu}^\lambda = \text{diag}(-1, 1, 1, 1)$.  We can restate the transformation rule for~$f_{\mu\nu m}$ by saying that~$T$ acts on the 2-form~$f^{(2)}_m = \half f_{\mu\nu m} dx^\mu \wedge dx^\nu$ as~$T(f^{(2)}_m) = - f^{(2)}_m$. 

Since~$T$ commutes with the hypermultiplet gauge charges and~$SU(2)_R$ transformations, its action on the hypermultiplet scalars is fixed (up to a gauge transformation),
\be
T: h_{i m} \to h^i_{m}~, \qquad \psi_{ \alpha  m}^\suppm \to - i \ep^{\a \b} \psi_{\b m}^\suppm~.
\ee
Here we have used~\eqref{eq:tonqbis} and~\eqref{eq:hypers2} to infer the~$T$-action on the fermions. Note that the relation~$T^2 = 1$, which holds on gauge-invariant fields, is modified to~$T^2 = -1$ for the gauge-charged monopole fields. This relation is meaningful:\footnote{~Since~$T$ commutes with the hypermultiplet gauge charges, it follows that~$T^2$ is in fact gauge invariant.} it shows that the monopole states organize into Kramers doublets.

\item{\it 1-Form Symmetry:} The Abelian Higgs model at the multi-monopole point possesses an accidental magnetic~$U(1)_D^{N-1}$ 1-form symmetry, since all particles that carry magnetic charge under the dual magnetic gauge group (i.e.~microscopic electric charge from the point of view of the~$SU(N)$ gauge theory in the UV) are massive there.\footnote{~By contrast, the massless hypermultiplet monopoles explicitly break all electric 1-form symmetries of the model.}  As long as there is no room for confusion, we will denote the~$U(1)_{Dm}$ gauge group and the associated magnetic 1-form symmetry by the same symbol. 

As reviewed in section~\ref{sec:symm}, the microscopic SYM theory has a~$\Z_N^{(1)}$ electric 1-form symmetry associated with the center of the~$SU(N)$ gauge group. At the multi-monopole point this~$\Z_N^{(1)}$ must be a subgroup of the~$U(1)_D^{N-1}$ 1-form symmetry. 

We claim that the~$\Z_{N}^{(1)}$ symmetry is a subgroup of the following (non-diagonal) linear combination of~$U(1)_D$'s at the multi-monopole point,\footnote{~Here the additive formula for~$\widetilde { U(1)}$ in terms of the~$U(1)_{Dm}$ should be read as applying to the associated integer-valued charges that we do not wish to introduce explicitly.} 
	\ba{
	 \label{utildefirst}
\Z_N^{(1)} \subset	\widetilde{U(1)} =\sum_{m=1}^{\lfloor \frac{N-1}{2} \rfloor} m \Big ( U(1)_{Dm} - U(1)_{D(N-m)} \Big ) +  \left\{ \begin{array}{cc}  \frac{N}{2} U(1)_{D \left( \frac{N}{2}\right)} & N \ \text{even} \\ 0 & N\ \text{odd}   \end{array} \right. 
	}
To check this, one needs to compute the~$\widetilde{U(1)}$ charges of all massive states at the multi-monopole point. It suffices to check the $W$-bosons, whose charges can be deduced from~\eqref{eq:Wcyc}. Indeed, one finds that their~$\widetilde{U(1)}$ charges are all~$0$ or~$\pm N$. 

Note that the~$\widetilde{U(1)}$ charge in~\eqref{utildefirst} is odd under charge-conjugation~$C$, which exchanges~$m \leftrightarrow N-m$. This is indeed the correct action of~$C$ on~$\Z_N^{(1)}$, since~$C$ exchanges~$SU(N)$ representations whose~$N$-alities sum to zero modulo~$N$ (such as the fundamental representation of~$N$-ality~$1$ and the anti-fundamental representation of~$N$-ality~$N-1$).  

\end{itemize}

In passing, we note that the renormalizable terms~\eqref{eq:Lbos} and~\eqref{eq:Lferm} of the effective Lagrangian are classically invariant under superconformal transformations, including a superconformal~$U(1)_r$ symmetry under which the gauginos~$\rho^i_{\alpha m}$ have charge~$+1$ and the hypermultiplet fermions~$\psi_{ \alpha m}^\suppm$ charge~$-1$. Both conformal and~$U(1)_r$ invariance is ruined by quantum anomalies. Moreover, these symmetries are explicitly broken by the irrelevant operators in the~$\CN=2$ effective Lagrangian, and also by the relevant SUSY-breaking terms discussed in section~\ref{sec:general} below. Thus we will not discuss them further here.\footnote{~See~\cite{Cordova:2018acb} for a detailed discussion in the case~$N = 2$, and the more recent~\cite{DelZotto:2024arv}.}

\subsection{Tracking the~$\CN=2$ stress-tensor multiplet from UV to IR}

\label{sec:stmult}

Here we discuss the stress-tensor supermultiplet of the~$\CN=2$ pure supersymmetric Yang-Mills theory with gauge group~$SU(N)$, whose UV Lagrangian is~\eqref{2.Lcomp}. A detailed discussion for~$SU(2)$ gauge group appears in section 5.2 of~\cite{Cordova:2018acb}. The generalization to~$SU(N)$ is essentially immediate, so we keep the discussion brief.  This multiplet was first discussed in~\cite{Sohnius:1978pk}, and further analyzed in~\cite{Antoniadis:2010nj,Cordova:2016xhm,Cordova:2016emh}. It is a short multiplet of~$\CN=2$ supersymmetry whose primary is a real, neutral scalar~$\CT$. Descendants of~$\CT$ include the~$SU(2)_R$ currents,\footnote{~Thus, this multiplet is not appropriate for~$\CN=2$ theories that explicitly break the~$SU(2)_R$ symmetry, e.g.~theories with~$\CN=2$ Fayet-Iliopoulos~$D$-terms.} the supersymmetry currents, and the stress tensor, all of which are conserved. The multiplet is defined by the following shortening conditions,
\be\label{eq:sohnmult}
Q^{\alpha i} Q_\alpha^j \CT = X^{(ij)}~, \qquad Q_\alpha^{(i} X^{jk)} = \bar Q_\alphadot^{(i} X^{jk)} = 0~.
\ee
Here~$X^{ij}$ is a complex~$\CN=2$ flavor current multiplet that gives rise to the complex central charge in the~$\CN=2$ supersymmetry algebra. It also contains the trace of the stress tensor and the spin-$\half$ traces of the supersymmetry currents. If~$X^{ij} = 0$ the theory is thus superconformal.

Classically the pure~$\CN=2$ SYM theory in~\eqref{2.Lcomp} is indeed conformal, and we can use the transformation rules in~\eqref{eq:n2susyt} to confirm that it has a superconformal stress-tensor multiplet with vanishing~$X^{ij}$, based on the primary
\be\label{eq:primTdef}
\CT = {2 \over g^2} \tr (\bar \phi \phi)~.
\ee
Quantum mechanically, the coupling~$g$ runs and conformal invariance is ruined. This generates the operator
\be\label{eq:stx}
X^{ij} \sim \beta \bar Q_\alphadot^{i} \bar Q^{\alphadot j} \bar u_2~, \qquad u_2 = \tr \phi^2~.\ee
Here~$\beta$ is the 1-loop beta function of the theory. 

Stress-tensor supermultiplets are generally not unique. For instance, given a well-defined stress-tensor multiplet primary~$\CT$ satisfying~\eqref{eq:sohnmult}, we can shift
\be\label{eq:imp}
\CT \to \CT + \CO + \bar \CO~, \qquad X^{ij} \to X^{ij} + Q^{\alpha i} Q_\alpha^j  \CO~, \qquad \bar Q_\alphadot^i \CO =0~,
\ee
while preserving the form of~\eqref{eq:sohnmult}. Here~$\CO$ is an~$\CN=2$ chiral multiplet, which must itself be well defined. The shift~\eqref{eq:imp} modifies the conserved currents in the multiplet by improvement terms -- well-behaved total derivatives that do not affect current conservation or the integrated charges. See~\cite{Komargodski:2010rb,Dumitrescu:2011iu} for a detailed discussion of such improvements in~$\CN=1$ theories, and~\cite{Antoniadis:2010nj,Cordova:2018acb} for~$\CN=2$ theories. An example of an improvement is~$\CO \sim \beta u_2$ to make~$X^{ij}$ real. This variant was studied in~\cite{Antoniadis:2010nj}; here we instead continue to study the multiplet whose primary is~\eqref{eq:primTdef}, and for which~$X^{ij}$ in~\eqref{eq:stx} is complex. 

Our analysis of supersymmetry-breaking below rests on our ability to track the~$\CN=2$ stress-tensor multiplet from the UV to the IR, where it is expressed in terms of the effective degrees of freedom on the Coulomb branch. We will need the following facts (verified, for instance, in appendix B of~\cite{Antoniadis:2010nj}; see also~\cite{Luty:1999qc, Cordova:2018acb}):  
\begin{itemize}
\item[(i)] At generic, smooth points of the Coulomb branch, where the low-energy effective action is~\eqref{2.LIR}, the stress-tensor primary~$\CT$ flows to
\be\label{eq:tcb}
\CT \to K = {1 \over 2 \pi} \sum_{m =1}^{N-1} \Im a_{Dm} \bar a_m~.
\ee
Here~$K$ is the low-energy K\"ahler potential of the Coulomb-branch sigma model in~\eqref{2.Kdef}. Since~$\CT$ is a well-defined operator in the UV, this must be true for~$K$ on the Coulomb branch. This was already argued around~\eqref{eq:fshift} from another point of view.\footnote{~Note that this statement is stronger than a similar statement for~$\CN=1$ theories, which states that the existence of a Ferrara-Zumino~\cite{Ferrara:1974pz} stress-tensor supermultiplet implies that~$dK$ must be a globally well-defined 1-form on the K\"ahler target manifold~\cite{Komargodski:2010rb,Dumitrescu:2011iu}.}

\item[(ii)] At the multi-monopole point  the stress-tensor primary~$\CT$ flows to
\be\label{eq:tmm}
\CT \to K^\text{eff} (a_D) - \half \sum_{m = 1}^{N-1} \bar h^{i}_m h_{im}~.
\ee
Here~$K^\text{eff} (a_D)$ is the Wilsonian effective K\"ahler potential for the magnetic vector multiplet scalars~$a_{Dm}$ in~\eqref{eq:keffah} at the multi-monopole point, while~$h_{i m}$ are the scalars in the massless monopole hypermultiplets at that point. Note that if we move away from the multi-monopole point, the hypermultiplets become massive and we can integrate them out. In this case~\eqref{eq:tmm} reduces to~\eqref{eq:tcb}. 
\end{itemize}
\noindent In principle~$\CT$ receives corrections from additional massless fields (such as the monopoles in~\eqref{eq:tmm}) at all singular loci on the Coulomb branch, but we will not need these explicitly.

\newpage

\section{RG flow from~$\CN=2$ SYM to adjoint QCD} 
\label{sec:general}

Here we elaborate on the discussion in section~\ref{intro:susy}. We explain how to flow from~$\CN=2$ SYM to adjoint QCD using a non-holomorphic scalar mass~$M^2 \tr (\bar\phi \phi)$ for the vector multiplet scalar~$\phi$. We then show how to analyze this deformation in the regime~$M \ll \Lambda$, by tracking~$\tr (\bar \phi \phi)$ onto the Coulomb branch of the~$\CN=2$ theory. Most importantly, we formulate a dual Abelian Higgs model of the~$M$-deformed~$\CN=2$ theory. The dual correctly reproduces the small-$M$ regime, but it can be analyzed for all values of~$M$, and in particular the large-$M$ regime relevant for adjoint QCD. In this section we focus on establishing general features of the dual, e.g.~its unbroken symmetries in compliance with the Vafa-Witten theory for adjoint QCD, while leaving a detailed analysis of its vacua and phase diagram to later sections.

\subsection{The SUSY-breaking scalar mass~$M$ in the UV}

We will analyze the family of RG flows that start from pure~$\CN=2$ SYM, with gauge group~$SU(N)$ and UV Lagrangian~$\SL_{SU(N)}$ in~\eqref{2.Lcomp}, and that are triggered by turning on the following non-holomorphic SUSY-breaking mass term for the adjoint scalar~$\phi$ in the~$\CN=2$ vector multiplet,
\begin{equation}\label{eq3:susybrdef}
\SL = \SL_{SU(N)} - \SV_\text{\cancel{SUSY}}~, \qquad \SV_\text{\cancel{SUSY}} = M^2 \CT~, \qquad \CT = {2 \over g^2} \tr \left( \bar \phi \phi\right)~.
\end{equation}
Given the supersymmetric kinetic terms in~\eqref{2.Lcomp}, the SUSY-breaking parameter~$M > 0$ is nothing but the mass of the adjoint scalar~$\phi$ in the~$\CN=2$ vector multiplet. It preserves all symmetries (and thus 't Hooft anomalies), except for supersymmetry itself. 

This family of RG flows is labeled by the dimensionless parameter~$M / \Lambda$, where~$\Lambda$ is the strong-coupling scale of the~$\CN=2$ gauge theory:
\begin{itemize}
\item When~$M \ll \Lambda$, the RG flow is nearly supersymmetric, only deviating from that of the~$\CN=2$ theory in the deep IR. It can therefore be analyzed by perturbing the IR effective theory describing the~$\CN=2$ Coulomb branch, provided we can track the SUSY-breaking mass deformation in~\eqref{eq3:susybrdef} onto the Coulomb branch. This can indeed be done, as we explain in section~\ref{sec:smallm} below. 

\item When~$M \gg \Lambda$, the scalar~$\phi$ decouples and the theory flows to adjoint QCD with~$N_f = 2$ adjoint Weyl fermions.\footnote{~In this regime, the strong-coupling scale~$\Lambda$ of~$\CN=2$ SYM, and its counterpart~$\Lambda_\text{adj.}$ in adjoint QCD, are related via
$$
\Lambda_\text{adj.}^{\beta_\text{adj.}} \sim M^{\beta_\text{adj.} - \beta} \Lambda^\beta~.
$$
Here~$\beta_\text{adj.} > \beta$ are (minus) the 1-loop beta functions of adjoint QCD and~$\CN=2$ SYM. Throughout, we will use the~$\CN=2$ strong-coupling scale~$\Lambda$.} Clearly, in this regime the small-$M$ analysis on the~$\CN=2$ Coulomb branch is no longer valid.
\end{itemize}

In this paper, we analyze the IR phases and the transitions between them as a function of~$M / \Lambda$. In order to push beyond the small-$M$ regime (which is the only regime that can be analyzed completely rigorously), we propose a useful dual description that extends to all values of~$M$. This dual description will be introduced and motivated in section~\ref{sec:gen1} below; the remainder of the paper is dedicated to exploring its consequences.

\subsection{IR analysis for small SUSY-breaking ($M \ll \Lambda$) and the origin of the Coulomb branch} 
\label{sec:smallm}

As explained above, for small values~$M \ll \Lambda$ we can study the effects of SUSY-breaking by tracking the operator~$\CT$ in~\eqref{eq3:susybrdef} to the low-energy theory on the Coulomb branch of the~$\CN=2$ theory and analyzing its effects  there. As reviewed in section~\ref{sec:stmult}, this can be done reliably because~$\CT$ is the primary (i.e.~bottom component) of the protected~$\CN=2$ stress-tensor supermultiplet. In the deep IR, $\cT$ flows to the K\"ahler potential on the Coulomb branch as in~\eqref{eq:tcb}, which we repeat here, 
\be\label{eq:tcbii}
\CT \to K = {1 \over 2 \pi} \sum_{m =1}^{N-1} \Im a_{Dm} \bar a_m~.
\ee
As explained before, this particular K\"ahler potential~$K$ is globally well defined on the Coulomb branch. Strictly speaking~\eqref{eq:tcbii} is only valid away from the singular points on the Coulomb branch, but for the purposes of our small-$M$ analysis it will be sufficient to work directly with~$K$ in~\eqref{eq:tcbii} (whose non-analyticities are rather mild).

Thus, to leading order in small~$M \ll \Lambda$, the only consequence of SUSY-breaking is the generation of a scalar potential on the~$\CN=2$ Coulomb branch, 
\begin{equation}\label{eq:smallMsBpot}
\SV_\text{\cancel{SUSY}} = M^2 K(u_I)~,
\end{equation}
where~$u_I$ are the gauge-invariant Coulomb-branch coordinates. In order to analyze its consequences, we recall from section~\ref{ssec:KPorigin} that~$K(u_I)$ is a convex function with a unique minimum at the origin of the Coulomb branch, where all~$u_I = 0$ and~$K(u_I = 0) < 0$ (see for instance figure~\ref{fig:kahler} for the cases of $SU(2)$ and $SU(3)$ gauge groups). This leads to the following conclusions about the vacuum in the small-$M$ regime:
\begin{itemize}
\item The vacuum is at the origin of the Coulomb branch, where all~$u_I$ vanish. All 0-form symmetries of the theory, i.e.~$SU(2)_R$, $\Z_{4N}$, $C$, and~$T$, are unbroken there. 
\item The positive curvature of the SUSY-breaking potential near the origin gives masses proportional to $M$ to all the scalars in the~$N-1$ Abelian~$\CN=2$ vector multiplets. By contrast, all~$N-1$ gauginos (each of which is an~$SU(2)_R$ doublet) and photons in these multiplets remain massless.  
\item The~$N-1$ massless photons imply that this vacuum describes a Coulomb phase. In particular, the microscopic~$\Z_N^{(1)}$ is spontaneously broken. 
\item The massless gauginos and photons match the 't Hooft anomalies of the UV theory in a highly non-trivial way (see~\cite{Cordova:2018acb} for a detailed discussion of the~$SU(2)$ case, with related discussions and generalizations in~\cite{Wan:2018djl,Hsin:2019fhf,Cordova:2019bsd,Cordova:2019jqi,Brennan:2023vsa}).
\end{itemize}

\noindent Note that the small-$M$ Coulomb phase is neither confining, nor does it break any chiral symmetries. While there may well be other phases with these features at larger~$M$ (as we will soon argue to be the case), we cannot reliably access this regime within the limitations of the approach we have pursued so far. 

In order to quantify these limitations, let us recall the BPS spectrum at the origin of the~$\CN=2$ Coulomb branch (see section~\ref{sec:BPSorigin}, and in particular figure~\ref{fig:BPSmassesbetter}). Up to an~$\CO(1)$ prefactor, the BPS masses at the origin are given by their simple large-$N$ spectrum~\eqref{2.BPSbigN}, which we recall here,
\begin{equation}\label{eq:BPSmasses3}
M_\text{BPS}(\mu_{km}) \sim \sqrt 2 \Lambda s_m~, \qquad k = 0, \ldots, 2N-1~,  \qquad m = 1, \ldots, N-1~.
\end{equation}
Recalling that~$s_m = \sin (\pi m/N)$, we see that the lightest of these BPS states has mass~$\sim \Lambda / N$, which is therefore also the natural cutoff of the IR effective theory at the origin. We should thus only trust our SUSY-breaking analysis above in the regime~$M \lesssim \Lambda / N$.

However, the existence of massive BPS states also suggests an opportunity for our SUSY-breaking analysis: if we manage to probe the regime where~$M$ is of order the BPS masses in~\eqref{eq:BPSmasses3}, we might activate these degrees of freedom and unveil new interesting phases and vacua. The challenge is that this requires a sufficiently tractable description of these states, which are not only massive but also mutually non-local.

\subsection{Dual description for all~$M$ via SUSY-breaking in the Abelian Higgs model at the multi-monopole point} 
\label{sec:gen1}

We shall now discuss a dual description of the~$N-1$ Abelian vector multiplets on the Coulomb branch, as well as the massive BPS states at its origin, that has the desirable features imagined at the end of the previous subsection while sidestepping the associated challenges outlined there. In particular, it will allow us to extend our SUSY-breaking analysis to all values of~$M$.

The dual originates at the~$N$ multi-monopole points of the~$\CN=2$ theory (see section~\ref{ssec:mmpts}). As reviewed in section~\ref{ssec:mmbps} (see in particular figures~\ref{fig:BPSmassesbetter} and \ref{fig:BPS}), all massive BPS states at the origin of the Coulomb branch become massless at one of the~$N$ multi-monopole points. They do so in groups of~$N-1$ mutually local full hypermultiplets -- precisely one from each of the~$N-1$ levels indexed by~$m$ in~\eqref{eq:BPSmasses3}. Because the~$N-1$ massless BPS states at every multi-monopole point are mutually local, they can (in a suitable duality frame) be described by a conventional~$\CN=2$ Abelian Higgs model (plus non-renormalizable terms), which we described in section~\ref{sec:AHM}. 

Our proposal is to describe the BPS states in the strong-coupling region surrounding the origin via the effective Abelian Higgs models at the~$N$ multi-monopole points. Importantly, we can only analyze one multi-monopole point at a time; the broken~$\Z_{4N}$ symmetry that relates these points implies that analyzing one of them is also sufficient. We will choose it to be {\it the} multi-monopole point, where all~$u_I$ are real, as in~\eqref{eq:unmmpt}. The price to pay is that this dual description does not have manifest~$\Z_{4N}$ symmetry. 

The Wilsonian effective Lagrangian for the Abelian Higgs model at the multi-monopole point was reviewed in section~\ref{sec:MMLag}. For most of our discussion we will focus on the renormalizable terms in~\eqref{eq:Lbos}, \eqref{eq:vsimple}, \eqref{eq:Lferm}, \eqref{eq:lyukbis}.\footnote{~The only exception, discussed in section~\ref{sec:aD3}, is to break some accidental degeneracies that arise in the renormalizable theory.} This amounts to only retaining terms up to and including~$\CO(a_D^2)$ in the effective K\"ahler potential~\eqref{eq:keffah} for the vector multiplet scalars~$a_{Dm}$ at the multi-monopole point,
\be
\label{eq:keffahii}
\begin{split}
& K^{\text{eff}}(a_D) = { N \Lambda \over \pi^2} \sum_{m=1}^{N-1}s_m \Im {a}_{Dm} 
	+\sum_{m, n=1}^{N-1} t_{mn} \, a_{Dm} \bar{a}_{Dn} + \CO(a_D^3)~.
\end{split}
\ee
Here the matrix~$t_{mn} = t_{mn}(\mu)$ is defined in~\eqref{eq:tdef1}, \eqref{eq:tdef1ii} and will be discussed further below.  The~$\CO(a_D)$ term in~\eqref{eq:keffahii} is a K\"ahler transformation and does not appear in the~$\CN=2$ Lagrangian, but it is needed to render~$K^{\text{eff}}(a_D)$ globally well-defined and will crucially enter our discussion of SUSY-breaking once we turn on~$M$. 

Much of our analysis will revolve around the scalar potential for the complex vector multiplet scalars~$a_{Dm}$ and the complex hypermultiplet scalars~$h_{im}$ describing the massless monopoles, i.e.~$h_{im}$ has unit electric charge under the dual magnetic gauge group~$U(1)_{Dm}$ (and is uncharged under the other~$U(1)_D$'s).\footnote{~Recall that~$m = 1, \ldots, N-1$ indexes the different~$U(1)_{Dm}$ gauge groups, and that~$i = 1, 2$ is an~$SU(2)_R$ doublet index. Furthermore~$\bar h_m^i = (h_{im})^\dagger$, and repeated upper-lower~$SU(2)_R$ indices are summed. See section~\ref{sec:MMLag} and appendix~\ref{app:conv} for further detail.} In the~$\CN=2$ theory, this is given by~\eqref{eq:vsimple}, which we recall here,
 \be 
\begin{split}\label{eq:vsimpleii}
\SV_\text{SUSY} = &  
\sum_{m = 1}^{N-1} 2 \left|a_{Dm}\right|^2 \bar h^i_m h_{i m} 
\\
& + \sum_{m, n = 1}^{N-1} \left(t^{-1}\right)_{mn} \left( 
\left(\bar h^i_m h_{i n}\right) \left(\bar h^j_n h_{j m}\right) 
- \half \left(\bar h^i_m h_{i m}\right) \left(\bar h^j_n h_{j n}\right)
\right)~.
\end{split}
\ee

The only marginal couplings in the~$\CN=2$ Abelian Higgs model that are not completely dictated by the matter content together with supersymmetry are the effective gauge couplings~$t_{mn}(\mu)$, which appear explicitly in~\eqref{eq:keffahii}, \eqref{eq:vsimpleii} and are discussed at length in section~\ref{sec:propt}.\footnote{~Note that these couplings were not taken into account completely or correctly in many previous discussions of the effective  theory at the multi-monopole point, starting with the influential~\cite{Douglas:1995nw}; see~\cite{DHoker:2020qlp} for a detailed survey of the literature on~$t_{mn}(\mu)$.} They are a threshold effect and arise from integrating out the massive BPS particles at the multi-monopole point, whose properties are thus reflected in the interactions of the massless fields in our dual.\footnote{~Of course, integrating out massive states also generates an infinite number of irrelevant couplings (many of them~$D$-terms not controlled by SUSY), including the higher-order terms in~$K$ (see~\eqref{eq:keffahii}).}  Through the detailed structure of~$t_{mn}(\mu)$, the dual is able to capture aspects of all massive BPS states at the origin --  or equivalently the massive $W$-bosons at the multi-monopole point (they are related by wall crossing, though the~$W$-bosons more naturally reflect the structure of the underlying~$SU(N)$ gauge theory) -- not just the ones that become massless at the multi-monopole point. 

A consequence of this fine structure was discussed around~\eqref{2.tinvneg}: $t_{mn}(\mu)$ is not only positive definite as required by unitarity, but also has the unexpected property that its off-diagonal elements are negative,\footnote{~It would be desirable to find an intuitive or elementary derivation of this key property (one that is perhaps simply related to the charges of the~$W$-bosons), but we have not found one.} 
\begin{equation}\label{eq:todneg3}
(t^{-1})_{mn}  < 0~, \qquad m \neq n~, \qquad \mu < \mu_\text{neg}~.
\end{equation}
This holds as long as the renormalization scale~$\mu$ of our effective theory satisfies~$\mu < \mu_\text{neg}$. Here~$\mu_\text{neg} \sim \Lambda/N^2$ is a natural cutoff that tracks the mass of the lightest~$W$-boson at the multi-monopole point. As already stated around~\eqref{2.tinvneg} we will assume~\eqref{eq:todneg3} throughout; but by no means is it sufficient to fully capture all detailed properties of~$t_{mn}(\mu)$. For this reason we always use the explicit formula for~$t_{mn}(\mu)$  in~\eqref{eq:tdef1}, \eqref{eq:tdef1ii}.

We now proceed to analyze the effect of the SUSY-breaking mass~$M$ in~\eqref{eq3:susybrdef} in the dual Abelian Higgs model at the multi-monopole point. As already shown in~\eqref{eq:tmm}, the operator~$\CT$ in~\eqref{eq3:susybrdef} flows to the primary of the~$\CN=2$ stress-tensor supermultiplet of the Abelian Higgs model,
\be\label{eq:tmmii}
\CT \to K^\text{eff} (a_D) - \half \sum_{m = 1}^{N-1} \bar h^{i}_m h_{im}~.
\ee
Substituting the effective K\"ahler potential in~\eqref{eq:keffahii} then leads to the following SUSY-breaking potential,
\begin{equation}\label{eq:s3sblagmm}
\SV_\text{\cancel{SUSY}} = M^2 \left({ N \Lambda \over \pi^2} \sum_{m=1}^{N-1}s_m \Im {a}_{Dm} 
	+\sum_{m, n=1}^{N-1} t_{mn} \, a_{Dm} \bar{a}_{Dn} - \half \sum_{m = 1}^{N-1} \bar h^{i}_m h_{im}\right)~. 
\end{equation}
Here, as in~\eqref{eq:keffahii}, we have only retained terms up to and including~$\CO(a_D^2)$, because these operators are related by supersymmetry.\footnote{~Note that this does not amount to using the full effective K\"ahler potential everywhere and subsequently truncating to the renormalizable terms, which would lead to cubic and quartic terms in~\eqref{eq:s3sblagmm}.} For instance, if we choose to include the leading non-renormalizable $\CO(a_D^3)$ terms in~\eqref{eq:keffahii}, we should correspondingly include these cubic terms (but no quartics) in the SUSY-breaking potential~\eqref{eq:s3sblagmm} (see section~\ref{sec:aD3}).  A more pragmatic reason for truncating to the quadratic terms in~\eqref{eq:s3sblagmm} is that it simplifies the challenging analysis of the scalar potential, without sacrificing substantial accuracy.\footnote{~The quadratic approximation to the K\"ahler potential~$K$ around the multi-monopole point is excellent, e.g.~extrapolating all the way to the origin of the Coulomb branch in the~$SU(2)$ theory only leads to a percent-level error. Loosely speaking, this reflects the convexity of~$K$, see~e.g.~figure~\ref{fig:kahler}.}

In the remainder of this paper we will explore the phases of the dual Abelian Higgs model described above as a function of the SUSY-breaking parameter~$M$, and the implications for adjoint QCD. Importantly, the dual enables us to explore all values of~$M$, well beyond the cutoff~$\sim \Lambda/ N^2$ set by the lightest~$W$-boson mass at the multi-monopole point, and we shall do so with impunity. The fact that we will be able to establish a consistent picture of the entire phase diagram, which beautifully matches onto the expected properties of adjoint QCD in the~$M \gg \Lambda$ limit, gives us hope that our approach is indeed justified (see section~\ref{sec:summary} for further discussion). 

We shall explore these phases semi-classically, by studying the vacua of the tree-level potential~$\SV$ of the dual, which is obtained by adding the~$\CN=2$ supersymmetric potential~$\SV_\text{SUSY}$ in~\eqref{eq:vsimpleii} and the SUSY-breaking potential~$\SV_\text{\cancel{SUSY}}$ in~\eqref{eq:s3sblagmm}, 
\bea
\label{defpot}
\SV & = &   \SV_\text{SUSY} + \SV_\text{\cancel{SUSY}}
\no \\ 
&=&  \sum_{m=1}^{N-1}\left( \frac{ M^2 N \Lambda   }{\pi^2} s_m \Im {a}_{Dm} 
	+  \left( 2 |a_{Dm}|^2 - \thalf M^2 \right )  \bar h^i_m h_{i m}  \right )
	+  \sum_{m, n=1}^{N-1}   M^2 \, t_{mn} \, {a_{Dm} \bar{a}_{Dn}} 
\no \\ &&
 + \sum_{m, n=1}^{N-1} 
		\left(t^{-1}\right)_{mn} \left [ 
\left(\bar h^i_m h_{i n}\right) \left(\bar h^j_n h_{j m}\right) 
- \half \left(\bar h^i_m h_{i m}\right) \left(\bar h^j_n h_{j n}\right )
\right ] ~.
\eea
In the remainder of this section, we initiate the study of the vacua of this potential. We will obtain a number of general results, valid for any~$N$, and show that our dual correctly recovers the small SUSY-breaking regime~$M \ll \Lambda$ already analyzed in section~\ref{sec:smallm}. The phase structure for all~$M$ will be studied analytically in sections~\ref{sec:su2} and~\ref{sec:su3}, for~$SU(2)$ and~$SU(3)$ respectively, before we move on to~$N \geq 4$ using a combination of analytical and numerical methods. A summary of our results, and the implications for adjoint QCD in the large-$M$ regime, are the subject of section~\ref{sec:summary}.

\subsection{Unbroken symmetries and vacuum alignment in the dual} 

As was already mentioned in section~\ref{intro:nf2adjoint}, adjoint QCD is subject to the constraints on symmetry breaking obtained by Vafa and Witten~\cite{Vafa:1983tf,Vafa:1984xg} in vector-like gauge theories. In particular, a~$U(1)_R$ subgroup of the~$SU(2)_R$ symmetry, as well as a suitably defined parity symmetry~$P$ (equivalently, by the~$CPT$ theorem, a suitable~$CT$ symmetry), cannot be spontaneously broken. 

It is therefore a reassuring fact that our dual Abelian Higgs model, with scalar potential~\eqref{eq:s3sblagmm}, only admits vacua that at most break $SU(2)_R \to U(1)_R$ (as well as vacua where~$SU(2)_R$ is not broken at all) and always preserve the symmetry~$C \t T$ in table~\ref{tab:symahm}. This holds for all values of the SUSY-breaking mass~$M$.

\subsubsection{Vacuum alignment and spontaneous~$SU(2)_R \to U(1)_R$ breaking}
\label{sec:vacal}

If a single hypermultiplet~$h_{im}$ gets a vev, it spontaneously breaks~$SU(2)_R \to U(1)_R$, leading to a single~$\C\P^1$ sigma model for the two massless Nambu-Goldstone bosons. The fact that the~$U(1)_R$ Cartan subgroup remains unbroken is due to mixing with the broken~$U(1)_{Dm}$ gauge symmetry acting on~$h_{im}$. 

If at least two hypermultiplets get vevs, they may in principle misalign and break~$SU(2)_R$ completely. We will now show that this does not happen in our dual Abelian Higgs model with scalar potential~\eqref{eq:s3sblagmm}: in any vacuum where at least two hypermultiplets get a vev, their vevs align in~$SU(2)_R$ space, leading to the symmetry-breaking pattern~$SU(2)_R \to U(1)_R$. We refer to this behavior as vacuum alignment. 

To prove this assertion, it suffices to examine the dependence of the scalar potential $\SV$ in~\eqref{eq:s3sblagmm} on the hypermultiplet scalars~$h_{im}$ and their complex conjugates~$\bar h^i_m$, for an arbitrary fixed value of $a_D$, 
\be
\begin{split}\label{partv}
\SV \big|_{h, \bar h}  =  &  \sum_{m=1}^{N-1}   \left( 2 |a_{Dm}|^2 - \thalf M^2 \right )  \bar h^i_m h_{i m} 
 \\
& + \sum_{m, n=1}^{N-1}
		\left(t^{-1}\right)_{mn} \left( 
\left(\bar h^i_m h_{i n}\right) \left(\bar h^j_n h_{j m}\right) 
- \half \left(\bar h^i_m h_{i m}\right) \left(\bar h^j_n h_{j n}
\right)\right)~.
\end{split}
\ee
This potential can be recast in an illuminating way be recalling that (taking into account the~$U(1)_{Dm}$ gauge transformations acting on~$h_{i m}$) all gauge invariant data is contained in the real~$SU(2)_R$ triplet vectors (or spins), 
\begin{equation}
\label{eq:Spindef}
\vec S_m = \bar h_m^i { \vec \sigma_i}^{\;\; j} h_{jm}~, 
\end{equation}
Here~${ \vec \sigma_i}^{\;\; j}$, with the indicated placement of~$SU(2)_R$ indices, denotes the three standard Pauli matrices (see appendix~\ref{sec:conventions}). Since these matrices are Hermitian, it follows that
\begin{equation}
\left(\vec S_m\right)^\dagger = \vec S_m~. 
\end{equation}
Using standard identities for Pauli matrices,\footnote{~In particular, we use
$$
{\left(\vec \sigma\right)_i}^j  \cdot {\left(\vec \sigma\right)_k}^\ell = 2 \delta_i^\ell \delta_k^j - \delta_i^j \delta_k^\ell~.
$$
}
we find
\begin{equation}
\vec S_m \cdot \vec S_n = 2 \left(\bar h_m^i h_{i n}\right) \left(\bar h_n^j h_{jm}\right) - \left(\bar h_m^i h_{im}\right) \left( \bar h_n^j h_{jn}\right)~.
\end{equation}
Using this formula, it is straightforward to check that~\eqref{partv} can be rewritten as follows,
\ba{
\label{partv2}
\SV \big|_{h,\bar{h}} =& \sum_{m=1}^{N-1} \left(\left(2 |a_{Dm}|^2 - \half  M^2   \right) \bar h_m^i h_{im} + \half (t^{-1})_{mm} (\bar h_m^i h_{im})^2 \right) + \sum_{m<n}   { (t^{-1})_{mn}} \, \vec S_m \cdot \vec S_n~.
}
\noindent Let us make some comments on this formula:
\begin{itemize}
\item The first, single-sum term in~\eqref{partv2} only depends on the~$SU(2)_R$ invariant magnitudes of the~$h_{im}$, or equivalently the magnitudes~$|\vec S_m|$ of the spin vectors~$\vec S_m$ defined in~\eqref{eq:Spindef}. Let us consider these magnitudes (along with~$a_{Dm}$) to be fixed, so that the first term in~\eqref{partv2} is also fixed.

\item The second, double-sum term in~\eqref{partv2} is a Heisenberg spin chain Hamiltonian for the~$\vec S_n$, with all-to-all couplings given by the off-diagonal matrix elements of~$(t^{-1})_{mn}$. Importantly, this term is invariant under simultaneous~$SU(2)_R$ rotations of the~$\vec S_m$, but it depends on their relative orientation. 

\item It is here that we use the assumption, spelled out in~\eqref{eq:todneg3} (see also the discussion around~\eqref{2.tinvneg}), that these off-diagonal matrix elements are negative, $(t^{-1})_{m \neq n} < 0$. This implies that the Heisenberg couplings in~\eqref{partv2} are all ferromagnetic, so that all non-vanishing spins~$\vec S_m$ must align in the vacuum. This perfect vacuum alignment implies that the symmetry-breaking pattern is indeed~$SU(2)_R \to U(1)_R$, with unbroken~$U(1)_R$ Cartan subgroup (as required by the Vafa-Witten theorem in adjoint QCD).\footnote{~The only exception to this is a vacuum where all~$\vec S_m$ vanish, leaving~$SU(2)_R$ unbroken.} This leads to exactly two massless Nambu-Goldstone bosons, which correspond to coherent spin wave oscillations around the aligned vacuum. They are described by a sigma model with target space~$\C\P^1 = SU(2)_R/U(1)_R$.

\item It is instructive to contemplate the role of relative, non-aligned oscillations of the~$\vec S_m$. The ferromagnetic Heisenberg couplings in~\eqref{partv2} ensure that exciting these costs more energy, leading to massive scalar particle excitations; only the two Nambu-Goldstone bosons discussed above are exactly massless.\footnote{~We will confirm this explicitly in section~\ref{sec:11}.} This is entirely due to the off-diagonal~$(t^{-1})_{m \neq n}$. If we were to (incorrectly) omit them, there would be~$N-1$ decoupled copies of~$SU(2)_R$, one for each spin~$\vec S_m$; moreover, there would be no vacuum alignment (with the diagonal~$SU(2)_R$ acting on all spins generically broken completely), and every non-zero spin would break its own copy of~$SU(2)_R$, leading to many Nambu-Goldstone bosons in decoupled copies of~$\C\P^1$. Clearly this would be a confounding scenario from the point of view of adjoint QCD; by supplying the correct, negative~$(t^{-1})_{m \neq n}$ the~$\CN=2$ SYM theory has elegantly absolved us from having to contemplate it.  
\end{itemize}

\sm

We can use vacuum alignment to simplify the form of the hypermultiplet vevs~$h_{im}$. Using a global $SU(2)_R$ rotation, we can choose the alignment direction of all non-zero spins~$\vec S_m \neq 0$ to be the~$3$-direction, i.e.~$S^1_m = S_m^2 = 0$, $S^3_m > 0$. Comparing with~\eqref{eq:Spindef}, we see that the first two conditions require $\bar{h}^1_m
 h_{2m}=0$, which combined with the third condition, requires $h_{2m}=0$. Finally, using a gauge transformation, one may align all $h_{1 m}>0$. Thus, as a consequence of~$SU(2)_R$ vacuum alignment, we can simplify our subsequent analysis of the scalar potential by only considering hypermultiplet vevs of the form
	\ba{
	\label{hpar}
	h_{im} = M h_m \delta_{i1}~, \qquad h_m \geq 0~, \qquad m=1,\ldots,N-1~. 
	}
Here we have scaled out a factor of the SUSY-breaking mass $M$ to render $h_m$ dimensionless. Note that, since we always take the~$h_m$ to be non-negative, we will use henceforth use~$h_m > 0$ and~$h_m \neq 0$ interchangeably throughout the paper.

\subsubsection{Invariance of all vacua under $C \t T$-symmetry}
\label{sec:CT}

The action of~$C \t T$ on the operator-valued fields of the Abelian Higgs model can be found in table 1. Here we are interested in the action on the c-number vevs (which we here, and only here, emphasize with the symbol~$\langle \cdots\rangle$), which are further subject to complex conjugation (which we here, and only here, denote by~$*$ for emphasis) because the symmetry is anti-unitary,
\begin{equation}\label{eq:cttvevs}
C \t T : \langle a_{Dm}\rangle  \to - \langle a_{Dm}\rangle^*~, \qquad \langle h_{im}\rangle \to \langle \bar h_m^i\rangle^* = \langle h_{im}\rangle~.
\end{equation}
Thus~$C \t T$ symmetry does not restrict the hypermultiplet vevs, but it requires the vevs of the~$a_{Dm}$ to be purely imaginary.

The total effective potential $\SV $ in (\ref{defpot}) is manifestly invariant under $C\t T$. It is also clear that $\Re a_{Dm}$ enters $\SV$ quadratically, with positive-definite coefficient matrix, in such a way that all extrema of~$\SV$ have $\Re a_{Dm} =0$, thereby preserving~$C\t T$.\footnote{~This conclusion remains valid upon the inclusion of certain higher-order corrections to the potential, discussed in section~\ref{sec:aD3}.} In the sequel, it will be convenient to introduce the following notation for $a_{Dm}$,
\bea
\label{3.xdef}
a_{Dm} = -i M x_m \qquad x_m \in \RR~,
\eea
where we have extracted a factor of the SUSY-breaking mass~$M$ from $a_{Dm}$ so that the variable $x_m$ is dimensionless. Note, however, that unlike the~$h_m$, which we have gauge fixed to be non-negative, the~$x_m$ are gauge invariant and can have either sign.

	
\subsection{Exploring the phases of the dual Abelian Higgs model} 
\label{sec:exphases}

Collecting the results of subsections~\ref{sec:vacal} and~\ref{sec:CT} above, we have reduced the study of the semi-classical vacua of the dual Abelian Higgs model to the special loci found in~\eqref{3.xdef} and \eqref{hpar}, parametrized in terms of the dimensionless variables~$x_m$ and~$h_m$,
\bea
\begin{cases} a_{Dm} = - i M x_m~, \qquad x_m \in \R \\ 
~ h_{im} = M h_m \delta_{i1}~, \qquad h_m \geq 0 \end{cases} \hskip 1in 
 m = 1, \ldots, N-1~.
\eea
With these restrictions, the scalar potential~$\SV$ in~\eqref{defpot} simplifies considerably. For the case~$M \neq 0$ that will occupy us throughout this paper,\footnote{~To restore supersymmetry, we must take~$M \to 0$ while holding~$a_{Dm}$ and~$h_{1m}$ fixed.} it is very convenient to express the scalar potential~$\SV$ in terms of a dimensionless potential, denoted by $V$,
\bea
 \SV=M^4 V
 \eea
where $V$ is given by,
	 \ba{\bs{
	\label{vscalea}
V= &  \sum_{m=1}^{N-1} \left(-\frac{ N  \Lambda  }{   \pi^2 M} s_m x_m +\frac{1}{2} \left( 4x_m^2- 1\right) h_m^2\right) +  \sum_{m,n=1}^{N-1}\left(  t_{mn} x_{m} x_{n} +    \frac{1}{2}  (t^{-1})_{mn}h_m^2 h_n^2  \right)
	}}
For future reference, it is very useful to parametrize the SUSY-breaking parameter~$M / \Lambda$ in terms of the dimensionless variable
\begin{equation}\label{eq:kappadefsec3}
\kappa = {N \Lambda \over 2 \pi^2 M}~,
\end{equation}
in terms of which the dimensionless scalar potential~$V$ in~\eqref{vscalea} reads
\ba{\bs{
	\label{vscaleawithk}
V= &  \sum_{m=1}^{N-1} \left(-2 \kappa s_m x_m  +\frac{1}{2} \left( 4x_m^2- 1\right) h_m^2\right) +  \sum_{m,n=1}^{N-1}\left(  t_{mn} x_{m} x_{n} +    \frac{1}{2}  (t^{-1})_{mn}h_m^2 h_n^2  \right)
	}}
Note that the small SUSY-breaking regime~$M \ll \Lambda$ corresponds to~$\kappa \to \infty$, while~$\kappa \to 0$ is the large-$M$ regime where we expect to make contact with adjoint QCD.

\subsubsection{Semi-classical analysis of the scalar potential} 	
\label{sec:symah2}	

The analysis of the semi-classical phase structure of the dual Abelian Higgs model will consist of three steps. Throughout, we assume that a value of~$N$ (which determines the UV~$SU(N)$ gauge group), the strong coupling scale~$\Lambda$, and the renormalization scale~$\mu$ (which enters the matrix~$t_{mn}(\mu)$) have been fixed. We only vary the SUSY-breaking mass~$M$, or equivalently the dimensionless parameter~$\kappa$ defined in~\eqref{eq:kappadefsec3}. 
\begin{itemize}
\itemsep=0in
\item[1.)] {\sl Existence} of solutions to the equations for the extrema of the dimensionless potential $V$ in~\eqref{vscalea} and~\eqref{vscaleawithk} as a function of real variables $x_m \in \R$ and non-negative variables $h_m \geq 0$ with $m=1,\ldots, N-1$,
\bea
\label{1aa}
-\kappa s_m  + 2 {h_m^2}  x_m +  \sum_{n=1}^{N-1} t_{mn} x_n & = & 0~, 
\hskip 0.7in \kappa = \frac{\Lambda N}{2\pi^2 M}
	  \\
\label{1ba}
h_m \left ( 4 x_m^2-1 + 2 \sum_{n=1}^{N-1} (t^{-1})_{mn} h_n^2 \right ) & = & 0 
\eea
For a given~$M$ (equivalently~$\kappa$), there may exist several solutions, whose stability must then be analyzed. 
	
\item[2.)] {\sl Local stability} of the solutions obtained in item 1.) above requires positivity of the Hessian matrix $\CH$ of second derivatives of $V$ in \eqref{vscalea},
	\ba{
	\CH = \left(  \begin{array}{cc}  \CH_{xx} & \CH_{xh} \\ \CH_{hx} & \CH_{hh}  \end{array}    \right)
	}
Here each block is an $(N-1)\times (N-1)$ matrix, whose components are given as follows,
\bea
\label{hess}
(\CH_{xx})_{ mn }&= & \frac{\partial^2 V}{\partial x_m \partial x_n } 
= 4h_m^2  \delta_{mn}   +2 t_{mn}
 \\
(\CH_{hx})_{ mn}&= &  \frac{\partial^2 V}{\partial h_m \partial x_n} 
= 8  x_m h_m \delta_{mn} 
\no \\
(\CH_{hh})_{mn}&= & \frac{\partial^2 V}{\partial h_m \partial h_n} 
= \delta_{mn} \left(  4 x_m^2 - 1+ \sum_{\ell=1}^{N-1} 2(t^{-1})_{m \ell} h_\ell^2 \right )+ 4  (t^{-1})_{mn} h_m h_n 
\no
\eea
The Hessian matrix is symmetric so that $\cH_{xh}= (\cH_{hx})^t$. Local stability, i.e.~positivity of the Hessian, only retains those solutions found in item~1.) that are free of tachyons.

\item[3.)] {\sl Global stability} of a solution satisfying the conditions in items 1.) and 2.) above must be decided by evaluating the potential $V$ on the solution. The globally stable solution -- and thus the ground state of the dual Abelian Higgs model in the semi-classical approximation -- is always the one with the lowest value of~$V$. If there is a unique solution to 1.) and 2.), it is automatically globally stable (since~$V$ is bounded from below), but in general there are multiple branches of locally stable solutions. Assessing global stability must generally be done numerically, since the full solutions are typically not available analytically.

\end{itemize}


\subsubsection{Recovering the Coulomb branch vacuum at the origin for~$M \ll \Lambda$} \label{sec:dualsmallM}

An important check of the dual Abelian Higgs model is whether it correctly reproduces the small-$M$ regime, because that regime was reliably analyzed in section~\ref{sec:smallm}. There we found that the theory is in a Coulomb phase at the origin~$u_I =0$ of the Coulomb branch, with~$N-1$ massless photons and gaugino~$SU(2)_R$ doublets, but no massless scalars.

In order to study the small-$M$ regime in the dual Abelian Higgs model, we must first solve the equations~\eqref{1aa} and~\eqref{1ba} for small~$M \ll \Lambda$, or equivalently for large~$\kappa$. It is easy to see that these equations always admit a Coulomb vacuum, with all~$h_m = 0$ and~$x_m$ given by
\begin{equation}\label{eq:xcb}
x_m = \kappa \sum_{n=1}^{N-1}(t^{-1})_{mn} s_n~, \qquad h_m = 0~.
\end{equation}
What is much less obvious, but true (as we will show in section~\ref{sec:largek}), is that this is the only stable solution when~$\kappa$ is sufficiently large, i.e.~when~$M$ is sufficiently small. 

With this in mind, we can check whether the solution~\eqref{eq:xcb} matches our expectations about the small-$M$ Coulomb vacuum at the origin established in section~\ref{sec:smallm}:
\begin{itemize}
\item Since all~$h_m = 0$, the~$U(1)_D^{N-1}$ gauge symmetry is not Higgsed and there are~$N-1$ massless photons. 

\item We see from the Yukawa couplings~\eqref{eq:lyukbis} of the dual Abelian Higgs model that the~$N-1$ gauginos~$\rho_m^i$ are massless, because all~$h_m = 0$. 

\item We see from the Hessian~\eqref{hess}, or directly from the full scalar potential~\eqref{defpot}, that the scalars~$a_{Dm}$ have positive-definite mass matrix~$M^2 t_{mn}$, i.e.~none of them are massless. 

\item The hypermultiplet scalars~$h_{im}$ (and their fermionic superpartners~$\psi_m^\suppm$)  acquire masses reminiscent of the BPS mass formula,
\be
\label{3.bpsdual}
(\text{hypermultiplet mass})  = \sqrt 2 |a_{Dm}| = {\sqrt 2 \Lambda N \over 2 \pi^2} \sum_{n=1}^{N-1} (t^{-1})_{mn} s_n~.
\ee
It can be checked numerically  that this formula is in good agreement with the exact BPS mass formula~\eqref{2.MBPS} at the origin of the Coulomb branch of the~$\CN=2$ theory. The agreement is most striking (and simplest to deduce analytically) in the large-$N$ limit of~\eqref{3.bpsdual}, which can be evaluated using~\eqref{useful},
\be
 \sqrt 2 |a_{Dm}| \rightarrow \sqrt 2 \Lambda s_m \qquad \text{as} \qquad N \to \infty~.
\ee
This is in perfect agreement with the large-$N$ BPS masses at the origin~\eqref{eq:BPSmasses3} in the same limit. Moreover, the hypermultiplets from all~$N$ multi-monopole points all have the same spectrum, and thus effectively restore the unbroken~$\Z_{4N}$ symmetry at the origin (which is not manifest in the dual). 

\end{itemize}

\noindent We see that the dual gives an excellent description of the small-$M$ Coulomb vacuum at the origin, modulo the fact that it obscures the unbroken~$\Z_{4N}$ symmetry there. We take this as encouragement to analyze the phases of the dual for all values of~$M$, the ultimate goal being adjoint QCD in the large-$M$ limit.

\subsubsection{Simplifications for maximal Higgs branches}
\label{sec:red}

We collect here some general simplifications that will be useful when analyzing maximal Higgs branches, which we define to be solutions of~\eqref{1aa} and~\eqref{1ba} for which all~$h_m \neq 0$. 
\sm

For any maximal Higgs branch the system of equations (\ref{1ba}) is solved as follows,
\bea
\label{3.maxH}
 h_m^2 = \half \sum_{n=1}^{N-1} t_{mn} (1-4x_n^2)
\eea
We note that a sufficient condition for the existence of solutions to this equation is given by $4x_m^2<1$ for all $m=1,\ldots, N-1$, but this condition is not necessary since the off-diagonal elements of $t^{-1}$ are negative. 

\sm

As a result of~\eqref{3.maxH}, the Hessian matrix~\eqref{hess} simplifies, and we have,
	\ba{
	\CH = \left(  \begin{array}{cc}  4 h_m^2 \delta_{mn}  + 2 t_{mn} & 8 x_m h_m \delta_{mn} \\  
	8 x_m h_m \delta_{mn} & 4 (t^{-1})_{mn} h_m h_n  \end{array} \right)
	}
The associated quadratic form $Q$ in the variables $\alpha_m,\beta_m\in \mathbb{R}$ is given by,
	\ba{
	Q = 
	\sum_{m,n=1}^{N-1} \Big (  
	\left( 4 h_m^2 \delta_{mn}  + 2 t_{mn}  \right)\alpha_m \alpha_n +  4 (t^{-1})_{mn} h_m h_n \beta_m \beta_n + 16 x_m h_m \delta_{mn}  \alpha_m \beta_n  \Big )
	}
Positive definiteness of the Hessian is equivalent to positive definiteness  of the quadratic form $Q$. We shall now reduce the criterion for positivity of $Q$ to a simplified criterion in half as many variables. To do so, we change variables from $\beta_m$ to $\gamma_m$ using the relation $2 h_m \beta_m=\sum_n t_{mn} \gamma_n$, and express $h_m^2$ in terms of $x_m$ using \eqref{3.maxH}. In terms of the variables $\alpha_m$ and $\gamma_m$, $Q$ then reduces as follows, 
\bea
\label{QQu}
Q &= & Q_\alpha + \sum_{m,n=1}^{N-1} t_{mn } (\gamma_m + 4 x_m \alpha_m) (\gamma_n + 4 x_n \alpha_n)
\no	\\
Q_\alpha &= & \sum_{m,n=1}^{N-1} t_{mn} \Big (   (1-4x_m^2) \alpha_n^2 + (1-4x_n^2) \alpha_m^2 
+ 2(1-8x_m x_n) \alpha_m \alpha_n    \Big )
\eea
In view of the positive definiteness of $t$, the quadratic form $Q-Q_{\alpha}$ is positive definite in $\gamma$ for arbitrary $\alpha$, and vanishes if and only if $\gamma_m=-4x_m\alpha_m$ for all $m=1,\ldots, N-1$. The remaining quadratic form $Q_\alpha$ depends only on the variables $\alpha$, and  positive-definiteness of the Hessian $\CH$ is equivalent to positive-definiteness of $Q_\alpha$, a simpler problem that will considerably facilitate the analysis of local stability for maximal Higgs branches.

\newpage

\section{Phase structure for  $SU(2)$ gauge group} 
\label{sec:su2}

In this section, we present an analysis of the dual Abelian Higgs model with soft supersymmetry breaking  for the case $N=2$ at the classical level, and use the results to infer the semi-classical phase structure of  the theory with $SU(2)$ gauge group first studied in \cite{Cordova:2018acb}. This is an essential prerequisite for the much more involved analysis of the saddle point equations \eqref{1aa}-\eqref{1ba}, the local stability conditions on the Hessian~\eqref{hess}, and global stability for the case~$N \geq 3$ that will occupy us for much of the paper.
\sm

Drastic simplifications take place when $N=2$:  the matrix $t_{k\ell}$ of effective~$U(1)^{N-1}$ gauge couplings has a single, positive entry $t_{11} > 0$,\footnote{~As explained around~\eqref{eq:tposmu}, this holds as long as~$\mu \leq \Lambda$, which we assume.} and vacuum alignment is automatic as there is only one hypermultiplet field.  Nevertheless, the phase structure is nontrivial, as we now review.  

\sm

We parameterize~$t_{11}$ in terms of the (dual magnetic) gauge coupling~$e$, as part of the following dimensionless variables appropriate to the~$N=2$ case,
	\ba{
	t_{11} = e^{-2} > 0\ ,\qquad x = x_1 \in \R \ ,\qquad h = h_1 \geq 0 \ ,\qquad s_1=1
	}
The dimensionless potential $V$ in \eqref{vscalea} then gives
	\ba{
	\label{pot2}
	V= - \frac{2  x \Lambda}{\pi^2 M} + \frac{x^2}{e^2} + \frac{1}{2} h^2 \left( 4x^2-1\right) + \frac{e^2 h^4}{2}~,
	}
so that the saddle point equations \eqref{1aa}-\eqref{1ba} reduce to
	\ba{
	\label{su2e}
		  { x}+ 2e^2{h^2}  x - \frac{ e^2\Lambda}{\pi^2 M}  = 0\ ,   \qquad h\left( 4 x^2-1 + 2 e^2 h^2 \right)= 0~,
	}
	and the Hessian $\CH$ in \eqref{hess} takes the form
	\ba{
	\label{hess2}
	\CH = \left( \begin{array}{cc} 2/e^2 + 4  h^2 & 8   h x \\ 8  h x & 4 x^2-1 + 6 e^2 h^2 \end{array}\right)
	}
To analyze the saddle point equations and stability conditions, we consider the branches $h=0$ (the Coulomb branch) and $h \not= 0$ (the Higgs branch) separately.\footnote{~Recall that~$h \geq 0$, so that~$h \neq 0$ implies~$h >0$.}

\subsection{The Coulomb branch (CB) with $h=0$} 
\label{sec:su2cb}

The saddle point equations (\ref{su2e}) always admit a solution with vanishing Higgs vev, $h = 0$, which we refer to as the Coulomb branch (CB). This solution exists for all values of the SUSY-breaking mass~$M$, and the first equation in~(\ref{su2e}) fixes the vev~$x$ of the vector-multiplet scalar as follows, 
	\ba{
	\label{xcb2}
	x = \frac{ e^2 \Lambda}{\pi^2 M}
	}	
The Hessian \eqref{hess2} is diagonal since $h=0$; the entry~$\CH_{xx}$ is always positive, while positivity of $\CH_{hh}$ requires $4x^2 > 1$. Thus, even though the Coulomb branch solution of the saddle point equations exists for all~$M$, it is only locally stable provided that~$x>x_{\text{CB}}=\frac{1}{2}$, which translates into the following upper bound for~$M$, 
	\ba{
	\label{mcb2}
	M < M_{\text{CB}} \hskip 1in M_{\text{CB}}  = \frac{2e^2 \Lambda }{\pi^2}
	}
When~$M > M_{\text{CB}}$, the Coulomb branch is not locally stable, i.e.~there are tachyons. It follows that there must be a phase transition (denoted by~$*$) to the Higgs branch that must occur for some~$M = M_* \leq M_{\text{CB}}$; as we will see below, the inequality turns out to be strict, $M_* < M_\text{CB}$, so that the transition occurs before the Coulomb branch becomes locally unstable. 

\begin{figure}
  \centering
\begin{tikzpicture}[scale=7.5]
\draw[->,line width=0.9] (0, 0) -- (0.9, 0) node[right] {\Large $x$};
\draw[->,line width=0.9] (0, 0) -- (0, 0.9) node[left] {\Large ${e^2\Lambda\over \pi^2 M}$};
\draw(0, 0.53) node[left] {\Large ${e^2\Lambda\over \pi^2 M_{\text{*}}}$};
\draw[scale=1, domain=0:0.5, densely dotted, variable=\x, red,line width=1] plot ({\x}, {\x});
\draw[scale=1, domain=0.5:0.89, smooth, variable=\x, red,line width=1.3] plot ({\x}, {\x});
\draw[scale=1, domain=0:0.408248, smooth, variable=\x, blue,line width=1.3] plot ({\x}, {2*\x-4*\x*\x*\x});
\draw[scale=1, domain=0.408248:0.5, densely dotted, variable=\x, blue,line width=1] plot ({\x}, {2*\x-4*\x*\x*\x});

  \draw[line width = 0.4,densely dotted] (-0.1,0.35) -- (0.9,0.35) node[right] {Higgs branch (HB): $M > M_*$};
  \draw[line width = 0.4,densely dotted] (-0.1,0.75) -- (0.9,0.75) node[right] {Coulomb branch (CB): $M < M_*$};
   \draw[line width = 0.4,densely dotted] (-0.03,0.53) -- (0.9,0.53) node[right] {Phase transition (first order)};
      \draw[line width = 0.4,gray,densely dotted] (1/2,0) -- (1/2,0.5);
      
    \draw (0.9,-0.15) node{\large  $ x_{\text{CB}} = \half $};
    \draw (1/2,-0.03) -- (1/2,0.03);
       \draw [->] (0.75, -0.12) -- (0.51, -0.04);
       
     \draw (0.45,-0.25) node{\large $ x_{\text{HB}} = {1 \over \sqrt 6}$};
         \draw (0.408248,-0.03) -- (0.408248,0.03);
         \draw[line width = 0.4,gray,densely dotted] (0.408248,0) -- (0.408248,0.54);
         \draw [->] (0.423, -0.2) -- (0.41, -0.05);
         
             \draw (0.02,-0.15) node{\large $x_{*} = {1 \over \sqrt 8}$};
         \draw (0.353553,-0.03) -- (0.353553, 0.03);
          \draw[line width = 0.4,gray,densely dotted] (0.353553,0) -- (0.353553,0.53);
          \draw [->] (0.18, -0.10) -- (0.34, -0.04);
         
         \draw[black,fill=blue] (0.19,0.35) circle (.06ex);
         \draw[black,fill=red] (0.75,0.75) circle (.06ex);
         \draw[ red] (0.7, 0.6) node{\large $x$};
         \draw[blue] (0.27, 0.63) node{\large $2x-4x^3$};
         \draw[black,fill=black] (0.353553,0.53) circle (.06ex);
         \draw[black,fill=black] (0.53,0.53) circle (.06ex);
         

\end{tikzpicture}
\caption{\label{fig:cubic} Plot of the dimensionless vector multiplet scalar vev~$x$ (horizontal axis) against~$e^2 \Lambda / \pi^2 M$ (vertical axis). As explained in the text, $M > 0$ implies that all solutions of the saddle point equations have~$x > 0$, so that we can restrict to the first quadrant. The~$h = 0$ Coulomb branch (CB), shown in red, is the graph of the function~$x$. The CB always exists, and is locally stable (indicated by the solid red line) when~$x>x_{\text{CB}} = \half$; the region of local instability is indicated by the dotted red line.   The~$h \neq 0$ Higgs branch (HB), shown in blue, is the graph of the function $2x - 4 x^3$; it only exists when~$0 < x < x_\text{CB} = \half$ (the two endpoints touch the Coulomb branch), and is locally stable (solid blue line) when~$0 < x < x_{\rm HB}=\frac{1}{\sqrt{6}}$; the region where the HB exists but is locally unstable is indicated by the dotted blue line. For~$M < M_*$, the CB is the globally stable vacuum (an example is the red dot intersecting a horizontal dotted black line), and for~$M > M_*$ the HB is the globally stable vacuum (e.g.~the blue dot intersecting a horizontal dotted black line). At~$M = M_*$ there is a first order phase transition between the two branches, where~$x$ jumps discontinuously from its Coulomb branch value~$e^2 \Lambda / \pi^2 M_* > x_\text{CB}= \half $ (indicated by the black dot on the solid red curve) to the strictly smaller Higgs branch value~$x_* = {1 \over \sqrt 8}$ (indicated by the black dot on the solid blue curve). }
\end{figure}
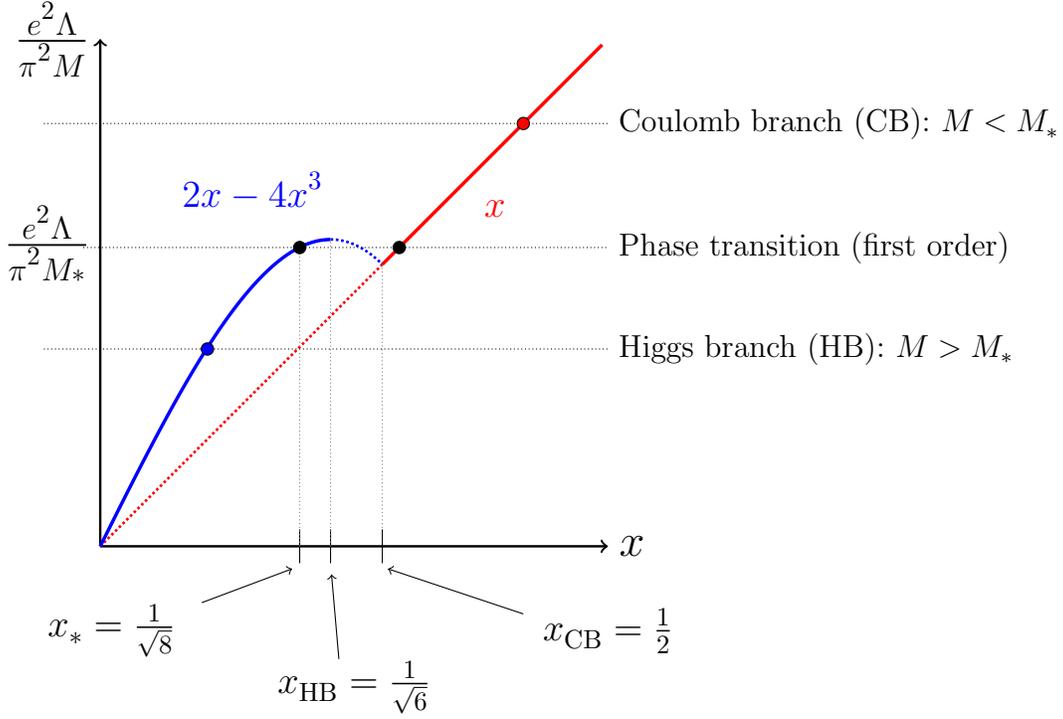

The Coulomb branch solution \eqref{xcb2} is indicated by the red line of unit slope in figure~\ref{fig:cubic}, where~$x > 0$ is plotted on the horizontal axis and~${e^2 \Lambda / (\pi^2 M)} > 0$ is plotted on the vertical axis.\footnote{~In principle, $x \in \R$, but~$M > 0$ implies that all solutions to the saddle point equations also have~$x > 0$.} The line is solid red in the region of local stability~$x > x_\text{CB} = \half$, and dotted red in the region $0 < x< x_{\text{CB}}$ where the solution exists but is locally unstable.

In order to analyze the global stability of the Coulomb branch relative to the Higgs branch (see below), we will need to know the value of the dimensionless potential~$V$ in~\eqref{pot2} evaluated on the Coulomb branch solution~\eqref{xcb2}, 
	\ba{
	\label{pe2a}
	V_{\text{CB}} =  - \frac{e^2 \Lambda^2}{\pi^4 M^2 } \hskip 1in M < M_{\text{CB}}
	}

\subsection{The Higgs branch (HB) with~$h\neq 0$} 
\label{sec:su2hb}

We refer to solutions of the saddle point equations~\eqref{su2e} with $h\neq 0$ as the Higgs branch~(HB). On this branch, the second equation in~\eqref{su2e} gives
	\ba{
	\label{hs2}
	h^2 = \frac{1}{2e^2} (1-4x^2)
	}
Since $h > 0$ and~$e> 0$, such solutions require $4x^2 < 1$, or equivalently $| x| < x_\text{CB} = \thalf$.   Substituting~\eqref{hs2} into the first equation in~\eqref{su2e} then gives a cubic equation for $x$,
	\ba{
	\label{cubic1}
	2x- 4 x^3 = \frac{ e^2\Lambda}{\pi^2 M}  \hskip 1in | x| < x_\text{CB} = \half
	}
Given that~$M > 0$ and the restriction on~$x$, it follows that all solutions are positive, $x > 0$. 
	
The graph of~$2x-4x^3$ is represented by the  blue curve in figure~\ref{fig:cubic}, where we restrict to the interval~$0 < x < x_\text{CB} = \half$ for reasons explained above. Next, we turn to analyzing the conditions for the existence, local stability, and global stability of the HB solutions. 

\subsubsection{Existence of HB solutions} 

The solutions to the cubic equation \eqref{cubic1} may be obtained graphically by intersecting the blue curve in figure~\ref{fig:cubic} (the graph of its left-hand side~$2x - 4 x^3$) by a horizontal line with intercept $e^2\Lambda/({\pi^2 M})$ (its right-hand side). As already explained around~\eqref{cubic1}, all physical HB solutions lie in the interval~$0<x<x_{\text{CB}} = \frac{1}{2}$. In this region, the function $2x - 4 x^3$ attains its unique maximum at $x_{\text{HB}}=1/\sqrt{6}$.  Thus HB solutions only exist for sufficiently large~$M$, when
\ba{
\label{su2hb}
M > M_{\text{HB}} \hskip 1in M_{\text{HB}} =  \frac{3 \sqrt{6} e^2\Lambda}{4\pi^2}
}

For $M > M_{\text{HB}}$ there are two solutions with $x >0$: one solution lies to the left of the maximum of the cubic, $x < x_{\text{HB}}$, while the other lies to its right, $x > x_{\text{HB}}$. The second solution only satisfies~$x < x_{\text{CB}} = \half$ as long as~$M_\text{HB} < M < M_\text{CB}$,\footnote{~Comparing~\eqref{mcb2} and~\eqref{su2hb}, we indeed see that~$M_\text{HB} < M_\text{CB}$.} and in this interval the CB and HB solutions co-exist.

\subsubsection{Local stability of HB solutions} 

 Local stability requires positivity of the Hessian \eqref{hess2}, which holds provided $\tr (\CH) >0$ and $\det (\CH) > 0$. Upon substituting the expression for $h$ from \eqref{hs2} into $\CH$, we see that the condition $\tr (\CH) > 0$ is automatically satisfied for $0< x < \frac{1}{2}$, while the determinant is given by,
	\ba{
	\det \CH = \frac{8}{e^2} (1-4x^2) (1-6x^2)
	}
Positivity of $\det \CH$ then restricts $0< x < x_{\text{HB}}=1/\sqrt{6}$; the solution in this range is locally stable (indicated by the solid blue curve in figure~\ref{fig:cubic}), while HB solutions with~$x > x_{\text{HB}}$ have tachyons (indicated by the dotted blue curve in figure~\ref{fig:cubic}). Therefore, a single locally stable HB solution exists when~$M > M_{\text{HB}}$, and it is given by the solution to \eqref{cubic1} that satisfies $0<x<x_{\text{HB}}= 1/\sqrt{6}$. The value of the dimensionless potential on the HB solutions\footnote{~Since we are using the field equations, this is an ``on-shell'' potential.} can be expressed as a simple function of~$x$ by substituting $h$ in \eqref{hs2} into the potential \eqref{pot2} and using the cubic~\eqref{cubic1}, 
	\ba{
	\label{vehb2}
V_{\text{HB}} = - \frac{1}{8e^2} \left( 1 + 16 x^2 - 48 x^4 \right) \hskip 1in 0< x < x_{\text{HB}} ={ 1\over \sqrt{6}}
	}

\subsection{Global stability of CB and HB solutions}

To investigate the global stability of the solutions and determine the true vacuum as a function of~$M$, we first summarize the existence and local stability properties of the Coulomb and Higgs branch solutions:
\begin{itemize}
\itemsep=0in
\item For $0 < M < M_{\text{HB}}$ only the CB solution with~$h = 0$ exists. It is therefore automatically locally stable (as verified above) and globally stable. 
\item For $M > M_{\text{CB}}$ the only solution that exists and is locally stable is the HB solution with $h \not=0$, which is therefore necessarily also globally stable. 
\item  In the interval~$M_{\text{HB}} < M < M_{\text{CB}}$ two locally stable solutions co-exist: a CB with~$h = 0$ and a HB with~$h \neq 0$ and~$x < x_\text{HB}$. 
Their global stability, as well as the phase transition between them, is determined by comparing the values of the potentials, to which we now turn.
\end{itemize}

\smallskip
\noindent In the coexistence region, we can use~\eqref{cubic1} to express the difference between the vacuum energies \eqref{pe2a} and \eqref{vehb2} in terms of the value of~$x$ on the Higgs branch,
	\ba{
	\label{deltavsu2}
	V_{\text{HB}} - V_{\text{CB}} = - \frac{1}{8e^2} (1-8x^2) (1-4x^2)^2
	\hskip 1in 2x- 4 x^3 = \frac{ e^2\Lambda}{\pi^2 M}
	}
This difference has a double zero at the upper end~$M = M_\text{CB}$ of the coexistence window, where~$x = x_\text{CB} = \half$ and the two branches touch, and another zero at the phase transition point~$M = M_*$ where
	\ba{
	\label{mss2}
	M_* = \frac{4 \sqrt{2} e^2\Lambda}{3\pi^2} \hskip 1in  x = x_* = \frac{1}{\sqrt{8}}
	}
Both~$x_* = 1/ \sqrt{8}$ and~$M_*$ are indicated in figure~\ref{fig:cubic}. 

Let us make a few comments about this phase transition:
\begin{itemize}
\item The transition occurs within the coexistence region, where both CB and HB solution are locally stable, as can be seen from
\bea
\label{mmmineq}
{M_{\text{HB}} \over e^2 \Lambda}  \approx 0.1861
~ < ~
{M_{*} \over e^2 \Lambda}  \approx 0.1910
~ < ~ 
{M_{\text{CB}} \over e^2 \Lambda}  \approx 0.2026
\eea

\item At~$M < M_*$ the CB is the true, globally stable vacuum (persisting down to~$M = 0$), while for~$M > M_*$ the HB is the true vacuum (persisting for all larger values of~$M$).  

\item The transition is first order: at the transition point~$M = M_*$ we can use~\eqref{xcb2} and~\eqref{mss2} to evaluate the discontinuous jump in~$x$ from CB to HB,\begin{equation}
\Delta x = x_* - {e^2 \Lambda \over \pi^2 M_*} = - {\sqrt 2 \over 8}~.
\end{equation}
The two distinct values of~$x$ at the transition point~$M = M_*$ are indicated by the two black dots in figure~\ref{fig:cubic}: the black dot on the solid blue (locally stable) HB curve indicates~$x_* = {1 \over \sqrt 8}$, while the black dot on the solid red (locally stable) CB curve is at~$x = {e^2 \Lambda \over \pi^2 M_*} = {3 \sqrt 2 / 8}$. Note that the transition occurs strictly before the CB reaches~$x = x_\text{CB} = \half$ and becomes locally unstable, as indicated in~\eqref{mmmineq}. 

\end{itemize}

\subsection{Graphical summary of the~$SU(2)$ phase diagram} 
\label{sec:graphsu2}

\begin{figure}[t!]
\centering
\includegraphics[width=0.52\textwidth]{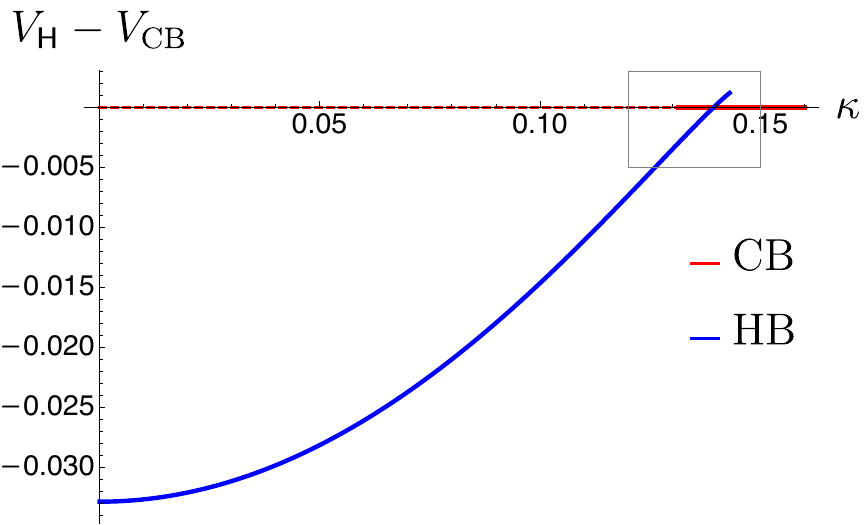} \qquad
\includegraphics[width=0.4\textwidth]{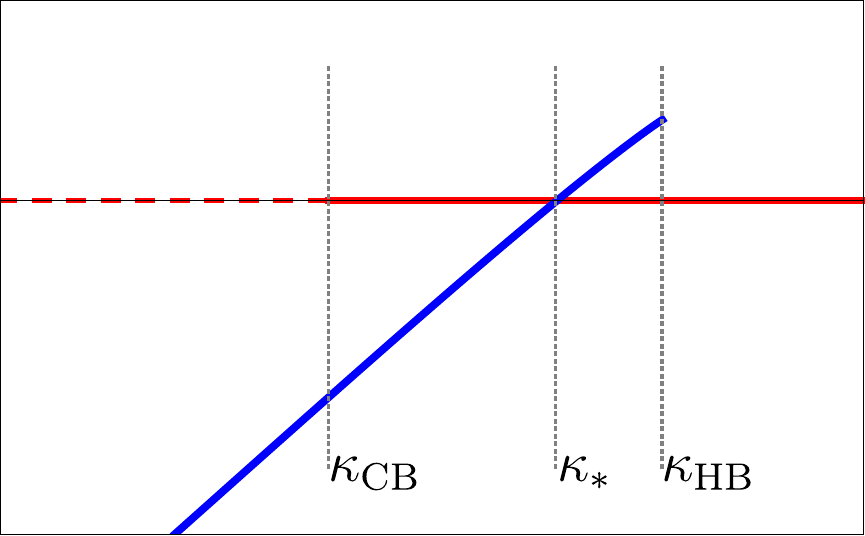}
\caption{In both panels, the dimensionless potential difference~$V_\text{HB}-V_\text{CB}$  is plotted against $\kappa=\Lambda/(\pi^2M)$ for the~$SU(2)$ theory. The locally stable part of the Coulomb branch (CB) is shown in the horizontal, solid red line (at zero relative potential), and the Higgs branch (HB) is shown in solid blue. The right panel zooms in on the boxed area of the left panel. All potentials are evaluated at the renormalization scale~$\mu = 10^{-3}\Lambda$, namely $e^2=3.8057$. \label{fig:su2}}\end{figure}

It is very convenient to have a compact graphical representation of the energetics of the CB and HB of the~$SU(2)$ theory, which in turn determines the phase diagram as a function of~$M$. Indeed, this becomes essential when we generalize to~$SU(N \geq 3)$, where there are many more branches. Throughout, we will adopt the following graphical conventions, which are implemented in figure \ref{fig:su2} for the~$SU(2)$ case:

\begin{itemize}
\item On the vertical axis, we will plot the dimensionless potential~$V$ of each branch, relative to the potential~$V_\text{CB}$ of the Coulomb branch. The CB will thus always be a horizontal line with vanishing intercept (i.e.~it is embedded within the horizontal axis). In~$SU(2)$ the only other branch is the HB, for which we plot the difference~$V_\text{HB} - V_\text{CB}$. The globally stable branch is always the one with the lowest potential. 

\item For any gauge group~$SU(N)$, we take the horizontal axis to be parametrized by the dimensionless variable
\bea\label{eq:kdef}
\kappa = { N \Lambda \over 2 \pi^2 M}
\eea
Plots in this variable are clearer and more concise than those obtained by plotting against $M/\Lambda \sim 1/\kappa$. For the~$SU(2)$ case we set~$N = 2$ and use~$\kappa_\text{CB} < \kappa_* < \kappa_\text{HB}$ corresponding to the values of~$M$ in~\eqref{mmmineq}. 
\item We use solid lines to indicate branches that exist and are locally stable. Different branches are distinguished by their color (e.g. the CB is always red and the HB is always blue). If needed, we will indicate locally unstable portions of a given branch using dotted lines of the appropriate color. 
\end{itemize}

It is now straightforward to infer the globally stable branches, and hence the phase diagram, by reading figure~\ref{fig:su2} from right to left while tracing the lowest-energy branch: 
\begin{itemize}
\item At small~$M$, $\kappa \sim {\Lambda /M}$ is large, and the CB (indicated by the solid red line) is the only stable branch, with zero potential (relative to itself).
\item As we increase~$M$, $\kappa \sim {\Lambda /M}$ decreases. At~$\kappa_\text{HB}$ the HB becomes  locally stable, but $V_\text{HB} - V_\text{CB} > 0$ so that the CB remains the true, globally stable vacuum.
\item At~$\kappa = \kappa_*$ there is a phase transition because~$V_\text{HB} - V_\text{CB}$ changes sign, rendering the HB globally stable. The CB remains locally stable until~$\kappa_\text{CB}$, but has higher energy than the HB. 
\end{itemize}
Thus, we recover the previously deduced semi-classical picture of the phase structure for the~$SU(2)$ theory: 
\bea
M < M_* & \qquad  & \text{Coulomb phase} \hskip 0.2in h=0 
\no \\
M > M_* & \qquad  & \text{Higgs phase} \hskip 0.44in h \not= 0
\eea
For a discussion of the mass spectrum in these phases, we refer to section \ref{sec:11} where the masses are obtained for arbitrary $SU(N)$ gauge group. For the special case of $N=2$ considered here, the massless spectrum is as follows: in the Coulomb phase, there is a massless photon and an~$SU(2)_R$ doublet of massless Weyl fermions; in the Higgs phase, the fact that~$h \neq 0$ leads to a mass for the photon and all fermions, and it also spontaneously breaks~$SU(2)_R \to U(1)_R$, leading to two massless Nambu-Goldstone bosons parameterizing a $\mathbb{CP}^1$ nonlinear sigma model. A detailed discussion of these phases, with an emphasis on symmetries and 't Hooft anomaly matching, can be found in~\cite{Cordova:2018acb}.

\newpage

\section{Phase structure for  $SU(3)$ gauge group} 
\label{sec:su3}

In this section we present a detailed analysis of the semi-classical phase structure for gauge group $SU(3)$. As in the case $N=2$, the results may still be obtained analytically and will provide further valuable information before we proceed to the case of arbitrary $N$. An interesting question that does not arise for~$SU(2)$ gauge group is the realization of charge-conjugation symmetry~$C$. 
\sm

\subsection{Adapted parameterization of~$t_{mn}$}

For $N=3$, the components of the symmetric~$2 \times 2$ matrix $t_{mn}$ (with~$m, n = 1, 2$) satisfy $t_{22}=t_{11}$ due to charge-conjugation symmetry. The remaining two independent components of $t_{mn}$ are then given as follows (see~\eqref{eq:tdef1} and~\eqref{eq:tdef1ii}), 
\ba{
\label{texp3}
t_{11} =t_{22} = \frac{1}{4\pi^2} \left( \log \frac{\Lambda}{\mu}  + \log (18\sqrt{3}) \right)
\hskip 1in 
t_{12}=t_{21} = \frac{1}{4\pi^2} \log 4
	}	
	
As we have $0<t_{12}<t_{11}$ for all $\mu < \Lambda$, we may conveniently (and without loss of generality) parametrize 
the matrix elements of $t$ and $t^{-1}$ in terms of two real gauge couplings~$e_1$ and $e_2$ that satisfy $0<e_1<e_2$,
	\ba{
	\bs{
	\label{tes3}
	t_{11} = t_{22} = \frac{1}{2} \left( \frac{1}{e_1^2} + \frac{1}{e_2^2} \right) > 0
	\qquad
	  (t^{-1})_{11} = (t^{-1})_{22} = \frac{1}{2} \left( e_1^2 + e_2^2\right) >0 \\
	t_{12} = t_{21} = \frac{1}{2} \left( \frac{1}{e_1^2} - \frac{1}{e_2^2} \right) > 0 
	\qquad
	 (t^{-1})_{ 12} = (t^{-1})_{21} = \frac{1}{2} \left( e_1^2 - e_2^2\right) < 0
	}}
Positivity of the matrix $t$ is automatic as both the trace and determinant are positive. The diagonal entries of $t$ and $t^{-1}$ are positive while the off-diagonal elements of $t^{-1}$ are negative. 

\subsection{Taxonomy of different branches}

The semi-classical vacuum solutions to the dual Abelian Higgs model, which solve the system of equations given in (\ref{1aa}) and (\ref{1ba}),  split into different branches according to whether (\ref{1ba}) is solved by setting $h_m=0$ or by setting the second factor to zero, for each value of $m=1,\ldots, N-1$. These $2^{N-1}$ branches correspond to partitions, as will be discussed in detail  for arbitrary $N$ in section \ref{sec:sun}. There we will develop a condensed notation that is useful in dealing with the exponential proliferation of branches for larger~$N$, which is further complicated by the action of charge-conjugation symmetry~$C$. Since the case~$N = 3$ that we are considering here is still fairly tame, we eschew this condensed notation here.

For~$SU(3)$ we have the following branches: 
\begin{itemize}
\item We always refer to the branch~$h_1=h_2=0$ where all~$h_m = 0$ as the Coulomb branch (CB). 

\item We always refer to the branch~$h_1, h_2 >0$ where all~$h_m \neq 0$ as the (maximal) Higgs branch (HB). When desired, this branch can be further subdivided by considering the realization of charge-conjugation~$C$:
\begin{itemize}
\item The~$C$-symmetric Higgs branch (HB$^+$) has~$h_1 = h_2 > 0$. 
\item The~$C$-non-symmetric Higgs branch (HB$^-$) has~both~$h_1, h_2 > 0$ but~$h_1 \neq h_2$. 
\end{itemize}

\item The two branches~$(h_1>0, h_2=0)$ and $(h_1=0, h_2>0)$ are mixed Coulomb-Higgs branches. In general we refer to a branch on which~$p$ Higgs fields are non-vanishing as a~$p$H branch (e.g.~the maximal HB has~$p  = N-1$). Thus $(h_1>0, h_2=0)$ and $(h_1=0, h_2>0)$ are both 1H, or single Higgs branches.  Since these two branches are exchanged by~$C$, it suffices to analyze one of them.
\end{itemize}

\noindent We will now analyze them in turn.

\subsection{The Coulomb branch (CB) with $h_1=h_2=0$}

The relations  $h_1=h_2=0$ trivially solve the equations \eqref{1ba} for all values of $M$. The solution to the remaining equations~\eqref{1aa} is given by,\footnote{~We have left this expression in terms of $s_1=\sqrt{3}/2$ for $N=3$ so that it will be easy to compare with the case of arbitrary $N$ to be investigated in later sections. }
	\ba{
	\label{xcb3}
	x_1=x_2 =\frac{3 e_1^2 s_1 \Lambda }{2 \pi^2 M} 
	}
Thus the CB is automatically~$C$-symmetric.  While this solution always exists, it is not always locally stable. The nonzero blocks of the Hessian \eqref{hess} are $\CH_{xx}$, which is always positive definite, and $\CH_{hh}$, whose positivity requires $4x_1^2 = 4x_2^2 > 1$. Therefore, the Coulomb branch solution is locally stable provided that,
	\ba{
	\label{mcb3}
	M < M_{\text{CB}} \hskip 1in M_{\text{CB}} = \frac{3  e_1^2 s_1\Lambda}{ \pi^2}	} 
The vacuum energy is given by substituting the CB solution~\eqref{xcb3} into~\eqref{vscaleawithk}, 	\ba{
	\label{vcb3}
V_{\text{CB}} = - \frac{9 s_1^2 e_1^2 \Lambda^2 }{2 \pi^4 M^2} 
	}

\subsection{The~$C$-symmetric Higgs branch (HB$^+$) with~$h_1 = h_2 > 0$}

On this branch the set of equations \eqref{1ba} imposes $x_1^2=x_2^2$ while \eqref{1aa} further imposes $x_1=x_2$, so that this branch is actually fully~$C$-symmetric. The reduced saddle point equations  \eqref{1aa} and \eqref{1ba} may be expressed in terms of the reduced variables,
	\ba{
	x = x_1 = x_2 \hskip 1in 
	h = h_1=h_2
	}
which satisfy the following reduced equations,
	\ba{
	\label{cubic3}
	2x- 4 x^3   = \frac{3e_1^2 s_1\Lambda}{2 \pi^2 M}  \hskip 1in 
	h^2 =\frac{1}{2e_1^2}\left( 1- 4 x^2 \right)
	}
These equations are identical to the equations \eqref{hs2} and \eqref{cubic1} for the $N=2$ Higgs branch solution, provided we identify the $N=2$ coupling $e^2$ with the $SU(3)$ coupling $3 e_1^2 s_1/2$ and rescale~$h$. Importing the corresponding results from our analysis of the $N=2$ case around~\eqref{su2hb}, we obtain the following existence conditions for~HB$^+$,
	\ba{
	\label{bound33}
	M > M_{\text{HB}}
	\hskip 1in 
	 M_{\text{HB}}  = \frac{9\sqrt{6} e_1^2 s_1\Lambda}{8\pi^2}
	 	}
or equivalently  $x< x_{\text{HB}} = 1/\sqrt{6}$.

Local stability requires positivity of the Hessian $\CH$. On this maximal Higgs branch we may use the results of section~\ref{sec:red} to reduce the condition of positivity of $\CH$ to the equivalent condition of positivity of the quadratic form $Q_{\alpha}$ in the real variables $\alpha _1, \alpha_2$ given in \eqref{QQu}, which takes the following form,
\ba{
	e_1^2 e_2^2 Q_{\alpha} = \left( e_1^2 + 3 e_2^2 - (8 e_1^2 + 16 e_2^2 ) x^2 \right) (\alpha_1^2 + \alpha_2^2)  + 2 (e_2^2 - e_1^2) (1-8x^2) \alpha_1 \alpha_2
	}
Positivity of $Q_\alpha$ is equivalent to positivity of both the trace and determinant of the matrix corresponding to the quadratic form $e_1^2 e_2^2 Q_\alpha$, which amount to,
	\ba{
	\bs{
	0  &<  (e_1^2 + 3 e_2^2) (1-6 x^2)  + 2 (e_2^2 - e_1^2) x^2 
	\\
	0 &< e_2^2 (1-6 x^2) \big [  (e_1^2 + e_2^2) (1-6 x^2) + 2 (e_2^2 - e_1^2) x^2 \big ]
	}}
Since $e_2^2 > e_1^2 > 0 $, both conditions are manifestly satisfied, and the solutions are therefore  locally stable, throughout their region of existence~$x < x_{\text{HB}}= \frac{1}{\sqrt{6}}$. 

In preparation for the study of global stability we evaluate the potential $V$ of (\ref{vscalea})  on the solution for $h$ given in the second equation of (\ref{cubic3}),  and obtain the following reduced potential in the Higgs branch HB, in the interval $0<x< x_{\text{HB}} = \frac{1}{\sqrt{6}}$, 
	\ba{
	\label{vhball3}
	V_{\text{HB}} = - \frac{1}{4 e_1^2} \left( 1+ 16 x^2 - 48 x^4\right)
	\hskip 1in 
	2x- 4 x^3   = \frac{3e_1^2 s_1\Lambda}{2 \pi^2 M}
	}

\subsection{The~$C$-non-symmetric Higgs branch (HB$^-$) with~$h_1 \neq h_2 > 0$}\label{sec:SU3noCstab}

For this maximal Higgs branch, the saddle point equations~\eqref{1ba} allow us to solve for $h_1$, $h_2$ in terms of $x_1$, $x_2$ and we obtain
\bea
	h_1^2 &= & \frac{1}{2e_1^2} - \frac{x_1^2+ x_2^2}{e_1^2} - \frac{x_1^2-x_2^2}{e_2^2} 
	\no \\ 
	h_2^2 & = & \frac{1}{2e_1^2} - \frac{x_1^2+ x_2^2}{e_1^2} + \frac{x_1^2-x_2^2}{e_2^2} 
\eea
The assumption $h_2\neq h_1$ implies $x_2^2\neq x_1^2$.  The existence of solutions with real $h_1, h_2$  requires the following restriction on the range of $x_1$ and $x_2$,
 	\ba{
	x_1^2 + x_2^2 + \frac{e_1^2}{e_2^2} \,  \big |x_1^2-x_2^2 \big | < \frac{1}{2}
	}
Eliminating $h_1$, $h_2$ from~\eqref{1aa} gives a set of reduced equations for $x_1$ and  $x_2$, which we  express in terms of $x_{\pm}=x_1\pm x_2$. Since $x_2^2 \not= x_1^2$ we may use $x_\pm\neq 0$   to simplify the resulting equations and obtain a relation expressing $x_-^2$ in terms of $x_+^2$,
	\ba{
	\label{xmm}
	x_-^2 = 1 + \frac{e_1^2}{e_2^2} - x_+^2 \left( 1 + 2 \frac{e_1^2}{e_2^2} \right) \quad \text{for}\quad x_+^2 < \frac{e_1^2 + e_2^2}{ 2 e_1^2 + e_2^2  }
	}
Here the inequality ensures that~$x_-^2 > 0$, and it is saturated when~$x_- = 0$. 	We also obtain a reduced equation for $x_+$ alone,
		\ba{
		\label{xpluseq}
		4 e_1^2 (e_1^2 + e_2^2) x_+^3 - (2 e_1^4 + 3 e_1^2 e_2^2 - e_2^4 )x_+ = \frac{  3 s_1\Lambda}{\pi^2 M} e_1^2 e_2^4
		}

We shall now show that the solutions of this type are never locally stable. To do so, we use the fact that positivity of the Hessian is equivalent to positivity of the quadratic form $Q_\alpha$ in \eqref{QQu}, since we satisfy its applicability condition that $h_1,h_2\neq 0$. Positivity of $Q_\alpha$ is equivalent to positivity of both the trace and determinant of the matrix $\CQ$ corresponding to the rescaled quadratic form~$e_1^2 e_2^2 Q_\alpha$. In terms of the variable $x_{+}$, the trace evaluates to
	\ba{
	\tr \CQ = \frac{2}{e_2^2} \left( 4 e_1^2 ( e_1^2 + 2 e_2^2 ) x_+^2 - \left( 2 e_1^4 + 5 e_1^2 e_2^2 + e_2^4 \right)   \right)
	}
where we used \eqref{xmm} to eliminate $x_-$ in favor of $x_+$. Note that~$\tr \CQ$ is a monotonically increasing function of $x_+^2 > 0$. At the upper bound allowed for $x_+$ in \eqref{xmm}, it evaluates to a negative value
\begin{equation}
\tr \CQ   =  {2 e_2^2 (e_1^2 - e_2^2) \over  2 e_1^2 + e_2^2} < 0~,
\end{equation}
because~$e_1 < e_2$. As a result, the trace is always negative in the region \eqref{xmm}, and the~HB$^-$ solutions with $h_1 \neq h_2 > 0$ are always locally unstable. Therefore, we shall not consider this branch any further.

\subsection{The  single Higgs (1H) branch with~$h_1=0, h_2> 0$}

Solutions for which one $h_m$ is nonzero while the other vanishes spontaneously break charge conjugation symmetry $C$. Since the solutions with $(h_1=0,h_2> 0)$ and $(h_1> 0,h_2=0)$ are exchanged by $C$, we restrict attention to the former without loss of generality. The set of equations \eqref{1ba} are solved by,
	\ba{
	h_1=0\ ,\qquad h_2^2 = \frac{1-4x_2^2}{e_1^2 + e_2^2} 
	}           
which has real solutions for $4x_2^2 < 1$. Eliminating $h_2$ in~\eqref{1aa} we obtain $x_1$ in terms of $x_2$,
	\ba{
	\label{xxx}
	x_1 = \frac{1}{e_1^2+ e_2^2} \left(   \frac{3 \Lambda}{\pi^2 M} s_1e_1^2 e_2^2    - (e_2^2 - e_1^2) x_2 \right)
	}
and a cubic for the remaining variable $x_2$, 
\ba{
\label{xxx1}
2x_2- 4 x_2^3  =  \frac{3 e_1^2 s_1\Lambda }{2 \pi^2 M} 
}	
The cubic in $x_2$ is precisely the cubic \eqref{cubic3} encountered in the case $h_1=h_2 \neq 0$ upon setting $x=x_2$. Thus, the bound $M > M_{\text{HB}}$ given in  \eqref{bound33} for the existence of the solution applies. 

\sm

To analyze local stability of the solution, we can no longer use the reduced stability conditions of subsection \ref{sec:red} because we are not considering a maximal Higgs branch. Instead,  we shall directly investigate the positivity of  the full Hessian,
	\ba{
	\label{hess33}
	\CH = \left( \begin{array}{cccc}  2 t_{11} & 2 t_{12} & 0 & 0 \\ 2 t_{12} & 2 t_{22} + 4 h_2^2 & 0 & 8 h_2 x_2 \\ 0 & 0 &  (\CH_{hh})_{11} & 0 \\ 0 & 8 h_2 x_2 & 0 & 4 (t^{-1})_{22} h_2^2 \end{array}  \right)
	}
where $(\CH_{hh})_{11} = 4 x_1^2 - 1 + 2 (t^{-1})_{12} h_2^2$ may be expressed entirely in terms of $x_2$ by eliminating $x_1$ with the help of  \eqref{xxx}, to obtain,
	\ba{
	\label{chh11}
	(\CH_{hh})_{11} = \frac{2e_2^2 (1 - 4 x_2^2)}{(e_1^2 + e_2^2)^2}\left( - 32 e_2^2 x_2^4 + 8 e_1^2 x_2^2 + 16 e_2^2 x_2^2 - e_1^2 - e_2^2\right)
	}
 The entry $(\CH_{hh})_{11}$ decouples from the other entries in the Hessian and must be positive by itself. Given the reality condition $4x_2^2<1$ derived earlier,  positivity of $(\CH_{hh})_{11}$ requires,
	\ba{
	\label{xcc}
x_{\text{1H}} < x_2 < \half 
\hskip 1in 	x_{\text{1H}}^2 = \frac{1}{8e_2^2} \left( 2e_2^2 + e_1^2 - \sqrt{e_1^4 + 2 e_1^2e_2^2 + 2e_2^4} \right) 
	} 
Here the subscript 1H indicates the single Higgs branch we are considering. Applying the Sylvester criteria for positivity of the remaining $3\times 3$ reduced matrix $\CH$, we see that the upper left entry and the determinant of the upper left $2\times 2$ matrix are automatically positive. This leaves the remaining condition that the determinant of the $3 \times 3$ matrix be positive,
	\ba{
	\text{det} \left(   \begin{array}{ccc} 2 t_{11} & 2 t_{12} & 0 \\ 2 t_{12} & 2 t_{22} + 4 h_2^2 & 8 h_2 x_2 \\ 0 & 8 h_2 x_2 & 4 (t^{-1})_{22} h_2^2  \end{array}   \right)  = \frac{16}{e_1^2 e_2^2} (1-4x_2^2 ) (1-6x_2^2) > 0
	}
Given the reality condition $4 x_2^2 <1$, the positivity of the determinant reduces to the condition $6x_2^2 < 1$. It may be readily verified that $6x_{\text{1H}}^2 < 1$ for all values of $e_1^2$ and $e_2^2$, with $x_{\text{1H}}$ defined in \eqref{xcc}. Therefore, the window of local stability for the $h_1=0, h_2\neq 0$ solution is given by,
		\ba{
		x_{\text{1H}} < x_2 < x_{\text{HB}}= \frac{1}{\sqrt{6}}
		}
Equivalently, since the curve $2x_2-4x_2^3 $ is monotonically increasing in the interval $[0,\frac{1}{\sqrt{6}}]$ (see figure \ref{fig:cubic}), local stability imposes the following conditions on $M$, 
	\ba{
	\label{mcc}
	M_{\text{HB}} < M < M_{\text{1H}} \hskip 1in
	 2x_{\text{1H}} - 4 x_{\text{1H}}^3 = \frac{3 e_1^2 s_1 \Lambda}{2\pi^2 M_{\text{1H}}}
	}
	where $M_{\text{HB}}$ is given in \eqref{bound33}. In particular, $M_{\text{1H}}$ is a monotonically decreasing function of $e_1^2/e_2^2$ whose minimum value  is realized as $e_1^2 \to e_2^2$ and coincides with the upper bound for local stability of the Coulomb branch solution $M_{\text{CB}}$, given in \eqref{mcb3}, while its maximum value is realized at $e_1^2\to 0$, 
	\ba{
	\label{mcbound}
M_{\text{CB}}	< \ M_{\text{1H}} \ <   \frac{3\sqrt{4 - 2 \sqrt{2}} \, e_1^2 s_1\Lambda}{\pi^2}  
	}
We conclude that the region of existence and local stability of the 1H branch at least consists of the range $M_{\text{HB}} < M < M_{\text{CB}}$ where both the Coulomb branch and the maximal Higgs branch exist and are locally stable, and at most extends slightly beyond this range to the larger value of $M$ given in \eqref{mcbound}. 

\sm

To investigate global stability in the next subsection, we will need the value of the potential evaluated on the 1H solution, which is readily evaluated as follows,
	\ba{
	\label{vh13}
V_{\text{1H}} =  -   \frac{128 e_2^2 x_2^6-16(8e_2^2+3e_1^2) x_2^4 + 16 (2e_2^2 + e_1^2)x_2^2 + e_1^2}{ 4 e_1^2(e_1^2+e_2^2)}  
	}
where $x_2$ is given in terms of $M$ by  \eqref{xxx1}.

\subsection{Global stability of~CB, 1H, and HB branches}

In this subsection we carry out the analysis of the global stability of the locally stable branches: the Coulomb branch (CB), single Higgs (1H) branch, and the~$C$-symmetric maximal Higgs branch (HB$^+$). It will be useful to recall the ordering of the various thresholds in $M$,
	\ba{
	M_{\text{HB}}   < M_{\text{CB}} <  M_{\text{1H}}
	}
as well as the results on existence and local stability  established above:
\begin{itemize}
\itemsep=0in
\item The Coulomb branch is locally stable for $M \! < \! M_{\text{CB}}$ with $M_{\text{CB}}$ given in \eqref{mcb3}.
\item The $C$-symmetric Higgs branch is locally stable for $M_{\text{HB}} \! < \! M$ with $ M_{\text{HB}}$ given in \eqref{bound33}.
\item  The single Higgs  branch  is locally stable for $M_{\text{HB}} \! < \! M \! < \!  M_{\text{1H}}$ with $M_{\text{1H}}$ given in \eqref{mcc}. 
\end{itemize}

\subsubsection{Coulomb versus maximal Higgs branch}

In the window $M_{\text{HB}} < M < M_{\text{CB}}$, both the Coulomb and $C$-symmetric Higgs branch exist and are locally stable. To compare the values of the potential in these branches we use equations \eqref{vcb3} and \eqref{vhball3} and  express the potential for the Coulomb branch for a given value of $M$ in terms of the value $x$ corresponding to $M$ in the Higgs branch, as given in \eqref{cubic3} and as we did for the $N=2$ case. The result for their difference is as follows,
	\ba{
	\label{vcbhb}
	V_{\text{HB}} - V_{\text{CB}} = - \frac{1}{4e_1^2} (1-4x^2)^2 (1-8x^2)
	}
The transition point is at $x_* = \frac{1}{\sqrt{8}}$ and corresponds to a value of $M$ given by
\ba{
M_* = \frac{2 \sqrt{2} e_1^2 s_1 \Lambda}{\pi^2}
\hskip 1in  
M_{\text{HB}} < M_* < M_{\text{CB}}
	}
	For $M < M_*$ we have $V_{\text{CB}} < V_{\text{HB}}$ so that the Coulomb branch has lower energy, and for $M > M_*$ we have $V_{\text{CB}} > V_{\text{HB}}$ so that the maximal Higgs branch has lower energy.
	
\subsubsection{Maximal versus single Higgs branches}
	
In the window  $M_{\text{HB}} \! < \! M \! < \!  M_{\text{1H}}$ both the maximal and single Higgs branches exist and are locally stable. To compare the potentials $V_{\text{1H}}$ given in \eqref{vh13}  and $V_{\text{HB}}$  given in \eqref{vhball3}, we express both in terms of $x=x_2$ since the relations between $M$ and $x$ in \eqref{cubic3} and $M$ and $x_2$ in \eqref{xxx1} are identical, and we obtain,  
	\ba{
	\label{vh1hb}
	V_{\text{1H}} - V_{\text{HB} }= \frac{ e_2^2}{4 e_1^2 (e_1^2 + e_2^2) } (1-4x^2)^2 (1-8x^2)
	}
	We conclude that for $x < x_*=\frac{1}{\sqrt{8}}$, the maximal Higgs branch has lower energy than the single Higgs branch. Since, in this range of $M$,  the maximal Higgs branch also has lower energy than the Coulomb branch by \eqref{vcbhb}, it is the globally stable branch for all $M > M_*$.

\subsubsection{Coulomb versus single Higgs branches}	
	
In the window $M_{\text{HB}} < M < M_{\text{CB}}$, both the Coulomb and the single Higgs branches are stable. The difference of their potentials is obtained from \eqref{vcb3} and \eqref{vh13}, where $M$ is eliminated in the formula for the Coulomb branch in favor of $x_2$ using the relation (\ref{xxx1}), and we find, 
		\ba{
		V_{\text{1H}} - V_{\text{CB}} = - \frac{1}{4 (e_1^2 + e_2^2) } (1-4x_2^2)^2 (1-8x_2^2)
		}
For $x_2 < x_*$, we have already established in the preceding subsections that the maximal Higgs branch is the globally stable solution. The above formula shows that for $x_2 > x_*$ we have $V_{\text{1H}} > V_{\text{CB}}$, so that the Coulomb branch is globally stable in this range. Therefore, the single Higgs branch is nowhere globally stable, except exactly at~$x = x_*$ where the three branches exactly cross. This is manifest in the graphical representation plotted in figure~\ref{fig:su3} below.

\subsection{Graphical summary of the~$SU(3)$ phase diagram} 

\begin{figure}[t!]
\centering
\includegraphics[width=0.53\textwidth]{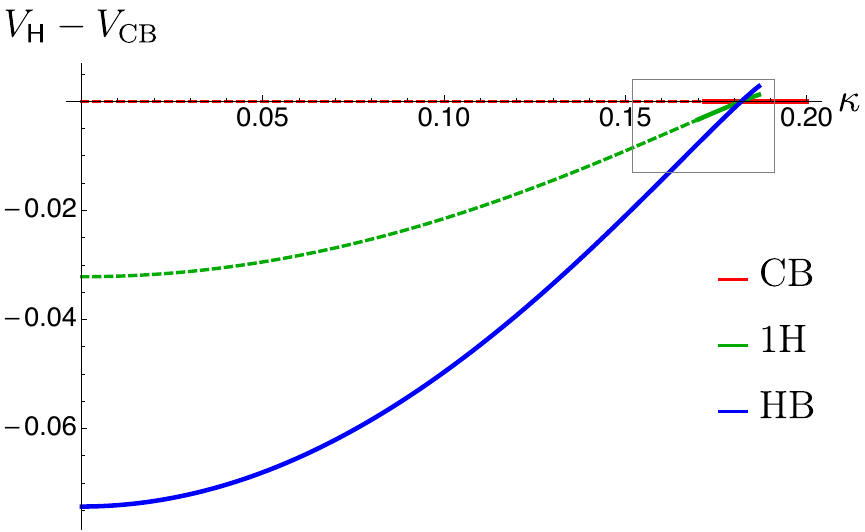} \qquad
\includegraphics[width=0.4\textwidth]{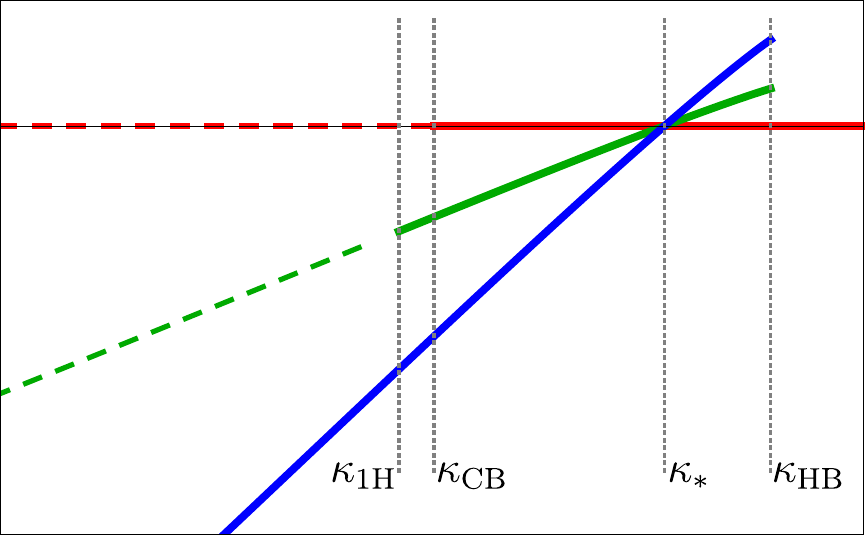}
\bigskip
\caption{In both panels, the potential differences $V_{\mathsf H}-V_\text{CB}$ are plotted against $\kappa$ for the various branches, denoted by $\BB$. The Coulomb branch (CB) is shown in the horizontal  red line; the maximal $C$-symmetric Higgs branch (HB) is shown in blue; and the $C$-non-symmetric single Higgs branch 1H is shown in green. Solid and dashed lines indicate regions of local stability and instability, respectively. The right panel zooms in on the boxed area of the left panel. Note that the three branches exactly cross at the transition point~$\kappa_*$. (See section~\ref{sec:1Hstabsu3}.) All potentials are evaluated for the renormalization scale $\mu = 10^{-3}\Lambda$ (corresponding to $e_1^2$ = 3.36452 and $e_2^2 = 4.40551$). \label{fig:su3}}\end{figure}

The global stability analysis above is summarized in table \ref{tab:su3sum} and figure \ref{fig:su3}. 

\begin{table}[htb]
\begin{center}
\begin{tabular}{|c||c|c|c|c|c|}
\hline
Branch & Higgs fields & existence & local stability & global stability & $C$ \\
\hline \hline
Coulomb		& $h_1=h_2=0$ 	& $0 < M < \infty$ 		& $M < M_{\text{CB}}$ 	& $M \leq M_*$ 	& yes \\ \hline
single Higgs		& $h_1=0, h_2\neq 0$ & $M_{\text{HB}} \! < \! M  $ & $M_{\text{HB}} \! < \! M \! < \! M_{\text{1H}}$  & $M = M_*$  &  no  \\ \hline 
maximal Higgs 	&  $h_1=h_2\neq 0$  & $M_{\text{HB}} < M  $  &  $M_{\text{HB}} < M $ & $M_* \leq M$ & yes \\ \hline
\end{tabular}
\caption{ The locally stable solutions for $N=3$. For each branch, we indicate whether~$C$ is spontaneously broken. The $C$-non-symmetric maximal Higgs branch~HB$^-$ with $h_1 \neq h_2>0$  is never locally stable, and  therefore we do not list it. We also only list one of the two 1H branches that are exchanged by~$C$. \label{tab:su3sum}}
\end{center}
\end{table}

At the transition point~$M = M_*$ there are three globally stable, exactly degenerate vacua: the CB, the 1H branch, and the HB, which are separated in field space since the value of the Higgs fields on the 1H and HB branches is nonzero at the transition point. As we dial~$M$ from~$0$ to $\infty$, we are on the CB for~$M < M_*$ and on the HB for~$M > M_*$, resulting in a first-order phase transition between them. Thus, the 1H branch is never actually realized as we dial~$M$ to~$M_*$ from the left or the right. We will make further comments on the fate of the accidental degeneracy between the three branches in section~\ref{sec:1Hstabsu3} below. 

We plot the~$SU(3)$ phase diagram in figure \ref{fig:su3}, following the graphical conventions established in section~\ref{sec:graphsu2} above: we plot the potential differences~$V_{\mathsf H} - V_\text{CB}$ of the various branches relative to that of the Coulomb branch (on the vertical axis) versus the dimensionless variable~$\kappa$ introduced in~\eqref{eq:kdef}, which we specialize here to~$N = 3$, 
\bea
\kappa = { 3 \Lambda \over 2 \pi^2 M}
\eea
with corresponding relations for $\kappa_\text{CB}$, $\kappa_\text{HB}$, $\kappa_\text{1H}$, and $\kappa_*$. Thus, the small-$M$ region of weak SUSY-breaking corresponds to large $\kappa$, while the large-$M$ region corresponds to small $\kappa$. At intermediate values, we have
\bea
\kappa_\text{1H}  < 
\kappa_\text{CB}  <
\kappa_* <
\kappa_\text{HB} 
\eea
These values are also illustrated in figure \ref{fig:su3}.

\bigskip

\subsection{Stability of the $C$-non-symmetric 1H branch}
\label{sec:1Hstabsu3}

As may be seen from figure \ref{fig:su3} and confirmed by inspection of (\ref{vcbhb}) and (\ref{vh1hb}), the potentials of all three branches exactly coincide $V_\text{CB}=V_\text{HB}=V_\text{1H}$ at the point $M=M_*$. One implication of this degeneracy is that the 1H single Higgs branch is never actually accessed as we dial~$M$. This accidental degeneracy is an exact prediction of the potential~\eqref{vscaleawithk} of our Abelian dual at the multi-monopole point. We shall see in section \ref{sec:oneH} that  similar degeneracies persist to arbitrary values of $N \geq 3$. 

Recall, however, that the analysis of the potential~\eqref{vscaleawithk} that we carried out above for~$N = 3$ involved two simplifying assumptions:
\begin{itemize}
\item We analyzed the problem classically. Quantum corrections (however small) are expected to break accidental degeneracies.
\item The potential in~\eqref{vscaleawithk}  was obtained by only retaining terms up to and including~$\CO(a_D^2)$ in the Seiberg-Witten effective K\"ahler potential, as in~\eqref{eq:keffahii}, the expectation being that the subleading~$\CO(a_D^3)$ do not generically change the leading-order answers in a qualitatively significant fashion. However, precisely this expectation breaks down when the leading-order answers have accidental degeneracies, as in our case.  We should therefore analyze whether these degeneracies are lifted by the~$\CO(a_D^3)$ corrections that we have omitted -- and that are known explicitly, see~\eqref{eq:keffah}. 
\end{itemize}

In section~\ref{sec:aD3} these~$\CO(a_D^3)$ corrections will be taken into account perturbatively. As expected, they correct the plots in figure \ref{fig:su3} to those displayed in figure~\ref{fig:aD3su3}, and they lift the accidental degeneracy. For the case~$N = 3$, we moreover see that the corrections lower the potential of the~$C$-non-symmetric 1H branch in the vicinity of~$\kappa_*$, so that there is a phase in which it is globally stable. Thus the~$SU(3)$ theory now has three phases: the CB, the 1H branch, and the HB, which are traversed in order of ascending~$M$ and separated by first-order transitions. The derivation of these results and their generalization to arbitrary $N$ will be discussed in detail in Section \ref{sec:aD3}.

\newpage

\section{Branch structure for  arbitrary $SU(N)$ gauge group} 
\label{sec:sun}

In this section we shall analyze the existence, local stability and global stability of semi-classical vacua for the dimensionless reduced effective potential~\eqref{vscaleawithk}, which emerges from the Abelian dual at the multi-monopole point after SUSY-breaking and vacuum alignment. We will study the problem for general~$SU(N)$ gauge group. Throughout we  make use of the assumption that $\mu/\Lambda$ is sufficiently small so that the matrix $t_{mn}$ in (\ref{eq:tdef1}) is positive definite and satisfies $(t^{-1})_{mn} (\mu)<0$ for all $n \not= m$, as was already postulated in (\ref{2.tinvneg}). 

\subsection{Review of the dual Abelian Higgs model}

The results obtained previously under these assumptions are as follows. All vacua have $\Re (a_{Dm})=0$, as shown in section \ref{sec:CT}, and are perfectly aligned in~$SU(2)_R$ space, as shown in section \ref{sec:vacal}. Summarizing the results of equations (\ref{3.xdef}) and (\ref{hpar}), we express $a_{Dm}$, $h_{im}$ and $\Lambda$ in terms of the dimensionless real-valued variables $x_m$, $h_m$ and $\kappa$, respectively, 
\bea
\left \{ \bma a_{Dm} & = & -i M \, x_m \cr h_{1m} &= & M \, h_m \geq 0 \cr h_{2m}  &= &0 \ema \right .
\hskip 1in 
\kappa = { N \Lambda \over 2 \pi^2 M} 
\eea
where $m=1,\ldots, N-1$. Note that $M,\Lambda>0$ so that $\kappa >0$ as well. The dimensionless reduced effective potential $V$ of (\ref{vscaleawithk})  is expressed in terms of $x_m, h_m, \kappa$ and the entries of the matrix $t_{mn}$ (see (\ref{eq:tdef1}) and~\eqref{eq:tdef1ii}), and is reproduced here for convenience,
\ba{\bs{
\label{Vred}
V =&  \sum_{m=1}^{N-1} \left(-  2 \kappa s_m x_m   +\frac{1}{2} \left( 4x_m^2	- 1\right) h_m^2\right) 
+  \sum_{m,n=1}^{N-1}\left(       t_{mn} \, x_{m} x_{n} +    \frac{1}{2}  (t^{-1})_{mn} \, h_m^2 \, h_n^2  \right) 
}}
We also recall the associated field equations: varying~$x_m$ leads to (\ref{1aa}), 
\bea
\label{1aaa}
2{h_m^2}  x_m +  \sum_{n=1}^{N-1} t_{mn} \, x_n = \kappa s_m   
\eea
while varying $h_m$ leads to (\ref{1ba}),
\bea
\label{1baa}
h_m \bigg ( 4 x_m^2-1 + 2 \sum_{n=1}^{N-1} (t^{-1})_{mn} \, h_n^2 \bigg )= 0
\eea
For each value of $m$, equation (\ref{1baa}) has two solutions, one corresponding to $h_m=0$, and the other corresponding to the vanishing of the expression inside the large parentheses. We refer to the latter branch as $h_m \not=0$, or equivalently $h_m >0$. To disentangle the vacuum structure of the resulting $2^{N-1}$ branches,  we shall now  introduce a convenient terminology and notation.

\subsection{Taxonomy of different branches}	
\label{sec:9.1}	
		
We parametrize  the different branches of solutions to \eqref{1aaa}, \eqref{1baa} in terms of the partitions of the set of indices $\{ 1, \ldots , N-1\}$ into two mutually disjoint subsets $\AA$ and $\BB$. Let $\AA$ denote the set of values $m$ for which $h_m=0$  and $\BB$ the  set of values $m$ for which $h_m \not=0$, 
\bea
\label{partition}
\left \{ \begin{array}{cc} m \in \AA: & h_m=0 \cr m \in \BB: & h_m \not= 0 \end{array} \right . 
\hskip 1in
\left \{ \begin{array}{cl} \AA \cap \BB & =  \emptyset   \\ \AA \cup \BB & =  \{1,\dots,N-1\}  \end{array} \right .
 \eea
One may denote such a partition by $\AA| \BB$ or, equivalently when the value of $N$ has been specified, simply by $\BB$. The letters $\AA$ and $\BB$ stand for Coulomb and Higgs respectively. The partition given by $\AA= \{ 1, \ldots , N-1\}$ and thus $\BB=\emptyset$ corresponds to the Coulomb branch (often abbreviated as CB) while the partition given by $\BB= \{ 1, \ldots , N-1\}$ and thus $\AA=\emptyset$ corresponds to the maximal Higgs branch (often abbreviated as HB). A partition $\AA|\BB$ for which neither $\AA$ nor $\BB$ is empty corresponds to a mixed Coulomb-Higgs branch as we shall see in more detail in the sequel.\footnote{~As will be explained in subsection \ref{sec:Csym}, certain partitions $\AA|\BB$ naturally exhibit further sub-structure of solutions depending on whether charge-conjugation symmetry~$C$ is preserved or spontaneously broken. } Henceforth, we shall prefer to label the partitions by $\BB$ when the value of $N$ has been specified. Finally, we will often refer to a branch with~$p$ non-zero Higgs fields (so that~$p = |\BB|$ is the cardinality of~$\BB$) as a $p$H branch, e.g.~$p = N-1$ for the maximal HB. 
 
 \sm
 
Inspection of (\ref{1baa}) reveals that the only matrix elements  $(t^{-1})_{mn}$ upon which this equation depends are those for which $m,n \in \BB$. We now define a square matrix $u$ whose dimension is the cardinality $p = |\BB|$ of $\BB$, and whose inverse is the restriction of  $t^{-1}$ to $\BB$,
\ba{
\label{uinvv}
(u^{-1})_{mn} = (t^{-1})_{mn}\ ,\qquad m,n \in \BB \ .
}
Since the matrix $t^{-1}$ is positive definite, so is its restriction $u^{-1}$, whose inverse is $u$.  The off-diagonal elements of $u^{-1}$ are also negative since those of $t^{-1}$ are in view of (\ref{2.tinvneg}). This property, combined with the positive definiteness of $u$, implies that all matrix elements of $u$ are positive. To prove this, we set $u^{-1} = D - L$ where $D$ is a positive definite diagonal matrix while $L$ has vanishing diagonal entries and positive off-diagonal entries.  We then obtain $u$ as a convergent geometric series $u=D^{-1} +D^{-1} L D^{-1} + \cdots$ which shows that all the entries of the matrix $u$ are positive.

\subsection{Reducing the field equations on each branch}

The reduced field equations corresponding to a given partition $\AA|\BB$ (or simply $\BB$, since we are working at fixed $N$) may be organized as follows. By definition of the partition $\AA|\BB$ in~(\ref{partition}), we have $h_m=0$ for all $m \in \AA$ and we may solve equation (\ref{1baa}) for all $h_m$ with $m \in \BB$ in terms of the variables $x_n$ with $n \in \BB$, making use of the matrix $u$ defined in (\ref{uinvv}), 
\bea
\label{k2a}
h_m^2 = \half \sum_{n \in \BB} u_{mn} (1-4x_n^2) \hskip 1in m \in \BB
\eea
For a given partition $\AA| \BB$, equation (\ref{k2a}) along with $h_m=0$ for all $m \in \AA$ provides the complete solution to equation (\ref{1baa}).  

To eliminate $h_m$ from the remaining equations (\ref{1aaa}) we proceed by treating the equations for $m \in \AA$ and $m \in \BB$ separately, 
\bea
\label{partx}
m \in \AA & \hskip 0.5in &   \sum_{k \in \AA} t_{mk} x_k + \sum _{n \in \BB} t_{mn} x_n = \kappa s_m 
\no \\
m \in \BB &&    \sum_{k \in \AA} t_{mk} x_k + \sum _{n \in \BB} t_{mn} x_n + x_m \sum_{n \in \BB} u_{mn} (1-4x_n^2)  
= \kappa s_m
\eea 
The equations for $m \in \AA$ are linear in $x_k$ for $k \in \AA$, and may be solved for $x_k $ with $k \in \AA$ in terms of the  $x_n$ for $n \in \BB$. To do so, we introduce an auxiliary matrix $\sigma$, whose  dimension is the cardinality $|\AA|$ of $\AA$, and whose inverse is the restriction of the matrix $t$  to $\AA$,
\bea
\sum_{k \in \AA} t_{m k} \, \sigma _{k \ell } = \delta_{m, \ell} \hskip 1in m,\ell \in \AA
\eea
Clearly, the values of the entries of the matrix $\sigma$ depend on the partition $\AA| \BB$. 
In terms of $\sigma$, the first equation of (\ref{partx}) is solved for $x_k$ with $k \in \AA$ as follows,
\bea
\label{xA}
x_k =  \sum_{\ell \in \AA} \sigma _{k \ell} \Big ( \kappa s_\ell - \sum _{n \in \BB}  t_{\ell n} \, x_n \Big )~, \qquad k \in \AA 
\eea
Eliminating $x_k$ for $k \in \AA$ from the second equation in (\ref{partx}) gives the following reduced equation for $x_m $ with $m \in \BB$, 
\bea
\label{xB}
\sum_{n \in \BB} u_{m n} \Big (  x_n   + x_m (1-4x_n^2) \Big )  =   
\kappa \sum_{n \in \BB} u_{m n}   (t^{-1} s)_n
\eea
Throughout, we shall use the following shorthand, for arbitrary values of $n=1, \ldots, N-1$, 
\bea
(t^{-1} s)_n = \sum _{p=1}^{N-1} (t^{-1})_{np} s_p
\eea
To simplify and consolidate the various contributions to (\ref{xB}),  we have used the following matrix algebra relation for $m,n \in \BB$, 
\bea
\label{tsigma}
t_{mn} - \sum_{k,\ell \in \AA} t_{mk} \, \sigma _{k\ell} \, t_{\ell n} = u_{mn}
\eea
which may be  proven by block decomposing the matrix $t$ and its inverse. 

\sm

An alternative form of the field equations (\ref{xB}), which will be particularly useful in the sequel, is obtained by moving the first term in the parentheses in (\ref{xB}) to the right side of the equation, and then matrix-multiplying by $u^{-1}$ on both sides of the equation. The resulting alternative for (\ref{xB}) for all $m \in \BB$ is given by,
\bea
\label{xBalt}
 \kappa (t^{-1} s)_m = x_m +  \sum _{p,q\in \BB} (u^{-1})_{mp} \, x_p \, u_{pq} \, (1-4x_q^2) ~, \qquad m \in \BB
\eea
Having solved this system of cubic equations for $x_m$ with $m \in \BB$, the solutions for $x_k$ with $k \in \AA$ and $h_m$ with $m \in \BB$ may be obtained by direct substitution into (\ref{xA}) and  (\ref{k2a}), respectively. Thus, the problem of solving the system of equations (\ref{1aaa}) and (\ref{1baa}) has been reduced to solving the system of cubics (\ref{xBalt}) for each partition $\AA|\BB$. Note that these cubics exactly reduce to equations we have previously encountered, e.g.~\eqref{cubic1} for the HB of~$N = 2$, and~\eqref{cubic3} for the 1H branch~$h_1 = 0, h_2 \neq 0$ (denoted by~$\AA = {1}$ and~$\BB = {2}$) for~$N = 3$.

\subsection{Reducing the effective potential  in each branch}

For a given partition $\AA| \BB$, the potential $V$ may be reduced by evaluating $V$ on the solution for the Higgs field $h_m$ for $m \in \BB$  given by (\ref{k2a}),  and the solution for $x_k$ with $k \in \AA$ given in (\ref{xA}). The result is a reduced potential $V_\BB$ that is specific to the partition $\AA|\BB$,  
\bea
V _\BB= V \Big |_{(\ref{k2a}) \& (\ref{xA})}
\eea
and whose dependence on the variables $x_m$ with $m \in \BB$ is given by, 
\bea
\label{redV}
V_\BB = V_\text{CB}  +
\sum_{m,n \in \BB} u_{mn} \bigg [ \left ( x_m- \kappa (t^{-1} s) _m \right ) \left ( x_n - \kappa (t^{-1} s) _n \right )
 - \tfrac{1}{8}  (1-4x_m^2) (1-4 x_n^2) \bigg ]
\eea
Here $V_\text{CB}$ is the potential of the Coulomb branch (for which $\BB = \emptyset$) given by,
\bea
\label{VA}
V_\text{CB} =  - \kappa ^2 \sum _{k,\ell=1}^{N-1} (t^{-1})_{k\ell} s_k s_\ell 
\eea 
For a given partition $\AA| \BB$, the equations (\ref{xB}) for $x_m$ with $m \in \BB$ may be derived  by applying the variational principle to $V_\BB$, which is therefore an \textit{off-shell potential} for these variables.

Using the alternative presentation of the field equations for $x_m$ with $m \in \BB$ given in~(\ref{xBalt}), we may evaluate the effective potential $V_\BB$ on a solution to these equations so as to eliminate the $\kappa$-dependence and obtain the following simplified form of the potential,
\bea
\label{redValt}
V^{\rm sol} _\BB = V_\text{CB} + \sum _{m,n \in \BB} \bigg ( - \tfrac{1}{8} u_{mn} + \sum _{p,q\in \BB} u_{m p} x_p (u^{-1})_{pq} x_q u_{qn} \bigg ) (1-4x_m^2) (1-4x_n^2)
\eea
It must be stressed that this potential, obtained by evaluating $V_\BB$  on a solution to the field equations, is now an \textit{on-shell potential}, i.e.~the field equations cannot be derived by varying~$V^{\rm sol} _\BB$ in~\eqref{redValt}.

Let us examine two important special cases:

\begin{itemize}
    
\item When $\BB = \emptyset$, i.e.~on the CB, the sums over $\BB$ in (\ref{redV}) and in (\ref{redValt}) are absent and the potential reduces to $V_\text{CB}$, which indeed was defined to be the value of the effective potential in the pure Coulomb branch (see~\eqref{VA}).

\item When $\AA = \emptyset$, i.e.~on the maximal HB, one may  substitute $t$ for $u$,   $t^{-1}$ for $u^{-1}$,  and the full range $m,n=1,\ldots, N-1$ for $m,n \in \BB$ in the above expressions  to obtain the field equations and the potential for the maximal Higgs branch.

\end{itemize}

\subsection{Reducing the local stability conditions}
\label{sec:red2}

The local stability conditions, formulated generally in terms of the Hessian matrix $\cH$  in~(\ref{hess}), may now be analyzed for each partition $\AA|\BB$ by eliminating the Higgs fields in terms of~$x_m$. To do so,  it will be convenient to recast the positivity conditions of the Hessian (\ref{hess}) in terms of an associated quadratic form $Q$ in real variables $\alpha_m, \beta_m$,
\bea
Q = \sum_{m,n =1}^{N-1} \Big ( (\cH_{xx})_{mn} \, \alpha_m \alpha_n + 2 (\cH_{xh})_{mn} \, \alpha_m \beta_n +
(\cH_{hh})_{mn} \,\beta_m \beta_n  \Big )
\eea
Positive definiteness of $\cH$ is equivalent to positive definiteness of the quadratic form $Q$. To simplify the latter condition, we decompose its contributions according to whether $m \in \AA$ or $m \in \BB$, and similarly for~$n$. For $m \in \AA$, the first term in $(\cH_{xx})_{mn}$, the last term in $(\cH_{hh})_{mn}$ and all of $(\cH_{xh})_{mn}$ vanish, while for $m \in \BB$, the first term in  $(\cH_{hh})_{mn}$ vanishes in view of (\ref{1baa}) and the fact that $h_m \not=0$. Taking these simplifications into account, and rearranging terms into absolute square combinations, we decompose $Q$  into a sum of four quadratic forms of $\a_m$ and $\b_m$, 
\bea
Q= Q_\a + Q_\b + Q_1 + Q_2
\eea
which are given by,
\bea
\label{stabB}
Q_\a & = & \sum_{m,n \in \BB} u_{mn} \Big ( 
(1-4x_m^2)\a_n^2 + (1-4x_n^2)\a_m^2 + 2 (1-8 x_m x_n) \a_m \a_n  \Big )
\no \\
Q_\b & = & \sum_{k \in \AA} \bigg ( 4 x_k^2 -1 
+  \sum _{m,n \in \BB} (t^{-1})_{kn} \, u_{n m} (1-4x_m^2)  \bigg ) \beta _k^2
\no \\
Q_1 & = & 2 \sum _{k,\ell \in \AA} t_{k\ell} \bigg ( \a_k + \sum _{m \in \BB} (\sigma t)_{km} \a_m \bigg  )
\bigg ( \a_\ell + \sum _{n \in \BB} (\sigma t)_{\ell n} \a_n \bigg )
\no \\
Q_2 & = & 4 \sum_{m,n \in \BB} u_{mn}
\bigg ( 2 x_m \a_m + \sum_{p \in \BB} (t^{-1})_{mp} h_p \b_p   \bigg ) 
\bigg ( 2 x_n \a_n + \sum_{q \in \BB} (t^{-1})_{n q} h_q \b_q   \bigg )
\eea
The quadratic  form $Q_1$ contains all the dependence of $Q$ on the variables $\a_k$ for $k \in \AA$, while $Q_2$ contains all the dependence on the variables $\beta _p$ for $p \in \BB$. Both $Q_1$ and $Q_2$ are positive definite since $t$ and  $u$ are positive definite. Thus, positive definiteness  of $\cH$ and $Q$ is equivalent to positive definiteness  of both $Q_\a$ and $Q_\b$,
\bea
\label{stabC}
\cH > 0 &  \Longleftrightarrow  & \bigg \{ Q_\a > 0 \, \hbox{ and } \,  Q_\beta > 0   \bigg \}
\quad
\eea
Positive definiteness of $Q_\beta$ is equivalent to the following  conditions for all $k \in \AA$, 
\bea
\label{stabbeta}
4 x_k^2 -1  +  \sum _{m,n \in \BB} (t^{-1})_{kn} \, u_{nm} (1-4x_m^2) >0~, \qquad k \in \AA
\eea
For arbitrary $N$ and an arbitrary partition $\AA| \BB$, the conditions (\ref{stabC}) are difficult to study analytically. However, some simple necessary conditions may be obtained, as we now do.

\subsection{Charge conjugation}
\label{sec:Csym}

The matrix of gauge couplings $t$, defined in (\ref{eq:tdef1}) and~\eqref{eq:tdef1ii}, and the values $s_m=\sin(\pi m/N)$ are invariant under charge conjugation~$C$,  
\bea
t_{(N-m)( N-n)} = t_{mn}  \hskip 1in s_{N-m} = s_m
\eea
The combined set of field equations (\ref{1aaa}) and (\ref{1baa}) and the reduced effective potential~(\ref{Vred}) are also $C$-invariant provided $x_m$ and $h_m$ transform as follows,
\bea
C : \ \begin{cases} x_m  \to  x_{N-m} \\ h_m  \to  h_{N-m} \end{cases} 
\eea
Furthermore, charge conjugation maps a partition $\AA| \BB$ into a partition $\AA^c| \BB^c$ where,
\bea
C: ~ \AA| \BB \to \AA^c| \BB^c \hskip 0.6in \begin{cases} 
\AA^c = \big \{ k \in \{1, \ldots, N-1\} ~  \hbox{ s.t. } N-k \in \AA \big \} \\
\BB^c = \big \{ k \in \{1, \ldots, N-1\} ~ \hbox{ s.t. } N-k \in \BB \big \} 
\end{cases} 
\eea
A partition that satisfies $\AA^c| \BB^c \not= \AA| \BB$ is not~$C$-invariant; the corresponding solutions spontaneously break~$C$ and are exchanged by it, but are otherwise physically identical. It is therefore sufficient to analyze just one of the two~$C$-non-symmetric partitions. A partition that satisfies  $\AA^c| \BB^c = \AA| \BB$ is $C$-invariant and can be further subdivided into two different branches of solutions: one corresponding to $C$-symmetric vacua, the other to vacua with spontaneously broken charge conjugation. The branch of $C$-symmetric solutions corresponding to the partition $\AA|\BB= \AA^c |\BB^c$  will be denoted by $\BB^+$ while the branch of $C$-non-symmetric solutions will be denoted $\BB^-$.

\subsubsection{$C$-invariant solutions for $C$-invariant partitions}

As explained above, $C$-invariant solutions can only occur for $C$-invariant partitions. We shall now consider the field equations and the effective potential for $C$-invariant solutions in a $C$-invariant partition $\AA| \BB$ with reduced matrix $u$.  Invariance of the solution requires, \footnote{~Actually, requiring $x_{N-m}=x_m$ for all $m$ is equivalent to requiring $h_{N-m}^2 = h_m^2$ for all $m$, as follows from~(\ref{1aaa}) and~(\ref{1baa}). Since all~$h_m \geq 0$ are non-negative, either condition in~(\ref{Cxh}) implies the other.}
\bea
\label{Cxh}
x_{N-m} = x_m \hskip 0.8in h_{N-m} = h_m
\hskip 1in
m = 1 , \ldots, N-1
\eea
Defining the set $\BB_* = \{ m \in \BB \hbox{ s.t. } m <N/2\} $, the reduced equations for $x_m$ are as follows,
\bea
\label{xBstar}
\sum_{n \in \BB_*} \hat u_{mn} \Big (  x_n - \kappa (t^{-1} s)_n  + x_m (1-4x_n^2) \Big )  = 0  \qquad
\hbox{ for all } m \in \BB_*
\eea
where the entries of $\hat u$ are given by,
\bea
\label{6.hatu}
\hat u_{mn} = 2u_{mn}  + 2u_{m(N-n)}  \hskip 1in m,n = 1 , \ldots, \left [ \tfrac{N-1}{ 2} \right ]
\eea
supplemented by the following relations when $N=2\nu$ is an even integer,
\bea
\label{6.hatusupp}
\hat u _{\nu  \nu} = u_{\nu \nu}
\hskip 0.8in 
\hat u_{m\nu} = u_{m \nu} + u_{(N-m) \nu} 
\hskip 0.8in 
m=1,\ldots, \nu-1
\eea
The reduced potential is given by,
\bea
\label{redVa}
V_\BB = V_\text{CB}  +
\sum_{m,n \in \BB_*} \hat u_{mn} 
\bigg [ \left ( x_m - \kappa (t^{-1} s) _m \right ) \left ( x_n- \kappa (t^{-1} s) _n \right )
 - {1 \over 8}  (1-4x_m^2) (1-4 x_n^2) \bigg ]
 \quad
\eea
It may be readily verified that these equations reproduce the reduced equations for gauge group $SU(3)$ on its~$C$-symmetric Higgs branch (HB$^+$) analyzed in section \ref{sec:su3}.

\subsubsection{$C$-non-invariant solutions for $C$-invariant partitions}
\label{sec:6.CH}

The analysis of $C$-non-invariant solutions to a $C$-invariant partition $\AA|\BB$ is more involved than that for $C$-invariant solutions because the number of independent variables is larger. Here we shall provide a set-up that simplifies the equations without actually solving them. 

The starting point is the set of reduced equations for $x_m$ with $m \in \BB$ given in~\eqref{xB}. Since $\BB^c=\BB$, the index $N-m$ also belongs to $\BB$. The issue is whether the differences $x_m - x_{N-m}$ vanish or not. If they all vanish, then the corresponding solution is $C$-invariant, while otherwise the solution spontaneously breaks $C$-symmetry. To analyze the possible branches that can appear for a given $C$-symmetric partition, we study the equations for the differences $x_m - x_{N-m}$ and the sums $x_m + x_{N-m}$ using the following parametrization,
\bea
\label{6.varxD}
x_m = y_m +\Delta_m \hskip 1in x_{N-m} = y_m - \Delta _m
\eea 
Taking the sums and differences of the $m$ and $N- m$ equations in (\ref{xBalt}) and using the fact that $(t^{-1}s)_{N-m} = (t^{-1}s)_m$ implies $(u^{-1}s)_{N-m} = (u^{-1}s)_m$  gives the sum equations, 
\bea
\label{6.sumxx}
\kappa \sum_{n \in \BB} u_{mn} (t^{-1} s)_n = 
 \sum _{ n \in \BB} u_{m n} 
\Big ( y_m (1-4y_n^2-4\Delta_n^2) + y_n - 8 y_n \Delta _m \Delta_n \Big ) 
\eea
and the difference equations, 
\bea
\label{6.diffxx}
\sum_{n \in \BB} U_{mn} \Delta _n=0
\eea
where the components of the matrix $U$ for $m,n \in \BB$ are given by,
\bea 
\label{6.coeffxx}
U_{mn} = u_{mn} (1-4y_m y_n) +  \delta_{m,n} \sum_{p \in \BB} u_{n p}  (1-4y_p^2-4\Delta_p^2)   
\eea
Under charge conjugation, $y_m$ and $U_{mn}$ are invariant, while $\Delta_n \to - \Delta_n$. Thus, a charge conjugation invariant solution is characterized by $\Delta_n=0$ for all $n \in \BB$ while otherwise the solution spontaneously breaks $C$.  For the latter case, the analysis of the $SU(3)$ case has shown that such a solution exists but that it is always locally unstable. In section \ref{sec:num} we shall establish numerically that a similar conclusion holds for the cases of gauge groups $SU(4), SU(5)$, and $SU(6)$, but we have not found an analytic proof.

\subsection{The Coulomb branch CB: {\rm $\BB=\emptyset$}}
\label{sec:largek}

The Coulomb branch CB has vanishing Higgs fields $h_k=0$ for all $k=1,\ldots, N-1$ while the solution for $x_k$ is given by (\ref{xA}), 
\bea
\label{xmcb}
x_k = \kappa (t^{-1} s)_k
\eea 
The conditions for local stability of the Coulomb branch may be read off from  (\ref{stabC}),
\bea
\thalf < |x_k|   \hskip 0.5in \hbox{ for all } \quad k =1,\ldots, N-1
\eea
and imply the bound $M<M_{{\rm CB}}$ or equivalently, 
\bea
\kappa _{{\rm CB}} < \kappa  
\hskip 1in 
 \kappa _{\rm CB}   = \left ( 2 \min _{1 \leq k \leq N-1} \left \{ (t^{-1} s )_k \right \} \right )^{-1} 
\eea
As $\kappa \to \infty$, the only solution is the Coulomb branch. Indeed, if we assumed that $\BB \not = \emptyset$ as $\kappa \to \infty$, we see from (\ref{1aaa}) and (\ref{1baa}) that $x_k$ cannot remain bounded  for at least one $k \in \BB$ since otherwise the left side would remain bounded while the right side diverges as $\kappa \to \infty$. But if  $x_k$ diverges for any $k \in \BB$, then $h_k^2 <0$ for all $k \in \BB$ in view of (\ref{k2a}), which is contradictory to our assumptions. Hence $\BB$ must be empty in the limit $\kappa \to \infty$ and, by continuity must remain empty for sufficiently large $\kappa$. Finally, the value of the potential in the Coulomb branch is given by $V_\text{CB}$ in (\ref{VA}).

\subsection{The maximal Higgs branch HB: {\rm $\AA=\emptyset$}}
\label{sec:smallk}

In this subsection, we analyze the existence and stability of solutions for small $\kappa$ which corresponds to the case where the supersymmetry breaking scale is large compared to $N \Lambda$. We shall show that the stable solutions necessarily lie on a maximal Higgs branch HB, where $\AA = \emptyset$. It is instructive to begin with the special case $\kappa=0$.

\subsubsection{The solution for $\kappa =0$}

To analyze the existence and stability of solutions for $\kappa = 0$, we begin by considering the equations for $x_m$ in (\ref{1aaa}), 
\bea
\sum_{n=1}^{N-1} \Big ( t_{mn} + 2 h_m^2 \delta _{m,n} \Big ) x_n = 0
\eea
Clearly, positive definiteness of $t$ implies  $x_n=0$  for all $n=1,\ldots, N-1$. The local stability condition (\ref{stabbeta}) for a solution corresponding to an arbitrary partition $\AA | \BB$ for $x_n=0$ for all $n=1,\ldots, N-1$ reduces to the following condition for all $k \in \AA$,
\bea
\sum_{m,n \in \BB} (t^{-1})_{kn} u_{nm} > 1
\eea
Since $k \in \AA$ and $n \in \BB$, the matrix elements $(t^{-1})_{kn}$ are all negative, while the matrix elements $u_{nm}$ are all positive. As a result,  the left side is negative and the inequality can never be satisfied, unless $\AA = \emptyset$ in which case the condition is simply absent. Thus we conclude that local stability eliminates all but the maximal Higgs branch. The remaining local stability condition $Q_\alpha \geq 0$ is satisfied for $x_m=0$ as $Q_\a$ reduces to, 
\bea
\label{6.locstab}
Q_\a = \sum_{m,n =1}^{N-1} t_{mn} (\a_m+\a_n)^2
\eea
Since the matrix elements of $t$ are strictly positive $Q_\a$ is positive. It is definite since the vanishing of all $\a_m+\a_n$ implies $\a_m=0$ for all $m$. We conclude that for $\kappa=0$ the maximal Higgs branch is the only locally stable solution and therefore it must also be globally stable, as is explicitly proven in Appendix~\ref{sec:maxHglob}.

\subsubsection{Series expansion of the solution for  small $\kappa $}

By continuity in $\kappa$,  the solution corresponding to the partition with $\AA=\emptyset$ will remain the global  minimum  of the potential for $\kappa$ non-zero but small. The solution may be constructed by Taylor series expanding equation (\ref{xB}) for $\AA=\emptyset$ in powers of $\kappa$ to obtain $x_m$, 
\bea
\label{6.TT}
\sum_{n=1}^{N-1} T_{mn} x_n   - 4  x_m \sum_{n=1}^{N-1} t_{mn} x_n^2 = \kappa s_m
\hskip 0.8in 
T_{mn} = t_{mn} + \delta _{m,n} \sum_{p =1}^{N-1} t_{n p}
\eea
for $m,n=1,\ldots, N-1$, and then using (\ref{k2a}) to obtain $h_m$. To leading order in $\kappa$, the equation reduces to the linear matrix equation $Tx=\kappa s$. Since the polynomial in $x$ on the left side of the first equation in (\ref{6.TT})  is odd in $x$, and the right side is odd in $\kappa$, the Taylor expansion of $x_m$ in powers of $\kappa$ involves only odd powers of $\kappa$,  
\bea
x_m = \kappa x_m^{(1)} + \kappa^3 x_m^{(3)} + \kappa^5 x_m^{(5)} + \cO(\kappa ^7)
\eea
It is immediate to obtain the first two coefficients recursively, 
\bea
\label{6.x1}
x_m^{(1)}  = \sum _{n=1}^{N-1} (T^{-1})_{m n} s_n
\hskip 1in
x_m ^{(3)} = 4 \sum _{n,p=1}^{N-1} (T^{-1}) _{mn} \, t_{n p} \, x_n ^{(1)}  \, (x_p^{(1)})^2
\eea
The Taylor series  solution is manifestly~$C$-invariant to all orders  in $\kappa$ since, by induction on the order~$(r)$ of the expansion,  the individual contributions satisfy, 
\bea
x_{N-m}^{(r)}  = x_m ^{(r)} \hskip 1in m=1,\ldots, N-1
\eea
using the relations $s_{N-m}=s_m$, $t_{(N-m)(N-n)}=t_{mn}$ and $T_{(N-m)(N-n)}=T_{mn}$. 
The radius of convergence of this expansion is finite, but its value depends on the detailed structure of the matrix  $t$. 
In the crude approximation where $t_{mn}$ is dominated by its diagonal entries,  the $N-1$ cubics decouple, and the condition for convergence for the expansion of each cubic becomes $27 \kappa ^2 s_m^2   <8 t_{mm}^2$ for all $m$. 
Equivalently this is the point at which the Hessian ceases to be positive definite. 

\sm

To investigate global stability, we expand the effective potential $V_\BB$, given in (\ref{redV}) for an arbitrary partition,  to second order in $\kappa$,
\bea\label{eq:smallKHB}
V_ \BB= V_\text{CB} 
- {1 \over 8} \sum_{m,n \in \BB} u_{mn} 
+ \kappa ^2 \sum_{m,n \in \BB} u_{mn}  \, ( x_m^{(1)} \big )^2 
+ \cO(\kappa^4)
\eea
The first term is the energy of the Coulomb branch, the sum of the first two terms gives the potential of the $\kappa=0$ HB, while the third term  systematically raises the value of the effective potential for every partition as $\kappa$ is increases away from 0. Note that the quadratic $\kappa$-dependence in~\eqref{eq:smallKHB} explains the universally parabolic shape at small $\kappa$ of the HB potential plotted in figures~\ref{fig:su2} and \ref{fig:su3} above for~$SU(2)$ and $SU(3)$, respectively, as well as in figures \ref{fig3a}, \ref{fig:7}, and~\ref{fig:77} below for $SU(4)$, $SU(5)$, and~$SU(6)$, respectively.

\subsection{Comparing the branches $\BB={\rm CB}, \{m\}, \{m, N-m\}$}
\label{sec:oneH}

In this subsection we shall compare the existence and local and global stability of the Coulomb branch CB with the following branches, for arbitrary $N\geq 3$ and $m <N/2$,\footnote{~Note that for $N=2 \nu$ an even integer and $m=\nu$, the branch $\BB_2^-$ does not exist, while the other two branches coincide $\BB_2^+=\BB_1$.}
\begin{align}
\BB_1 & = \{ m \} & h_m&  >0 && h_k =0, \,  k \not= m
\no \\ 
\BB_2^+ & = \{m, N-m \}^+ & h_m&=h_{N-m}   >0 && h_k =0, \,  k \not= m, N-m
\no \\
\BB_2^- & = \{ m, N-m \}^- & h_m &\not =h_{N-m}   >0 && h_k =0, \, k \not= m, N-m
\end{align} 
We will show that when the potentials for the branches  $\BB_1$ and $\BB_2^+$ are lower than the potential of the Coulomb branch, then the potential of the branch $\BB_2^+$ is always lower than the potential of the branch $\BB_1$. Furthermore, the branch $\BB_2^-$ is always locally unstable. This simple case exemplifies many salient features of the cascade that we will explore below for general~$SU(N)$. 

The reduced equations for $x_m$ in branch $\BB_1$ and for $x_m=x_{N-m}$ in branch $\BB_2^+$ are given by (\ref{xB}) in terms of the same relation between $x_m$ and $\kappa$,  
\bea
\label{oneB}
2 x_m - 4 x_m^3 = \kappa (t^{-1} s) _m 
\eea
The equations are those of the $SU(2)$ case studied in section~\ref{sec:su2hb}, with an adapted parameter on the right side of the cubic relative to the~$SU(2)$ case (see~\eqref{cubic1}), and we may therefore directly import the results from $SU(2)$.  Existence of the solutions requires $x_m<1/\sqrt{6}$ and imposes the following condition on $\kappa$, 
\bea
\label{6.kapbound}
 \kappa (t^{-1} s)_m < { 4 \over 3 \sqrt{6}}
\eea
The reduced potentials on the branches are given as follows, 
\begin{align}
\label{oneV}
V_{\BB_1} & = V_\text{CB}  - \tfrac{1}{8} \, { (1-8x_m^2) (1-4x_m^2)^2  \over  (t^{-1})_{mm} }  
& \BB_1 & =\{m\}
\no \\
V_{\BB_2^+} & = V_\text{CB} - \tfrac{1}{4} \, {(1-8x_m^2) (1-4x_m^2)^2  \over (t^{-1})_{mm} + (t^{-1})_{m,N-m}} 
& \BB_2^+ & =\{m, N-m\}^+
\end{align}
Since $(t^{-1})_{m,N-m}<0$ for $m<[N/2]$, the potential for the branch $\BB_2^+$ is lower than the potential of the branch $\BB_1$ as soon as $8 x_m^2 <1$ and both potentials are lower than the potential $V_\text{CB}$ in the Coulomb branch, as announced earlier. For $N$ even and $m=N/2$, the two potentials coincide, in agreement with the fact that the branches $\BB_1$ and $\BB_2^+$ coincide.

\subsubsection{Local instability of the branch $\BB_2^-$}\label{sef:h2mininstab}

To analyze the local stability for the branch $\BB_2^-$, we make use of the formalism developed in subsection \ref{sec:6.CH} for $C$-non-symmetric solutions to branches corresponding to $C$-symmetric partitions. The only non-vanishing variables here are $y_m=y_{N-m}$ and $\Delta _m = - \Delta _{N-m}$, defined in (\ref{6.varxD}), where $\Delta_m \not=0$ by definition of the branch $\BB_2^-$. Equation (\ref{6.diffxx}) then reduces to the condition $U_{m,m} - U_{m , N-m} =0$ which is solved in $\Delta_m$ by,
\bea
\label{6.Delta}
\Delta_m^2 = { u_{m,m} (1-4 y_m^2) \over 2 ( u_{m,m} + u_{m, N-m})}
\eea
A cubic equation for $y_m$ may be obtained by eliminating $\Delta_m$ between (\ref{6.Delta})  and the remaining equation (\ref{6.sumxx}), but it will not be needed here. Instead we shall right away pass to the analysis of the local stability condition $Q_\a >0$ given in (\ref{stabB}). Positivity of the quadratic form $Q_\a$ requires that its trace be positive. Expressing the trace of $Q_\a$ in terms $y_m$ and $\Delta_m$, we find the following $C$-symmetric expression, 
\bea
\tr \, Q_\a = 8 u_{m,m} ( 1 - 6 y_m^2 - 6 \Delta_m^2) + 4 u_{m, N-m} ( 1 - 4 y_m^2 - 4 \Delta _m^2) 
\eea
Eliminating $\Delta_m$ from the trace, using (\ref{6.Delta}), we obtain, 
\bea
\tr \, Q_\a = { -4 u_{m,m}^2  (1-3y_m^2) + 4 (1-4 y_m^2) u_{m, N-m}^2 + 4 (1-8 y_m^2) u_{m,m} u_{m, N-m}
\over u_{m,m} + u _{m, N-m}}
\eea
An immediate rearrangement of the terms in the numerator gives,
\bea
\tr \, Q_\a = - 4 {  (u_{m,m}^2- u_{m,N-m}^2 - u_{m,m} u_{m, N-m})   (1-3y_m^2) 
+  ( u_{m, N-m}^2  +5 u_{m,m}  u _{m, N-m} ) y_m^2
\over u_{m,m} + u _{m, N-m}}
~
\eea
Positivity of $h_m^2$ requires $4y_m^2<1$ so that both terms in the numerator of the above expression are positive and the trace is negative. Hence the branch $\BB_2^-$ is always locally unstable. As a special case, we recover the result of section~\ref{sec:SU3noCstab} for~$SU(3)$.

\subsubsection{Local stability of the branch $\BB_2^+$}

Local stability requires the conditions of (\ref{stabC}). One verifies that $Q_\a$ is positive definite for $8 x_m^2<1$ by diagonalizing the quadratic form. To enforce the condition $Q_\b >0$, we first solve  for $x_k$ with $k \in \AA$ using (\ref{xA}), 
\bea
\label{oneA}
x_k = \sum_{\ell \in \AA} \sigma _{k\ell} \Big ( \kappa s_\ell -  (t_{\ell,m} +t_{\ell,N-m}) x_m \Big )
\eea 
Positivity of $Q_\b$ then requires the following inequalities for all $k \in \AA$, 
\bea
x_k^2  > \tfrac{1}{4} - \tfrac{1}{8} \Big ( (t^{-1})_{k,m} + (t^{-1})_{k,N-m} \Big )   
\hat u_{mm}  (1-4 x_m^2) 
\eea
Since we have $0< \hat u_{mm}$, $8x_m^2<1$, and $(t^{-1})_{km}<0$ for $k \in \AA$ in view of (\ref{2.tinvneg}),  the above bound requires the following necessary condition, 
\bea
\label{oneloc}
x_k^2  > \tfrac{1}{4} + \tfrac{1}{16} \hat u_{mm} \, \Big | (t^{-1})_{km} + (t^{-1})_{k,N-m} \Big |   
\quad \hbox{ for all } k \in \AA
\eea
As $\kappa \to 0$, equation (\ref{oneB}) implies that also $x_m \to 0$ as is familiar from the $SU(2)$ case. Clearly, the values of $x_k$ obtained by solving (\ref{oneA}) then also tend to zero and the bound (\ref{oneloc}) will not be satisfied. This means that the solution in the branch $\BB=\{m,N-m\}$ is locally stable only for a limited range of $\kappa$ below the upper bound (\ref{6.kapbound}). Note that these constraints trivialize for~$N= 3$ with~$m = 1$, so that~$\BB_2^+ = $ HB is the maximal Higgs branch, which is stable down to arbitrarily small~$\kappa$.

\subsubsection{Relative structure of the branches $\BB_1$ and $\BB_2^+$ for different values of $m$}

\begin{figure}[t!]
\centering
\includegraphics[width=0.45\textwidth]{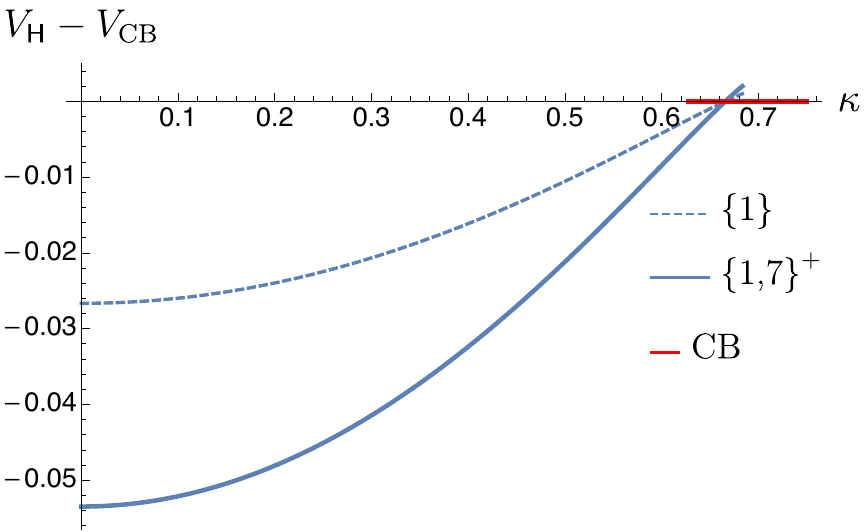}
\includegraphics[width=0.45\textwidth]{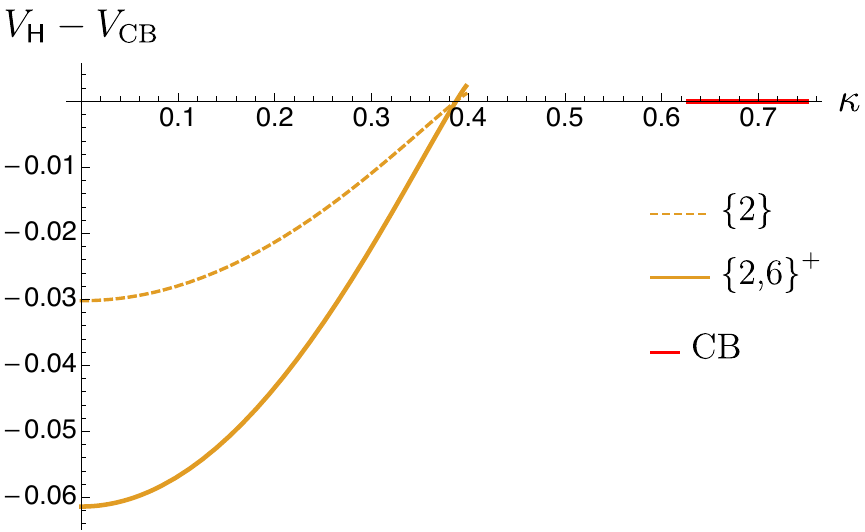}
\caption{The case of gauge group $SU(8)$: the potential for the Coulomb branch is shown  in red on the horizontal axis; the potentials for the branches $\BB=\{m,N-m\}$ are drawn in solid lines, those for the branches  $\BB=\{m\}$ are drawn in dashed lines; in blue for $m=1$; in orange for $m=2$.  The renormalization scale is~$\mu = 10^{-3}\Lambda$.
\label{fig:6}}\end{figure}

The analytical expressions for $s_n = \sin(n\pi/N)$ and $t_{mn}$ in (\ref{eq:tdef1}), \eqref{eq:tdef1ii} imply the following ordering of their entries, independently of the value of $\mu$,
\bea
s_m < s_n \hskip 0.8in t_{mm} < t_{nn} \hskip 0.8in 1\leq m < n \leq \tfrac{N}{2}
\eea
For $\mu$ sufficiently small, so that the matrix $t$ is positive definite and satisfies the inequality $(t^{-1})_{k\ell} <0$ for all $k\not= \ell$,  numerical analysis reveals the following ordering of $(t^{-1}s)_m$,\footnote{~See for instance figures~\ref{fig:skm}, \ref{fig:tinvs}, and~\ref{fig:tvTnequal50} in appendix~\ref{app:appt}. See also the analytic discussion around~\eqref{useful}.}
 \bea\label{6.tinvsineq}
 (t^{-1} s)_m < (t^{-1}s)_n 
 \hskip 0.7in 
 1\leq m < n \leq \tfrac{N}{2}
\eea
It follows from~\eqref{6.tinvsineq} and ~\eqref{6.kapbound} that, as $\kappa$ is decreased from $\infty$ (where only the CB exists), the first one of the branches $\BB=\{m, N-m\}$ to be allowed is $m=1$, then $m=2$ and so on. (See figure~\ref{fig:6} for the case~$N = 8$.) This observation lies at the heart of the cascading phase structure that we will uncover below. 

Note that the branches CB, $\BB = \{m\}$, and~$\BB = \{m, N-m\}$ have a accidental triple intersection, as we already encountered for~$SU(3)$ (see figure~\ref{fig:su3} and section~\ref{sec:1Hstabsu3}). The fate of these degeneracies is discussed numerically in section~\ref{sec:num}, and analytically in section~\ref{sec:cascade}.

\newpage

\section{Numerical phase diagrams for $N=4,5,6$} 
\label{sec:num}

In this section, we shall carry out a detailed analysis of the existence and stability of the various branches for the case of $SU(4)$ gauge group, and plot the potentials $V_\BB$ versus $\kappa$ for the various branches $\AA|\BB$. Since the analysis for the $SU(5)$ and $SU(6)$ cases is very similar, we shall only present the final results, i.e.~a plot of the potentials $V_\BB$ for the various branches.

As was explained in subsection \ref{sec:Csym} for arbitrary $N$, specifying a partition $\AA|\BB$ does not always suffice to specify a branch of solutions completely. This is not an issue if the set~$\BB$ is not~$C$-invariant, in which case~$\BB$ and its charge-conjugate~$\BB^c$ are physically identical branches, exchanged by the spontaneously broken~$C$-symmetry. However, a $C$-symmetric partition $\AA|\BB$, with~$\BB^c = \BB$, must be further refined into two different sub-branches:  one containing only $C$-symmetric solutions, which we denote by $\BB^+$, and the other with solutions that spontaneously break~$C$, which we denote by~$\BB^-$. As before, we continue to use the abbreviations CB and HB for the Coulomb branch and the maximal Higgs branch, respectively.

The numerical analysis presented here was carried out with the help of two different numerical methods, whose results were found to be in perfect agreement with one another within the prescribed precisions, and to match with the analytical results wherever they are available. The methods are as follows. 
\begin{itemize}
\itemsep=0in
\item A direct numerical method by which all possible solutions to the reduced field equations (\ref{k2a}), (\ref{xA}), (\ref{xBalt}) are found for a given value of $\kappa$ (which is then incremented in steps of $\Delta \kappa = 0.0001$);  only those solutions are retained for which the Hessian (3.27) is positive; and the effective potential  for those solutions is plotted. The plots in figures~\ref{fig3a}, \ref{fig:7}, and \ref{fig:77} have been drawn using this method. 
\item An algorithmically  simpler method which proceeds by scanning the entire parameter space of the variables $x_m$ for $m \in \BB$, enforcing the field equations of (\ref{k2a}), (\ref{xA}) exactly, while solving the field equations (\ref{xBalt}) within a prescribed precision $\ep$, and then retaining only those solutions for which the Hessian is positive.  This method is presented in detail in appendix~\ref{app:B} for the case of gauge group $SU(4)$.
\end{itemize}

The numerics in this section depend  on the value of the renormalization scale~$\mu$. Positivity of the matrix $t$ and negativity of the off-diagonal entries of its inverse $t^{-1}$ require $\mu < 0.451 \Lambda$, as listed in~\eqref{tnegvaluesbis}. Unless stated otherwise, our numerical estimates will be carried out for $\mu = 10^{-3} \Lambda$.

\subsection{Phase Diagram for $SU(4)$}

\begin{figure}[t!]
\centering
\includegraphics[width=0.6\textwidth]{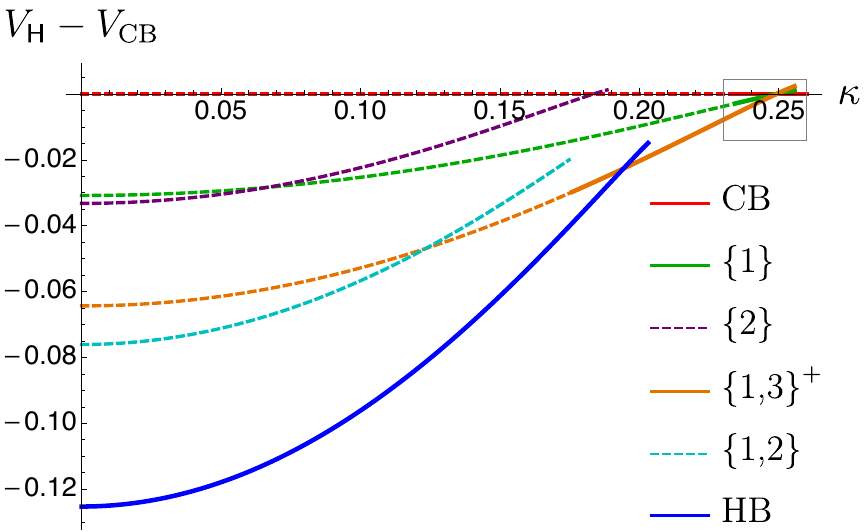}
\includegraphics[width=0.35\textwidth]{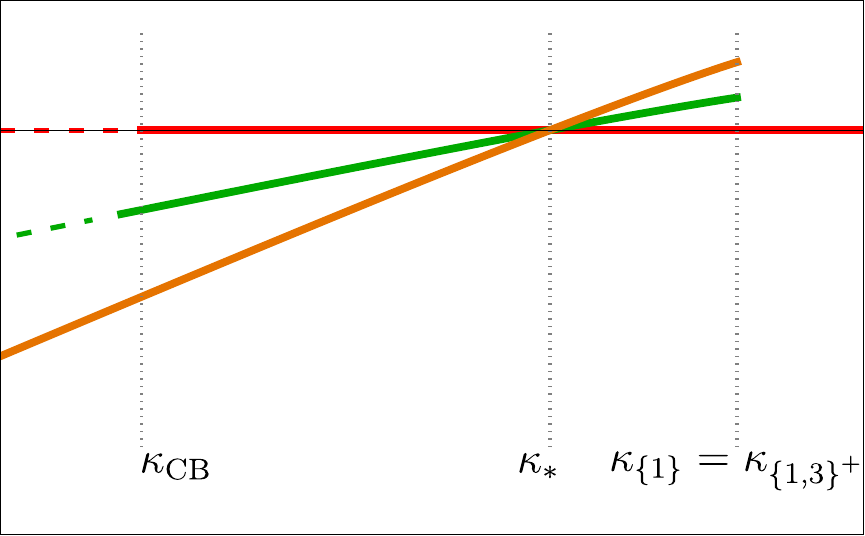}
\caption{The potential $V_\BB - V_\text{CB}$ is plotted for gauge group $SU(4)$ as a function of $\kappa$ for the different branches labeled by the partition~$\AA | \BB$. Solid and dotted lines correspond to locally stable and locally unstable vacua, respectively.  The Coulomb branch~CB is drawn in red on the horizontal axis and extends to $\kappa \to \infty$; the branch $\BB=\{1\}$ is drawn  in green; $\BB=\{2\}$ in purple; $\BB=\{1,3\}^+$ in orange; $\BB= \{ 1,2\}$ in cyan, and the maximal Higgs branch HB=$\{ 1,2,3\}^+$ in blue. The branches $\{ 1,3\}^-$ and $\{ 1,2,3\}^-$ are empty and are not shown. The right panel is a magnification of the boxed area of the left panel; note the accidental degeneracy at the triple intersection of the~CB, $\BB = \{1\}$ and~$\BB = \{1, 3\}^+$ branches. See section~\ref{sec:su4phases} for more detail.   \label{fig3a}}\end{figure}

The phase diagram of~$SU(4)$ is summarized in figure~\ref{fig3a}, where the potential difference~$V_\BB-V_\text{CB}$ of each branch~$\BB$ relative to the Coulomb branch CB are plotted as a function of the dimensionless~$\kappa = {N \Lambda / (2 \pi^2 M)}$ (with~$N = 4$) defined in~\eqref{eq:kappadefsec3}. The phase diagram as a function of increasing SUSY-breaking mass~$M$, equivalently decreasing~$\kappa$, is obtained by tracing the curve of lowest potential, starting with the CB at the top right of the figure. See section~\ref{sec:su4phases} for a more detailed description of the~$SU(4)$ phase diagram and its various transitions. 

Since~$N = 4$ is fixed, we label branches by the index set~$\BB$ of non-zero Higgs vevs; if needed, we also use~$\pm$ to indicate the~$C$-parity of~$C$-even partitions. This leads to the following distinct branches (for the~$C$-non-invariant partitions~$\BB$, we only list one of the two degenerate partitions~$\BB$, $\BB^c$): 
\bea\label{eq:su4branches}
C\text{-invariant} \, \begin{cases} \text{CB} \\  \{2\} \\ \{ 1,3\}^+ \\  \text{HB}= \{ 1,2,3\}^+ \end{cases}
\hskip 0.4in
C\text{-non-invariant} \, \begin{cases}  \{1\} \\  \{1,2\}  , \, \{ 1,3 \}^- \\   \{1,2,3\}^- \end{cases}
\eea
We shall now discuss each one of these branches, and compare their relative global stability.

\subsubsection{Analytic and numerical results for the branches of SU(4)}
\label{sec:su4branches}

Our prior discussion for general~$SU(N)$ gauge group already covered the CB in section~\ref{sec:largek}, the maximal HB in section~\ref{sec:smallk}, as well as a discussions of the branches~$\BB = \{m\}, \{m, N-m\}^\pm$ for any~$m$. We shall briefly recall these results specialized to~$SU(4)$, and then focus on the new branches in~\eqref{eq:su4branches}.

\subsubsection*{$\bullet $ The Coulomb branch CB with $h_1=h_2=h_3=0$}

Using the results of subsection \ref{sec:largek} we have,
\bea
x_1=x_3 = \kappa (t^{-1} s)_1 
\hskip 1in 
x_2 = \kappa (t^{-1}s)_2 
\eea 
Numerical analysis shows that throughout the range~$\mu < \mu_\text{neg}$, we have $(t^{-1}s)_1 < (t^{-1}s)_2$. 
Thus, the branch exists for the following range of $\kappa$, 
\bea
\kappa > \kappa_\text{CB} =  { 1 \over 2 (t^{-1}s)_1} \approx 0.234855
\eea
The solution is locally stable in this range and its potential is given  by $V_\text{CB}$. 
Here and below, the subscripts on $\kappa$ and  on $V$ refer to the branch names.

\subsubsection*{$\bullet $ The single Higgs branch $\BB = \{ 1\}$  with $h_1 > 0, \, h_2=h_3=0$}

Using the general results of subsection \ref{sec:oneH}, the variables $x_1$ and $h_1$ are given by,
\bea
2x_1 - 4 x_1^3 = \kappa (t^{-1}s)_1 
\hskip 0.6in 
2 h_1^2 = u_{11} (1-4x_1^2)
\hskip 0.6in 
\kappa < \kappa_{\{1\} } 
\eea
where $\kappa_{ \{ 1\} }$ is given by,
\bea
\label{7.kappa1}
\kappa_{ \{ 1 \} }  = \frac{4}{3\sqrt{6} (t^{-1}s)_1} \approx 0.255678
\eea
The variables $x_2,x_3$ are given in terms of $x_1$ by
\bea
x_2 & = & \Big ( t_{33}(\kappa s_2 - t_{21} x_1) - t_{23} (\kappa s_3 - t_{31} x_1) \Big ) / (t_{22} t_{33} - t_{23}^2)
\no \\
x_3 & = & \Big (  t_{33} (\kappa s_3 - t_{31} x_1) - t_{23}(\kappa s_2 - t_{21} x_1)  \Big ) / (t_{22} t_{33} - t_{23}^2)
\eea
Local stability requires the conditions, 
\bea
6x_1^2 < 1 &  \hskip 1in  &
4x_k^2 > 1- u_{11} (t^{-1})_{k1} (1-4x_1^2) 
\eea
for $k=2,3$. The value of the reduced effective potential is given by
\bea
\label{7.V5}
V_{ \{ 1 \} } = V_\text{CB} - \tfrac{1}{8} u_{11} (1-8x_1^2)   (1-4x_1^2)^2
\eea
Numerical analysis shows that, in the interval $x_1 \in [0,1/\sqrt{8}]$ where the potential $V_{ \{ 1  \} } $  is lower than the potential $V_\text{CB}$ of the Coulomb branch, the solution in branch $\{ 1 \}$  is locally stable for a small region below $x_1=1/\sqrt{8}$, and is locally unstable throughout the remaining interval, as indicated by the solid and dotted green lines in figure \ref{fig3a}.

\subsubsection*{$\bullet $ The single Higgs branch $\BB=\{ 2 \}$   with $h_1=h_3=0, \, h_2 > 0$}

Using the general results of subsection \ref{sec:oneH} the variables $x_2$ and $h_2$ are given by,
\bea
2x_2 - 4 x_2^3 = \kappa (t^{-1}s)_2 
\hskip 0.6in 
2 h_2^2 = u_{22} (1-4x_2^2)
\eea
where we recall the relations $(t^{-1})_{22} = (u^{-1})_{22}= (u_{22})^{-1}$ that are applicable here. The branch exists provided $\kappa $ satisfies, 
\bea
\kappa  < \kappa _{ \{ 2 \} }  = \frac{4}{3\sqrt{6} (t^{-1}s)_2} \approx 0.188408
\eea
The remaining variables $x_1,x_3$ are given in terms of the solution for $x_2$  by,
\bea
x_1=x_3= {\kappa s_1 - t_{12} x_2 \over t_{11} + t_{13}}
\eea
Local stability requires the conditions, 
\bea
6x_2^2 < 1 
\hskip 1in
4x_1^2 > 1- u_{22} (t^{-1})_{12} (1-4x_2^2)
\eea
and the value of the reduced effective potential is given by,
\bea
V_{ \{ 2 \} } = V_\text{CB} - \tfrac{1}{8} u_{22} (1-8x_2^2) (1-4x_2^2)^2
\eea
Numerical analysis shows that, in the interval $x_2 \in [0,1/\sqrt{8}]$ where the potential $V_{ \{ 2 \} }$  is lower than the potential $V_\text{CB}$ of the Coulomb branch, the solution in branch $\{ 2 \} $ is  locally unstable throughout the interval, as indicated by the dotted purple line in figure \ref{fig3a}.

\subsubsection*{$\bullet $ The 2H branch $\BB= \{ 1, 3 \} ^+$ with $h_1=h_3>0$ and $h_2= 0$}

Using the general results of \ref{sec:oneH}, we find that the variables $x_1$ and $x_3$ must be equal to one another along with $h_1=h_3$, and are given by,
\bea
\begin{cases} 2x_1 - 4 x_1^3 = \kappa (t^{-1}s)_1 \\ 
2 h_1^2 = (u_{11}+u_{13})  (1-4x_1^2) \end{cases} 
\hskip 0.6in 
\kappa< \kappa_{ \{ 1,3 \} ^+} = \kappa _{ \{ 1 \} }
\eea
and $\kappa_{ \{ 1 \} }$ was given in (\ref{7.kappa1}).
The remaining variable $x_2$ is given in terms of $x_1$  by,
\bea
x_2 = {1 \over t_{22}} \Big ( \kappa s_2-(t_{11}+t_{13}) x_1 \Big )
\eea
Local stability requires the conditions, 
\bea
6x_1^2 < 1 &  \hskip 1in  &
4x_2^2 > 1- 2  (t^{-1})_{21} (u_{11}+u_{13})  (1-4x_1^2)
\eea
and the value of the reduced effective potential is given by,
\bea
\label{7.V3}
V_{ \{ 1,3 \}^+}  = V_\text{CB} - \tfrac{1}{4} (u_{11}+u_{13})  (1-8x_1^2) (1-4x_1^2)^2
\eea
This potential is shown in orange in figure \ref{fig3a}. 

\sm

Comparing the potentials  in (\ref{7.V3}) and  (\ref{7.V5}) we observe that $V_{ \{ 1,3 \} ^+} < V_{ \{ 1 \} }$ for any value $0 \leq x_1^2 < 1/8$ using the fact that $|u_{13} |< u_{11}$. The latter follows from the positivity of the matrices $t$ and $u$, while the former guarantees that both potentials are smaller than the potential $V_\text{CB}$ of the Coulomb branch. Hence the branch $\{ 1 \}$ is not globally stable (except at the triple intersection with the CB and the~$\{1,3\}^+$ branch).

\subsubsection*{$\bullet $  The double Higgs branch $\BB=\{ 1,3 \}^-$ with  $h_1 \not= h_3 >0, \, h_2=0$}

The equations for this branch reduce to two coupled cubics in two variables that can be analyzed numerically. Our general methodology is explained in Appendix \ref{app:B}. The result is that the branch $\{ 1,3 \}^-$ is empty or, in other words, that starting out with $h_2=0$ and arbitrary $h_1, h_3 >0$ invariably leads to $h_1=h_3$, which characterizes the branch $\{ 1,3 \}^+$. Independently, we have already shown in section~\ref{sef:h2mininstab} that the branch~$\BB = \{1, 3\}^-$ is always locally unstable (were it to exist).

\subsubsection*{$\bullet $ The double Higgs branch $\BB=\{ 1,2\}$ with $h_1, h_2 >0, \, h_3=0$}

The study of the branch $\BB=\{ 1,2\}$ also requires numerical analysis. While this branch is found to be non-empty, its reduced potential, for a given value of $\kappa$,  is always larger  than the potential in branch HB=$\{ 1,2,3\}^+$, so that the branch $\BB=\{ 1,2\}$ is not globally stable. In fact, it is not even locally stable, as indicated by the dotted cyan line in figure~\ref{fig3a}.

\subsubsection*{$\bullet$ The maximal HB, $\BB=\{1,2,3\}^+$, with $h_1=h_3, h_2>0$}
\label{7.maxHa}

The analysis of the maximal Higgs branch HB involves two coupled cubics, as may be seen by eliminating $h_1=h_3>0$ and $h_2>0$ from the field equations for $x_1=x_3$ and $x_2$,  
\bea
\label{7.maxEq}
2x_k h_k^2 + \sum _{\ell=1}^3 t_{k\ell} x_\ell = \kappa s_k
\hskip 1in 
2 h_k^2 = \sum_{\ell=1}^3 t_{k \ell } (1-4x_\ell^2)
\eea
The reduced potential, evaluated on the solutions of the maximal Higgs branch with $x_1=x_3$, is conveniently expressed as follows,
\bea
\label{redVmaxother}
V_\text{HB} ^\text{sol} = 
V_\text{CB} - {1 \over 8}  \sum _{k, \ell, m,n =1}^3 (t^{-1})_{k\ell} \, t_{km} \, t_{\ell n} (1-8x_k x_\ell)(1-4x_m^2) (1-4x_n^2)
\eea
The plot of the potential $V_\text{HB}$ versus $\kappa$ is obtained by the methods explained in Appendix \ref{app:B} and produces the blue curve in figure \ref{fig3a}.

\subsubsection*{$\bullet$ The~$C$-odd maximal Higgs branch $\BB=\{ 1,2,3\}^-$ with $h_1 \not =h_3, h_2>0$}
\label{7.maxHb}

Using the charge conjugation invariance of $t_{k\ell}$ and $s_k$, and eliminating $h_k^2$ using the left equation of (\ref{7.maxEq}), we see that the difference of the left equations in (\ref{7.maxEq})  for $k=1$ and $k=3$ factorizes. One solution, namely $x_1=x_3$ gave the maximal Higgs branch HB. The other solution is given by the following relation, 
\bea
4t_{11}(x_1^2 + x_3^2) + 4(t_{11} - t_{13}) x_1 x_3 + 4t_{12} x_2^2 = 2 t_{11} +  t_{12} 
\eea
and governs the $\BB=\{ 1,2,3\}^-$ branch.  The numerical analysis of the $\BB=\{ 1,2,3\}^-$ branch proceeds along the same lines as that of the HB branch and is detailed in Appendix \ref{app:B}.  Our numerical analysis shows that the  branch $\BB=\{ 1,2,3\}^-$ is empty.

\subsubsection{Details of the~$SU(4)$ Phase Diagram}
\label{sec:su4phases}

Expanding on the preceding numerical analysis, we now give a more detailed description of the different (non-empty) branches~$\BB$, whose potentials $V_\BB - V_\text{CB}$ are plotted against $\kappa$ in figure \ref{fig3a}, and the resulting phase diagram.

Above, we have numerically established the following ordering:
\bea
0 <  \kappa_{ \{ 2 \} } < \kappa_\text{HB}   < \kappa_\text{CB}  < \kappa_{ \{ 1 \} }  = \kappa_{ \{1,3 \}^+ }
\eea
The Coulomb branch (CB) exists for $\kappa > \kappa_\text{CB}$ while all other branches exist for $\kappa$ smaller than the corresponding critical value of $\kappa$, e.g.~$\kappa < \kappa_{ \{ 1 \} }= \kappa_{ \{ 1,3 \} ^+}$ for  the branches $\{ 1 \}$ and $\{ 1,3\}^+$; $\kappa < \kappa_{ \{ 2 \} }$ for branch $\{ 2 \}$; and $\kappa < \kappa _\text{HB}$ for the maximal Higgs branch HB. The most important inequality is $\kappa_{\text{HB}} < \kappa_{\text{CB}}$, which indicates that the maximal Higgs branch and the Coulomb branch can never coexist. This forces the existence of at least one (and generally several) intermediate phases -- a new phenomenon for~$SU(N)$ with~$N \geq 4$. 

Let us summarize the branch structure in more detail: 
\begin{itemize}
\itemsep=-0.02 in
\item For $\kappa_{\{ 1 \} } = \kappa_{ \{ 1,3 \} ^+}  < \kappa$, none of the Higgs branches exist and the system must be on the Coulomb branch, which is globally stable in this range of $\kappa$;

\item For $\kappa_\text{CB} < \kappa < \kappa_{ \{ 1,3\}^+}$, there are three locally stable branches: CB, $\{ 1 \}$, and $\{ 1,3\}^+$. The right panel of figure~\ref{fig3a} shows that the three branches intersect at the point 
\bea
\kappa _* =  {3 \over 4\sqrt{2} \, (t^{-1}s)_1 }~, \qquad \kappa _{\text{CB}} < \kappa _* < \kappa_{\{ 1,3\}^+}~.
\eea 
The~$\{ 1,3\}^+$ branch has lower potential for~$\kappa < \kappa_*$, while the CB has the lowest potential for~$\kappa > \kappa_*$. Thus there is a phase transition~CB$ \to \{1,3\}^+$ as we dial through~$\kappa_*$ from right to left, while the single Higgs branch~$\{1\}$ is passed over. This is identical to the situation for~$SU(3)$ (see section~\ref{sec:1Hstabsu3}).

\item For $\kappa _{\text{HB}}  < \kappa < \kappa _{\text{CB}}$, neither the Coulomb branch nor the maximal Higgs  branch exists. The $\{ 1,3\}^+$ branch is globally stable with the lowest potential. The~$\{1\}$ branch ceases to be locally stable in this range of~$\kappa$. 

\item For $\kappa < \kappa _{\text{HB}}$ we have a competition between the maximal Higgs branch HB and the branch $\{ 1,3\}^+$. (All other branches that exist are locally unstable.) We know analytically that the HB is the globally stable vacuum for sufficiently small~$\kappa$. Thus there must be a phase transition~$\{1,3\}^+ \to $HB, whose precise location ($\kappa \simeq 0.194$ in figure~\ref{fig3a}) must be settled numerically. 
\end{itemize}

\subsection{Phase Diagram for $SU(5)$}

\begin{figure}[t!]
\centering
\includegraphics[width=0.9\textwidth]{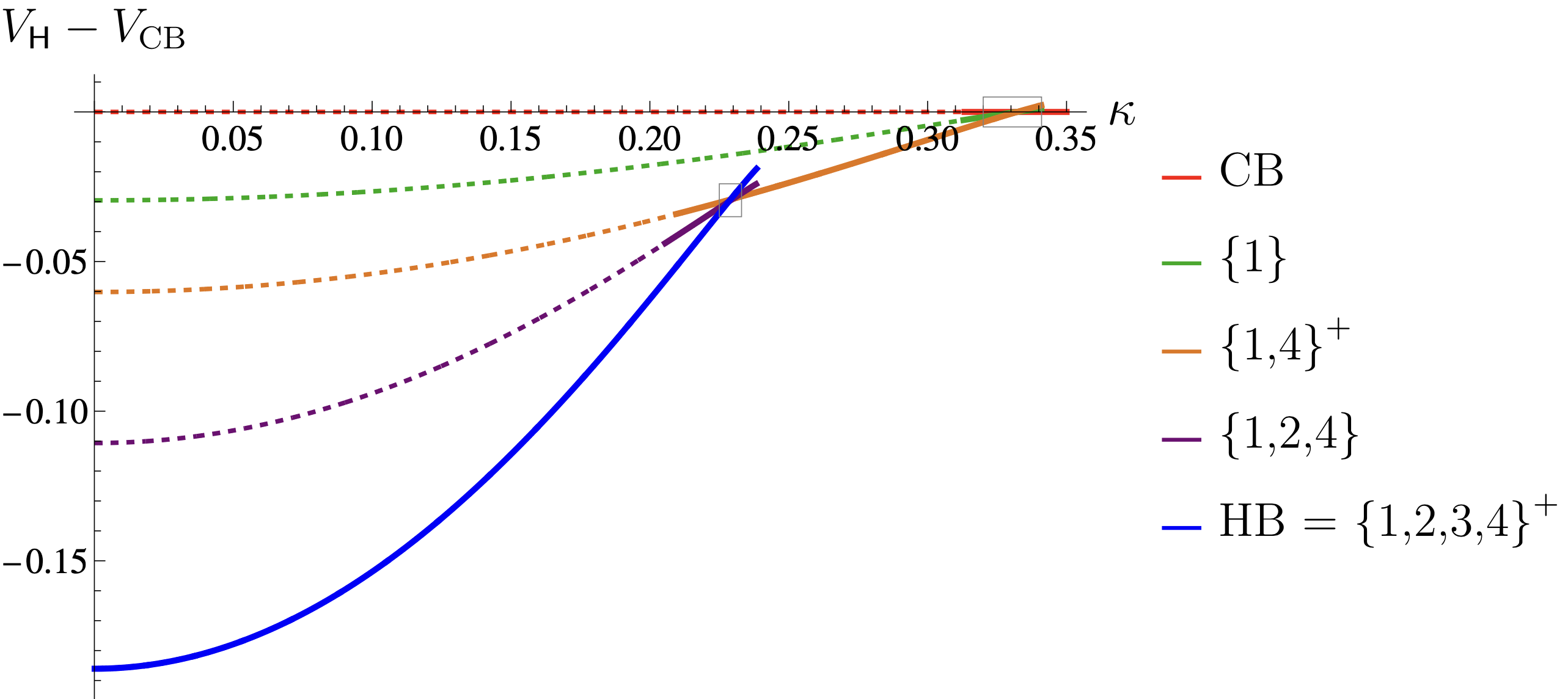}
\includegraphics[width=0.4\textwidth]{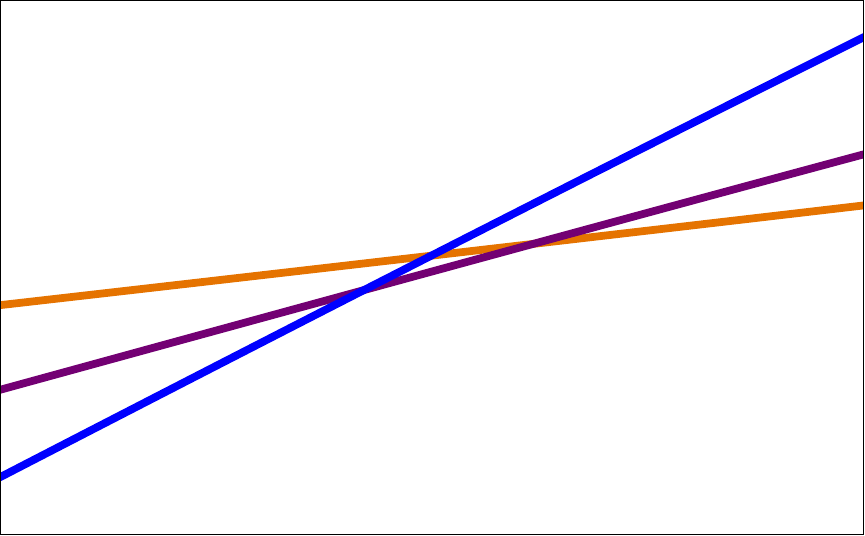}
\includegraphics[width=0.4\textwidth]{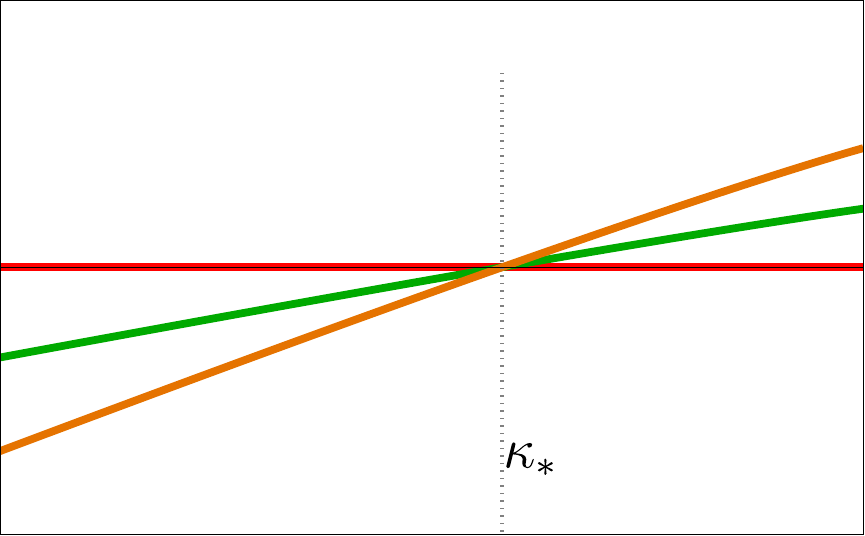}
\caption{The potentials $V_\BB - V_\text{CB}$ for gauge group $SU(5)$ as a function of $\kappa$ for different branches $\BB$. Solid/dotted lines indicate locally stable/unstable solutions. The Coulomb branch (CB) is drawn in red on the horizontal axis and extends to $\kappa \to \infty$;  the branch $\BB=\{1 \}$ is drawn in green; $\BB=\{ 1,4\}^+$ in orange; $\BB=\{ 1,2,4\}$ in purple; and the maximal HB in blue. To avoid clutter, we do not show the remaining branches, which are either empty or never locally stable. The bottom right panel zooms in on the (exact) triple intersection between the CB, $\{1\}$, and~$\{1,4\}^+$ branches at~$\kappa = \kappa_*$; the bottom left panel zooms in on the (approximate) crossing of the~$\{1,4\}^+, \{1,2,4\}$ and HB branches, and exhibits the brief existence of a globally stable $C$-odd~$\{1,2,4\}$ phase around~$\kappa \simeq 0.23$. \label{fig:7}}
\end{figure}

Here we will be brief and present only results, since the approach is identical to the~$SU(4)$ case described in detail above (see also appendix~\ref{app:B} for more details on the numerical analysis). 

For $SU(5)$ the distinct partitions $\AA \cup \BB$ are given by 
\bea
\label{7.branchesSU5}
C\text{-inv.} \, \begin{cases} \text{CB} \\  \{ 1,4\}^+ , \,  \{ 2,3\}^+ \\ \text{HB}= \{ 1,2,3,4\}^+ \end{cases}
\hskip 0.4in
C\text{-non-inv.} \, \begin{cases}  \{1\}, \,    \{2\} \\  \{1,2\}  , \, \{ 1,3 \} , \, \{ 1,4\}^- , \, \{ 2,3\}^-\\   
\{1,2,3\}, \{ 1,2,4\} \\ \{ 1,2,3,4\}^- \end{cases}
\eea

In figure~\ref{fig:7} we plot the potentials~$V_\BB - V_\text{CB}$, for the different branches~$\BB$, but in order to avoid cluttering the figure we omit branches that do not exist, or that exist but are never locally stable.

In accord with general expectations (see section~\ref{sec:sun}), the CB is the true vacuum for large~$\kappa$, and the HB for small~$\kappa$. Reading the figure from large to small~$\kappa$, the first phase transition CB$ \to \{1,4\}^+$ proceeds as for~$SU(3)$ and~$SU(4)$ above, with the transition point~$\kappa_*$ also being the exact triple intersection with the single Higgs~$\{1\}$ branch. 

A new feature of the~$SU(5)$ case not previously encountered, is that the second phase transition $\{1,4\}^+ \to \{1,2,4\}$ briefly opens up a phase that spontaneously breaks~$C$. This phase only persists for a short range of~$\kappa$, before further transitioning $\{1,2,4\} \to $HB. The range in~$\kappa$ over which this happens is so small that the three branches~$\{1,4\}^+, \{1,2,4\}$ and HB look almost degenerate in the top panel of figure~\ref{fig:7}. However, the magnification in the bottom left panel clearly shows that (unlike the exact branch crossing of CB, $\{1\}, \{1,4\}^+$) this is a near miss, with the $C$-odd $\{1,2,4\}$ branch actually being the true vacuum for a small range of~$\kappa$. We will revisit this phenomenon analytically in section~\ref{sec:cascade}.

\newpage

\subsection{Phase diagram for $SU(6)$}

\begin{figure}[t!]
\centering
\includegraphics[width=0.9\textwidth]{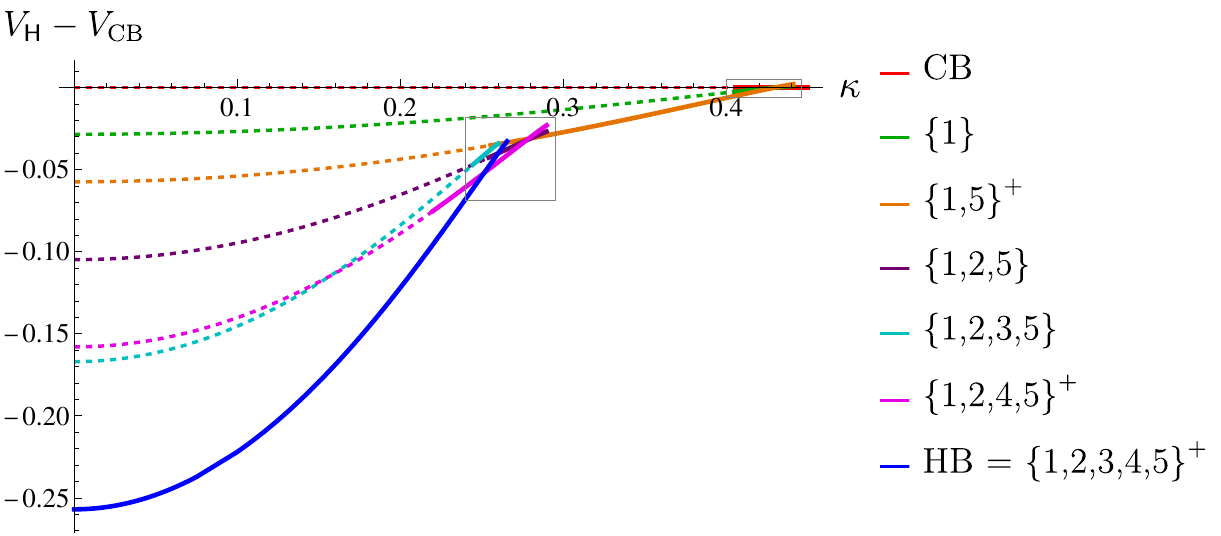}
\includegraphics[width=0.4\textwidth]{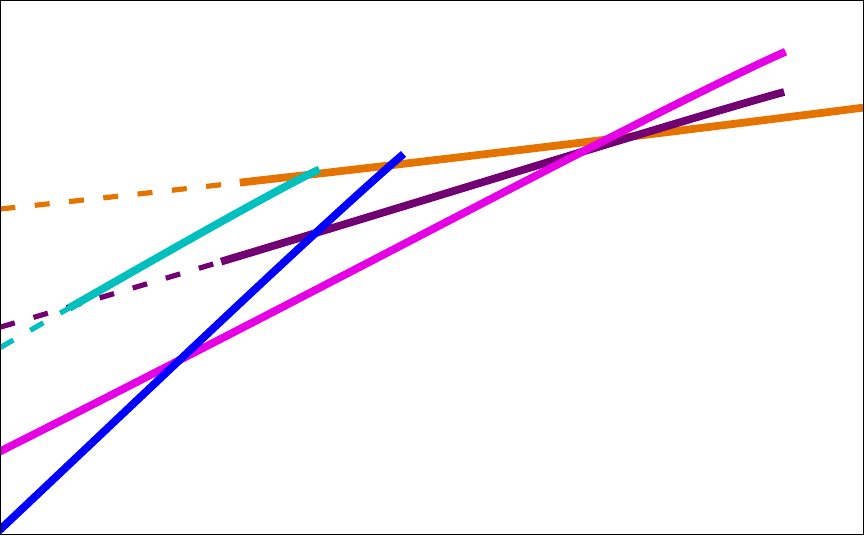}
\includegraphics[width=0.4\textwidth]{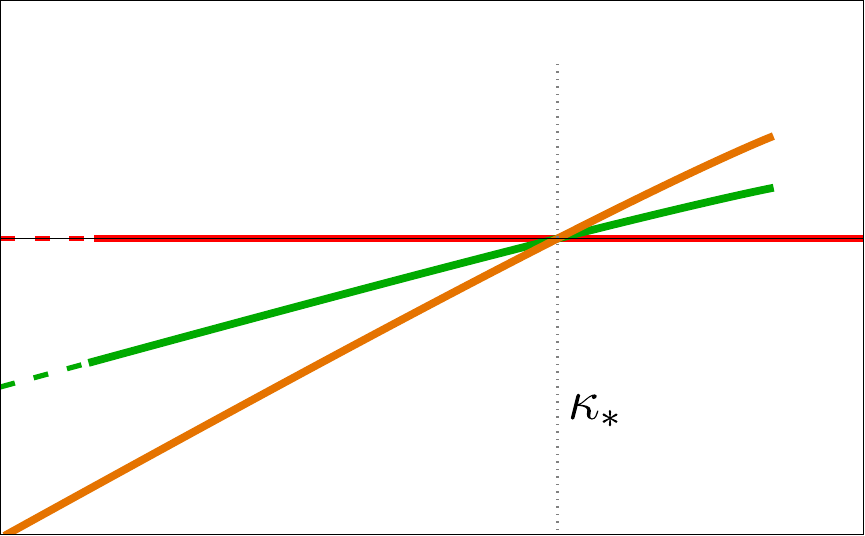}
\caption{The potentials $V_\BB - V_\text{CB}$ for gauge group $SU(6)$ as a function of $\kappa$ for different branches $\BB$. Solid/dotted lines indicate locally stable/unstable solutions. The CB is drawn in red and extends to $\kappa \to \infty$; the branch $\BB=\{1 \}$ is drawn in green; $\BB=\{ 1,5\}^+$ in orange; $\BB=\{ 1,2,5\}$ in purple; $\BB= \{1,2,3,5\}$ in cyan; $\BB=\{1,2,4,5\}^+$ in pink; and the maximal HB in blue. To avoid clutter, we do not show the remaining branches, which are empty or never locally stable. The bottom right panel zooms in on the (exact) triple intersection between the CB, $\{1\}$, and~$\{1,5\}^+$ branches at~$\kappa = \kappa_*$; the bottom left panel zooms in on the (approximate) crossing of the~$\{1,5\}^+, \{1,2,5\}$, and~$\{1,2,4,5\}^+$ branches, and exhibits the brief existence of a globally stable $C$-odd~$\{1,2,5\}$ phase around~$\kappa \simeq 0.28$. \label{fig:77}}
\end{figure}

We plot the potentials~$V_\BB - V_\text{CB}$ for the different~$SU(6)$ branches~$\BB$ in figure~\ref{fig:77}. Again we only show branches that exist and are locally stable for at least some of their existence.

\subsection{Evidence for a cascade of phase transitions}\label{sec:numcascade}

In the previous subsections we have explored the phase diagram of the~$SU(4)$, $SU(5)$, and~$SU(6)$ theories as a function of the SUSY-breaking parameter~$\kappa = {N \Lambda / (2 \pi^2 M)}$ defined in~\eqref{eq:kappadefsec3}, by numerically minimizing the classical scalar potential~\eqref{vscaleawithk} obtained from the Abelian dual at the multi-monopole point. This analysis is captured by the effective potentials plotted in figures~\ref{fig3a}, \ref{fig:7}, and~\ref{fig:77} respectively. By starting in the large-$\kappa$ regime at the top-right of these figures and tracing the envelope of lowest potential as~$\kappa$ decreases, we obtain the phase diagrams. Here we summarize these phase diagrams, emphasizing common structural aspects that naturally generalize to all~$SU(N)$ gauge groups:

\begin{table}[htp]
\begin{center}
\begin{tabular}{|c||c|c|c|c|} \hline
$N$ & $\kappa_\text{CB}$ & $\kappa_{\{1\}} $  & $\kappa_\star$ & $\kappa_\text{HB}$ 
\\ \hline
2 & 0.1314 	& 		&  0.1394	& 0.1430		\\ \hline
3 & 0.1716	& 0.1700	&  0.1820	& 0.1868		\\ \hline
4 & 0.2349	& 0.2557	&  0.2491	& 0.2031		\\ \hline
5 & 0.3133	& 0.3409 	&  0.3322	& 0.2384		\\ \hline
6 & 0.4052	& 0.4410	&  0.4297 	& 0.2655		\\ \hline
\end{tabular}
\end{center}
\caption{Various threshold values of~$\kappa$, including its value~$\kappa_*$ at the phase transition out of the Coulomb branch, for~$N =2$ through~$N = 6$. All values are obtained with~$\mu = 10^{-3} \Lambda$.}
\label{tab:kappastar}
\end{table}

\begin{itemize}
    \item At large~$\kappa$ we always find the Coulomb branch. The first phase transition out of the CB occurs at~$\kappa = \kappa_*$, where the three branches CB, $\{1\}$ and~$\{1, N-1\}^+$ are exactly (and accidentally) degenerate. (The values of~$\kappa_*$ for~$N = 2$ through~$N=6$ are listed in table~\ref{tab:kappastar}.) Thus dialing~$\kappa$ through~$\kappa_*$ leads to the phase transition CB$ \to \{1, N-1\}^+$, bypassing the~$C$-odd single Higgs branch~$\{1\}$. Exactly this feature was already discussed for~$SU(3)$ in section~\ref{sec:1Hstabsu3}, and it persists for~$N = 4,5,6$. 

    \item At small~$\kappa$ we always find the maximal HB, which is necessarily  separated by the CB by at least one intermediate phase for~$N \geq 4$ (and unlike what happens for~$N = 2, 3$). 

    \item The additional phases and transitions can be roughly -- but in general incorrectly (see below) -- characterized by saying that the Higgs fields turn on in the following~$C$-symmetric, cascading pattern of phase transitions,
    \begin{equation}\label{eq:naivecascade}
 \text{CB} \; \to \; \{1, N-1\}^+ \; \to \; \{1, 2, N-2, N-1\}^+ \; \to  \; \cdots \; \to \; \text{HB}~.
    \end{equation}
This picture in fact correctly describes the phase diagram of~$SU(4)$ in figure~\ref{fig3a}, but it is not correct for~$SU(5)$: there the transition~$\{1,4\}^+ \to $ HB is interrupted by the brief appearance of a~$C$-breaking intermediate phase, as shown in the bottom-left panel of figure~\ref{fig:7}, so that
    \begin{equation}
     SU(5) \; : \qquad   \{1,4\}^+ \to \{1,2,4\} \to  \text{HB}
    \end{equation} 
    Something similar happens for~$SU(6)$, where the naive transition~$\{1,5\}^+ \to \{1,2,4,5\}^+$ is very briefly interrupted by a~$C$-breaking~$\{1,2,5\}$ phase (barely visible in the bottom-left panel of figure~\ref{fig:77}). Note, however, that for~$SU(6)$ (as was the case for~$SU(4)$) this phenomenon does not happen for the final transition into the HB, because this only involves turning on a single~$C$-even Higgs field. 

We see that although the~$C$-even phases~\eqref{eq:naivecascade} dominate the cascade (in the sense that they occupy most of the phase diagram in~$\kappa$-space), a $C$-odd interpolating phase very briefly appears in between generic~$C$-even phases. Thus most transitions in~\eqref{eq:naivecascade}, which involve the simultaneous activation of two Higgs fields, are actually split into two transitions involving one Higgs field at a time,  
    \begin{equation}\label{eq:truecascade}
        \BB = \{1, \ldots, m, N-m, \ldots, N-1\}^+ \to \BB \cup \{m+1\} \to \BB \cup \{m+1, N-m-1\}^+~. 
    \end{equation}
However, in all examples the intermediate~$C$-odd phase~$\BB \cup \{m+1\}$ only persists very briefly in~$\kappa$. We have also uncovered two exceptions to~\eqref{eq:truecascade}:
\begin{itemize}
    \item The first transition~CB$ \to \{1, N-1\}^+$ is unmodified. 
    \item If~$N = 2\nu$ is even and~$m + 1 = \nu = N-m-1$, then all branches in~\eqref{eq:truecascade} are~$C$-even, and the last two branches coincide. Thus the last transition~$\AA = \{\nu\} \to $ HB is also unmodified in this case. 
\end{itemize}

\item Finally, all transitions are first order.

\end{itemize}

\smallskip

 We refer to the structure uncovered above as a cascade of first-order phase transitions interpolating between the CB and HB. In more detail:
\begin{itemize}
\item[(i)] We refer to the approximate pattern of~$C$-even phases~\eqref{eq:naivecascade} that dominates the cascade when one zooms out far enough in~$\kappa$ as its {\it coarse structure}.

\item[(ii)] We refer to the exact pattern~\eqref{eq:truecascade}, where the~$C$-even phases are briefly interrupted by~$C$-odd ones, as the {\it fine structure} of the cascade. 

\end{itemize}
We conjecture that this structure persists for all values of~$N$. (At large-$N$ the number of transitions is~$\CO(N)$.) This is strongly supported by the approximate analytic approach to the cascade that we will develop in section~\ref{sec:cascade} below, which is valid for all~$SU(N)$ gauge groups and explains many of its qualitative (and even quantitative) features.

\newpage

\section{Cascading phase transitions in perturbation theory} 
\label{sec:cascade}

In the preceding sections we have analyzed the phase structure of the~$\CN=2$ SYM theory with gauge group~$SU(N)$ and SUSY-breaking mass~$M$, i.e.~$\kappa = N \Lambda / (2 \pi^2 M)$ in~\eqref{eq:kappadefsec3}, by analyzing the semi-classical vacua of the Abelian dual at the multi-monopole point. These are found by minimizing the potential~\eqref{vscaleawithk} as a function of~$\kappa$. Above we have done this for~$N \leq 6$; we have also obtained results for general~$N$ in the large- and small-$\kappa$ limits corresponding to the Coulomb and maximal Higgs branch, respectively. As described in section~\ref{sec:numcascade}, these two regimes are connected by a cascade of first-order phase transitions, with a two-tier structure that we referred to as the coarse structure (in~\eqref{eq:naivecascade}) and the fine structure (in~\eqref{eq:truecascade}) of the cascade. 

\sm

In this section, we will introduce a perturbative approximation scheme, applicable for all values of~$N$ and~$\kappa$, that establishes this cascade -- including its coarse structure (at leading order) and its fine structure (at higher orders) -- and is in excellent agreement with the results obtained for~$N \leq 6$ in previous sections. 

\sm

This approximation involves Taylor expanding the field equations, their solutions, and the potential in powers of the off-diagonal entries of the matrix $t$ or its inverse $t^{-1}$.\footnote{~Although roughly equivalent, these expansions differ in the details, including in the assumptions under which they are convergent, as discussed in section~\ref{sec:8.1}.} This approximation is clearly justified if the RG scale~$\mu$ in~\eqref{eq:tdef1} is sufficiently small, so that the diagonal entries of $t$ and $t^{-1}$ become large and dominate the off-diagonal entries. Reassuringly, even at larger values of~$\mu$ its predictions are in good agreement with our previous results. 

\sm

We stress that the approximations obtained by expanding in powers of the off-diagonal entries of $t$ or $t^{-1}$ will be applied to the reduced field equations of (\ref{xB}) or (\ref{xBalt}) and the reduced effective potential (\ref{redV}), whose derivation already accounts for the vacuum alignment established in section~\ref{sec:vacal}. This in turn relied on the fact that the off-diagonal entries of the matrix $t^{-1}$ are negative definite.

\subsection{Defining perturbation theory around diagonal $t^{-1}$}
\label{sec:8.1}

The inverse of the diagonal part of the matrix $t$ is not equal to the diagonal part of the matrix $t^{-1}$. Thus, one may naturally define different expansions: either around the diagonal of $t$, or around the diagonal of $t^{-1}$. We will choose the latter, and expand the matrix $t^{-1}$ around its diagonal. This has both practical and conceptual advantages. 

Practically, the decomposition of the matrix $t^{-1}$ and the matrix $u^{-1}$ in an arbitrary partition $\AA|\BB$ may then be parametrized as,
\begin{align}
\big ( t^{-1} \big )_{mn} & =   \delta _{mn} \, \big ( t^{-1} \big )_{mm} - P_{mn} 
& m,n & \in \{ 1,2, \ldots, N-1 \}
\no \\
\big ( u^{-1} \big )_{k \ell} & =   \delta _{k \ell} \, \big ( t^{-1} \big )_{kk} - P_{k \ell}  & k, \ell & \in \BB
\end{align} 
where the non-negative matrix~$P$ is minus the (negative definite) off-diagonal part of $t^{-1}$, 
\bea\label{eq:pdef}
P_{mn}  = - \big ( 1- \delta_{mn}  \big ) \big ( t^{-1} \big )_{mn}
\eea
The matrices $t$ and $u$ themselves are then obtained by geometric series,
\begin{align}
\label{9.tu}
t  & =  t^{(0)}   + t^{(0)} P t^{(0)} + t^{(0)} P t^{(0)} P t^{(0)} + \cO(P^3)  
& t^{(0)} _{mn} & = \delta_{mn} / \big ( t^{-1} \big ) _{mm} 
\no \\
u & =  u^{(0)} +  u^{(0)} P u^{(0)} + u^{(0)} P u^{(0)} P u^{(0)} + \cO(P^3)
& u^{(0)} _{k\ell} & = \delta_{k\ell} / \big ( t^{-1} \big ) _{kk} 
\end{align}
Here~$m,n \in \{ 1,2,\ldots, N-1 \}$ and~$k, \ell \in \BB$; the matrix $P$ in the expression for $u$ is restricted to $\BB$. The diagonal matrix $u^{(0)}$ is the restriction to $\BB$ of the diagonal matrix $t^{(0)}$.

\sm

The conceptual advantage of the expansion of the matrix $t^{-1}$ stems from the fact that all its off-diagonal entries are negative (see~(\ref{2.tinvneg})), and hence the matrix~$P$ in~\eqref{eq:pdef} is non-negative. (More precisely, its diagonal entries vanish, while the off-diagonal ones are strictly positive.) Thus both series in (\ref{9.tu})  are absolutely convergent.  

\subsubsection{A modified perturbation expansion}
\label{sec:exp}

Throughout, we will use a modified prescription for carrying out the perturbative expansion in the off-diagonal elements~of~$t^{-1}$, i.e.~in the matrix~$P$ defined in~\eqref{eq:pdef}:

\begin{itemize}
    \item The matrix~$u$ that appears in the field equations (\ref{xB}) or (\ref{xBalt}) and in the effective potential (\ref{redV}), is treated perturbatively, by expanding in~$P$ using (\ref{9.tu}).
    \item However, the full matrix $t^{-1}$ is retained in the combination $(t^{-1} s)$ that appears in these same equations.
\end{itemize}
This modified expansion leads to simple analytic formulas, and we have found that it converges much more rapidly to the the numerical results obtained in section~\ref{sec:num}.

\subsection{Cascading transitions to leading order: coarse structure} 

In this subsection, we shall solve the reduced field equations (\ref{xB}) or (\ref{xBalt}) to leading order in the perturbation expansion described in  subsection \ref{sec:exp} above: this involves retaining the matrix~$u$ to leading~$u^{(0)}$ order, while keeping the exact values of $(t^{-1}s)$. At this order, the system of cubics decouples, and each cubic reduces to the $SU(2)$ case, with modified parameters, which was already  solved in section \ref{sec:su2}. The results are as follows. 

\sm

To leading order in the expansion,  the field equations (\ref{xA}) and (\ref{xB})  for a given partition $\AA | \BB $ reduce as follows,\footnote{~Note that the solution for~$x_k \; (k \in \AA)$ precisely agrees with the Coulomb branch solution~\eqref{xmcb}. }
\begin{align}
\label{diagsol}
k & \in \AA &  x_k & = \kappa (t^{-1}s)_k   & h_k & =0
\no \\
k & \in \BB &  2x_k - 4 x_k^3 &= \kappa (t^{-1}s)_k  &  2h_k^2 & =  u_{kk}^{(0)}  (1-4x_k^2)
\end{align}
and the reduced off-shell potential of (\ref{redV}) takes the following form, 
\bea
\label{redVdiagoff}
V_\BB = V_\text{CB}  +
\sum_{k \in \BB} u_{kk}^{(0)}  \bigg [ \left ( x_k- \kappa (t^{-1}s)_k \right )^2   - {1 \over 8}  (1-4x_k^2)^2  \bigg ]
\eea
where $V_\text{CB}$ is given by (\ref{VA}). Evaluating the effective potential on a solution to (\ref{diagsol}) gives the on-shell effective potential $V_\BB^\text{sol}$. It may be usefully expressed by eliminating $x_k - \kappa (t^{-1}s)_k$ in favor of $4x_k^3-x_k$ using the field equation for $x_k$ with $k \in \BB$, and we obtain, 
\bea
\label{redVdiagon}
V_\BB ^\text{sol} = V_\text{CB} -
{ 1 \over 8} \sum_{k \in \BB} u_{kk}^{(0)}   \left ( 1 - 8 x_k^2 \right )   (1-4x_k^2)^2  
\eea
where $x_k$ is a solution to $2x_k - 4 x_k^3 = \kappa (t^{-1}s)_k$ and $u^{(0)}_{kk}$ is given by (\ref{9.tu}) in terms of the diagonal entries of $t^{-1}$,  $u^{(0)}_{kk}=1/(t^{-1})_{kk}$.

\subsubsection{Existence of solutions and their stability to leading order}

Next, we analyze the existence of these solutions and their local and global stability as a function of $\kappa$, to leading order in the expansion of subsection \ref{sec:exp}. 

\sm

\begin{itemize}
    \item \textit{Existence} of a solution corresponding to a partition $\AA | \BB$ requires that, for every $k \in \BB$,  we have $4x_k^2<1$ by the positivity of $h_k^2$, and $0 < x_k$ by the positivity of $\kappa$. These conditions  impose  restrictions on the values of $\kappa$ for which solutions corresponding to the partition $\AA | \BB$ can exist.  Since the maximum of the function $2x_k - 4 x_k^3$ is attained for $x_k=1/\sqrt{6}$ in the allowed interval $[0, \thalf]$ for $x_k~(k \in \BB)$, the solution only exists when~$\kappa$ is bounded by
\bea
\kappa < \frac{4}{3\sqrt{6}} { 1 \over (t^{-1}s)_k}~, \qquad \frac{4}{3\sqrt{6}} \approx  0.54433105~, \qquad \text{for all } k \in \BB~.
\eea
On the other hand, for $k \in \AA$, the solution $x_k = \kappa (t^{-1}s)_k$ exists  for all $\kappa$.

\item  \textit{Local stability} of a solution corresponding to a partition $\AA|\BB$ requires $x_k < 1/\sqrt{6}$ for all $k \in \BB$ and $\thalf < x_k$ for all $k \in \AA$.  

\item \textit{Global stability} of a (locally stable) solution corresponding to a partition $\AA|\BB$ for a given value of $\kappa$,  requires that solution to have the lowest potential relative to all other partitions and solutions that exist for that value of~$\kappa$. As we will now explain, this induces an ordering of the partitions that gives rise to the coarse structure of the cascade. 
\end{itemize}

\subsubsection{Ordering of partitions and the coarse structure of the cascade}

 To compare the potentials for different partitions, we begin by clarifying the structure of the different combinations  $(t^{-1}s)_k$ that occur in the field equations. The charge conjugation relations $s_{N-k}=s_k$ and $t_{N-k, N-\ell}  = t_{k,\ell}$ imply that all solutions satisfy $x_{N-k}=x_k$ and $h_{N-k} = h_k$, and are therefore charge conjugation symmetric. Furthermore, we have, 
\bea
(t^{-1}s)_k < (t^{-1}s)_\ell \hskip 0.5in \hbox{ for } 1\leq k < \ell  \leq \tfrac{N}{2}
\eea
Thus,  the coefficients of $\kappa$ on the right side of the field equations for $x_k$ in (\ref{diagsol})  are ordered. See also the closely related discussion around~\eqref{6.tinvsineq}.

\sm

The global stability conditions may be read off from the reduced potential evaluated on the solutions, as given in (\ref{redVdiagon}). Clearly, any solution for which $8 x_k^2>1$ for some $k \in \BB$ has higher potential than the corresponding solution where the same $k \in \AA$, namely for which the Higgs field $h_k$  is turned off. Thus, a necessary condition that any globally stable solution must satisfy is $x_k < 1/\sqrt{8}$ for all $k \in \BB$, whenever such a solution to the equation $2x_k - 4 x_k^3 = \kappa (t^{-1}s)_k$ exists.  Putting all together, we obtain the following picture for the globally stable solutions for a given value of $\kappa>0$,
\bea\label{eq:diagkbounds}
  { 3  \over  4 \sqrt{2} }   { 1 \over (t^{-1}s)_k} < \kappa & \qquad \Longrightarrow \qquad & k \in \AA
\no \\
 { 3 \over  4 \sqrt{2} }  {  1 \over (t^{-1}s)_k} > \kappa & \Longrightarrow & k \in \BB
\eea

 \begin{figure}[t!]
  \centering
\begin{tikzpicture}[scale=8]
  \draw[->,line width=0.9] (-0.1, 0) -- (0.9, 0) node[right] {\large $x$};
  \draw[->,line width=0.9] (0, -0.4) -- (0, 0.9) node[left] {\large $\kappa (t^{-1} s)_n$};
  
 \draw[scale=1, domain=-0.1:0.5, densely dotted, variable=\x, red,line width=1] plot ({\x}, {\x});
 \draw[scale=1, domain=0.5:0.89, smooth, variable=\x, red,line width=1.3] plot ({\x}, {\x});
    \draw[scale=1, domain=-0.07:0.408248, smooth, variable=\x, blue,line width=1.3] plot ({\x}, {2*\x-4*\x*\x*\x});
    \draw[scale=1, domain=0.408248:0.5, densely dotted, variable=\x, blue,line width=1] plot ({\x}, {2*\x-4*\x*\x*\x});

  \draw[line width = 0.5,densely dotted, blue] (-0.1,0.35) -- (0.9,0.35) node[right] {\footnotesize $n=1$ };
   \draw[line width = 0.5,densely dotted, blue] (-0.1,0.42) -- (0.9,0.42) node[right] {\footnotesize $n=2$ };
 \draw[line width = 0.5,densely dotted, blue] (-0.1,0.47) -- (0.9,0.47)  node[right] {\footnotesize $\cdots$ };
 
  \draw[line width = 0.5,densely dotted, red] (-0.1,0.6) -- (0.9,0.6) node[right] {\footnotesize $\cdots$ };
   \draw[line width = 0.5,densely dotted, red] (-0.1,0.7) -- (0.9,0.7) node[right] {\footnotesize $\cdots$ };
 \draw[line width = 0.5,densely dotted, red] (-0.1,0.8) -- (0.9,0.8)  node[right] {\footnotesize $\cdots$ };
  \draw[line width = 0.5,densely dotted, red] (-0.1,0.75) -- (0.9,0.75) node[right] {\footnotesize $\cdots$ };
 \draw[line width = 0.5] (-0.2,0.53) -- (1.2,0.53)  node[right] {Transition point};
 \draw [color=red] (1.42, 0.68) node{Coulomb branch};
 \draw[color=blue]  (1.38,0.38 ) node{Higgs branch};   
 \draw[black] (-0.3, 0.53) node{\large $\frac{3}{4 \sqrt{2}}$};
   
      \draw[line width = 0.4,gray,densely dotted] (1/2,0) -- (1/2,0.5);     
    \draw (0.7,-0.25) node{\large $\hskip 0.1in x_{\text{CB}} $};
    \draw (1/2,-0.03) -- (1/2,0.03);
       \draw [->] (0.65, -0.2) -- (0.52, -0.05);
       
     \draw (0.45,-0.26) node{\large $ x_{\text{HB}}$};
         \draw (0.408248,-0.03) -- (0.408248,0.03);
         \draw[line width = 0.4,gray,densely dotted] (0.408248,0) -- (0.408248,0.54);
         \draw [->] (0.423, -0.2) -- (0.41, -0.05);
         
             \draw (0.22,-0.26) node{\large $x_{*} \hskip 0.1in $};
         \draw (0.353553,-0.03) -- (0.353553, 0.03);
          \draw[line width = 0.4,gray,densely dotted] (0.353553,0) -- (0.353553,0.53);
          \draw [->] (0.25, -0.2) -- (0.34, -0.05);
         
         \draw[fill=blue,line width=0] (0.189,0.35) circle (.04ex);
          \draw[fill=blue,line width=0] (0.235,0.42) circle (.04ex);
           \draw[fill=blue,line width=0] (0.277,0.47) circle (.04ex);
         \draw[fill=red,line width=0] (0.7,0.7) circle (.04ex);
          \draw[fill=red,line width=0] (0.75,0.75) circle (.04ex);
           \draw[fill=red,line width=0] (0.6,0.6) circle (.04ex);
            \draw[fill=red,line width=0] (0.8,0.8) circle (.04ex);
         \draw[ red] (0.95, 0.85) node{\large $x$};
         \draw[blue] (-0.2, 0.2) node{\large $2x-4x^3$};
         \draw[black,fill=black] (0.353553,0.53) circle (.04ex);
         \draw[black,fill=black] (0.53,0.53) circle (.04ex);
         
\end{tikzpicture}
\caption{Graphical representation of globally stable solutions to leading order in the perturbative expansion defined in subsection~\ref{sec:exp}. Horizontal lines  intersecting the solid red curve~$x$ at a red dot give solutions $x_n~(n \in \AA)$, while horizontal lines intersecting the solid blue curve~$2x - 4x^4$ at a blue dot give solutions $x_n~(n \in \BB)$. As $\kappa$ is decreased the horizontal lines sweep the figure from top to bottom, eventually crossing the horizontal black line that indicates the first order phase transition where the Higgs field~$h_n$ turns on. This figure is a rescaled version of the~$SU(2)$ figure~\ref{fig:cubic}.
\label{fig:7.1}}
\end{figure}
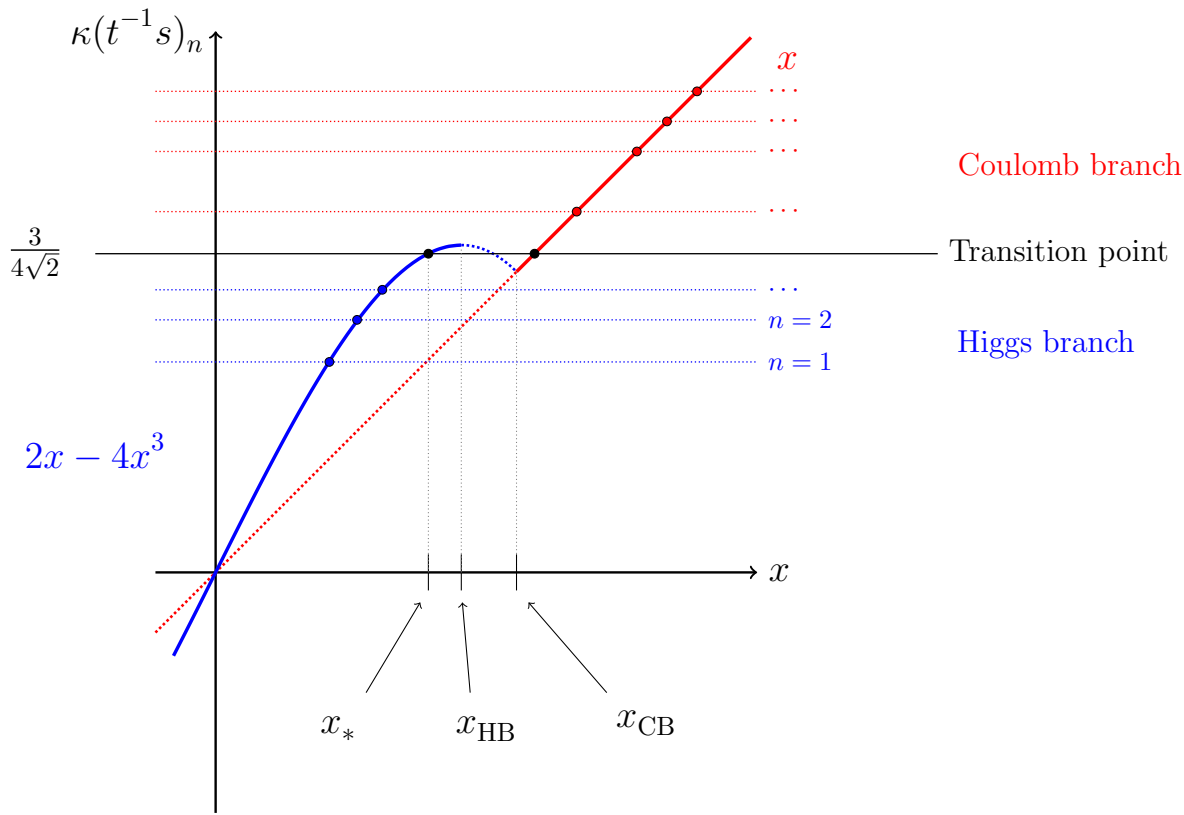

We conclude from~\eqref{eq:diagkbounds} that the phase transitions at which~$k \in \AA$ moves to~$k \in \BB$ occurs at the following value of~$\kappa$,
\begin{equation}\label{eq:PTkappastar}
    \kappa_{*k} = {3 \over 4 \sqrt{2}} { 1 \over (t^{-1}s)_k} ~, \qquad {3 \over 4 \sqrt{2}} \approx 0.53033008
\end{equation}
Due to the~$C$-symmetry of this expression, we have the ordering
\begin{equation}\label{eq:PTkordering}
\kappa_{*1} = \kappa_{*, N-1} > \kappa_{*2} = \kappa_{*, N-1} > \cdots 
\end{equation}
The existence of the cascade, and its coarse structure, immediately follow:
\begin{itemize}
    \item For  large values of $\kappa$, which exceed the transition points $\kappa_{*k} = {3 / ( 4 \sqrt{2} (t^{-1}s)_k})$ in~\eqref{eq:PTkappastar} for all $k =1, \ldots, N-1$, we find the Coulomb branch corresponding to the partition $\BB=0$, a result already obtained in subsection \ref{sec:largek}. 

\item For small values of $\kappa$, less than the transition values $\kappa_{*k} = {3 / ( 4 \sqrt{2} (t^{-1}s)_k})$ in~\eqref{eq:PTkappastar} for all $k =1, \ldots, N-1$, we find the maximal Higgs branch $\AA= \emptyset$, $\BB = \text{HB}$, a result already obtained  in subsection \ref{sec:smallk}. 

\item As $\kappa$ is decreased from larger  to smaller values, the Higgs vevs are successively turned on in a charge-conjugation symmetric pattern that follows from~\eqref{eq:PTkordering}, starting with $h_1=h_{N-1}$, followed by $h_2=h_{N-2}$ and so forth, as depicted in figure~\ref{fig:7.1}. This precisely leads to the coarse structure of the cascade~\eqref{eq:naivecascade},
\begin{equation}
    \text{CB} \to \{1, N-1\}^+ \to \{1, 2, N-2, N-1\}^+ \to \cdots \text{HB}~.
\end{equation}

\item These transitions are all first order: the square Higgs vev $h_k^2$ jumps from $h_k^2 = 0$ for  $ k \in \AA$ to the strictly positive value $h_k^2 = t_{kk}^{(0)}/4 > 0$ for  $k \in \BB$. Correspondingly, $x_k$ jumps discontinuously from $x_k = 3/\sqrt{32} \approx 0.53033$ for $k \in \AA$ to $x_k = 1/\sqrt{8} \approx 0.35355$ for~$k \in \BB$, as indicated by the two black dots on the horizontal black line in figure \ref{fig:7.1}. 
\end{itemize}

\subsubsection{Comparison with exact numerics}

\begin{figure}[h!b]
    \centering
    \includegraphics[width=0.49\linewidth]{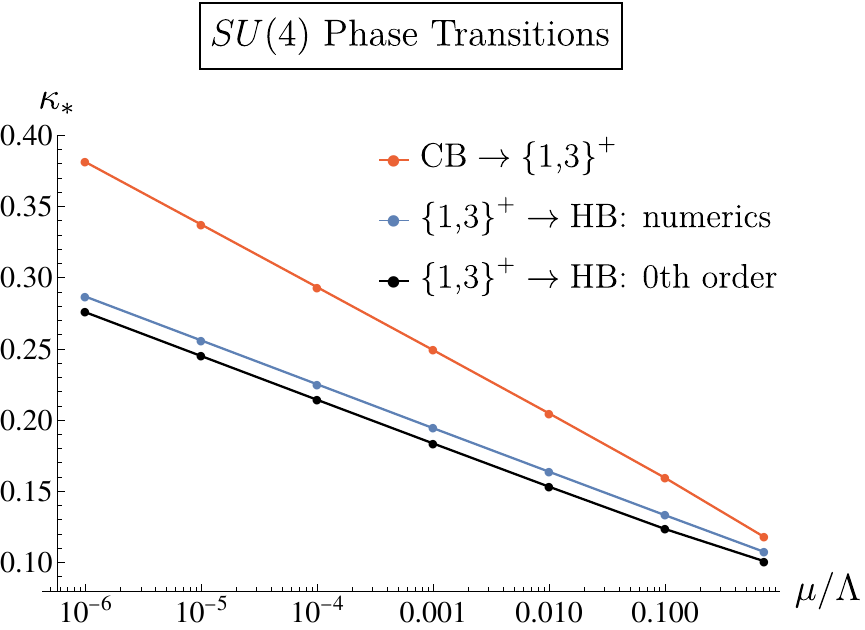}
        \includegraphics[width=0.49\linewidth]{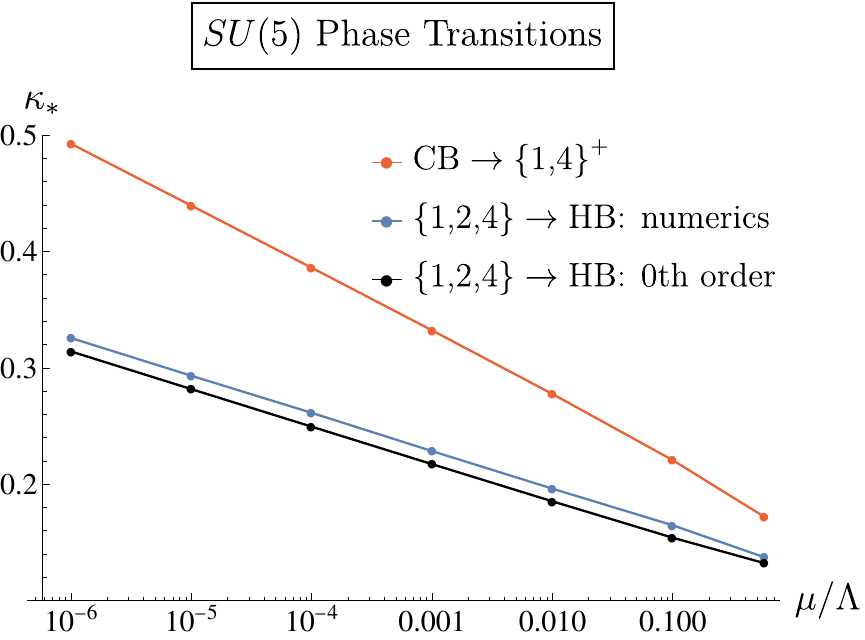}
    \caption{Phase transition values~$\kappa_*$ for~$SU(4)$ and~$SU(5)$ gauge groups, as a function of RG scale~$\mu$. The red dots indicate the first transition into the Coulomb branch (for which perturbation theory is exact); the blue dots (exact numerics) and the black dots (leading order perturbation theory) depict the last phase transition into the maximal Higgs branch.}
    \label{fig:comparingk}
\end{figure}

In the previous subsection we have analytically established the coarse structure of the cascade, by working to leading order in the perturbative scheme explained in section~\ref{sec:exp}. This coarse structure is also observed numerically, as summarized in section~\ref{sec:numcascade}. We would now like to quantitatively compare leading order perturbation theory and the exact numerics in section~\ref{sec:num}. 

To this end, we compare the phase transition values~$\kappa = \kappa_{*m}$ obtained at leading perturbative order in~\eqref{eq:PTkappastar}, with the exact numerical values. The resulting comparison is shown in figure~\ref{fig:comparingk}, for gauge groups~$SU(4)$ and~$SU(5)$, as a function of the RG scale~$\mu$, and all the way to the largest possible values~$\mu \lesssim \mu_\text{neg}$. 
 For both gauge groups, we plot the value of~$\kappa$ for the first transition out of the Coulomb branch (for which perturbation theory and the exact numerics agree, see section \ref{sec:aD3} below), and the value of~$\kappa$ for the last transition into the maximal Higgs branch. We see that even the leading order perturbative expression~\eqref{eq:PTkappastar} for the transition values tracks the exact answers quite closely, over a large range of~$\mu$, and is therefore a good quantitative approximation. The approximation can be improved by including higher-order perturbative corrections, as we will do below.

\subsection{Perturbative corrections}

We now proceed to investigate how the above leading-order picture is altered as we include subleading corrections in the perturbative expansion of subsection \ref{sec:exp}. First, we will obtain the relevant equations for $x_k$, and then study the evolution of the transitions as the perturbation is tuned on. This requires evaluating~$V_\BB - V_\text{CB}$ for the solution given by a partition $\AA | \BB$ in this approximation, and then comparing the results from different partitions. As we will see in subsection~\ref{sec:liftdegen} below, the perturbative corrections are especially important in the vicinity of the phase transitions, because of accidental degeneracies at leading order in perturbation theory. Here we begin with a general discussion of the subleading terms of the perturbative expansion. 

The starting point is the exact reduced off-shell potential,
\bea
V_\BB = V_\text{CB} + \sum_{k, \ell \in \BB} u_{k \ell} \, \Big [
\left (x_k - \kappa (t^{-1} s)_k \right ) \left (x_\ell - \kappa (t^{-1} s)_\ell \right ) 
- \tfrac{1}{8} (1-4 x_k^2)(1-4 x_\ell^2) \Big ]
\eea
and the corresponding exact reduced field equations of (\ref{xB}), 
\bea
\sum_{\ell \in \BB} u_{k \ell} \, \Big [ x_\ell - \kappa (t^{-1} s)_\ell + x_k (1-4 x_\ell^2) \Big ] =0
\eea
Recall that the matrix $u$ is defined in terms of the matrix $t^{-1}$ and the partition $\BB$ by (\ref{uinvv}), and thus explicitly depends on $\BB$. An economical formulation of the expansion is obtained in terms of the solutions $y_k$ to the following system of decoupled equations, 
\bea
\label{9.yt}
2 y_k - 4 y_k^3 = \kappa (t^{-1} s)_k \hskip 0.6in k \in \BB
\eea
We stress that, in this equation, the full matrix $t^{-1}$ is always retained on the right side, regardless of the order of perturbation theory, as explained in  subsection \ref{sec:exp}.  Thanks to the fact that the combinations $(t^{-1} s)_k$ remains the same to all orders in the expansion, the function $y_k$ similarly remains the same to all orders.\footnote{~If we had also expanded  $(t^{-1}s)$, the values of $y_k$ would similarly require expansion, which would significantly complicate the calculations.} In fact, the~$y_k$ are nothing but the solutions of the leading-order equations~\eqref{diagsol}.  

\sm

Next, we eliminate $\kappa (t^{-1} s)_k$ from the potential and the field equations in favor of the variables $y_k$ which are given in terms of $\kappa$ by (\ref{9.yt}).  The potential becomes, 
\bea
V_\BB = V_\text{CB} + \sum_{k, \ell \in \BB} u_{k \ell} \, \Big [
\left (x_k - 2y_k + 4 y_k^3 \right ) \left (x_\ell - 2y_\ell + 4 y_\ell^3 \right ) 
- \tfrac{1}{8} (1-4 x_k^2)(1-4 x_\ell^2) \Big ]
\eea
while the field equations are given by, 
\bea
\sum_{\ell \in \BB} u_{k \ell} \, \Big [ x_k + x_\ell - 2y_\ell + 4 y_\ell^3 -4 x_k x_\ell^2 \Big ] =0
\eea
Up to this point, the expressions for the potential and for the field equations are exact and, given the solutions $y_k$ to (\ref{9.yt}), depend only on $u_{k\ell}$ and $x_k$. Both of these functions may now be expanded  around diagonal $t^{-1}$, as shown in (\ref{9.tu}) for $u$,
\bea
\label{9.xu}
x_k & = & y_k + x_k^{(1)} + x_k ^{(2)} + \cdots 
\no \\
u_{k\ell} & = & u^{(0)} _{kk} \delta _{k\ell} + u^{(1)}_{k\ell} + u^{(2)}_{k\ell} + \cdots
\eea
where the ellipses stand for terms of order three and higher. Comparison with (\ref{9.tu}) gives the following explicit expressions for the leading and first order corrections to~$u$, for $k, \ell \in \BB$,
\bea
\label{8.u1}
u^{(0)} _{k\ell} = { \delta _{k\ell} \over (t^{-1})_{kk}} 
\hskip 0.6in
u^{(1)}_{kk}=0 \hskip 0.6in u^{(1)}_{k \not= \ell} = { - (t^{-1})_{k \ell} \over  (t^{-1})_{k k} (t^{-1})_{\ell \ell} } 
\eea
Higher order corrections may be evaluated analogously, but will not be needed explicitly  here.

\sm

Carrying out the expansion to first and second order, we obtain the following expressions for the corrections $x_k ^{(1)}$ and $x_k ^{(2)}$ with $k \in \BB$, 
\bea
2 u_{kk}^{(0)}  (1-6 y_k^2) \, x_k^{(1)} &  = & \sum _{\ell  \not = k } u_{k\ell}^{(1)} (1-4y_\ell^2) (y_\ell-y_k) 
\no \\
2 u_{kk}^{(0)}  (1-6y_k^2) \, x_k ^{(2)}
& = &  
12 u_{kk}^{(0)}  y_k \big ( x^{(1)}_k \big )^2 
- \sum _{k,\ell \in \BB} u^{(2)} _{k\ell} (y_k-y_\ell) ( 1  -4 y_\ell^2)
\no \\ &&
- \sum _{\ell \not= k} u^{(1)} _{k\ell} \Big [  
 x_k ^{(1)} (1- 4 y_\ell^2) + x_\ell^{(1)} (1- 8 y_k y_\ell)  \Big ]
\eea
while the on-shell potential evaluates as follows to this order,
\bea
\label{9.potred}
V_\BB ^\text{sol} - V_\text{CB}  &=&
- \tfrac{1}{8} \sum _{k, \ell \in \BB} \Big ( u^{(0)} _{kk} \delta _{k\ell} + u^{(1)}_{k\ell} + u^{(2)}_{k\ell} \Big ) 
 ( 1- 8 y_k y_\ell ) (1-4y_k^2)(1-4y_\ell^2) 
\no \\ && \quad
- 2 \sum_{k \in \BB} u_{kk}^{(0)} \big (1-6y_k^2 \big ) \big ( x^{(1)}_k \big )^2 
\eea
We stress again that the matrices $u^{(0)}, u^{(1)}$ and $u^{(2)}$ in this expression depend explicitly on the partition $\BB$. Note that, as expected, the second order correction $x_k ^{(2)}$ does not enter into the corrections to the potential at that order.

\subsection{Lifting approximate degeneracies:  fine structure of the cascade}
\label{sec:liftdegen}

In this subsection, we analyze the behavior of the potentials for the three branches
\bea
\label{8.branch}
\BB_0, \qquad \BB_1= \BB _0\cup \{ p \}, \qquad  \BB_2 = \BB_0 \cup \{ p, N-p \} 
\eea
where $2 p \not= N$ and $p, N-p \not \in \BB_0$. As we shall show below, the potentials for the three branches as a function of $\kappa$ intersect at a single point to leading order in the perturbative expansion. These are also the only degeneracies. For $\BB_0=\emptyset$, this degeneracy persists to higher orders in the expansion, and is in fact exact. By contrast, for $\BB_0 \not= \emptyset$ the degeneracy is lifted by first order terms in the perturbative expansion. 

\subsubsection{Leading (zeroth) order} 

The leading order in the expansion  reveals an exact triple intersection of the three branches in (\ref{8.branch}).  The result follows from the expression of (\ref{redVdiagon})  for the corresponding potential evaluated on the solutions to leading order in the expansion. Although the matrix $u$ in general depends on the partition $\BB$, its leading order expression is diagonal and its entries are actually independent of the partition $\BB$.  As  a result, all terms in~\eqref{redVdiagon} with $k \in \BB_0$ in fact cancel in the following differences,  
\bea\label{eq:vmiusvleadingPT}
V^\text{sol} _{\BB_1} - V^\text{sol} _{\BB_0} & = & - \tfrac{1}{8} u_{pp}^{(0)}  (1-8y_p^2) (1-4y_p^2)^2
\no \\
V^\text{sol} _{\BB_2} - V^\text{sol} _{\BB_0} & = & - \tfrac{1}{4} u^{(0)} _{pp} (1-8y_p^2) (1-4y_p^2)^2
\eea
Here~$u^{(0)}_{pp} $ is given by the first equation in (\ref{8.u1}). Thus, the three branches intersect at $x_p = 1/\sqrt{8}$, namely at $\kappa (t^{-1}s) _{p} = 3/(4\sqrt{2})$. Comparing with~\eqref{eq:PTkappastar} this is precisely the value of~$\kappa$ at which~$h_p = h_{N-p}$ turn on. This triple intersection is an accidental degeneracy of leading-order perturbation theory, i.e.~it is not protected by any symmetry, so we expect that it is generically lifted. 

\subsubsection{Subleading corrections at first order}

To establish the fate of the above degeneracy as the expansion is carried out to first order, we record the expression for the  potential (\ref{9.potred}) to this order,\footnote{~We drop the second order terms in~\eqref{9.potred}, which will not be needed.}
\bea\label{eq:vminusvfirstPT}
V_\BB ^\text{sol} - V_\text{CB}  & = & 
- \tfrac{1}{8} \sum _{k \in \BB}  u_{kk}^{(0)}  ( 1- 8 y_k^2 ) (1-4y_k^2)^2
\no \\ &&
- \tfrac{1}{8} \sum _{k, \ell \in \BB} u^{(1)} _{k\ell} ( 1- 8 y_k y_\ell ) (1-4y_k^2)(1-4y_\ell^2) 
\eea
where $u^{(0)}_{k\ell} $ and $u^{(1)}_{k\ell} $ for $k, \ell \in \BB$ are given by (\ref{8.u1}) for each partition $\BB$. 

We distinguish the following two cases: 
\begin{itemize}
    \item When $\BB_0= \emptyset$, the only off-diagonal entry  is $u^{(1)}_{p, N-p}$ with $y_{N-p}=y_p$, so that,
\bea
V^\text{sol} _{\BB_1} - V_{\BB_0}^\text{sol} & = & 
- \tfrac{1}{8} u_{pp}^{(0)} (1-8y_p^2) (1-4y_p^2)^2
\no \\
V^\text{sol} _{\BB_2} - V^\text{sol} _{\BB_0} & = & 
- \tfrac{1}{4} \big ( u^{(0)} _{pp} + u^{(1)} _{p, N-p} \big ) (1-8y_p^2) (1-4y_p^2)^2
\eea
These expressions are identical to the leading order result~\eqref{eq:vmiusvleadingPT}, so that the three potentials remain degenerate at $y_p = 1/\sqrt{8}$. In fact, the degeneracy is exact: the exact equations that govern the Coulomb branch, as well as the~$\BB = \{1\}$ and~$\BB = \{1, N-1\}$ branches  are simply rescaled versions of the equations for~$SU(3)$, already analyzed in section~\ref{sec:su3}. As discussed there (see especially subsection~\ref{sec:1Hstabsu3}), the accidental degeneracy of the three branches is an exact property of the classical potential~\eqref{vscaleawithk}. In section~\ref{sec:aD3} we will show how this degeneracy is broken once we take into account additional subleading terms in the effective field theory at the multi-monopole point, which are not included in~\eqref{vscaleawithk}.

\item When $\BB_0 \not= \emptyset$, the non-trivial couplings of $y_k$ to $y_p$ and $y_{N-p}$ for $k \in \BB_0$ and $k \not= p, N-p$ remove any good reason for the degeneracy. To see this, we use the first-order formula~\eqref{eq:vminusvfirstPT} to compute the potential differences between the three branches to first order in perturbation theory,
\bea\label{eq:firstPTtriplepot}
V ^\text{sol} _{\BB_1} - V^\text{sol} _{\BB_0} & = &
- \tfrac{1}{8} u_{pp}^{(0)} (1-8y_p^2) (1-4y_p^2)^2
 \\ &&
- \tfrac{1}{4} \sum_{k \in \BB_0} u^{(1)} _{pk} (1-8y_p y_k) (1-4y_p^2) (1-4y_k^2)
\no \\
V^\text{sol} _{\BB_2} - V^\text{sol} _{\BB_0} & = &
- \tfrac{1}{4} \left ( u_{pp}^{(0)} + u_{p,N-p}^{(1)} \right )  (1-8y_p^2) (1-4y_p^2)^2
\no \\ &&
- \tfrac{1}{4} \sum_{k \in \BB_0} \left ( u^{(1)} _{pk} + u^{(1)}_{N-p,k} \right )  (1-8y_p y_k) (1-4y_p^2) (1-4y_k^2)
\no
\eea
Here the components of $u^{(0)}$ are given by the first equation in (\ref{8.u1}), while the other components (for~$k \neq p, N-p$) are given as follows,
\bea
u^{(1)} _{pk} = { - (t^{-1})_{pk} \over  (t^{-1})_{pp} (t^{-1})_{kk} }~, \quad 
u^{(1)} _{N-p,k} = { - (t^{-1})_{N-p, k} \over  (t^{-1})_{pp} (t^{-1})_{kk} }~, \quad 
u^{(1)} _{p, N-p} = { - (t^{-1})_{p, N-p} \over  (t^{-1})_{pp} (t^{-1})_{pp} } 
\eea
where we have used the bi-symmetry relation $(t^{-1})_{N-p,N-p}=t^{-1}_{pp}$ to simplify the expressions. 
\end{itemize}

While we generically expect the first-order potential differences in~\eqref{eq:firstPTtriplepot} to remove the accidental degeneracy between the~$\BB_0 \neq \emptyset$, $\BB_1 = \BB_0 \cup \{p\}$, and~$\BB_2 = \BB_0 \cup \{p, N-p\}$ branches, this is by no means obvious analytically. The reason is that perturbation theory shifts the values of~$\kappa$ at which the branches cross, so that we cannot simply substitute the leading order values into~\eqref{eq:firstPTtriplepot}. Moreover, it is also not obvious which branch has the lowest energy after the degeneracy is broken. 

To answer these questions we will plot the potential differences~\eqref{eq:firstPTtriplepot} obtained in first-order perturbation theory numerically, starting with the cases~$N = 4,5,6$ that we can compare with the (much more time-consuming) exact numerics in section~\ref{sec:num}. As we will see, first-order perturbation theory is sufficient to account for the fine structure~\eqref{eq:truecascade} of the cascade observed numerically at low values of~$N$. We will then confirm using perturbation theory that the same fine structure also persists to higher values of~$N$, for which we do not have exact numerics:

\begin{figure}[t!]
    \centering
    \includegraphics[width=0.49\linewidth]{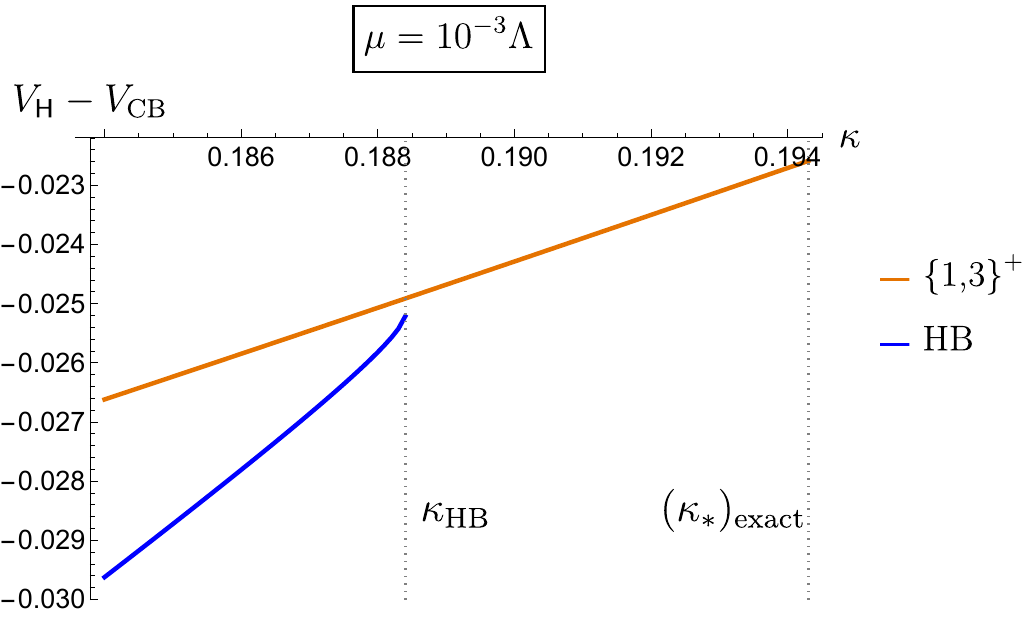}
        \includegraphics[width=0.49\linewidth]{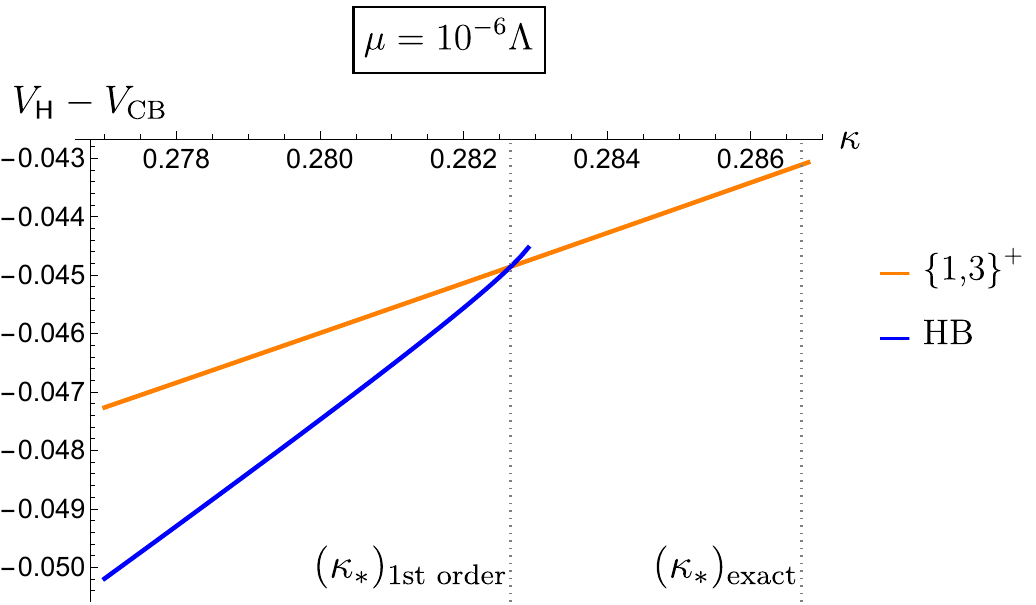}
    \caption{First-order perturbative potentials~\eqref{eq:vminusvfirstPT} for the branches~$\BB = \{1, 3\}^+$ (orange) and the maximal HB (blue), for~$SU(4)$ gauge group and RG scale~$\mu = 10^{-3} \Lambda$ (left panel) and~$\mu = 10^{-6}\Lambda$ (right panel). For both cases we indicate the exact numerical value~$(\kappa_*)_\text{exact}$ at which the transition occurs. On the left, perturbation theory breaks down at~$\kappa > \kappa_\text{HB}$ because the HB ceases to exist at leading order. This happens before the branches can cross, so that we cannot infer where -- or even if -- a transition occurs. This is solved on the right by going to smaller~$\mu$, where perturbation theory predicts a transition at~$(\kappa_*)_\text{1st order}$, just shy of the exact answer.}
    \label{fig:su4pt}
\end{figure}

\begin{itemize}
\item In figure~\ref{fig:su4pt} we study the second transition~$\{1, 3\}^+ \to \text{HB}$ for~$SU(4)$. Even though there are no accidental degeneracies here, this example is useful to illustrate how perturbation theory works, and how it compares to the exact numerics. In both panels of figure~\ref{fig:su4pt} we show the first-order perturbative potentials~\eqref{eq:vminusvfirstPT} for the branches~$\BB = \{1, 3\}^+$ (orange) and the maximal HB (blue). 

The left panel has RG scale~$\mu = 10^{-3} \Lambda$, just as the exact~$SU(4)$ potential plotted in figure~\ref{fig3a}. Note that the two branches never cross in perturbation theory. This is because the leading HB solution only exists when~$\kappa < \kappa_\text{HB}$, and hence the fate of this branch for larger~$\kappa > \kappa_\text{HB}$ cannot be determined in perturbation theory. This breakdown occurs before the two branches actually cross at the larger value~$\kappa = (\kappa_*)_\text{exact}$ taken from figure~\ref{fig3a}. 

We can circumvent this problem by lowering the RG scale, which improves the quality of the perturbative expansion. (Recall that it is exact in the limit~$\mu \to 0$.) This is shown in the right panel of figure~\ref{fig:su4pt}, where we take~$\mu = 10^{-6} \Lambda$. Now first-order perturbation theory reliably predicts a transition at the value~$(\kappa_*)_\text{1st order} < \kappa_\text{HB}$, where the HB already exists. Note that~$(\kappa_*)_\text{1st order}$ is only slightly smaller (by about~$\sim 1\%$) than the true value~$(\kappa_*)_\text{exact}$ from the exact~$SU(4)$ numerics at this value of~$\mu$. 

The lesson is that the effects of perturbation theory can always be ascertained by working at sufficiently small~$\mu$. This will be important below.

\begin{figure}[t!]
    \centering
    \includegraphics[width=0.8\linewidth]{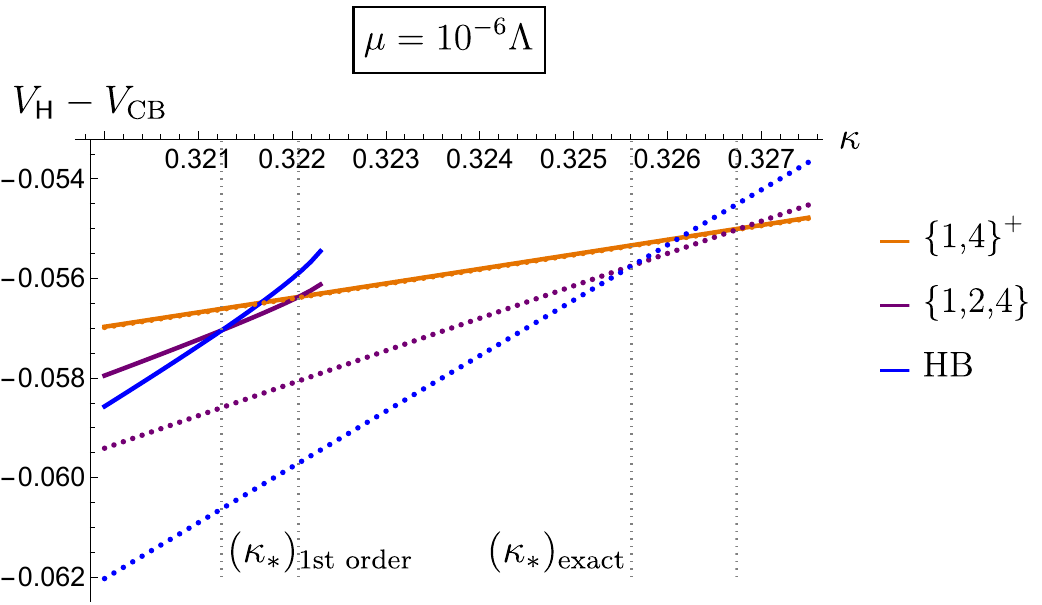}
    \caption{ Perturbative results, including first order corrections in the expansion of subsection \ref{sec:exp}, are shown as solid lines for gauge group~$SU(5)$ and RG scale~$\mu = 10^{-6} \Lambda$ for the transitions~$\{1, 4\}^+ \to \{1,2,4\} \to \text{HB}$.  The exact numerics are shown in dotted lines. The transition value~$(\kappa_*)_\text{1st order}$ obtained in first-order perturbation theory slightly underestimates the exact value~$(\kappa_*)_\text{exact}$. However, the qualitative features of the phase diagram are correctly reproduced by perturbation theory. }
    \label{fig:su5pt}
\end{figure}

\item In figure~\ref{fig:su5pt}, we study the first non-trivial example that can be explained by lifting an accidental degeneracy in perturbation theory: the second pair of closely-spaced~$SU(5)$ transitions $\{1, 4\}^+ \to \{1,2,4\} \to \text{HB}$, established numerically in figure~\ref{fig:7} at $\mu = 10^{-3} \Lambda$. Note that the~$C$-odd~$\{1,2,4\}$ branch briefly has the lowest potential. 

As for the~$SU(4)$ case discussed in the previous bullet point, perturbation theory fails to predict a phase transition at~$\mu = 10^{-3} \Lambda$, but does so reliably at the lower value~$\mu = 10^{-6} \Lambda$ depicted in figure~\ref{fig:su5pt}. There we see that first order perturbation theory (solid lines) reliably predicts the correct transition pattern~$\{1, 4\}^+ \to \{1, 2,4\}\to \text{HB}$, including the brief appearance of the~$C$-odd~$\{1,2,4\}$ phase. The exact numerical potentials at this value of~$\mu$ are indicated by dotted lines. We see that the qualitative arrangement of the three branches, and the resulting phase diagram as a function of~$\kappa$, agree with the perturbative prediction. The only difference is that perturbation theory slightly underestimates the transition value: $(\kappa_*)_\text{1st order} < (\kappa_*)_\text{exact}$, by an amount~$\sim 1\%$.

This example gives us confidence that first-order perturbation theory (as defined in subsection~\ref{sec:exp}), at sufficiently small~$\mu$, correctly predicts the way in which the degeneracies of the leading-order cascade solution are split. We have similarly  verified that it  captures the transitions~$\{1,5\}^+ \to \{1,2,5\} \to \{1,2,4,5\}^+$ shown for~$SU(6)$ in figure~\ref{fig:77}. Thus first-order perturbation theory correctly predicts (and indeed explains) the fine-structure~\eqref{eq:truecascade} of the cascade, when compared to the cases~$N = 4,5,6$ for which we have exact numerics from section~\ref{sec:num}.

\begin{figure}[t!]
\centering
\includegraphics[width=0.49\textwidth]{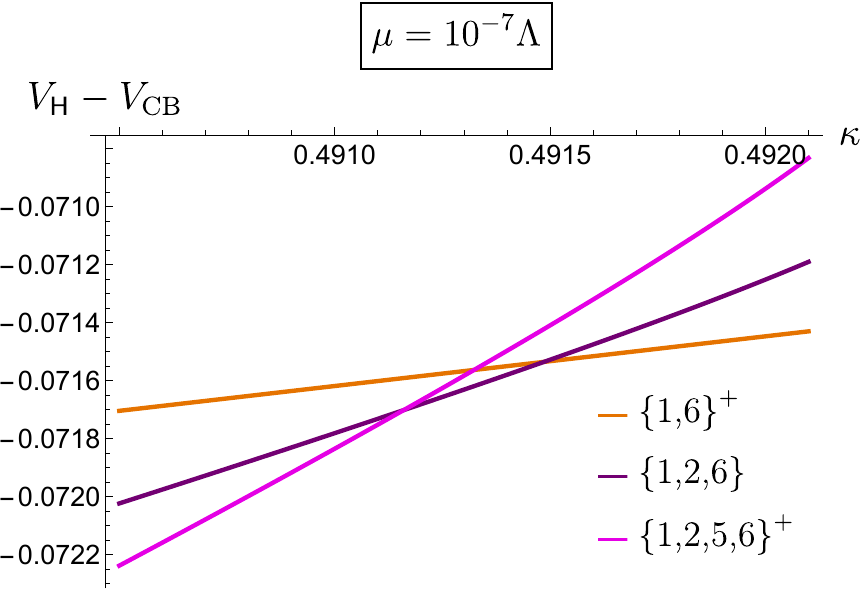}
\includegraphics[width=0.49\textwidth]{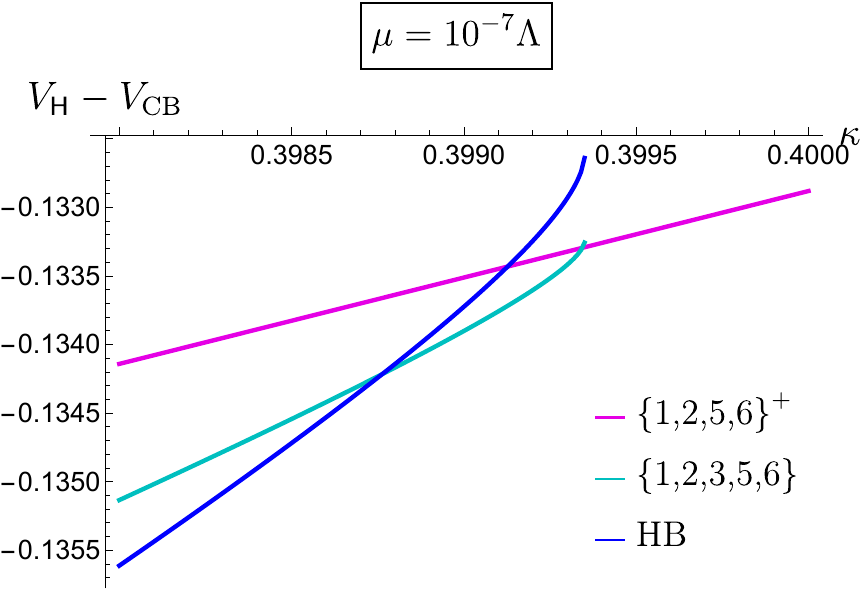}
\caption{Perturbative results, including first order corrections in the expansion of subsection \ref{sec:exp}, are shown as solid lines for gauge group~$SU(7)$ and RG scale~$\mu = 10^{-7} \Lambda$ for the transitions~$\{1, 6\}^+ \to \{1,2,6\} \to \{1,2,5,6\}^+$ (left panel), and~$\{1,2,5,6\} \to \{1,2,3,5,6\} \to \text{HB}$ (right panel).  In both cases perturbation theory confirms the fine structure of the cascade, with the~$C$-odd phases~$\{1,2,6\}$ and~$\{1,2,3,5,6\}$ coming down in energy and briefly appearing in between the~$C$-even phases that make up the cascade at leading perturbative order (i.e.~its coarse structure).\label{fig:su7pt} }
\end{figure}

\item We have carried out the perturbative analysis for~$SU(7)$ and~$SU(8)$, for which there are no exact numerical results to compare with. In both cases we confirm that the fine structure of the cascade takes the form~\eqref{eq:truecascade}, as we already found for~$N \leq 6$. This strongly suggests that the same structure persists to higher values of~$N$ as well, that we have not explicitly checked. 

The results for~$SU(7)$ are shown in figure~\ref{fig:su7pt}. We see that the second and third leading-order transitions split, at first order, into two pairs of closely space transitions,
\begin{equation}
\begin{split}
  &   \{1,6\}^+ \to \{1,2,6\} \to \{1, 2, 5, 6\}^+~, \cr
  & \{1, 2, 5, 6\}^+ \to \{1, 2, 3, 5, 6\} \to \text{HB} = \{1, 2, 3, 4, 5, 6\}^+
    \end{split}
\end{equation}
in exact agreement with the fine structure~\eqref{eq:truecascade} of the cascade. Note that the low RG scale~$\mu = 10^{-7} \Lambda$ in figure~\ref{fig:su7pt} is required to see the phase transitions in perturbation theory. 
\end{itemize}

\subsection{Lifting degeneracies via~$\CO(a_D^3)$ terms in~$K^\text{eff}$: the first transition}
\label{sec:aD3}

Throughout our analysis so far, we have restricted the expansion of the~$\CN=2$ effective K{\"a}hler potential $K^{\text{eff}}$ at the multi-monopole point to quadratic order in the magnetic periods~$a_{Dm}$; see  the discussion in section \ref{sec:gen1}, and in particular equation~(\ref{eq:keffahii}) for $K^\text{eff}$ in this approximation. For consistency, we must then also truncate the SUSY-breaking potential at the same order in the~$a_{Dm}$, as was done in~\eqref{eq:s3sblagmm}. This was done (a) for simplicity, and (b) because it is reasonable to expect that the effect of the higher-order~$\CO(a_D^3)$ corrections is small -- especially if we are working suitably close to the multi-monopole point.

We now discuss a question for which the inclusion of such higher-order terms is essential. This concerns the fate of the accidental degeneracy between the Coulomb branch ($\BB = \emptyset$), the~$C$-odd single Higgs branch~$\BB_1 = \{1\}$, and the~$C$-even double Higgs branch~$\BB_2^+ = \{1, N-1\}^+$. As we have seen from numerous points of view (analytically, in sections~\ref{sec:su3} and~\ref{sec:liftdegen}, and numerically in section~\ref{sec:numcascade}), the classical potential~\eqref{vscaleawithk} predicts an exact triple intersection of these three branches. 

Since this is an accidental degeneracy, not enforced by an exact symmetry of the problem, it should be lifted by higher-order corrections. Once source of these are quantum corrections in the dual, which we will not consider here. 

Instead, we will consider the effect of retaining the previously neglected~$\CO(a_D^3)$ term in the~$\CN=2$ effective K{\"a}hler potential $K^{\text{eff}}$ at the multi-monopole point. These in turn require the~$\CO(a_D^3)$ corrections to the prepotential $\cF_D(a_D)$, which were obtained in \cite{DHoker:1997mlo}, and already appear in~(\ref{fdfull}) above. They were carried over to the effective prepotential $\cF_D^\text{eff}(a_D)$ in (\ref{fdefffull}), to the effective electric period $a^\text{eff}_m(a_D)$ in (\ref{aMM}), to the gauge coupling matrix 
$\tau^\text{eff} _{Dmn}(a_D)$ in (\ref{eq:taueff}),  and to the effective K\"ahler potential $K^\text{eff}(a_D)$ in (\ref{eq:keffah}). Additionally, the corrections to $\tau^\text{eff} _{Dmn}(a_D)$ induce corrections to the effective K\"ahler metric $g_{mn}^\text{eff}  = \Im(\tau^\text{eff}_{Dmn})/2 \pi$ of (\ref{2.Kdef}), which are given by,
\bea
\label{8.g}
g_{mn}^\text{eff}  & = & t_{mn}(\mu) 
 -{ \delta_{m,n}  \over 8 \pi^2 N \Lambda}  \left [ { - 3 \, \Im a_{Dm} \over 4 s_m^3} 
 + \sum_{p\not= m} { s_p \, \Im a_{Dp} \over (c_p-c_m)^2} \right ]
\no \\ &&  \hskip 0.5in
- { 1- \delta_{m,n} \over 8 \pi^2 N \Lambda} \, { s_m \, \Im a_{Dn} + s_n \, \Im a_{Dm} \over (c_m-c_n)^2}
+ \cO(a_D^2)
\eea

The arguments relating to vacuum alignment of the hypermultiplet Higgs scalars and unbroken~$C\tilde T$ symmetry that led us to set $h_{im} = M \delta _{i1} h_m$ with $h_m>0$ in (\ref{hpar}) and $a_{Dm} = - i M x_m$ with $x_m \in \RR$ in (\ref{3.xdef}) remain in full force. Thus we can continue to work with a dimensionless reduced potential~$V$ for these variables. Previously this potential was given by (\ref{Vred}), which we repeat here, 
\bea
\label{8.Vrep}
V =  \sum_{m=1}^{N-1} \left(-  2 \kappa s_m x_m   +\frac{1}{2} \left( 4x_m^2	- 1\right) h_m^2\right) 
+  \sum_{m,n=1}^{N-1}\left(       t_{mn} \, x_{m} x_{n} +    \frac{1}{2}  (t^{-1})_{mn} \, h_m^2 \, h_n^2  \right) 
\eea

The corrections to this potential that stem from the inclusion of the~$\CO(a_D^3)$ in the effective K\"ahler potential are two-fold: 
\begin{itemize}
\item[(1.)] The sum over $t_{mn} x_mx_n$ in the potential above, which was given by the effective K\"ahler potential to quadratic order in $a_D$, is now replaced by the effective K\"ahler potential $K^\text{eff} /M^2$ given in terms of (\ref{eq:keffah}). This correction to~$V$ will be denoted by~$\Delta_K V$.

\item[(2.)]  The sum over $\thalf (t^{-1})_{mn} h_m^2 h_n^2$, which involved the inverse of the K\"ahler metric to quadratic order in $a_D$, is now replaced by the inverse of the effective K\"ahler metric (\ref{8.g}). This correction to~$V$ will be denoted by $\Delta _h V$.  
\end{itemize}
The two corrections~$\Delta_K V$ and $\Delta _h V$ are given by the following formulas, 
\bea
\label{8.Delta12}
\Delta_K V & = &   
\frac{3}{32\pi^4 \kappa}  \sum_{m=1}^{N-1}   
\left( - \frac{  x_m^3}{4 s_m^3} +  \sum_{n \neq m} \frac{s_m \, x_{m} x_{n}^2 }{(c_n - c_m)^2}  \right) 
\no \\
\Delta_h V & = & 
{ 1 \over 32 \pi^4 \kappa} \sum_{m,n,p,q=1}^{N-1}  (t^{-1})_{mp} (t^{-1})_{nq} \, h_m^2 \, h_n^2 \left \{ 
- ( 1- \delta_{p,q} ) { s_p x_q + s_q x_p \over (c_p-c_q)^2} \right .
\no \\ && \hskip 2in \left . 
+ \delta_{p,q} \left [  { 3 x_p \over 4 s_p^3} - \sum_{r \not= p} { s_r x_r \over (c_p-c_r)^2} \right ] 
  \right \}
\eea
Here we have eliminated the dimensionless ratio $M/\Lambda$ in favor of $\kappa=N\Lambda/(2 \pi^2 M)$ that was introduced already earlier in (\ref{eq:kappadefsec3}).

\begin{figure}[t!]
	\centering
	\includegraphics[width=0.9\textwidth]{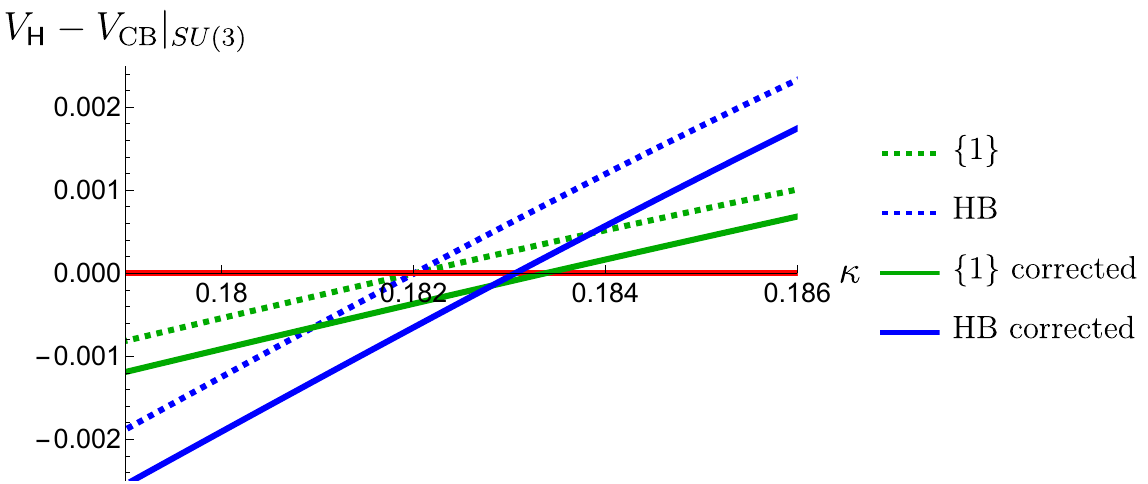}
\caption{Coulomb branch (horizontal red line), as well as~$\BB = \{1\}$ and~$\BB = \text{HB}$ for~$SU(3)$ gauge group, with RG scale~$\mu = 10^{-3} \Lambda$. The unperturbed potential~$V$ in~\eqref{8.Vrep} gives rise to the dotted lines (with a triple intersection at~$\kappa \simeq 0.182$). The potential~$V + \Delta_K V + \Delta_h V$ corrected by~$\CO(a_D^3)$ terms in the effective K\"ahler potential gives rise to the solid lines. 
	\label{fig:aD3su3}}
	\end{figure}

\begin{figure}[h]
	\centering
	\includegraphics[width=0.51\textwidth]{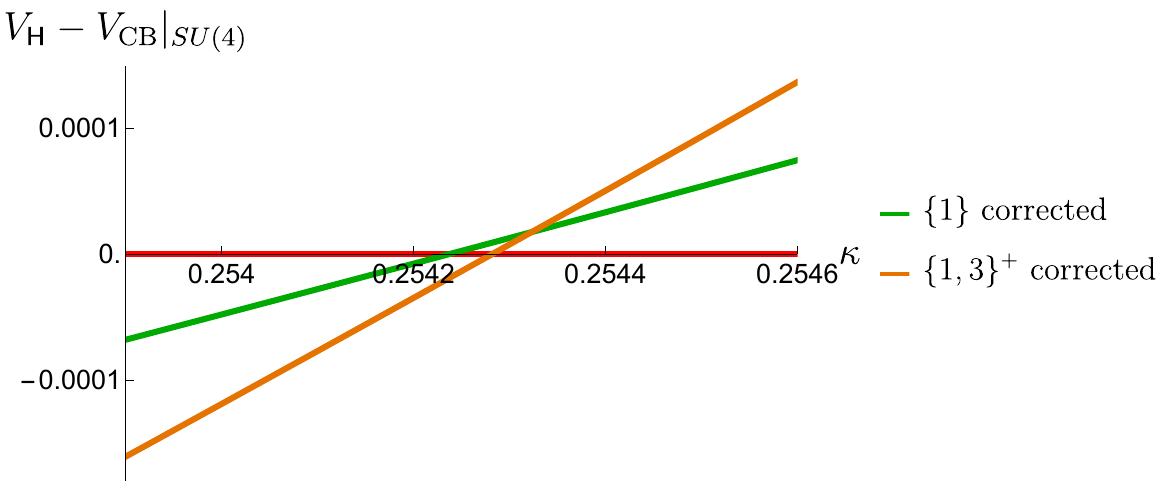}\\
    \includegraphics[width=0.49\textwidth]{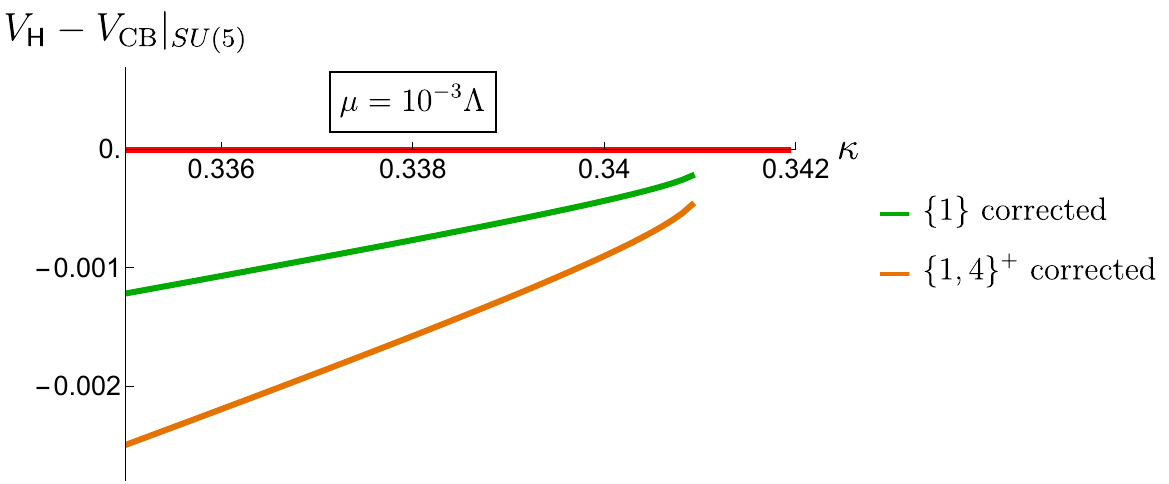}
	\includegraphics[width=0.49\textwidth]{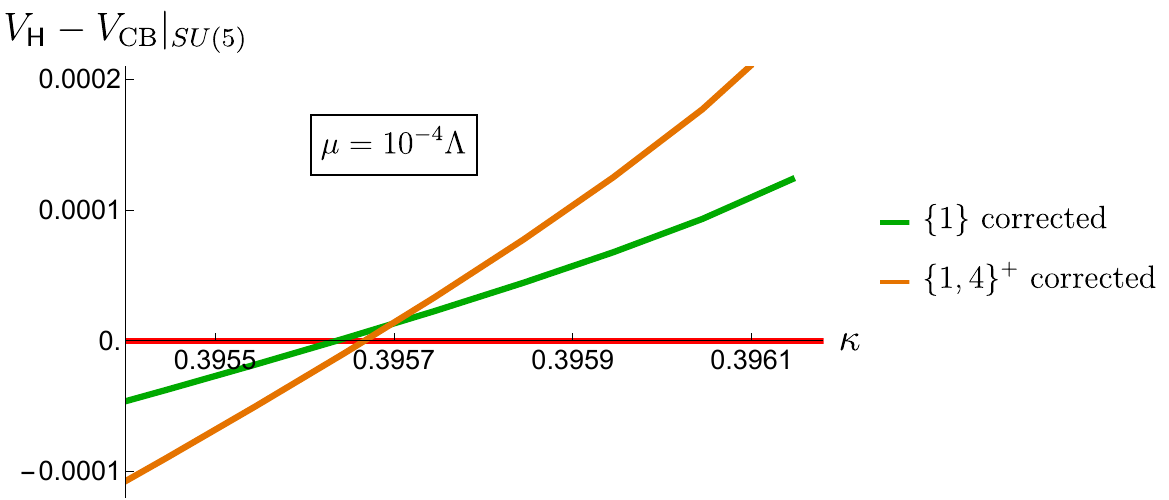}
\caption{Coulomb branch (red), as well as the~$C$-odd ~$\BB = \{1\}$ branch (green) and the~$C$-even~$\BB = \{1, N-1\}^+$ branch (orange) for~$SU(N)$ gauge group, with~$N = 4$ (top panel) and~$N = 5$ (bottom panels). We only show the corrected potentials~$V + \Delta_K V + \Delta_h V$ for the various branches. For~$SU(4)$ perturbation theory is reliable at~$\mu = 10^{-3} \Lambda$ and shows that the degeneracy is split by raising the~$\{1\}$ branch in energy. For~$SU(5)$, perturbation theory fails at~$\mu = 10^{-3} \Lambda$ (bottom left panel), but reliably shows the lifting of the~$\{1 \}$ branch at~$\mu = 10^{-4}\Lambda$ (bottom right panel). }
	\label{fig:aD3su4and5}
	\end{figure}

We must now evaluate the perturbations $\Delta_KV$ and $\Delta_hV$ on the solutions to the unperturbed field equations derived from $V$ in (\ref{8.Vrep}), and discussed  in sections \ref{sec:largek} and \ref{sec:oneH}. Here we restrict to the branches CB, $\BB_1 = \{ 1 \}$ and $\BB_2^+ = \{ 1,N-1\}^+$. The pertinent unperturbed solutions are as follows:
\begin{itemize}
\itemsep=0in
\item 
For the Coulomb branch CB we have $h_m=0$ and $x_m=\kappa (t^{-1} s)_m$  for $m=1,\ldots,  N-1$, so that 
$\Delta_h V=0$ and $\Delta_K V$ equals $\kappa^2 $ times a factor that depends only on the matrix $t$. 
\item 
For the single Higgs branch $\BB_1=\{1 \}$ we have $h_1 >0$ and $h_m=0$ for $m=2,\ldots, N-1$, while the variables $x_m$ are given by $x_{1} = x$ with $2 x-4x^3 = \kappa (t^{-1}s)_1$ and,
\bea
x_k = \sum_{\ell=2}^{N-1} (\sigma_1)_{k \ell} \Big ( \kappa s_\ell - t_{\ell, 1} x \Big ) 
\hskip 0.8in 
\sum_{\ell=2}^{N-1} (\sigma _1)_{k\ell} \, t_{\ell m} =\delta_{k,m}
\eea
for $k,m =2,\ldots, N-1$. The non-vanishing Higgs field is $h_1^2 = \thalf (1-4x^2)/(t^{-1})_{11}$.
\item 
For the double Higgs branch $\BB_2^+= \{ 1, N-1\}^+$, we have $h_1=h_{N-1}$ and $h_m=0$ for $m=2,\ldots, N-2$, while the variables $x_m$ are given by $x_1=x_{N-1}=x$ with $2 x-4x^3=\kappa (t^{-1}s)_1$ and,
\bea
x_k = \sum_{\ell=2}^{N-2} (\sigma_2)_{k \ell} \Big ( \kappa s_\ell - (t_{\ell , 1} + t_{\ell, N-1}) x\Big ) 
\hskip 0.8in 
\sum_{\ell=2}^{N-2} (\sigma _2)_{k\ell} \, t_{\ell m} =\delta_{k,m}
\eea
for $k,m =2,\ldots, N-2$. The non-vanishing Higgs fields are given by
\bea
h_1^2 = h_{N-1}^2 = \half \, { 1-4x^2 \over (t^{-1})_{1,1} + (t^{-1})_{1,N-1} }
\eea
\end{itemize}
\noindent We now numerically evaluate the perturbations~$\Delta_K V$ and~$\Delta_h V$ on these unperturbed solutions, and plot the results for~$SU(3)$, $SU(4)$, and~$SU(5)$ in figures~\ref{fig:aD3su3} and~\ref{fig:aD3su4and5}. 

\bigskip

\begin{itemize}
\item The results for~$SU(3)$ are shown in figure~\ref{fig:aD3su3}. The Coulomb branch is shown as a horizontal red line. The potentials for the~$\{1\}$ and~$\text{HB} = \{1, 2\}^+$ branches are shown in green and blue -- dotted lines for the unperturbed potential~$V$ in ~\eqref{8.Vrep}; solid lines for the potential~$V$ plus the corrections~$\Delta_K V$ and~$\Delta_h V$ in~\eqref{8.Delta12}. Clearly the accidental degeneracy of the unperturbed potential is lifted, and the~$C$-odd~$\{1\}$ branch briefly comes down in energy, leading to the phase transitions~$\text{CB} \to \{1\} \to \text{HB}$. This is reminiscent, but distinct from, the fine structure of the cascade analyzed in the previous subsection, which was a feature of the unperturbed potential~$V$. In particular, it did not arise for~$SU(3)$ gauge group. 

\item Repeating the analysis for~$N \geq 4$, we find that the degeneracy between the branches is broken in the opposite way: the~$C$-odd single Higgs branch~$\{1\}$ is raised in energy and thus never globally stable, so that the transition remains~$\text{CB} \to \{1, N-1\}^+$. This is shown explicitly for~$SU(4)$ and~$SU(5)$ gauge group in figure~\ref{fig:aD3su4and5}. Note that for~$SU(5)$ we must reduce the RG scale to~$\mu = 10^{-4}\Lambda$ in order to reliably analyze the phase structure in perturbation theory.
\end{itemize}

\newpage

\section{Mass spectra}
\label{sec:11}

In this section we analyze the mass spectrum predicted by the dual Abelian Higgs model at the multi-monopole point as the supersymmetry breaking scale $M$ (equivalently~$\kappa$)  is varied and the system cascades through the various phases obtained in previous sections. Since we are analyzing the dual semiclassically, the masses may be read off  from the Lagrangian of the model given in equations (\ref{eq:Lbos}) through (\ref{eq:lyukbis}). Phases are labeled by the partitions $\AA|\BB$ introduced in subsection \ref{sec:9.1}, possibly with further refinements due to charge-conjugation symmetry~$C$. We will explicitly confirm that there are no tachyons in any stable phase of our model. (This is just a sanity check, given our previous extensive stability analysis.) We shall also determine the spectrum of massless particles in any phase labeled by the partition $\AA| \BB$. In particular, we will show that the only massless particles are  
\begin{itemize}
\itemsep=0in	
\item[(i)] $|\AA|$ massless vector bosons~$b_{\mu m}$ with~$m \in \AA$; the vector bosons in~$\BB$ are Higgsed and thus massive. 
\item[(ii)] $2 |\AA|$ massless Weyl fermions; they are the~$|\AA|$ gaugino doublets~$\rho_m^i$ for which~$m \in \AA$.
\item[(iii)] 2 Nambu-Goldstone bosons for $|\BB| >0$, and no massless scalars for $|\BB|=0$.
\end{itemize}
Here, $|\AA|$ and $|\BB|$ are the cardinalities of the sets $\AA$ and $\BB$, respectively, with $|\AA|+|\BB|=N-1$. Note that the two Nambu-Goldstone boson are precisely the degrees of freedom parameterizing the~$\C\P^1$ non-linear sigma model described below~\eqref{chiSB}. 

The fact that there are exactly two Nambu-Goldstone bosons no matter how many Higgs fields are turned on reflects the vacuum alignment discussed in section~\ref{sec:vacal}.  One consequence of this is that the complex-valued scalar fields $a_{Dm}$ and $h_{i m}$ have the following vacuum expectation values,
\bea
\label{11.vevs}
\< a_{Dm} \> = - i M x_m \hskip 1in \< h_{i,m} \> = M h_m \delta_{i,1}
\eea
with $M, x_m, h_m$ real and $h_m \geq 0$. By definition, $m \in \AA$ if $h_m=0$ while  $m \in \BB$ if $h_m>0$. 
 In the remainder of this section, we shall establish these results and obtain general formulas for the masses but we shall refrain from fully diagonalizing the fermion and scalar mass matrices as this can be done only numerically.

 \subsection{Mass spectrum of the fermions}
 
The quadratic part of the fermion Lagrangian may be obtained from (\ref{eq:Lferm}) and (\ref{eq:lyukbis}) and the expectation values of (\ref{11.vevs}). While the kinetic term of the fields $\psi_{\pm , m}$ is canonically normalized, the  normalization of the kinetic term of the fields $\rho^i_m$ is not canonical and involves the matrix $t_{mn}$ of $U(1)$ gauge couplings and mixings. It is convenient to collect all the Fermi fields into a column vector $\Psi$  and  recast the quadratic fermion Lagrangian in the following form, 
 \bea
 \label{eq:allfermions}
 \cL _\text{fermion}  =   - i \, \bar \Psi \, W^2 \, \bar \sigma^\mu \p_\mu \Psi  - \Psi ^t \, Y \, \Psi - \bar \Psi \, Y^* \, \bar \Psi
\eea
The $4(N-1)$-dimensional column vector of Weyl spinor fields $\Psi$ and the $4(N-1) \times 4(N-1)$ matrix $Y$ of Yukawa couplings are given as follows, 
\bea
\Psi = \left ( \bma \psi_{+,m} \cr \psi_{-,m} \cr \rho^1_m \cr \rho ^2 _m \ema \right )
\hskip 0.8in
Y = {M \over \sqrt{2}}\left ( \bma 
0 & i x_m \delta_{mn} & 0 &  h_m \delta _{mn} \cr
 i x_m \delta _{mn} & 0 & -h_m \delta _{mn} & 0 \cr
0 & -h_m \delta _{mn} & 0 & 0 \cr
h_m \delta _{mn} & 0 & 0 & 0 \ema \right )
\eea
The $4(N-1) \times 4(N-1)$ matrix $W^2$ that enters  the kinetic term for $\Psi$ accounts for the non-canonical normalization of the Fermi fields $\rho^i_m$ and is given in terms of the matrix $t_{mn}$ as follows: $W^2={\rm diag} ( \delta_{mn}, \delta _{mn} , t_{mn}, t_{mn})$. Since the matrix $t_{mn}$ is symmetric and positive definite the matrix $W$ is uniquely defined by requiring it to be symmetric and positive. The fermion mass matrix is then given by $W^{-1}Y W^{-1}$. 

\sm

The number of massive fermions is given by the rank of the mass matrix $W^{-1} Y W^{-1}$.  The rank of the matrix $W$ is maximal since the matrix $t$ is positive definite. Therefore, the rank of the mass matrix $W^{-1}Y W^{-1}$ equals the rank of $Y$ which also equals the rank of $Y^\dagger Y$.  The eigenvalues $y_{\pm, m}^2$ of $Y^\dagger Y$ are readily evaluated and we have, 
\bea
y_{\pm , m}^2 = {M^2 \over 4} \Big ( x_m^2 + 2 h_m^2 \pm \sqrt{ (x_m^2+2 h_m^2)^2 - 4 h_m^4} \Big )
\eea
each eigenvalue occurring with multiplicity 2. Manifestly, the eigenvalues $y_{+,m}$ never vanish, while $y_{-,m}$ vanishes if and only if $h_m=0$. Taking the multiplicity into account, the number of massless fermions is $2|\AA|$, thereby establishing point (i) above.

\subsection{Mass spectrum of the scalars and gauge bosons}

The scalar fields are the complex-valued fields $a_{Dm}$ and $h_{i m}$ and their complex conjugates, whose expectation values are given in (\ref{11.vevs}), while the~$U(1)_D^{N-1}$ gauge fields are $b_{\mu m}$.  We shall expand the scalar fields around their vacuum expectation values using the following notation for their real and imaginary parts,
\bea
a_{Dm} & = & - i M x_m + (a_{Dm}^R+ i a_{Dm}^I )/\sqrt{2}
\no \\
h_{1 m} & = & M h_m + (\eta_m + i \xi_m)/\sqrt{2}
\eea
and leave the field $h_{2 m}$ alone since it has vanishing expectation value. The quadratic part of the scalar and gauge fields is then,
\bea
\label{11.quad1}
\cL_\text{quad} & = & - \sum_{m=1}^{N-1} \Big [ 
\thalf  \p^\mu \eta_m \p_\mu \eta _m 
+ \thalf (\p^\mu \xi _m - \sqrt{2} Mh_m b^\mu_m)(\p_\mu \xi_m - \sqrt{2} M h_m b_{\mu  m}) 
+ \p^\mu \bar h_{2 m} \p_\mu h_{2 m} \Big ]
\no \\ &&
- \sum_{m,n=1}^{N-1} t_{mn} \Big [ 
\thalf \p^\mu  a_{Dm}^R \p_\mu a_{Dn}^R 
+ \thalf \p^\mu  a_{Dm}^I \p_\mu a_{Dn}^I 
+\tfrac{1}{4} f^{\mu \nu}_m f_{\mu \nu m} \Big ]  -  V_\text{quad}
\eea
The quadratic terms in the potential may be expressed using the Hessians,
\bea
\label{11.quad2}
V_\text{quad} & = & M^2 \sum_{m,n=1}^{N-1} \Big [ 
\tfrac{1}{4}  (\cH_{xx})_{mn} a_{Dm}^I a_{Dn}^I 
 - \thalf (\cH_{xh})_{mn} \eta_m a_{Dn}^I
 + \tfrac{1}{4} (\cH_{hh})_{mn} \eta_m \eta _n \Big ]
\no \\ &&
+ M^2 \sum_{m,n=1}^{N-1} \Big [  
\tfrac{1}{4}  (\cH_{xx})_{mn} a_{Dm}^R a_{Dn}^R
+ \tfrac{1}{4} (\cH_{\beta})_{mn} \xi_m \xi _n
+ \thalf (\cH_\alpha)_{mn} \bar h_{2 m} h_{2 n} \Big ]
\eea
where $\CH_{xx}, \CH_{xh}$ and $\CH_{hh}$ were given in (\ref{hess}), while $\CH_\alpha$ and $\CH_\beta$ are given as follows,
\bea
\label{11.hab}
(\CH_{\alpha})_{mn} &= & \delta_{mn} \bigg (  4 x_m^2 - 1- \sum_{p=1}^{N-1} 2(t^{-1})_{mp} h_p^2 \bigg )
+ 4  (t^{-1})_{mn} h_m h_n 
\no \\
(\CH_{\beta})_{mn} &=& \delta_{mn} \left ( 4 x_m^2 -1 + \sum_{p=1}^{N-1} 2 (t^{-1})_{mp} h_p^2 \right )
\eea
We note that the kinetic terms for the scalar fields $\eta_m, \xi_m$ and $h_{2,m}$ have canonical normalization, while those for $a_{Dm}$ and the gauge fields are non-canonical and set by the matrix $t$.  The corresponding canonically normalized fields are~$\tilde a _{Dm} = \sum_n \tilde W_{mn} a_{Dn}$ and~$b_{\mu m} = \sum_n \tilde W_{mn} b_{\mu  m}$, where the positive symmetric~$(N-1) \times (N-1)$ matrix $\tilde W$ is defined by $\tilde W^2 =(t_{mn})$.\footnote{~The matrix $\tilde W$ is therefore similar, but not identical, to the matrix~$W$ that appeared in~\eqref{eq:allfermions}.}

\subsubsection{Mass spectrum of gauge bosons}

The square of the mass matrix for the $U(1)$ gauge bosons may be read off from the quadratic part of the bosonic Lagrangian given in (\ref{11.quad1}) and (\ref{11.quad2}) and the expression for the Hessian $\cH_\beta$ in (\ref{11.hab}). In view of the field equations (\ref{1baa}) for the vacuum expectation value $h_m$, we see that for $m \in \BB$, namely when $h_m \not=0$, we have $(\cH_\beta)_{mm}=0$ and the corresponding massless scalar $\xi_m$  is eaten by $b_{\mu m}$ to render this gauge boson massive by the standard Abelian Higgs mechanism.   The components of the square of the full mass matrix for the fields $\tilde b_{\mu  m}$ with canonical kinetic terms are given as follows,
\bea
2M^2 \sum_{p=1}^{N-1} (\tilde W^{-1})_{mp} \, h_p^2 \, (\tilde W^{-1})_{pn}
\eea
For $m \in \AA$ the field $b_{\mu  m}$ is massless, thereby giving $|\AA|$ massless gauge bosons and establishing point (i) above.   The rank of the square of the mass matrix is $|\BB|$, and the positivity of its non-zero eigenvalues follows from the positivity of the quadratic form $Q_\beta$ in (\ref{stabB}) which was already demonstrated in section \ref{sec:red2}.

\subsubsection{Mass spectrum of scalars}

We have already established above that the scalar fields $\xi_m$ for $m \in \BB$ are massless and get eaten by the gauge field $b_{\mu  m}$ via the Higgs mechanism. The remaining scalar fields $\xi_m$ for $m \in \AA$ are massive and their mass square is simply given by the entries $(\CH_\beta)_{mm}$ of the diagonal matrix $\CH_\beta$. These entries are all strictly positive in view of the stability condition derived in subsection \ref{sec:red2}. 

\sm

Since the Hessian $\CH_{xx}$ of (\ref{hess}) is manifestly positive, the fields $a_{Dm}^R$ are all massive, and the square of their mass matrix is given by $\thalf M^2 \tilde W^{-1} \CH_{xx} \tilde W^{-1}$. Similarly, the square of the mass matrix for the fields $a_{Dm}^I, \eta_m$ is given by the matrix,
\bea
\half M^2 \left ( \bma \tilde W^{-1} & 0 \cr 0 & I \ema \right ) \CH  \left ( \bma \tilde W^{-1} & 0 \cr 0 & I \ema \right ) 
\hskip 1in 
\CH = \left ( \bma \CH_{xx} & \CH_{xh} \cr \CH_{hx} & \cH_{hh} \ema \right )
\eea
Positive definiteness of the Hessian $\CH$ is one of the basic stability conditions of the solutions to the field equations for the vacuum expectation values. Thus, the masses of the fields $a_{Dm}^I, \eta_m$ are all non-vanishing and their squares are strictly positive. 

\sm

It remains to analyze the properties of the matrix $\CH_\a$ which directly gives the square of the mass matrix $\thalf M^2 \cH_\a$ for the field $h_{2 m}$. To do so, we decompose the matrix $\CH_\alpha$ into four blocks according to whether the indices of the components $(\CH_\alpha)_{mn}$ belong to $\AA$ or to $\BB$.   The off-diagonal blocks vanish since we have $(\CH_\alpha)_{mn}=0$ whenever $m \in \AA$ and $n \in \BB$ (or $m \in \BB$ and $n \in \AA$). 
To study the spectrum of the diagonal blocks, we use \eqref{k2a} to simplify the $\BB$ block and recast the result for the $\AA$ block with the help of $\cH_\beta$, 
\begin{align}
\label{chalphab}
(\CH_{\alpha})_{mn} &= - 4\delta_{mn} \sum_\ell (u^{-1})_{n\ell} h_{\ell}^2 + 4(u^{-1})_{mn} h_m h_n
	 &  m,n & \in \BB \\
(\CH_{\alpha})_{mn} &=  (\cH_\beta)_{mn} - 4  \delta_{mn}  \sum_{\ell \in B} (t^{-1})_{m\ell} h_\ell^2 
	&  m,n& \in \AA
\no
\end{align}
To analyze the block on the first line of (\ref{chalphab}) we consider the associated quadratic form, 
\bea
\sum_{m,n \in \BB} (\cH_\a)_{mn} \alpha_m \alpha_n = 	
- 2 \sum_{m,n \in \BB} (u^{-1})_{mn} (h_m \alpha_n -  h_n \alpha_m)^2 
\eea
Since the contributions with $m=n$ vanish, and the off-diagonal elements of $(u^{-1})_{mn}= (t^{-1})_{mn}$, given by (\ref{uinvv}),  are all negative, we see that every term on the right side sum is non-negative and conclude that the block of $\cH_\a$ restricted to $\BB$ is non-negative. 

The quadratic form vanishes when each term vanishes, which requires $\alpha_m$ to be proportional to $h_m$. Thus, the eigenspace with zero eigenvalue is one-dimensional and generated by $h_m$. 
Since the matrix $\CH_\alpha$ multiplies $|h_{2 m}|^2$, the zero eigenvalue actually produces one complex, or two real, massless scalar fields. These are precisely the two expected Nambu-Goldstone bosons  associated with the spontaneous symmetry breaking $SU(2)_R\to U(1)_R$. 
	
\sm

Finally, positivity of the $\AA$ block of $\CH_\alpha$ on the second line in  \eqref{chalphab} may be established as follows. The matrix is diagonal, we have $(\CH_\beta)_{mm} >0$ for all $m \in \AA$ as required by local stability of the solution, only off-diagonal elements of $t^{-1}$ appear in this sum, since $m\in \AA$ and $\ell \in \BB$, and these matrix elements are all negative in view of the assumption (\ref{2.tinvneg}). Therefore, we have $(\CH_\alpha)_{mm} > (\CH_\beta)_{mm} >0$ so that the $\AA$ block of $\cH_\a$ is positive definite for any locally stable solution.

\newpage
	
\section{The cascade of phase transitions to adjoint QCD}
\label{sec:summary}

In this section we will give a detailed account of the cascade of phase transitions interpolating between the Coulomb branch (CB) at small SUSY-breaking~$M \ll \Lambda$ (equivalently, large~$\kappa$) to the maximal Higgs branch (HB) at sufficiently large~$M \gtrsim \Lambda$ (equivalently, sufficiently small~$\kappa$). Building on the earlier analytic and numerical explorations of the cascade in sections~\ref{sec:su2} through~\ref{sec:cascade}, we give a detailed account of the intermediate phases, how they realize the global symmetries, and the resulting massless spectrum (see section~\ref{sec:11}). On the HB, we find detailed agreement with the confining and chiral symmetry breaking scenario for adjoint QCD -- consisting of $N$ disconnected vacuum sectors, each with a~$\C\P^1$ sigma model -- reviewed in section~\ref{sec:confchisbIntro}. We elaborate on this by computing various physical observables, such as the vev of the chiral-symmetry breaking order parameter~\eqref{eq:vecOdef} or the radius of the~$\C\P^1$ sigma models, using our Abelian dual at the multi-monopole point. Importantly, and rather non-trivially, we find perfect agreement with the large-$N$ scaling that is expected for these quantities from adjoint QCD.

\subsection{Summary of the cascade: coarse and fine structure} 

The coarse and fine structure of the cascade are well captured by the perturbative analysis in section~\ref{sec:cascade}, which involves treating the off-diagonal elements of~$(t^{-1})_{mn}$ as a small perturbation to the diagonal, while keeping the full matrix $t^{-1}$ in the combination $(t^{-1}s)_n = \sum_{m = 1}^{N-1} (t^{-1})_{nm} s_m$, as explained in subsection \ref{sec:exp}. This approach, which is valid for all~$SU(N)$ gauge groups, is in good agreement with the exact analytic and numerical results obtained  for~$2 \leq N \leq 6$ in sections~\ref{sec:su2}, \ref{sec:su3}, and~\ref{sec:num}. Moreover, it gives a rather intuitive physical picture for these results:
\begin{itemize}
\item {\it Coarse structure from diagonal~$t^{-1}$:} Where we ignore the off-diagonal elements of~$t^{-1}$, the cascade proceeds by turning on pairs of Higgs fields, leading to the following~$C$-symmetric sequence of first-order phase transitions as we dial from large to small~$\kappa$,\footnote{~Note that for even~$N = 2\nu$, the last transition only involves turning on a single~$C$-even Higgs field~$h_\nu$.}
\begin{equation}\label{eq:diagtinvcascade}
\text{diagonal } t^{-1} \; : \; \text{CB} \to \{1, N-1\}^+ \to \{1,2, N-2, N-1\}^+ \to \cdots \to \text{HB}
\end{equation}
In this approximation, the transition
\begin{equation}
    \BB \to \BB \cup \{m, N-m\}^+~,
\end{equation}
which involves turning on~$h_m = h_{N-m} > 0$, occurs at~\eqref{eq:PTkappastar},
\begin{equation}\label{eq:ksdiagbis}
    \kappa = \kappa_{* m} = {3 \over 4 \sqrt 2} {1 \over (t^{-1} s)_m}~, \qquad M_{*m} = {2 \sqrt 2 N \Lambda  \over 3 \pi^2} (t^{-1} s)_m 
\end{equation}
This should be compared to the masses of the BPS states at the origin of the Coulomb branch (as computed using the dual Abelian Higgs model), which are given by~\eqref{3.bpsdual},
\begin{equation}
    M_\text{BPS}(\mu_{km}) = {\sqrt 2  N \Lambda\over 2 \pi^2} (t^{-1} s)_m~.
\end{equation}
This only differs from the transition points in~\eqref{eq:ksdiagbis} by~$4/3$. Thus, to leading order in the expansion, the naive idea that the BPS masses at the origin actually determine the thresholds in~$M$ at which a transition occurs is essentially born out.

Another feature of the leading order result is that each transition in~\eqref{eq:diagtinvcascade} actually occurs at an accidental triple degeneracy between the three branches\footnote{~Again, the case~$N = 2\nu$ and~$m + 1 = \nu = N-m-1$ is an exception; in that case only two of the branches in~\eqref{eq:tripdegbranch} are distinct.} 
\begin{equation}
\label{eq:tripdegbranch}
    \BB = \{1, \ldots, m,  N-m, \ldots, N-1\}^+~, \qquad \BB \cup \{m+1\}~, \qquad \BB \cup \{m+1, N-m-1\}^+
\end{equation}

\item {\it Fine structure from perturbative corrections in off-diagonal~$t^{-1}$:} these are well-behaved and produce moderate -- but qualitatively important -- changes to the diagonal $t^{-1}$ cascade above:
\begin{itemize}
\item The accidental degeneracy between the three branches~\eqref{eq:tripdegbranch} is generically lifted. 

\item Closely related to the previous point, the degeneracy is lifted in such a way that that the~$C$-odd phase~$H \cup \{m+1\}$ in~\eqref{eq:tripdegbranch} comes down in energy. It therefore briefly appears as an interpolating phase between the~$C$-even phases in~\eqref{eq:diagtinvcascade}. Thus, most transitions in~\eqref{eq:diagtinvcascade} split into two nearby transitions,\footnote{~The transitions occur at $\kappa \simeq \kappa_{*m} \pm \Delta$, with~$\Delta$ much smaller than the difference between successive~$\kappa_{*m}$'s.} each of which only involves a single Higgs field,
\begin{equation}\label{eq:intermed1hphase}
    \BB = \{1, \ldots, m, N-m, \ldots, N-1\}^+ \; \to  \; \BB \cup \{m+1\} \; \to \;  \BB \cup \{m+1, N-m-1\}^+
\end{equation}

\item An exception is the first transition CB$ \to \{1, N-1\}^+$, where there is no intermediate~$C$-odd~$\{1\}$ phase, and the three branches CB, $\{1\}$, and~$\{1, N-1\}^+$ remain exactly degenerate at the transition point. 

\item The central value~$\kappa = \kappa_{*m}$ around which the two closely-spaced phase transitions in~\eqref{eq:intermed1hphase} occur is raised somewhat, because the off-diagonal elements of~$t^{-1}$ account for the back-reaction from Higgs fields in~$\BB$, which have already condensed.\footnote{~For instance, the last term in the potential~\eqref{vscaleawithk} is such that the Higgs fields in~$\BB$, which already have vevs, favor the condensation of those in~$\AA$ that do not have vevs, once the off-diagonal matrix elements~$(t^{-1})_{m \neq n} < 0$ are taken into account. This naturally pushes the transitions to larger~$\kappa$.} 
\end{itemize}

\end{itemize}

As explained in section~\ref{sec:aD3}, the degeneracy between the Coulomb branch, as well as the branches~$\{1\}$ and~$\{1, N-1\}^+$, is lifted by including the (previously omitted) effects of the~$\CO(a_D^3)$ terms in the effective~$\CN=2$ K\"ahler potential~\eqref{eq:keffah}.\footnote{~Of course there could be additional contributions, e.g.~from quantum effects, that we do not consider.} The upshot of that analysis is that the $C$-odd single Higgs branch~$\{1\}$ comes down in energy for~$SU(3)$, but is lifted for higher~$SU(N)$, leading to the following picture for the first phase transition, out of the Coulomb branch,\footnote{~For~$SU(2)$, all branches are~$C$-even and the CB directly transitions to the maximal HB given by~$\BB = \{1\}$. }
\begin{align}
SU(2) \; & : \; \text{CB} \quad \to \quad \text{HB} =  \{1\}  \\
          SU(3) \; & : \; \text{CB} \quad \to \quad \{1\} \quad \to \quad \{1, N-1\}^+   \\
      SU(N \geq 4) \; & : \; \text{CB} \quad \to  \quad \{1, N-1\}^+ 
\end{align}

\begin{figure}[t!]
    \centering
    \includegraphics[width=\linewidth]{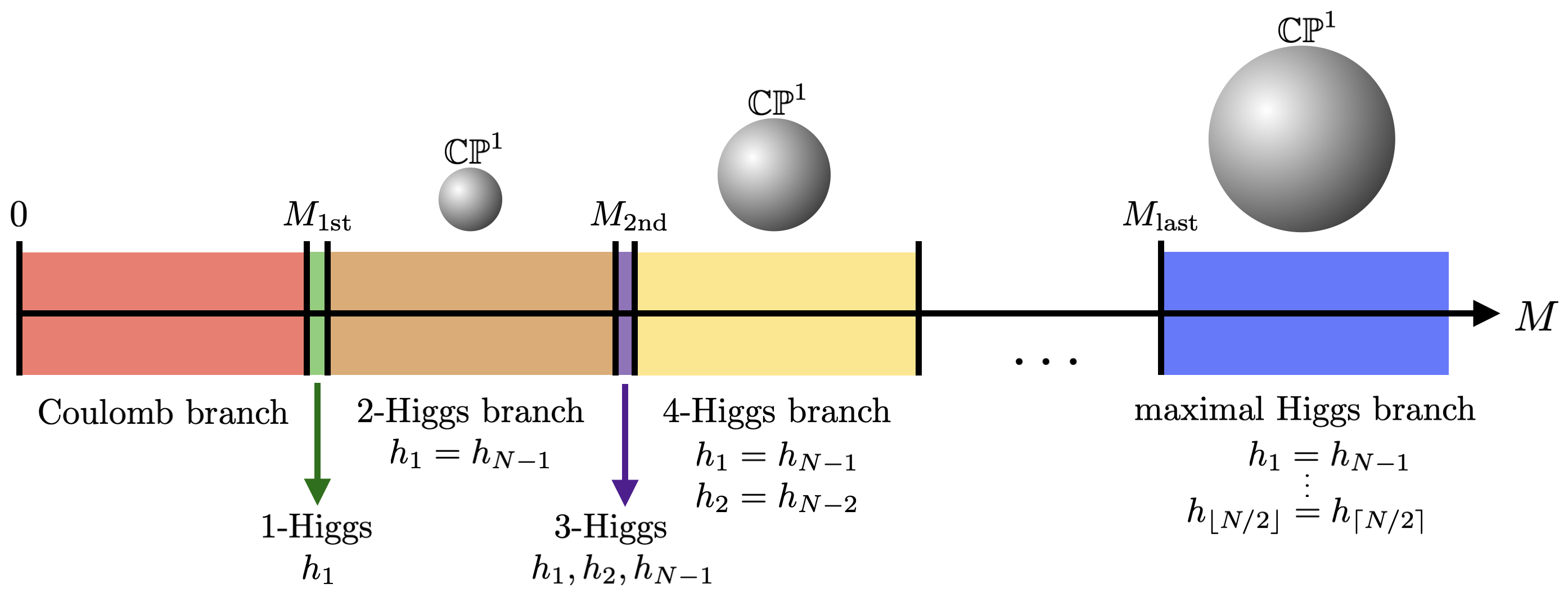}
    \caption{Cascade of phase transitions interpolating between the Coulomb branch at small~$M$ and the maximal Higgs branch/adjoint QCD at large~$M$.}
    \label{fig:phasesbis}
\end{figure}

The cascade reviewed above is depicted in figure~\ref{fig:phasesbis}, albeit as a function of increasing~$M$, rather than decreasing~$\kappa$. Note in particular, that the last transition, to the maximal Higgs branch (HB), occurs at
\begin{equation}
    M = M_\text{last} \sim \Lambda~,
\end{equation}
where~$\sim$ denotes an~$\CO(1)$ factor, without strong~$N$ dependence. This is consistent with the zero-order perturbative result in~\eqref{eq:aDonCB} (with~$m \sim N/2$), and the fact that higher orders in perturbation theory only shift the transition values by small amounts.    

\subsection{Massless fields along the cascade} 

Here we use the results of section~\ref{sec:11} to give a description of the massless degrees of freedom along the cascade reviewed above.

\begin{itemize}
\item On the Coulomb branch (CB) at large~$\kappa$, all scalars are massive; there are~$N-1$ massless photons~$b_{\mu m}$, and~$N-1$ massless~$SU(2)_R$ gaugino doubles. The latter precisely match the~$\Z_2$-valued Witten anomaly~\cite{Witten:1982fp} associated with~$SU(2)_R$, which counts~$SU(2)_R$ doublets modulo~$2$. Thus the anomaly is present if and only if~$N$ is even.

\item The moment the first Higgs field condenses, spontaneously breaking~$SU(2)_R \to U(1)_R$, there are two massless Nambu-Goldstone bosons parametrizing a~$\C\P^1$ non-linear sigma model with radius~$f_\pi$,
\begin{equation}\label{eq:cp1lag}
    \SL_{\mathbb C \mathbb P^1}= -{f_\pi^2 \over 2} \p^\mu \vec n \cdot \p_\mu \vec n~, \qquad \vec n^2 = 1~.
\end{equation}
The radius, or decay constant, $f_\pi$ depends on~$\kappa$, and jumps discontinuously across the first order phase transitions along the cascade. We will compute it in section~\ref{sec:cp1fpi}, where we show that it grows along the cascade, reaching the value appropriate to adjoint QCD on the maximal Higgs branch (HB). 

\item On a branch characterized by a partition~$\AA |\BB$, where~$|\BB|$ Higgs fields have condensed, there are~$|\BB|$ massive vector bosons, and~$|\AA| = N-1 - |\BB|$ massless ones. There are also~$|\AA|$ massless~$SU(2)_R$ gaugino doublets, which implies that the~$\C\P^1$ sigma model in~\eqref{eq:cp1lag} requires a discrete~$\theta$-angle to match the~$SU(2)_R$ Witten anomaly if and only if~$|\BB| = N-1 - |\AA|$ is odd. Since the number~$|\BB|$ of activated Higgs fields increases monotonically along the cascade, the number of massless fields monotonically decreases. 

\item On the maximal Higgs branch (HB) at small~$\kappa$, the only massless degrees of freedom that remain are the two Nambu-Goldstone bosons parametrizing the~$\C\P^1$ sigma model in~\eqref{eq:cp1lag}, with a discrete~$\theta$-angle if~$N$ is even. This is precisely the IR description expected in each of the~$N$ disconnected vacuum sectors of the confining, chiral-symmetry breaking  phase of adjoint QCD summarized in section~\ref{intro:nf2adjoint}.

\end{itemize}

\subsection{Order parameters along the cascade and the large-$N$ limit}\label{sec:ordparlargeN}

\subsubsection{Vacuum energy}

Since the~$\CN=2$ theory we start with has zero vacuum energy, it is meaningful to compute the vacuum energy of the deformed theory as a function of the SUSY-breaking mass~$M$. This results in the effective potentials we have been discussing throughout (mostly as a function of~$\kappa = N \Lambda / (2 \pi^2 M)$, rather than~$M$).

Here we would like to comment on the scaling of these effective potentials as we change the~$SU(N)$ gauge group, and especially in the large-$N$ limit. To this end, it will suffice to examine the vacuum energy on the large-$\kappa$ Coulomb branch, and on the small-$\kappa$ maximal Higgs branch, where we can make contact with adjoint QCD. 

\sm

\noindent {\it Coulomb Branch:} Recall from section~\ref{sec:largek} that
\begin{equation}\label{eq:cbsolbis}
    x_k = \kappa (t^{-1} s)_k~, \qquad h_k = 0~, \qquad k = 0, \ldots, N-1~.
\end{equation}
The dimensionless vacuum energy on that solution is given by~\eqref{VA},
\begin{equation}
    V_\text{CB} = - \kappa^2 \sum_{k, \ell = 1}^{N-1} (t^{-1})_{k\ell} s_k s_\ell~,
\end{equation}
and the physical one by
\begin{equation}\label{eq:CBpotphys}
    \SV_\text{CB} = M^4 V_\text{CB} = - {N^2 M^2 \Lambda^2 \over 4 \pi^4} \sum_{k, \ell = 1}^{N-1} (t^{-1})_{k\ell} s_k s_\ell~.
\end{equation}
It is interesting to examine this quantity in the standard large-$N$ limit~$N \to \infty$, with~$\Lambda$ fixed. A useful formula, which is valid at large $N$ and derived in appendix~\ref{app:appt}, is~\eqref{useful}, which we repeat here, 
\begin{equation}\label{eq:usefulbis}
    (t^{-1} s)_m \simeq {2 \pi^2 \over N} s_m \qquad \text{as} \qquad N \to \infty~.
\end{equation}
This can be used to show that the double sum in~\eqref{eq:CBpotphys} evaluates to~$\pi^2$, and thus
\begin{equation}\label{eq:largenvcb}
    \SV_\text{CB} \to - {N^2 M^2 \Lambda^2 \over 4 \pi^2}  \qquad \text{as} \qquad N \to \infty~.
\end{equation}
Note that the~$\CO(N^2)$ scaling is generically expected in a theory with adjoint fields in the large-$N$ limit.

\bigskip
    
\noindent {\it Maximal Higgs Branch:} Recall from sections~\ref{sec:smallk} and~\ref{sec:red} that
\begin{equation}\label{eq:hmonHBbis}
    h_m^2 = \half \sum_{n=1}^{N-1} t_{mn} (1-4 x_n^2)~, \qquad m = 1, \ldots, N-1~,
\end{equation}
where the small-$\kappa$ behavior of the~$x_m$ is given by~\eqref{6.x1},
\begin{equation}\label{eq:xonHBbis}
    x_m = \kappa \sum_{n=1}^{N-1} (T^{-1})_{mn} s_n + \CO(\kappa^3)~, \qquad m = 1, \ldots, N-1~,
\end{equation}
where the matrix~$T_{mn}$ was defined in~\eqref{6.TT}, which we repeat here,
\begin{equation}\label{bigTmatbis}
T_{mn} = t_{mn} + \delta_{mn} v_n~, \qquad v_n = \sum_{p = 1}^{N-1} t_{np}~.
\end{equation}
Evaluating the dimensionless vacuum energy to this order leads to~\eqref{eq:smallKHB} with~$u = t$,
\begin{equation}\label{eq:VphysHB}
    V_\text{HB} = V_\text{CB} - {1 \over 8} \sum_{m, n = 1}^{N-1} t_{mn} + \kappa^2 \sum_{m, n=1}^{N-1} t_{mn} (T^{-1} s)_m^2 + \CO(\kappa^4)~.
\end{equation}
The physical potential on the maximal Higgs branch is
\begin{equation}
    \SV_\text{HB} = \SV_\text{CB} - {M^4 \over 8} \sum_{m, n = 1}^{N-1} t_{mn} + { N^2 M^2 \Lambda^2 \over 4 \pi^2}  \sum_{m, n=1}^{N-1} t_{mn} (T^{-1} s)_m^2 + \CO(\Lambda^4)~,  
\end{equation}
where the physical Coulomb branch potential was already evaluated in~\eqref{eq:CBpotphys}. 

In the limit~$M \gg \Lambda$, deep in the regime appropriate to adjoint QCD, we have
\begin{equation}\label{eq:svonHB}
    \SV_\text{HB} \simeq - {M^4 \over 8} \sum_{m, n = 1}^{N-1} t_{mn} \to -  {7 \zeta(3) N^2 \over 16 \pi^4} M^4 \qquad \text{as} \qquad N \to \infty~.
\end{equation}
Here we have used~\eqref{eq:tmnlargeN} to evaluate the double sum over the matrix elements~$t_{mn}$ in the large-$N$ limit. Thus, in this limit, the leading large-$N$ vacuum energy also scales correctly as~$\CO(N^2)$. 

Interestingly, the two subleading~$\CO(M^2 \Lambda^2)$ terms in~\eqref{eq:VphysHB} also have~$\CO(N^2)$ scaling: the first one is simply the Coulomb-branch energy, already shown to be~$\CO(N^2)$ in~\eqref{eq:largenvcb}. To show that the third term in~\eqref{eq:VphysHB} is also~$\CO(N^2)$ in the large-$N$ limit, we must approximate~$(T^{-1} s)_n$, where~$T_{mn}$ is the matrix defined in~\eqref{bigTmatbis}. According to the numerical experiments in appendix~\ref{sec:app.C.5} the difference between~$(T^{-1} s)_n$ and~$(t^{-1} s)_n$ is an~$\CO(1)$ factor that ranges between roughly 1 and 3, without any pronounced~$N$-dependence. Since we are interested in the~$N$-scaling of this term, it is thus entirely sufficient to approximate
    \begin{equation}\label{eq:Tstsapprox}
       \sum_{p = 1}^{N-1} (T^{-1})_{mp}s_p \simeq  \sum_{p = 1}^{N-1} (t^{-1})_{mp}s_p \simeq {2 \pi^2 \over N} s_m~,
    \end{equation}
    where we have also used~\eqref{eq:usefulbis}, which applies at large $N$. It follows that
    \begin{equation}\label{eq:o1scaling}
        \sum_{m, n = 1}^{N-1} t_{mn} (T^{-1} s)_m^2 \simeq {4 \pi^4 \over N^2} \sum_{m, n= 1}^{N-1} t_{mn} s_m^2  = {4 \pi^4 \over N^2} \sum_{m = 1}^{N-1} v_m s_m^2 = \CO(1) \qquad \text{as} \qquad N \to \infty~.  
    \end{equation}
Here~$v_m$ is defined in~\eqref{bigTmatbis}. The large-$N$ scaling is obtained by converting the sum to an integral over~$\rho = {m / N}$, via~${1 \over N} \sum_m \to \int_0^1 d \rho$. It is shown in~\eqref{eq:vmlargeN} that~$v_m \sim N f(\rho)$ at large $N$, where~$f(\rho)$ is an~$\CO(1)$ function of~$\rho$ only, just as~$s_m \to \sin \rho$. Since the integral is convergent, this establishes~\eqref{eq:o1scaling}.

\subsubsection{Coulomb branch coordinates}

Here we would like to examine the gauge-invariant Coulomb branch moduli~$u_I$, or equivalently the dual vector multiplet scalars~$a_{Dm}$, along the cascade. We start by examining them on the Coulomb and maximal Higgs branches.

\begin{figure}[t!]
\centering
\includegraphics[width=0.49\textwidth]{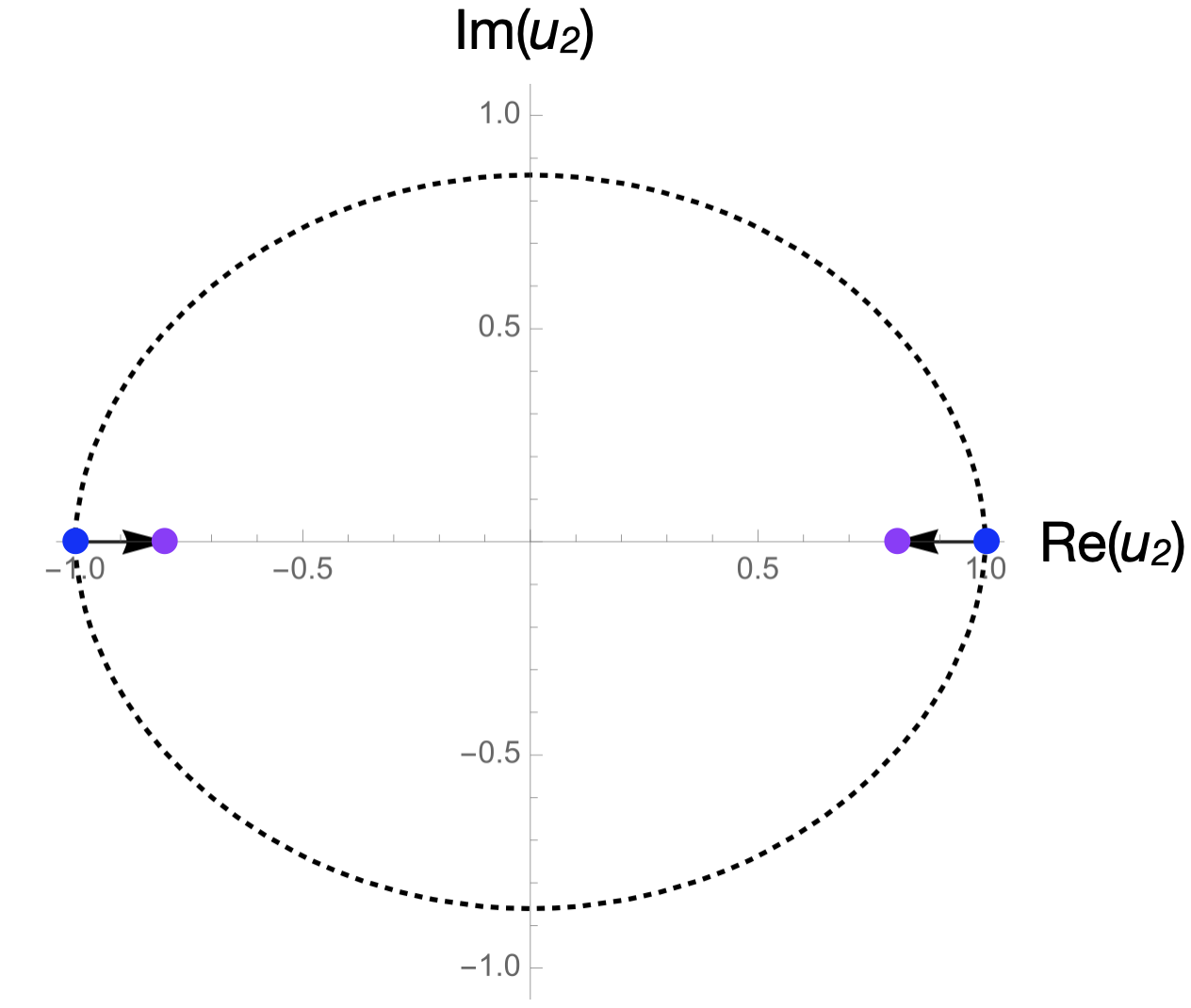}\includegraphics[width=0.5\textwidth]{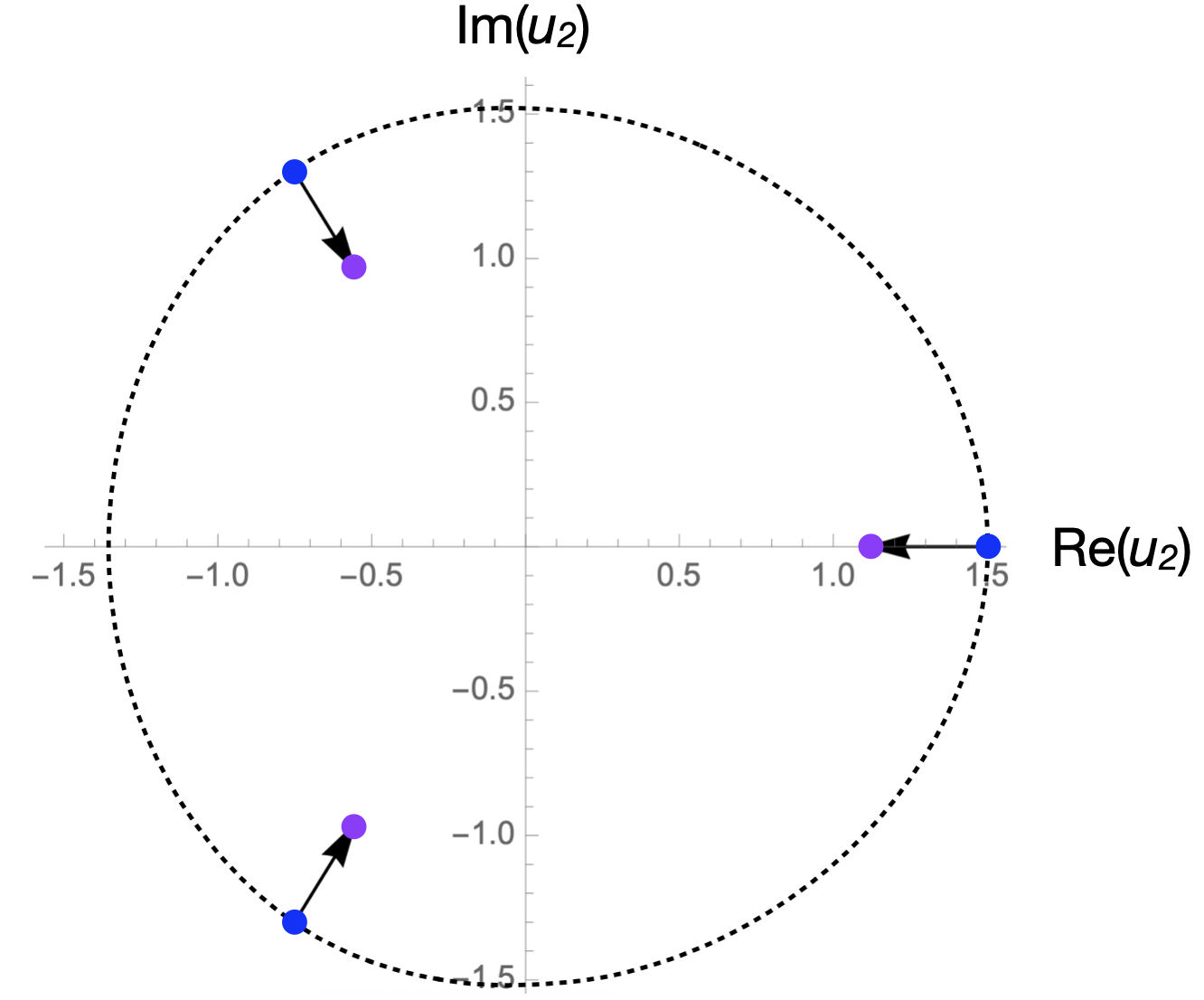}
	\caption{Complex~$u_2$ plane for~$SU(2)$ (left) and~$SU(3)$ (right), in units where~$\Lambda = \half$. The multi-monopole points are indicated by blue dots; the maximal Higgs branches at small~$\kappa$ (corresponding to adjoint QCD) are indicated by purple dots. (The renormalization scale is~$\mu=10^{-3} \Lambda$.) We see that they are close to each other, with the Higgs branch vacua slightly displaced into the strong-coupling region (indicated by the black dotted line), and pointing toward the origin~$u_2 = 0$, as indicated by the black arrows.}	\label{fig:sunmmpoints}
\end{figure}

\bigskip

\noindent {\it Coulomb branch:} At large~$\kappa$, it was shown in section~\ref{sec:smallm} that the gauge-invariant moduli~$u_I$ vanish for all~$I = 2, \ldots, N$. This point is special, because the discrete~$\Z_{4N}$ $R$-symmetry is unbroken there -- something that is not manifest in the description of this phase using the dual Abelian Higgs model at the multi-monopole point. In that description, the Coulomb branch vacuum is given by~\eqref{eq:cbsolbis}. To test this formula, we can use~\eqref{eq:u2ofad} to compute~$u_2$ in the dual (omitting~$\CO(a_D^3)$ terms),
\begin{equation}\label{eq:u2viaaDbis}
    u_2(a_D) = 2 N \Lambda^2 + \sum_{m=1}^{N-1} \left(-4 i \Lambda s_m a_{Dm} - {1 \over 2N} a_{Dm}^2\right)~.
\end{equation}
Recalling that~$a_{Dm} = - i M x_m$ from~\eqref{3.xdef}, and using~\eqref{eq:cbsolbis}, we obtain
\begin{equation}\label{eq:aDonCB}
a_{Dm} \Big |_{\text{CB}} 
	= - \frac{i  N\Lambda }{2\pi^2} (t^{-1} s)_m~.
\end{equation}
Substituting into~\eqref{eq:u2viaaDbis}, we find that the first two terms, both of which are~$\CO(N\Lambda^2)$ at large~$N$, cancel up to an~$\CO(\Lambda^2)$ remainder, which is also the scaling of the third term. 

\bigskip

\noindent {\it Higgs Branch:} At small~$\kappa$, the~$x_m$ are given by~\eqref{eq:xonHBbis}, with the matrix~$T$ defined in~\eqref{bigTmatbis}. Using~$a_{Dm} = - i M x_m$, we thus obtain
	\begin{equation}\label{eq:aDonHB}
     a_{Dm} \Big |_{\text{HB}} 
	=- \frac{i N \Lambda }{2\pi^2} (T^{-1} s)_m~.
	\end{equation}
Thus the only difference between the~$a_{Dm}$ on the Coulomb and Higgs branch is the matrix~$t$ versus~$T$ that appears in~\eqref{eq:aDonCB} and~\eqref{eq:aDonHB}. Note in particular that both results are~$\CO(N \Lambda)$ and do not scale with~$M$, even at very large~$M$. As already reviewed around~\eqref{eq:Tstsapprox}, the difference between~$(T^{-1} s)_n$ and~$(t^{-1} s)_n \simeq {2 \pi^2 / N} s_n$ is rather mild: they are of the same order, have the same symmetries (both look roughly proportional to~$s_n$), and are within a factor of~$1$ to~$3$ of one another, so that~$(T^{-1} s)_n \lesssim (t^{-1} s)_n$. Thus the~$a_{Dm}$ on the HB are somewhat smaller, and thus somewhat closer to the multi-monopole point.

This can be expressed using the gauge-invariant moduli~$u_I$. Here we focus on~$u_2$, which is given in~\eqref{eq:u2viaaDbis} for the multi-monopole point (where~$u_2>0$), for the case of $SU(2)$ and~$SU(3)$. The value of~$u_2$ at the~$N = 2,3$ multi-monopole points is indicated by blue dots in figure~\ref{fig:sunmmpoints}, while the value of~$u_2$ on the maximal Higgs branch at small~$\kappa$ (corresponding to adjoint QCD) is indicated by purple dots. We see that they are close to each other, with the Higgs branch points slightly displaced into the strong-coupling region, and pointing toward the origin~$u_2 = 0$. This closely reflects the shape of the~$\CN=2$ K\"ahler potential, which is convex with a unique minimum at the origin -- a fact that is captured in the dual via the~$\CO(a_D)$ tadpole and the~$\CO(a_D^2)$ terms in the SUSY-breaking potential~\eqref{eq:s3sblagmm}. 

We can now summarize how the~$u_I$ evolve along the cascade: they start at the origin~$u_I$ at large~$\kappa$, and once the Higgs fields start turning on they climb out of the potential well centered at the origin, and (roughly) towards the multi-monopole points.\footnote{~Linearly interpolating between the multi-monopole points and the origin in~$u_I$ space only leads to~$C$-even vacua; by contrast the cascade contains short phases where~$C$ is spontaneously broken, which cannot lie on these lines. Of course, the maximal Higgs branch shown in figure~\ref{fig:sunmmpoints} is~$C$-even.} As summarized in figure~\ref{fig:sunmmpoints}, they fall somewhat short, even at very small~$\kappa$.

\subsubsection{Gaugino bilinear}

Let us consider the gaugino bilinear~\eqref{eq:vecOdef}, which was introduced as an order parameter for chiral symmetry in adjoint QCD, 
\begin{equation}\label{eq:bilinearbis}
    \vec \CO = i \tr\left( \lambda^{\alpha i} {\vec \sigma_i}^{\;\; j} \lambda_{\alpha j}\right)
\end{equation}
In the~$\CN=2$ SYM theory, this operator resides in the~$\CN=2$ chiral multiplet whose bottom component is~$u_2 = \tr \phi^2$ defined in~\eqref{2.moduli}. To see this explicitly, we use the~$\CN=2$ SUSY-variations~\eqref{eq:n2susyt} to compute
\begin{equation}\label{10.llqqu}
    \tr \left(\lambda^{\alpha(i} \lambda_\alpha^{j)}\right) = {1 \over 4} \ep^{\alpha\beta} Q_\alpha^i Q_\beta^j u_2~.
\end{equation}
Here we have use the fact that the non-Abelian~$\CN=2$ $D$-term vanishes on-shell, $D^{ij} = 0$, in the pure~$SU(N)$ gauge theory. 

We can track this computation to the multi-monopole (MM) point, by using the expression~\eqref{eq:u2viaaDbis} for~$u_2$, together with~\eqref{10.llqqu} and the~$\CN=2$ SUSY-transformations~\eqref{eq:qonvect} at the multi-monopole point, 
\begin{equation}\label{eq:IRll}
 \tr \left(\lambda^{\alpha(i} \lambda_\alpha^{j)}\right) \Big|_\text{MM point} =   \sum_{m=1}^{N-1}\left(D_m^{ij} \Big( -2 \sqrt 2 \Lambda s_m + {i \sqrt 2 \over 2N} a_{Dm}\Big) - {1 \over 2N} \rho_m^{\alpha i} \rho_{\alpha m}^j\right)~.    
\end{equation}
This formula is valid in the~$\CN=2$ theory prior to SUSY-breaking, so it can receive explicit~$\CO(M^2)$ corrections, which we are not able to compute. However even the leading-order formula gives very sensible results, as we will now show.

Unlike their non-Abelian UV counterparts, the~Abelian~$\CN=2$ $D$-terms at the multi-monopole point are non-trivial functions of the hypermultiplet scalars given by~\eqref{app:IRDterms}, which we repeat here,
\be\label{app:IRDtermsbis}
D^{ij}_m = i \left(t^{-1}\right)_{mn} \left( h^i_{n} \bar h^j_{n} + h^j_{n} \bar h^i_{n}  \right)~. 
\ee
Substituting into~\eqref{eq:IRll}, we can then compute the vev of the triplet gaugino bilinear~\eqref{eq:bilinearbis},\footnote{~Note that the Abelian gauginos~$\rho_{m\alpha}^i$ are weakly coupled in the IR, so that the fermionic terms in~\eqref{eq:IRll} have vanishing vev.}
\begin{equation}\label{eq:OvevIR}
    \langle \vec \CO\rangle = \sum_{m,n = 1}^{N-1} (t^{-1})_{mn} \; \bar h^i_n {\vec \sigma_i}^{\;\; j} h_{jn} 
    \;\left( 4 \sqrt2 \Lambda s_m - {i \sqrt 2 \over N} a_{Dm}\right)~.
\end{equation}
Recall from~\eqref{eq:Spindef} that~$\vec S_n = \bar h^i_n {\vec \sigma_i}^{\;\; j} h_{jn}$, so that~\eqref{eq:OvevIR} is nothing but a linear combination of the different~$SU(2)_R$ spins~$\vec S_m$ arising from each hypermultiplet. Since all of these are perfectly aligned along the~$\vec e_3$ direction in~$SU(2)_R$ triplet space, as discussed around~\eqref{hpar}, the same is true of the vev of~$\vec \CO$,
\begin{equation}\label{eq:Ovevnodim}
    \langle \vec \CO\rangle = \vec e_3 M^2 \Lambda \sum_{m,n = 1}^{N-1} (t^{-1})_{mn} \; h_n^2
    \;\left( 4 \sqrt2  s_m - { \sqrt 2  \over 2 \pi^2 \kappa} x_m\right)~.
\end{equation}
Here we have switched to the dimensionless variables~$h_n$ and~$x_m$ introduced in~\eqref{hpar} and~\eqref{3.xdef}, and to~$\kappa = N \Lambda / (2 \pi^2 M)$ from~\eqref{eq:kappadefsec3}.  

\begin{figure}[t!]
\centering
\includegraphics[width=0.8\textwidth]{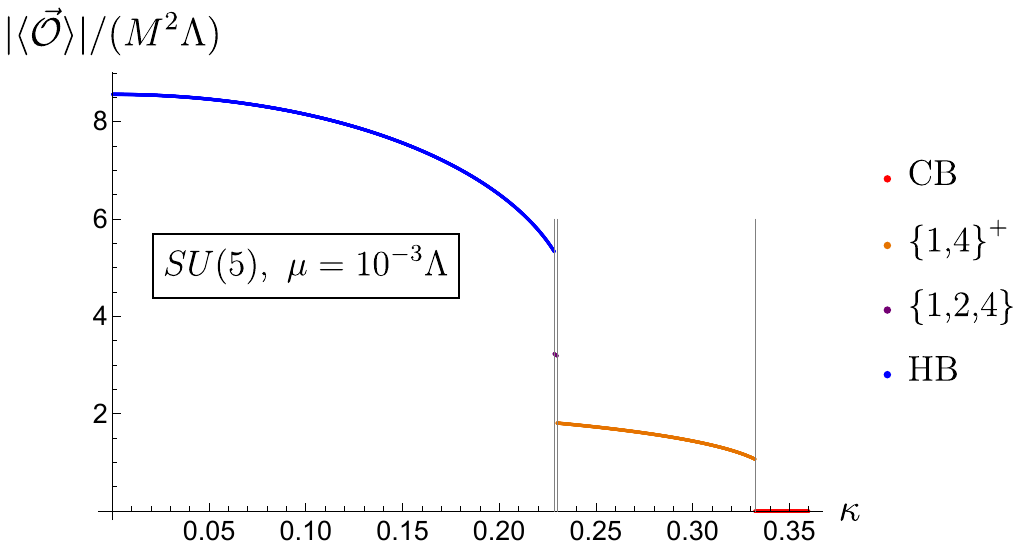}
	\caption{Vev of the gaugino bilinear~$\vec \CO$ in~\eqref{eq:bilinearbis} as a function of~$\kappa$ for~$SU(5)$.}	\label{fig:condensate}
\end{figure}

Of course~\eqref{eq:Ovevnodim} vanishes on the Coulomb branch, but once the first Higgs field turns on it is non-zero, and grows along the cascade until it reaches its asymptotic form appropriate to adjoint QCD on the maximal HB. This is shown for~$SU(5)$ gauge group in figure~\ref{fig:condensate}.

Let us  evaluate~\eqref{eq:Ovevnodim} on the Higgs branch HB, at leading order in small~$\kappa\ll 1$, or equivalently large~$M \gg \Lambda$. Due to the explicit~$1 / \kappa$ in~\eqref{eq:Ovevnodim}, we must work with the~$\CO(\kappa)$ contribution for the~$x_m$ from~\eqref{6.x1} (see also~\eqref{eq:xonHBbis}), with~$T$ in~\eqref{bigTmatbis},
\begin{equation}
    x_m = \kappa (T^{-1} s)_m~, \qquad  T_{mn} = t_{mn} + \delta_{mn} v_n~, \qquad v_n = \sum_{p=1}^{N-1} t_{np}
\end{equation}
However, we can take the leading~$\kappa =0$ order solution for the Higgs fields~$h_m$, by substituting~$x_m =0$ in~\eqref{eq:hmonHBbis}, so that
\begin{equation}\label{eq:smallkappahm}
    h_m^2 = \half \sum_{n = 1}^{N-1} t_{mn} = \half v_m~.
\end{equation}

We conclude that
\begin{equation}\label{eq:smallkOvev}
    \langle \vec \CO\rangle = \vec e_3 M^2 \Lambda \sum_{m,n = 1}^{N-1} (t^{-1})_{mn} \; v_n
    \;\left( 2 \sqrt2  s_m - { \sqrt 2  \over 4 \pi^2} (T^{-1}s)_m\right)
\end{equation}
Let us examine this quantity in the large-$N$ limit, where it can be simplified further: thanks to~\eqref{eq:Tstsapprox} the second term inside the big parentheses in~\eqref{eq:smallkOvev} is~$\CO(1/N)$ suppressed relative to the first one, and can be dropped. The remaining terms can be evaluated at large~$N$ using~\eqref{eq:usefulbis}, leading to
\begin{equation}\label{eq:largeNllfinal}
    \langle \vec \CO \rangle = {4 \sqrt 2 N \over \pi} M^2 \Lambda \vec e_3~.
\end{equation}
This is precisely the expected~large-$N$ dependence, given the normalization of~$\vec \CO$ in~\eqref{eq:bilinearbis}.\footnote{~The leading diagram is a single closed fermion loop, which gives a factor of~$N^2-1$ from the adjoint gauginos in the loop, and also a factor of the~$SU(N)$ Yang-Mills gauge coupling~$g^2$ from the propagator (see~\eqref{2.Lcomp}). Since~$g^2 N$ is the (fixed) 't Hooft coupling at large $N$, this diagram is~$\CO(N)$, just like~\eqref{eq:largeNllfinal}.} Note that~\eqref{eq:largeNllfinal} is an increasing function of~$M$, but on physical grounds we expect it to stabilize once we push deep enough into the HB. Precisely in that regime we expect the adjoint scalar of~$\CN=2$ SYM to decouple, leading to adjoint QCD. Since the transition to the maximal HB happens at~$M \sim \Lambda$ (without any strong~$N$-dependence), we see that the gaugino bilinear stabilizes at the scale~$\CO(N \Lambda^3)$ expected from adjoint QCD.\footnote{~The same comment applies to the vacuum energy in~\eqref{eq:svonHB}, which should stabilize at~$M \sim \Lambda$ and scale as~$\CO(N^2 \Lambda^4)$.}

\subsubsection{Radius~$f_\pi$ of the~$\C\P^1$ sigma model}\label{sec:cp1fpi}

\begin{figure}[t!]
\centering
\includegraphics[width=0.8\textwidth]{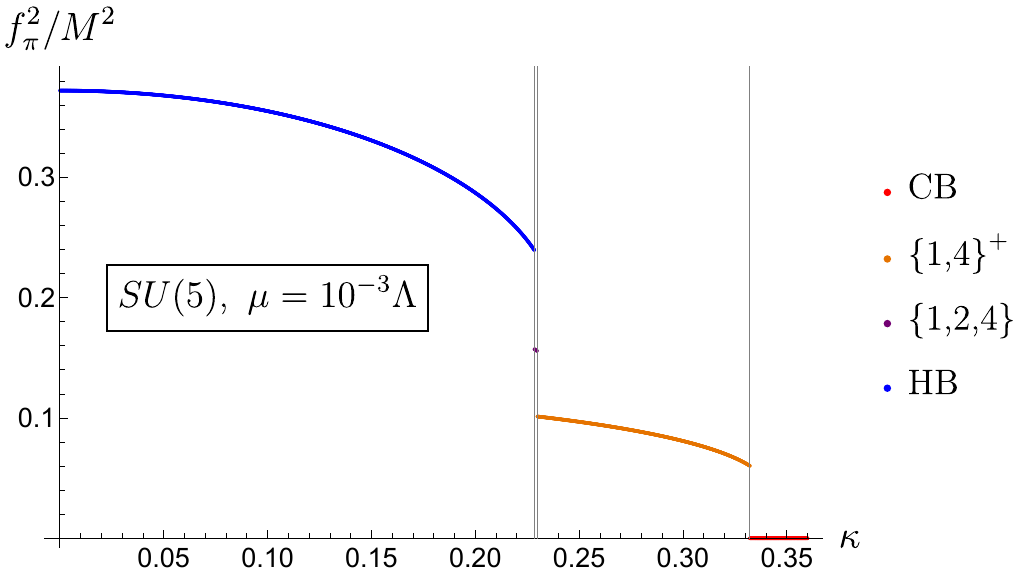}
	\caption{$\C\P^1$ radius-squared~$f_\pi^2$ in~\eqref{eq:cp1lag} as a function of~$\kappa$ for~$SU(5)$.}	\label{fig:c1radius}
\end{figure}

We would now like to compute the radius of the~$\C\P^1$ sigma model along the cascade. This radius is nothing but the pion decay constant~$f_\pi$, which appears explicitly in the~$\C\P^1$ Lagrangian~\eqref{eq:cp1lag}. 
Comparing the canonical kinetic terms for the~$h_{im}$ in~\eqref{eq:Lbos} to 
\begin{equation}
    \vec S_m = \bar h_m^i {\vec \sigma_i}^{\;\; j} h_{mj} = M^2 h_m^2 \vec n~,
\end{equation}
we can deduce the formula
\begin{equation}\label{eq:fpifmla}
    f_\pi^2 = \half M^2 \sum_{m \in \BB} h_m^2
\end{equation}
It is natural to switch to dimensionless variables
\begin{equation}
    {f_\pi^2 \over \Lambda^2} = {N^2 \over 4 \pi^4 \kappa^2} \sum_{m \in \BB} h_m^2~, \qquad \kappa = {N \Lambda \over 2 \pi^2 M}
\end{equation}
Thus~$f_\pi > 0$ from the moment the first Higgs field turns on, and it grows along the cascade until it reaches the small-$\kappa$ maximal Higgs branch and the adjoint QCD regime. This has been plotted for~$SU(5)$ gauge group in figure~\ref{fig:c1radius}.

We can evaluate~\eqref{eq:fpifmla} at leading order in small~$\kappa$ on the HB, using $h_m^2 = v_m / 2$ from~\eqref{eq:smallkappahm}, so that
\begin{equation}\label{eq:fpiHB}
        f_\pi^2 = {1 \over 4} M^2 \sum_{m,n = 1}^{N-1} t_{mn}~.
\end{equation}
We have already encountered the same double sum over the matrix elements of~$t_{mn}$ when computing the vacuum energy~\eqref{eq:svonHB}, so that we obtain the same~$\CO(N^2)$ scaling and the same transcendental pre-factor that we found there,
\begin{equation}\label{eq:finalfpiHB}
    f_\pi^2 \to  {7 \zeta(3) \over 8\pi^4} N^2 M^2~, \qquad N \to \infty~.
\end{equation}
This has precisely the expected large-$N$ scaling, since the Lagrangian should scale like~$N^2$ in a theory with adjoints.\footnote{~The~$N$-scaling~$f_\pi^2 \sim N^2$ can be obtained by examining the~$SU(2)_R$ current~$\vec j_\mu$ in the UV and IR,
\begin{equation}
\vec j_\mu \sim {1 \over g^2} \tr \bar \lambda \, \vec \sigma \, \bar \sigma_\mu \lambda \sim f_\pi^2 \p_\mu \vec n~, 
\end{equation}
The leading diagrams contributing to its connected 2-point function (a one-loop diagram in the UV, and a tree level one in the IR) scale like~$f_\pi^4/ f_\pi^2 \sim g^4 (N^2-1)/ g^4$, with some factors of~$f_\pi$ or~$g$ coming from the definition of the currents, and others from propagators.} As we already observed for the gaugino condensate in~\eqref{eq:largeNllfinal}, the HB formula~\eqref{eq:finalfpiHB} reduces to the expected~$\CO(N^2 \Lambda^2)$ answer in adjoint QCD if it saturates at~$M \sim \Lambda$, roughly where the phase transition into the maximal HB takes place.

\subsection{Realization of symmetries along the cascade}\label{sec:symcascade}

We conclude our analysis by determining the broken and unbroken symmetries along the cascade. Their detailed UV definition appears in section~\ref{sec:symm}, and their action on the magnetic Abelian dual at the multi-monopole point in section~\ref{sec:symmahm}. 

\subsubsection{Zero-form symmetries}

Here we examine the fate of the zero-form symmetries acting on local operators: the discrete and continuous~$R$-symmetries,
\begin{equation}
    {\Z_{4N} \times SU(2)_R \over \Z_2}~,
\end{equation}
with~$\Z_{4N}$ generator~$r$, charge-conjugation~$C$ (for~$N \geq 3$) and time-reversal symmetry~$T$.

\begin{itemize}
\item On the Coulomb branch at large~$\kappa$, where all Higgs fields vanish, all zero-form symmetries are unbroken. This is clear in the large~$\kappa$ analysis of section~\ref{sec:smallm}, but it is partially obscured in the magnetic dual at the multi-monopole point, because the~$\ZZ_{N} = {\ZZ_{4N} / \ZZ_{4}}$ quotient permutes the different multi-monopole points. 

The symmetries that fix the multi-monopole point are listed in table~\ref{tab:symahm}. Since~$r^N$ acts on the~$a_{Dm}$ like charge-conjugation, both are unbroken on the Coulomb branch, as can be seen from the manifestly~$C$-invariant expressions~\eqref{eq:xcb} or~\eqref{xmcb} for the~$a_{Dm} = - i M x_m$ on the Coulomb branch. Recall that the symmetry~$C \t T = C r^N T$ is always unbroken in the dual Abelian Higgs model at the multi-monopole point, and thus~$T$ is also unbroken.

\item The moment the first Higgs field turns on, and for the remainder of the cascade down to small~$\kappa$, both the continuous and the discrete~$R$-symmetries are spontaneously broken:
\begin{itemize}
    \item[(i)] The~$SU(2)_R$ symmetry is spontaneously broken as follows,
\begin{equation}
    SU(2)_R \to U(1)_R~,
\end{equation}
leading to two massless Nambu-Goldstone bosons parametrizing a~$\C\P^1$ sigma model as in~\eqref{eq:cp1lag}.
\item[(ii)] The discrete~$\Z_{4N}$ $R$-symmetry is spontaneously broken. The~$\Z_N = \Z_{4N} /\Z_4$ quotient permuting the~$N$ distinct multi-monopole points is always spontaneously broken (only the CB vacuum at the origin is an exception), but the precise unbroken subgroup depends on the details of the branch. 

By examining table~\ref{tab:symahm}, we see that combining~$\t C = r^N C$ with an~$SU(2)_R$ Weyl reflection~$W$ associated with the unbroken~$U(1)_R$ Cartan leads to a symmetry~$\t C W$ that is always unbroken. If we are on a branch with unbroken~$C$-symmetry, then~$r^N W$ is also preserved. Similarly, the time-reversal symmetry~$C\t T$ is always preserved, while~$\t T = r^N T$ is preserved on~$C$-even branches.

\end{itemize}

\item Since the maximal Higgs branch (HB) at small~$\kappa$ is~$C$-even, as shown in section~\ref{sec:smallk}, we conclude from point (ii) above that the unbroken discrete symmetries on the HB are~$C$, $r^N W$, and~$\t T = r^N R$. Together with the~$U(1)_R$ Cartan, these are precisely the unbroken symmetries of the confining and chiral symmetry breaking scenario for adjoint QCD spelled out in section~\ref{sec:confchisbIntro}. Note that the~$\Z_2$ symmetry extending~$U(1)_R$ to~$O(2)_R$ in~\eqref{chiSB} is precisely given by~$r^N W$. 

\end{itemize}

\subsubsection{One-form symmetries}

The~$\Z_N^{(1)}$ one-form symmetry associated with the center of the~$SU(N)$ gauge group is embedded into the emergent magnetic~$U(1)^{N-1}$ one-form symmetry of the Abelian dual at the multi-monopole point via~\eqref{utildefirst}. This shows that the fundamental, unit-charge 't Hooft line in the gauge group~$U(1)_{Dm}$ (with~$m = 1, \ldots, N-1$) has~$\Z_N^{(1)}$ charge~$m \text{ mod } N$. 

On a branch labeled by the partition~$\AA|\BB$, the Higgs fields in~$\BB$ have condensed and the corresponding 't Hooft loops have area law. By contrast the~$U(1)_{Dm}$ gauge groups with~$m \in \AA$ are still in the Coulomb phase and the associated 't Hooft loops have perimeter law, i.e.~vev. It thus follows that~$\Z_N^{(1)}$ is spontaneously broken as follows:
\begin{equation}
    \Z_N^{(1)} \to \Z_p^{(1)}~, \qquad p = \text{gcd}(N, m \in \AA)~.
\end{equation}
It can be checked that the greatest common divisor (gcd) that determines the unbroken subgroup evaluates to~$p = 1$ for almost all branches along the cascade, so that~$\Z_N^{(1)}$ is generically completely broken. There are only two exceptions:
\begin{itemize}
\item On the maximal Higgs branch, $\AA$ is empty and~$p = N$, so the entire~$\Z_N^{(1)}$ symmetry is unbroken. This is the fully confined phase expected for adjoint QCD, as discussed in section~\ref{intro:nf2adjoint}. 
\item If~$N = 2\nu$ is even and~$C = \{\nu\}$, then~$p = \text{gcd}(2\nu, \nu) = \nu$, so that there is an unbroken~$\Z_\nu^{(1)} \subset Z_N^{(1)}$ subgroup.
\end{itemize}

It is noteworthy that full confinement only occurs at the very end of the cascade, on the maximal Higgs branch. By contrast, the chiral symmetry breaking pattern of adjoint QCD is essentially locked in once the first Higgs field condenses, leading to a small~$\C\P^1$ sigma model, whose radius~\eqref{eq:fpifmla} grows along the cascade until it reaches the size expected in adjoint QCD (see the discussion below~\eqref{eq:finalfpiHB}).

\newpage

\appendix


	\section{Conventions}
	\label{sec:conventions}\label{app:conv}

	\subsection{Lie algebra and gauge fields}

Denote the $SU(N)$ generators in the defining representation by Hermitian~$N \times N$ matrices $T^a$ satisfying
\begin{equation}
[T^a,T^b]=if^{abc}T^c~, \qquad \tr (T^a T^b)  = \delta^{ab}/2~.
\end{equation}
For $N=2$, in the defining representation we have $T^a = \sigma^a/2$, where~$\sigma^a~(a = 1, 2, 3)$ are the standard Pauli matrices.

Any field in the adjoint representation of~$SU(N)$ is denoted by~$\chi = \chi^a T^a$. The covariant derivative in the adjoint representation is 
\begin{equation}
D_\mu \chi = \partial_\mu \chi - i [v_\mu,\chi] \quad \longleftrightarrow \quad  D_\mu \chi^a = \partial_\mu \chi^a + f^{abc} v_\mu^b \chi^c~,
\end{equation}
where~$v_\mu$ is the adjoint-valued~$SU(N)$ gauge field. Its field strength is given by 
\begin{equation}
v_{\mu\nu} = \partial_\mu v_\nu - \partial_\nu v_\mu - i [v_\mu,v_\nu]  \quad \longleftrightarrow \quad v_{\mu\nu}^a = \partial_\mu v_\nu^a - \partial_\nu v_\mu^a + f^{abc} v_\mu^b v_\nu^c~.
\end{equation}

\subsection{Weyl spinors and~$SU(2)_R$ symmetry}

The right-handed Hermitian conjugate of the left-handed Weyl fermion $\lambda_\alpha^{i}$ is
	\ba{
	\bar{\lambda}_{\dot{\alpha} i} = \left( \lambda_\alpha^{i} \right)^\dagger\ .
	}
All spinor conventions follow Wess and Bagger, including the use of bars for Hermitian conjugation. Spinor indices are raised and lowered by left action of $\ep^{\alpha\beta}$ and~$\ep_{\alpha\beta}$, where $\ep^{12}=\ep_{21}=1$. Similarly, we raise and lower $SU(2)_R$ doublet indices~$i,j, \ldots = 1,2,$ from the left with~$\ep^{ij}$ and~$\ep_{ij}$. 

We will denote the standard traceless, Hermitian~$SU(2)_R$ Pauli matrices as follows,
\begin{equation}
{{\vec \sigma}_i}^{\;\; j} = {\left(\sigma^1, \sigma^2, \sigma^3\right)_i}^{\;\; j}~.
\end{equation}
Note that Hermitian conjugation (indicated by bars) exchanges raised and lowered~$SU(2)_R$ doublet indices. 

\subsection{$\CN=2$ Supersymmetry}

The supercharges are 	
	\ba{
	Q_\alpha^i \ ,\qquad \bar{Q}_{\dot{\alpha}i} = (Q_\alpha^i)^\dagger\ ,\qquad i = 1,2 \ ,
	}
which satisfy the $\CN=2$ supersymmetry algebra 
	\ba{
	\{Q_\alpha^i,\bar{Q}_{\dot{\alpha}j}\}= 2\delta^{i}_{\ j} \sigma^\mu_{\alpha \dot{\alpha}} P_\mu\ , \qquad \{ Q_\alpha^i Q_\beta^j \}= 2 \ep_{\alpha\beta} \ep^{ij} \bar{Z},\qquad \bar{Z} = Z^\dagger\ . \label{eq:n2algebra}
	}
In this convention, the BPS bound reads~$M \geq |Z|$, so that BPS saturated particles have masses~$M_\text{BPS} = |Z|$.

We can embed the $\CN=1$ supersymmetry algebra into this $\CN=2$ algebra with 
	\ba{
	Q_{\alpha}^{\CN=1} = Q_\alpha^1\ ,\qquad \bar{Q}_{\dot\alpha}^{\CN=1} = (Q_\alpha^{\CN=1})^\dagger= \bar{Q}_{\dot{\alpha} 1} \ .
	}

	\subsubsection{$\CN=2$ Supersymmetry transformations of~$SU(N)$ pure Yang-Mills theory }

The $\CN=2$ supersymmetry transformations of the $\CN=2$ non-Abelian~$SU(N)$ vector multiplet are then the same as in \cite{Cordova:2018acb}, with $\ep_{abc}\to f_{abc}$,
	\ba{\bs{
	&Q_\alpha^i \phi^a = i \sqrt{2} \lambda_\alpha^{ia}\ ,\\
	& \bar{Q}_{\dot{\alpha}}^i \phi^a = 0 \ ,\\
	&Q_\alpha^i \lambda_\beta^{ja} = - \ep^{ij} (\sigma^{\mu\nu})_{\alpha \beta} v_{\mu\nu}^a + \ep_{\alpha\beta} \left( D^{ija} - \ep^{ij} f^{abc} \bar{\phi}^b \phi^c \right)\ ,\\
	& \bar{Q}_{\dot{\alpha}}^i \lambda_\alpha^{ja} = \ep^{ij}\sqrt{2} \sigma^\mu_{\alpha\dot{\alpha}} D_\mu \phi^a \ , \\
	&Q_\alpha^i v_\mu^a = i \sigma_{\mu \alpha \dot{\alpha}} \bar{\lambda}^{\dot{\alpha} i a}\ ,\\
	& \bar{Q}_{\dot{\alpha}}^i v_\mu^a = - i \sigma_{\mu \alpha \dot{\alpha}} \lambda^{\alpha i a} \ ,\\
	&Q_\alpha^i D^{jk a} = i \left( \ep^{ij} \sigma^\mu_{\alpha \dot{\alpha}}  D_\mu \bar{\lambda}^{\dot{\alpha} k a}  + \ep^{ik} \sigma^\mu_{\alpha \dot{\alpha}} D_\mu \bar{\lambda}^{\dot{\alpha} j a}\right) + i \sqrt{2} f^{abc} \bar{\phi}^b \left( \ep^{ij} \lambda_\alpha^{kc} + \ep^{ik}\lambda_\alpha^{jc} \right)\ ,\\
	&\bar{Q}_\alpha^i D^{jk a} =- i \left( \ep^{ij} \sigma^\mu_{\alpha \dot{\alpha}}  D_\mu {\lambda}^{{\alpha} k a}  + \ep^{ik} \sigma^\mu_{\alpha \dot{\alpha}} D_\mu {\lambda}^{{\alpha} j a}\right) + i \sqrt{2} f^{abc} {\phi}^b \left( \ep^{ij} \bar{\lambda}_{\dot{\alpha}}^{kc} + \ep^{ik} \bar{\lambda}_{\dot{\alpha}}^{jc} \right)\ .
	}\label{eq:n2susyt}}
These satisfy the algebra \eqref{eq:n2algebra} with $Z=0$, modulo gauge transformations.

\subsubsection{The~$\CN=2$ Abelian Higgs model at the multi-monopole point}\label{app:ahmlag}

Here we summarize the renormalizable terms in the~$\CN=2$ supersymmetric Abelian Higgs model at the multi monopole point, spelling out the component Lagrangian and supersymmetry transformations in detail. We closely follow appendix~B of~\cite{Cordova:2018acb}, which we generalize from rank one to rank~$N-1$. The dual magnetic gauge group is
\be
U(1)_D^{N-1} = \prod_{m = 1}^{N-1} U(1)_{Dm}~.
\ee 
In general we use~$m, n, \ldots$ to index~$U(1)_D^{N-1}$. When needed, we separate these indices from other ones (e.g.~Lorentz or~$SU(2)_R$ indices) by a comma for clarity.

Each~$U(1)_{Dm}$ gauge group factor gives rise to one~$\CN=2$ vector multiplet, with off-shell component fields
\be
a_{Dm}~, \quad \rho^i_{\alpha m}~, \quad f_{\mu\nu m} = \p_\mu b_{\nu m} - \p_\nu b_{\mu m}~, \quad D^{ij}_m~, \qquad m = 1, \ldots, N-1~.
\ee
Here~$a_{Dm}$ is a complex scalar, $\rho^i_{\alpha m}$ is the~$\CN=2$ gaugino (with~$SU(2)_R$ doublet index~$i = 1, 2$), and~$b_{\mu m}$ is the~$U(1)_{Dm}$ gauge field, with field strength~$f_{\mu\nu m}$. Finally, the auxiliary fields are real~$SU(2)_R$ triplets satisfying
\be
D^{ij}_m = D^{(ij)}_m = \left(D_{ij, m}\right)^\dagger~.
\ee
Since we will use~$\CN=1$ superspace to construct the Lagrangian, we choose an~$\CN=1$ supercharge~$Q_\alpha = Q_\alpha^{i= 1}$, under which the~$\CN=2$ vector multiplet decomposes into an~$\CN=1$ vector multiplet,\footnote{~Here we are using Wess and Bagger~\cite{Wess:1992cp} conventions~, with~$v_\mu$ the gauge field, $\lambda_\alpha$ the gaugino, and~$D$ the real auxiliary field in the~$\CN=1$ vector superfield.}
\be
\CV_{Dm} = \left(v_\mu = b_{\mu m}~, \quad \lambda_\alpha = i \rho_{\alpha m}^2~, \quad D = i D^{12}_m\right)~,
\ee
and into an~$\CN=1$ chiral multiplet,\footnote{~Here~$\phi$, $\psi_\alpha$, and~$F$ are the complex scalar bottom component, the fermion, and the complex auxiliary field in an~$\CN=1$ chiral superfield.} 
\be
A_{Dm} = \left(\phi = a_{Dm}~, \quad \psi_\alpha = \rho_\alpha^1~, \quad F = {i \over \sqrt 2} D^{11}\right)~.
\ee

There is precisely one~$\CN=2$ hypermultiplet of unit charge in every~$U(1)_{Dm}$ gauge group. On shell, every such hypermultiplet has the following component fields
\be
h_{i m}~, \qquad \psi_{\alpha m}^\suppm~, \qquad m = 1, \ldots, N-1~. 
\ee
Here~$h_{i m}$ is a complex~$SU(2)_R$ doublet of unit~$U(1)_{Dm}$ charge, while the fermions~$\psi_{\alpha m}^\suppm$ are neutral under~$SU(2)_R$ and carry~$U(1)_{Dm}$ charges~$\pm1$. (All these fields are neutral under the other gauge groups~$U(1)_{D, n \neq m}$.) We denote Hermitian conjugation by bars, so that 
\be
\bar h^i_m = \left(h_{i m}\right)^\dagger~, \qquad \bar h_{i m} = - \left(h^i_m\right)^\dagger~. 
\ee

With respect to the~$\CN=1$ supercharge~$Q_\alpha = Q_\alpha^1$ chosen above, the hypermultiplet decomposes into a pair of~$\CN=1$ chiral multiplets,
\be
\begin{split}
\CM_{m}^\supp & = \left(h_{1 m}~, \quad \psi_{\alpha m}^\supp~, \quad F_{m}^\supp \right)~, \\
\CM_{m}^\supm  & = \left(\bar h^2_m~, \quad \psi_{\alpha  m}^\supm~, \quad F_{m}^\supm \right)~. 
\end{split}
\ee
Here~$F_{m}^\suppm$ are~$\CN=1$ auxiliary fields that enable an~$\CN=1$ off-shell formulation. 

We can now write the Lagrangian in~$\CN=1$ superspace: 
\be\label{eq:ahmssapp}
\begin{split}
\SL = & \int d^4 \theta \, \left(\sum_{m, n = 1}^{N-1} t_{mn} \bar A_{Dm} A_{Dn} 
+ \sum_\pm \sum_{m =1}^{N-1}  \bar \CM_{m}^\suppm e^{\mp 2 \CV_{Dm}} \CM_{m}^\suppm \right)  
\\
& + \int d^2 \theta \, \left({1 \over 4} \sum_{m, n = 1}^{N-1} t_{mn} W^\alpha{} _m W_{\alpha n} 
+ \sqrt 2 \sum_{m =1}^{N-1} A_{Dm} \CM_{m}^\supp \CM_{m}^\supm \right) + \left(\text{h.c.}\right)~. 
\end{split}
\ee
Here~$W_{\alpha m} = -{1 \over 4} \bar D^2 D_\alpha \CV_{Dm}$ is the~$U(1)_{Dm}$ chiral field strength supermultiplet. Expanding this in terms of components and integrating out~$F_{m}^\suppm$ (which have no straightforward~$\CN=2$ version), but not~$D^{ij}_m$, we finally arrive at the following component Lagrangian,
\be\label{eq:applsum}
\SL = \SL_\text{kinetic} + \SL_\text{Yukawa}  + \SL_\text{scalar}~.
\ee 
The kinetic terms are given by
\be\label{eq:applkin}
\begin{split}
\SL_\text{kinetic} = &  - \sum_{m, n  = 1}^{N-1} t_{m n} \left(
\p^\mu \bar a_{Dm} \p_\mu a_{Dn} 
+{1 \over 4} f^{\mu\nu}_m f_{\mu\nu n} 
+ i \bar \rho_{i m} \bar \sigma^\mu \p_\mu \rho_n^i   \right) \\
& - \sum_{m = 1}^{N-1} \left( D^\mu \bar h^i_m D_\mu h_{i m} 
+ \sum_\pm i \bar \psi_{m}^\suppm \bar \sigma^\mu D_\mu \psi_{m}^\suppm \right)~,
\end{split}
\ee
with~$D_\mu$ is the gauge-covariant derivative. Since the charges of the hypermultiplet fields are diagonal, we have
\be
D_\mu h_{i m} = \left(\p_\mu - i b_{\mu m} \right) h_{i m}~, \qquad 
D_\mu \psi_{m}^\suppm = \left(\p_\mu \mp i b_{\mu m}\right) \psi_{m}^\suppm~.
\ee
The Yukawa terms take the form
\be\label{eq:applyuk} 
\begin{split}
\SL_\text{Yukawa} =&  \sqrt 2 \sum_{m = 1}^{N-1} \left( 
\bar h_{i m} \rho_m^i \psi_{m}^\supp - h_{i m} \rho^i_m \psi_{m}^\supm 
- h^i_m \bar \rho_{i m} \bar \psi_{m}^\supp - \bar h^i_m \bar \rho_{i m} \bar \psi_{m}^\supm \right) 
\\
& - \sqrt 2 \sum_{m = 1}^{N-1}  \left(a_{Dm} \psi_{m}^\supp \psi_{m}^\supm 
+ \bar a_{Dm} \bar \psi_{m}^\supp \bar \psi_{m}^\supm \right)~, 
\end{split}
\ee
and the scalar potential reads 
\be\label{eq:applscal}
\SL_\text{scalar} = {1 \over 4} \sum_{m, n=1}^{N-1} t_{mn} D_m^{ij} D_{ij n} 
- \sum_{m = 1}^{N-1} \left(i D_m^{ij} h_{i m} \bar h_{j m} + 2 \left| a_{Dm} \right|^2 \bar h^i_m h_{i m} \right)~.
\ee
Integrating out the~$\CN=2$~$D$-terms gives
\be\label{app:IRDterms}
D_{ij m} = 2 i \left(t^{-1}\right)_{mn} h_{(i| n} \bar h_{|j) n} 
=  i \left(t^{-1}\right)_{mn} \left( h_{i  n} \bar h_{j  n} + h_{j  n} \bar h_{i  n}  \right)~. 
\ee
Substituting back into~$\SL_\text{scalar}$ and using~$SU(2)_R$ Fierz identities gives the following supersymmetric scalar potential, 
\be
\SL_\text{scalar} = -\SV_\text{SUSY}~, 
\ee
where
\be
\begin{split}\label{eq:appscalpot}
\SV_\text{SUSY} = &  \sum_{m = 1}^{N-1} 2 \left|a_{Dm}\right|^2 \bar h^i_m h_{i m} 
\\
& + \sum_{m, n = 1}^{N-1} \left(t^{-1}\right)_{mn} \left( \left(\bar h^i_m h_{i n}\right) \left(\bar h^j_n h_{j m}\right) 
- \half \left(\bar h^i_m h_{i m}\right) \left(\bar h^j_n h_{j n}\right)\right)~.
\end{split}
\ee
Note that the two distinct contractions on the second line of this equation become identical in the rank-one case, where we reproduce equation (B.19) of~\cite{Cordova:2018acb} with~$t^{-1} = e^2$.  

Finally, we reproduce (in Wess-Zumino gauge) the~$\CN=2$ supersymmetry transformations of the components fields from equations (B.20) and (B.21) in \cite{Cordova:2018acb}.\footnote{~We supply each field in those equations with a~$U(1)_D^{N-1}$ gauge-group label~$m = 1, \ldots, N-1$.} For the~$\CN=2$ vector multiplets these are
\ba{\bs{\label{eq:qonvect}
	& Q_\alpha^{i} a_{D m} =  i \sqrt{2} \rho_{\alpha  m}^{i} \ ,\qquad\qquad\qquad\qquad\qquad  
	\bar{Q}_{\dot{\alpha}}^i a_{Dm} = 0\ , 
	\\
&  Q_\alpha^{i } \rho_{\beta  m}^{j} =  \ep_{\alpha\beta} D^{ij}_m - \ep^{ij} (\sigma^{\mu\nu})_{\alpha\beta} f_{\mu\nu  m}\ ,\hskip35pt 
	\bar{Q}^i_{\dot{\alpha}} \rho^j_{\alpha  m} = \ep^{ij}\sqrt{2}\sigma^\mu_{\alpha\dot{\alpha}} \partial_\mu a_{Dm}\ , 
	\\
& 		Q_\alpha^i D_m^{jk} = i \left( \ep^{ij}\sigma^\mu_{\alpha \dot{\alpha}} \partial_\mu \bar{\rho}^{\dot{\alpha}k}_m + \ep^{ik}\sigma^\mu_{\alpha \dot{\alpha}} \partial_\mu \bar{\rho}^{\dot{\alpha} j}_m  \right) \ , \hskip8pt 
		\bar{Q}_{\dot{\alpha}}^i D_m^{jk} = -i \left( \ep^{ij}\sigma^\mu_{\alpha \dot{\alpha}} \partial_\mu {\rho}^{{\alpha}k}_m + \ep^{ik}\sigma^\mu_{\alpha \dot{\alpha}} \partial_\mu {\rho}^{{\alpha} j}_m  \right) \ ,
		  \\
&		Q_\alpha^i f_{\mu\nu  m} = - i \left( \sigma_{\mu\alpha\dot{\alpha} }\partial_\nu \bar{\rho}^{\dot{\alpha} i}_m- \sigma_{\nu \alpha \dot{\alpha}}   \partial_\mu \bar{\rho}^{\dot{\alpha} i}_m \right)\ ,\hskip18pt 	
	\bar{Q}_{\dot{\alpha}}^i f_{\mu\nu  m}=  i \left( \sigma_{\mu\alpha\dot{\alpha} }\partial_\nu {\rho}^{{\alpha} i}_m- \sigma_{\nu \alpha \dot{\alpha}}   \partial_\mu {\rho}^{{\alpha} i}_m \right)\ .
	} }
These close off shell since we have not integrated out~$D_m^{ij}$. By contrast, the hypermultiplet transformations only close on shell, since we have integrated out~$F_{m}^\suppm$, 
	\ba{ \bs{
& 	Q_\alpha^{i} h^j_m = -  i \sqrt{2} \ep^{ij} \psi_{\alpha , m}^\supp \  , 
\qquad  
	\bar{Q}_{\dot{\alpha}}^{i}h^j_m=  i \sqrt{2} \ep^{ij} \bar{\psi}_{\dot{\alpha} m}^\supm \  , 
	\\
& 	Q_\alpha^{i} \bar{h}^j_m =   i \sqrt{2} \ep^{ij} \psi_{\alpha  m}^\supm \ , 
\hskip34pt
	\bar{Q}_{\dot{\alpha}}^{i} \bar{h}^j_m=   i \sqrt{2} \ep^{ij} \bar{\psi}_{\dot{\alpha}  m}^\supp \  ,   
	\\
&	Q_\alpha^{i} \psi_{\beta  m}^\supp = 2 i \ep_{\alpha\beta} \bar{a}_{Dm} h^i_m  \ , 
\hskip19pt  
	\bar{Q}_{\dot{\alpha}}^{i} \psi_{\alpha m}^\supp = \sqrt{2}\sigma^\mu_{\alpha \dot{\alpha}} D_\mu h^i_m  \ ,   
	\\
&	Q_\alpha^{i} \psi_{\beta  m}^\supm =  2 i \ep_{\alpha\beta} \bar{a}_{Dm} \bar{h}^i_m  \ , 
\hskip20pt   
	\bar{Q}_{\dot{\alpha}}^{i} \psi_{\alpha m}^\supm =  -\sqrt{2} \sigma^\mu_{\alpha \dot{\alpha}} D_\mu  \bar{h}^i_m ~  . 
	} \label{eq:hypers2}}\ 


\newpage

\newpage
\section{Numerical analysis of multiple-Higgs branches}
\label{app:B}

In this appendix, we present the methods used to carry out a streamlined numerical analysis for the mixed Coulomb-Higgs and maximal Higgs branches for which several independent Higgs fields are non-vanishing. In particular, these numerical methods were used to obtain the reduced potentials for the branches $\BB= \{1,3\}^-, \{1,2\}, \{1,2,3\}^\pm$ for the case of gauge group $SU(4)$, as well as corresponding results for~$SU(5)$ and~$SU(6)$. To be concrete, we shall explain the method here for the maximal Higgs branches $\BB=\{1,2,3\}^\pm$ of $SU(4)$, its generalization to the other cases being conceptually straightforward. 

\subsection{The branches $\BB= \{1,2,3\}^+$ and $\BB=\{ 1,2,3\}^-$ for $SU(4)$}
\label{7.maxH}

The branches $\BB= \{1,2,3\}^+$ and $\BB=\{ 1,2,3\}^-$ correspond to  the partition $h_1, h_2, h_3 >0$, in which case the reduced field equations are obtained by eliminating $h_k$ between the equations of (\ref{7.maxEq}), and are given as follows,
\bea
\label{B.1}
\sum_{\ell=1}^3 t_{k\ell} \Big ( x_\ell + x_k (1-4 x_\ell^2) \Big ) = \kappa s_k \hskip 1in
k=1,2,3
\eea 
The reduced potential, evaluated on these solutions, takes the form, 
\bea
\label{redVmax}
V_{\{1,2,3\}^\pm}^\text{sol} = V_\text{CB} 
- {1 \over 8}  \sum _{k, \ell,m,n =1}^3 (t^{-1})_{k\ell} \, t_{km} \, t_{\ell n} \, (1-8x_k x_\ell) (1-4x_m^2) (1-4x_n^2)
\eea
Using the charge conjugation invariance of $t_{k\ell}$ and $s_k$, the difference of the equations in (\ref{B.1})  for $k=1$ and $k=3$ factorizes as follows, 
\bea
(x_1-x_3) \Big ( 4t_{11}(x_1^2 + x_3^2) + 4(t_{11} - t_{13}) x_1 x_3 + 4t_{12} x_2^2 - 2 t_{11} -  t_{12}  \Big ) =0
\eea 
Therefore, the maximal Higgs branch is actually the union of two subbranches,
\bea\label{eq:B.subbranches}
 \BB=\{1,2,3\}^+  & \hbox{ with } & 
x_3=x_1
\no \\
 \BB=\{1,2,3\}^-   & \hbox{ with } & 
4t_{11}(x_1^2 + x_3^2) + 4(t_{11} - t_{13}) x_1 x_3 + 4t_{12} x_2^2 = 2 t_{11} +  t_{12} 
\eea
where $x_3=x_1$ implies $h_3=h_1$ corresponding to the $C$-invariant maximal Higgs branch. We now discuss several aspects of the numerical analysis used to examine each branch.

\subsection{Constraints on the range of the variables $x_k$}

The range of the variables $x_1, x_2, x_3$ for which any solutions to the field equations (\ref{7.maxEq}) exist is constrained  by the positivity conditions $h_k^2>0$ for all $k=1,2,3$ with $h_k^2$  given by (\ref{7.maxEq}). Any point in the cube $\{ 0< x_1, x_2, x_3< \thalf \}$ satisfies these conditions. However, the full range allowed by the conditions $h_k^2>0$ generally extends beyond this cube. It will be convenient to further constrain the allowed range of the variables $x_k$ by making use of the local stability conditions. For the maximal Higgs branch we have $u=t$ and the local stability conditions reduce to the positivity of the quadratic form $Q_\a$ given in (\ref{stabB}),
\bea
Q_\a = \sum _{k,\ell=1}^3 t_{k\ell} \Big ( (1-4x_k^2) \a_\ell^2 + (1-4 x_\ell^2 ) \a_k^2 + 2 (1-8x_k x_\ell) \a_k\a_\ell \Big )
\eea
A necessary (but not sufficient) condition is $Q_\a>0$ when only a single $\a_m \not=0$.   After some rearrangements this condition may be expressed as follows,
\bea
\label{7.Qm}
\cQ_m= 
3 \, t_{mm} (1-4x_m^2) + \sum _{\ell \not= m} t_{m\ell} (1-4x_\ell^2) - t_{mm} >0
\hskip 1in 
m=1,2,3
\eea
We shall now prove that this inequality implies $4x_k^2<1$ for $k=1,2,3$ by showing that all other options are excluded. The case where $4x_k^2>1$ for $k=1,2,3$ is excluded in view of the fact that $|t_{m \ell}|\ll t_{mm}$ for $\ell \not=m$ for sufficiently small $\mu/\Lambda$. The cases where $4x_k^2, 4 x_\ell^2 >1$ while $4x_n^2<1$ for $k,\ell,n$ mutually distinct are eliminated by observing that the left side of the sum of the inequalities for $m=k,\ell$  in (\ref{7.Qm})  is negative. Finally, the cases where $4x_k^2>1$ while $4 x_\ell^2, 4 x_n^2<1$ for $k, \ell, n$ mutually distinct  are excluded by observing that the left side of the inequality for $m=k$  in (\ref{7.Qm})  is negative. Thus local stability requires that the range of $x_m$  be contained in the cube $\{ 0< x_1, x_2, x_3< \thalf \}$.

\subsection{Maximum value of $\kappa$ for the maximal Higgs branches}

We shall numerically determine the largest value $\kappa_{\text{HB}}$  of $\kappa$ for which the maximal Higgs branches can exist. To proceed, we introduce the following three functions, 
\bea
K_\ell(x_1,x_2,x_3) = {1 \over s_\ell} \sum _{m=1}^3 t_{\ell m} \Big ( x_m + x_\ell ( 1-4x_m^2) \Big )
\eea
In terms of these functions, the reduced field equations for the variables $x_1, x_2, x_3$, for a given value of $\kappa$, reduce to the following relations,
\bea
\kappa = K_1 (x_1,x_2,x_3) =K_2 (x_1,x_2,x_3) =K_3 (x_1,x_2,x_3) 
\eea
for both branches HB$= \{1,2,3\}^+$ and $\BB=\{ 1,2,3\}^-$.  Thus, $\kappa_{\text{HB}}$ may be defined as follows,
\bea
\kappa _{\text{HB}} = \max \Big \{ K_1 (x_1,x_2,x_3) =K_2 (x_1,x_2,x_3) =K_3 (x_1,x_2,x_3)  , ~ 0< x_1,x_2, x_3 <  \thalf \Big \}
\eea
Since the functions $K_\ell$ are polynomials in the variables $x_\ell$ and the domain is bounded, one may simply scan through the cube to find the maxima. Numerical analysis shows that the value of $\kappa_{\text{HB}}$ and the corresponding point in the cube are given as follows,\footnote{~The numerical uncertainty for each result lies in the 7-th significant digit which has been omitted. } 
\bea
\label{7.maxNum}
\kappa_{\text{HB}} = 0.203100
\hskip 1in
(x_1,x_2,x_3) = (0.257919, 0.425976, 0.257919)
\eea
A number of remarks are in order. First, the point at which the maximum value of $\kappa$ is attained does indeed lie inside the cube $\{ 0< x_1, x_2, x_3< \thalf \}$.   Second, the value $\kappa_{\text{HB}}$ is attained for a solution that is {\sl charge conjugation invariant} since (\ref{7.maxNum}) obeys  $x_1=x_3$ and thus lies on the branch HB.  In other words, the branch HB exists for all $\kappa \leq  \kappa _{\text{HB}}$ while the branch $\BB=\{ 1,2,3\}^-$ can exist only for smaller values of $\kappa$.

\subsection{Numerical solution for $\BB=\{1,2,3\}^+$}

Numerically solving for the roots to pairs or triplets of coupled cubic equations in two or three variables is slow. In this subsection, we shall adopt a different method, which we explain here for the branch $\BB=\{1,2,3\}^+$, and adapt to branch $\{1,2,3\}^-$ in the next subsection. We have verified that the results of this method manifestly match those from the slower brute-force analysis. 

\begin{figure}[t!]
\centering
\includegraphics[width=0.23\textwidth]{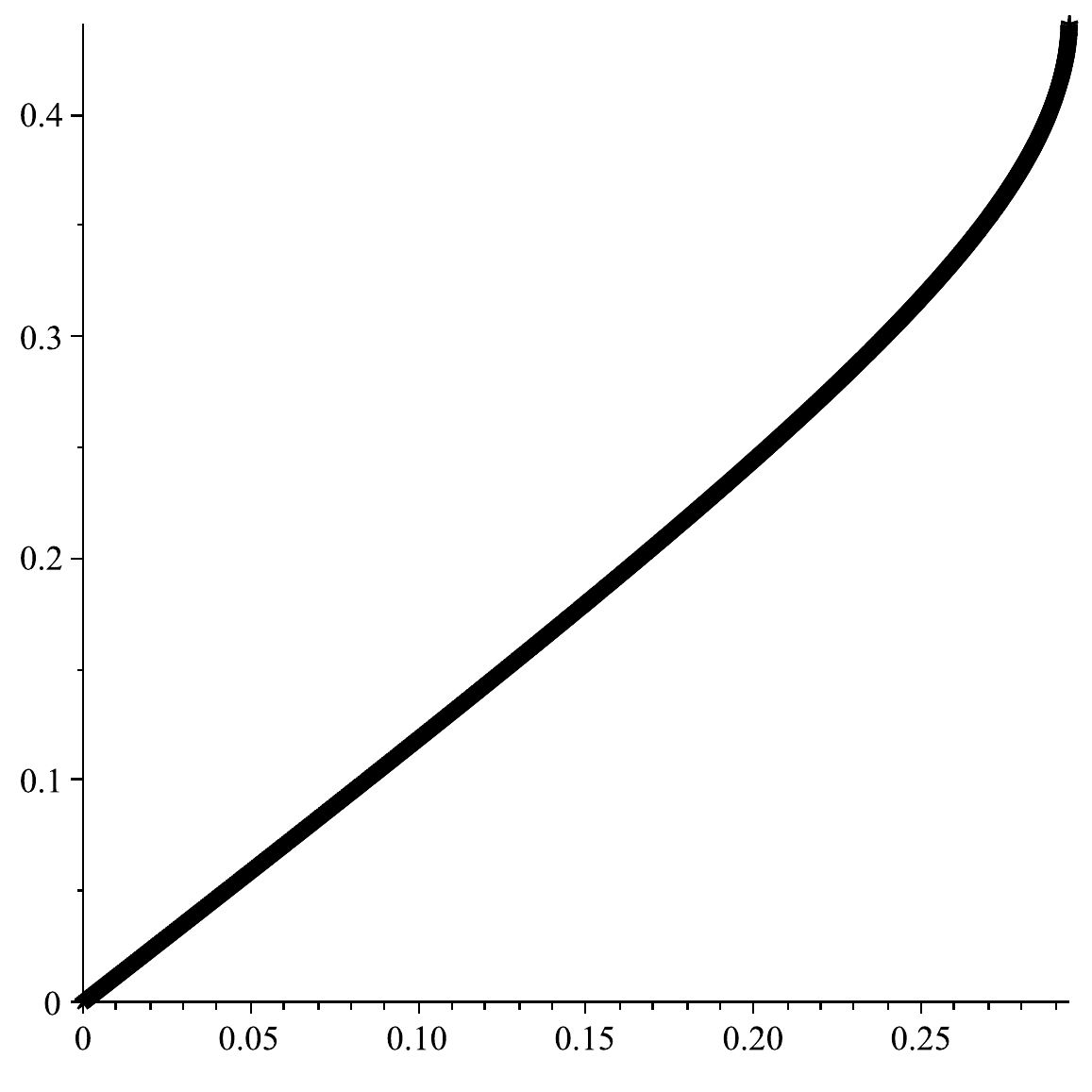}
\caption{Plot of the set $(x_1, x_2 ) \in \cS^\ep_\text{HB}$ for $\ep=10^{-6}$. \label{fig4a}}\end{figure}

Due to charge-conjugation symmetry, we have~$x_3=x_1$  and $K_3(x_1,x_2,x_1) = K_1(x_1,x_2,x_1)$. We proceed by defining the following $\kappa$-independent set,
\bea
\cS_\text{HB}^\ep = \Big \{ (x_1,x_2) \in \left [ 0,\thalf \right ]^2 \hbox{ s.t. } \big | K_2(x_1,x_2,x_1) - K_1(x_1,x_2,x_1) \big |< \ep, \, \cQ_m>0  \Big \}
\eea
where $\cQ_m$ was defined in (\ref{7.Qm}) and the condition $\cQ_m>0$ imposes a necessary (but not sufficient) condition for local stability.  A numerical plot of the set $\cS_\text{HB}^\ep$ is shown in figure~\ref{fig4a}. Finally,  the plot of the potential $V_{\{1,2,3\}^+}$ as a function of $\kappa$ is obtained by evaluating both of these quantities  on all the elements of $\cS^\ep_\text{HB}$  and is given by the solid blue curve in figure \ref{fig3a}. Note that we have independently checked that all points on that curve are actually locally stable (as indicated by the fact that the blue line in the figure is solid), even though the analysis above only imposed necessary conditions for local stability. 

\subsection{Numerical solution for $\BB=\{1,2,3\}^-$}

For the branch $\BB=\{1,2,3\}^-$ we proceed analogously, but instead of imposing $x_3=x_1$, we now impose the second solution to equation (\ref{eq:B.subbranches}),  
\bea
\label{7.bb}
4t_{11}(x_1^2 + x_3^2) + 4(t_{11} - t_{13}) x_1 x_3 + 4t_{12} x_2^2 = 2 t_{11} +  t_{12} 
\eea
It will be convenient to change variables from $x_1, x_3$ to $x, \delta$ related by $x_1= x+\delta$ and $ x_3 = x -\delta$ in terms of which the condition (\ref{7.bb}) becomes, 
\bea
\label{7.delta}
4 ( t_{11} + t_{12}) \delta ^2 = 2t_{11} (1-6x^2) + 4 t_{13} x^2 + t_{12}(1-4x_2^2)
\eea
Denoting the solutions for $\delta$ to this equation as a function of $x$ and $x_2$ by $\pm \delta(x,x_2)$ the combinations $x_1=x+\delta(x,x_2)$ and $x_3=x-\delta(z,x_2)$ parametrize all the solutions to the equation $K_1=K_3$, where the sign reversal on $\delta$ corresponds to swapping $x_1$ and $x_3$. The remaining independent components of the field equations are $K_+= \half (K_1+K_3)$ and $K_2$, both evaluated for $x_1= x+\delta$ and $ x_3 = x -\delta$. We proceed as before, by introducing the set,
\bea
\cS_{ \{ 1,2,3\}^-}  ^\ep = \Big \{  \big | K_+(x+\delta,x_2,x-\delta)-K_2(x+\delta,x_2,x-\delta) \big |< \ep,  \, \cQ_m>0 \Big \}
\eea
where~$\delta = \delta(x,x_2)$ is given by (\ref{7.delta}) and that $0 \leq x_2 \leq \thalf$ and $0 \leq x\pm \delta \leq \thalf$. Our numerical analysis shows that the set $\cS_{ \{ 1,2,3\}^-}  ^\ep$ is empty, and thus so is the entire branch $\BB=\{1,2,3\}^-$.

\newpage

 \section{Properties of the matrix $t(\mu)$}
 \label{app:appt}

In this appendix we establish various useful properties of the matrix~$t_{mn}(\mu)$ of effective~$U(1)_D^{N-1}$ gauge couplings in the dual Abelian Higgs model at the multi-monopole point, introduced in~\eqref{eq:tdef1}, \eqref{eq:tdef}, and~\eqref{eq:tdef1ii}, which we repeat here,
	\begin{align}
	\label{tklapp}
	t_{mn}(\mu)  =  \frac{1}{(2\pi)^2} \left( \delta_{mn} \log \frac{\Lambda}{\mu} + \log  \Lambda_{mn}\right)~, \qquad
	\Lambda_{mm} = 16 N s_m^3~, \qquad  
	\Lambda_{m \neq n} = \frac{1 - c_{m+n}}{1-c_{m-n}}~. 
	\end{align}
We always use the shorthand
\begin{equation}
    s_m = \sin \frac{m \pi}{N}~, \qquad c_m = \cos \frac{m\pi}{N}~, \qquad m = 1, \cdots, N-1~.
\end{equation}
The definition of $t$ in \eqref{tklapp} makes the following properties evident:
\begin{itemize}
	\item The matrix $t$ is symmetric, $t_{mn}=t_{nm}$, and furthermore is invariant under  charge conjugation, $t_{mn} = t_{N-m,N-n}$. Such a matrix is called bisymmetric.
	\item The off-diagonal elements of $t$ are all positive.  This follows from the fact that $c_{m \pm n} <1 $ and $c_{m -n} - c _{m+n} = 2 s_m s_n >0$ for all $m, n=1, \ldots, N-1$ and $m \not=n$. 
	\end{itemize}

\subsection{Eigenvalues of $t$}

As a real symmetric matrix, $t$ can be diagonalized by a real orthogonal matrix $O$ satisfying $O^T=O^{-1}$, so that the eigenvalue equation for $t$ takes the form,
	\begin{align}
	\label{exacteigs}
	\sum_{n = 1}^{N-1} t_{mn} O_{nk} = \lambda_k O_{mk}
	\end{align}
The $N-1$ real eigenvalues $\lambda_k$, $k=1,\dots,N-1$ furnish the inverse gauge couplings~$\lambda_k=1/e_k^2$ of the Abelian Higgs model in a basis in which the Maxwell kinetic terms are diagonal, without off-diagonal kinetic mixing. The diagonal one-loop running due to the massless monopoles leads to the following scale-dependence of the eigenvalues/gauge couplings, 
	\begin{align}
	\label{running}
	\lambda_k(\mu) = \lambda_k(\mu_0) - \frac{1}{4\pi^2} \log \frac{\mu}{\mu_0} \qquad 
	\Longleftrightarrow \qquad 
	e^2(\mu) = \frac{e^2(\mu_0)}{1- \frac{e^2(\mu_0)}{4\pi^2} \log \frac{\mu}{\mu_0}}
	\end{align}
We choose an ordering $e_1^2 < e_2^2 < \dots < e_{N-1}^2$, so that $\lambda_1$ is the largest eigenvalue of $t$ and $\lambda_{N-1}$ the smallest.

\subsection{Bounds on the range of $\mu$}

	\begin{figure}[t!]
	\centering
	\hspace{-1.2cm}
	\includegraphics[width=0.7\textwidth]{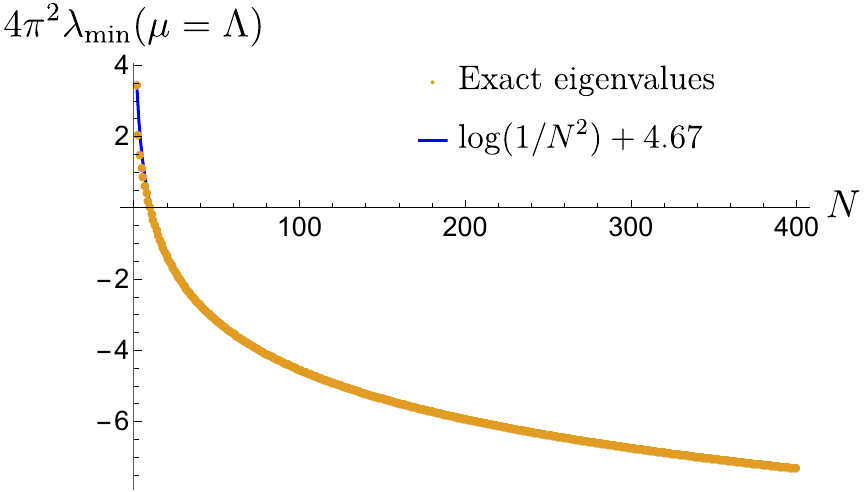}
	\caption{~The smallest eigenvalue~$\lambda_\text{min} = \lambda_{N-1}$ of the matrix~$t(\mu = \Lambda)$, rescaled by $(2\pi)^2$, plotted as a function of $N$ up to $N=400$. The best fit line  (largely not visible beneath the data points) is depicted in blue, and numerically establishes the large-$N$ scaling \eqref{pdbound}.
    \label{fig:mineig}}
	\end{figure}

The matrix~$t_{mn}(\mu)$ has two important properties that restrict the range of the RG scale~$\mu$:

\begin{itemize}
\item[(i)] Since the eigenvalues of~$t(\mu)$ are~$\lambda_k = 1/e_k^2 > 0$, they must all be positive, i.e.~$t(\mu)$ must be a positive definite matrix. We have shown explicitly that this holds for~$N \leq 10$ as long as~$\mu \leq \Lambda$, but for larger~$N$ we find a more stringent constraint:
\begin{align}
	\label{pdbound}
	t_{mn}(\mu) \quad \text{ positive definite if } \quad \mu < \mu_{\text{pos}}~, \qquad \mu_{\text{pos}} \approx \frac{107\, \Lambda}{N^2} \qquad 
	\text{as} \qquad
	N\to \infty
	\end{align}
We have numerically determined the scaling of $\mu_{\text{pos}}$ with $N$ for large values of $N$ by fitting the smallest eigenvalue $\lambda_{N-1}(\mu)$ at the reference scale $\mu=\Lambda$ as a function of $N$, and then applying \eqref{running} to run down to the critical value of $\mu$ for which the eigenvalues become positive. The resulting fit up to $N=400$ is shown in figure  \ref{fig:mineig}.

\begin{figure}[t!]
	\centering
	\hspace{-1cm}
	\includegraphics[width=0.7\textwidth]{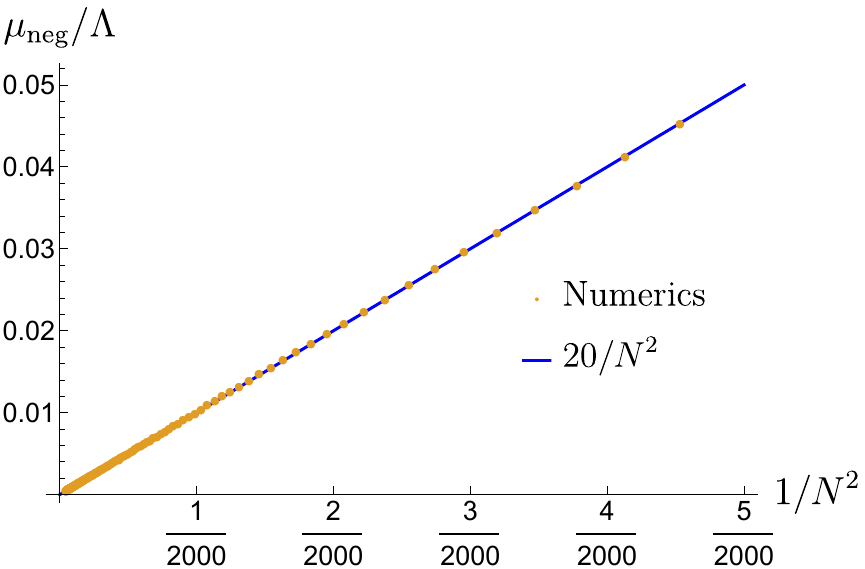}
	\caption{The data points depict the maximum value of $\mu \lesssim \mu_{\text{neg}}$ such that all off-diagonal elements of $t^{-1}$ are negative, plotted as a function of $1/N^2$  for values of $N$ ranging between $20$ and $200$.  The best fit line in blue establishes the large-$N$ scaling in~\eqref{negbound}. 
    \label{fig:muneg}}
	\end{figure}

\item[(ii)] We also restrict to sufficiently small~$\mu$ so that all off-diagonal matrix elements of the inverse matrix, $(t^{-1})_{m \neq n} < 0$ are negative. In general, this leads to the more stringent bound
\begin{equation}
    \mu < \mu_\text{neg} \leq \mu_\text{pos}~.
\end{equation}
For small values of $N$ this upper bound~$\mu_\text{neg}$ is readily computed as,  
	\begin{align}
	\label{tnegvalues}
	\begin{array}{c|cccccccccc}
	N & 2 & 3 & 4 & 5 & 6 & 7 & 8 & 9 & 10 \\ \hline
	\mu_{\text{neg}}/\Lambda  &1 & 1&  0.723& 0.577 & 0.451& 0.353 & 0.281& 0.228 & 0.188
	\end{array}
	\end{align}
     We numerically obtained the scaling of this bound with $N$ for large values of $N$ as,
	\begin{align}
	\label{negbound}
	\mu_{\text{neg}} \approx \frac{20  \Lambda}{N^2}\qquad\quad 
	\text{as}\qquad\quad 
	N\to \infty
	\end{align}
This behavior is demonstrated in figure \ref{fig:muneg}, which plots $\mu_{\text{neg}}$ as a function of $1/N^2$ over a large range of $N$.

\end{itemize}

\subsection{Approximations for the largest eigenvalues of $t$}

\begin{figure}[t!]
	\centering
	\includegraphics[width=0.48\textwidth]{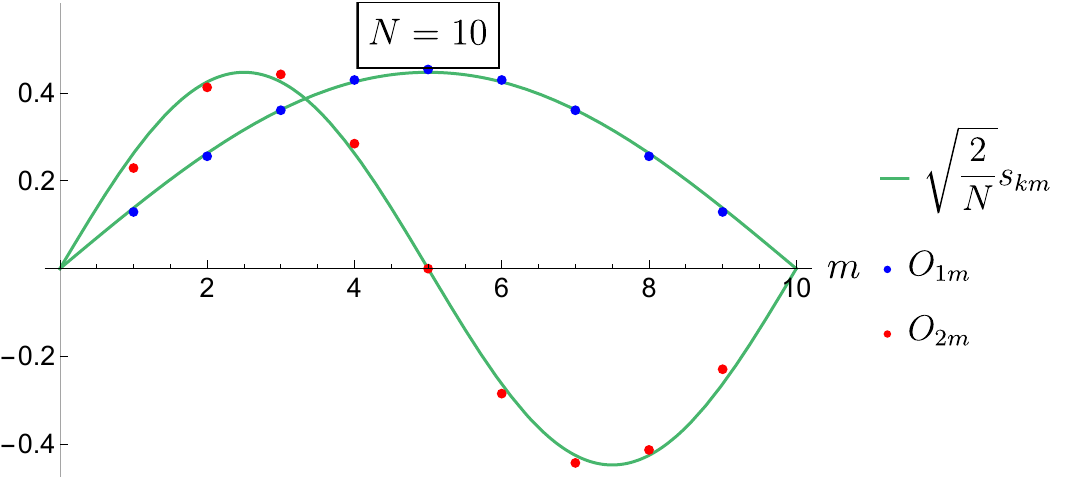}
    \bigskip
	\includegraphics[width=0.48\textwidth]{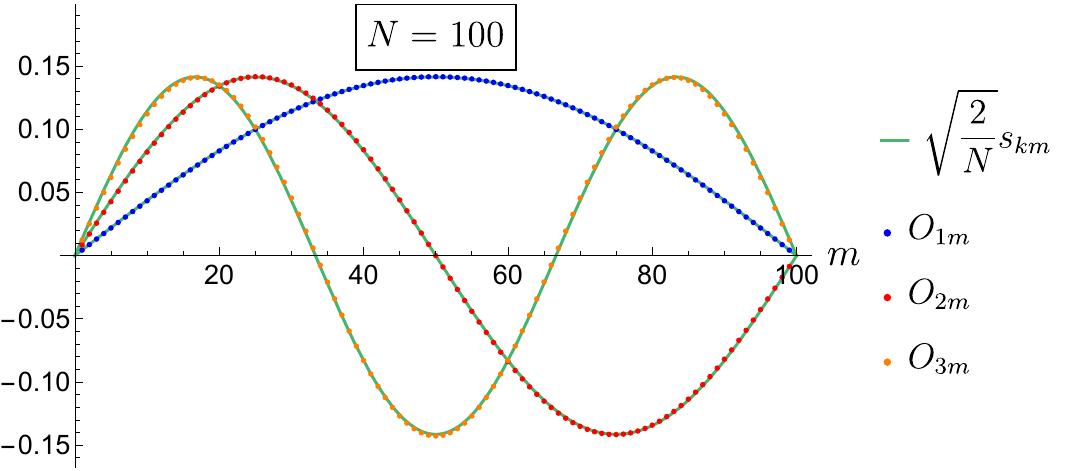}
    \includegraphics[width=0.5\textwidth]{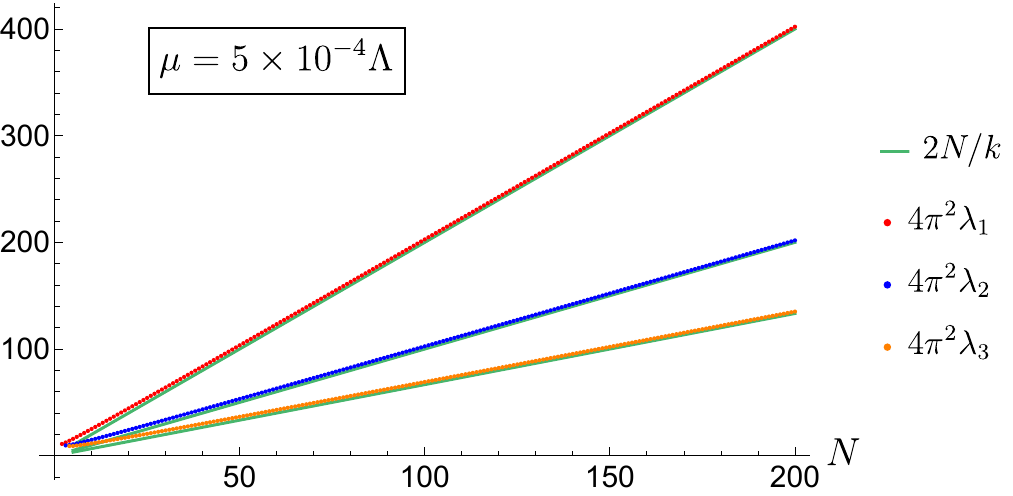}
	\caption{Exact eigenvectors $O_{km}$ corresponding to the largest eigenvalues of $t$, plotted against the approximate eigenvectors $\sqrt{2/N} s_{km}$. For $N=10$ (top left panel) we plot the $k=1,2$ eigenvectors, and for $N=100$ (top right panel) we plot the $k=1,2,3$ eigenvectors. The agreement improves for smaller $k$ and larger~$N$. The bottom panel shows the largest eigenvalues of $t$ (rescaled by $4\pi^2$) plotted as a function of $N$ up to $N=200$. We have taken $\mu=5\times 10^{-4}\Lambda$ to be consistent with the bound \eqref{negbound} for the largest values of $N$ depicted. 
	This plot numerically establishes the approximation \eqref{approxeigs}.\label{fig:skm}}
	\end{figure}

Here we establish useful analytic approximations for the largest eigenvalues~$\lambda_k$ of~$t$, and their corresponding eigenvectors, in the large-$N$ limit with~$k\ll N$. In that limit, the eigenvectors $O_{km}$ and eigenvalues $\lambda_k$ of $t$, defined in \eqref{exacteigs}, are well approximated by,\footnote{~As an aside, we note that the $s_{km}$ eigenvectors also appeared in \cite{Douglas:1995nw} as a basis for diagonalizing $t$ at large-$N$ on the special slice of moduli space that connects the multi-monopole point with semiclassical infinity (see also \cite{DHoker:2020qlp} for a review).}
	\begin{align}
	\label{approxeigs}
	O_{km} = \sqrt{\frac{2}{N}} s_{km} \qquad \quad 
	\lambda_k = \frac{N}{ 2\pi^2 k}\qquad\quad
	 \text{for}\qquad\quad 
	 k \ll N\,,\quad N\to \infty
	\end{align}
Taking~$k = 1$ in these formulas leads to the following particularly useful approximations, valid at leading large-$N$ order, which are used throughout the paper, 
	\begin{align}
	\label{useful}
	\sum_{n=1}^{N-1} t_{mn} s_{n} =  \frac{N}{ 2 \pi^2} s_{m} 
	\hskip 1in
	 \sum_{n=1}^{N-1} (t^{-1})_{mn} s_{n} =  \frac{ 2\pi^2}{ N} s_m
	\end{align}
To verify~\eqref{approxeigs}, we can replace the sum in~\eqref{exacteigs} by an integral in the large-$N$ limit and estimate the error via the Euler-Maclaurin formula. We omit the details, in part because it is straightforward to establish the same result numerically, as shown in figure~\ref{fig:skm}.

	\begin{figure}[t!]
	\centering
	\includegraphics[width=0.49\textwidth]{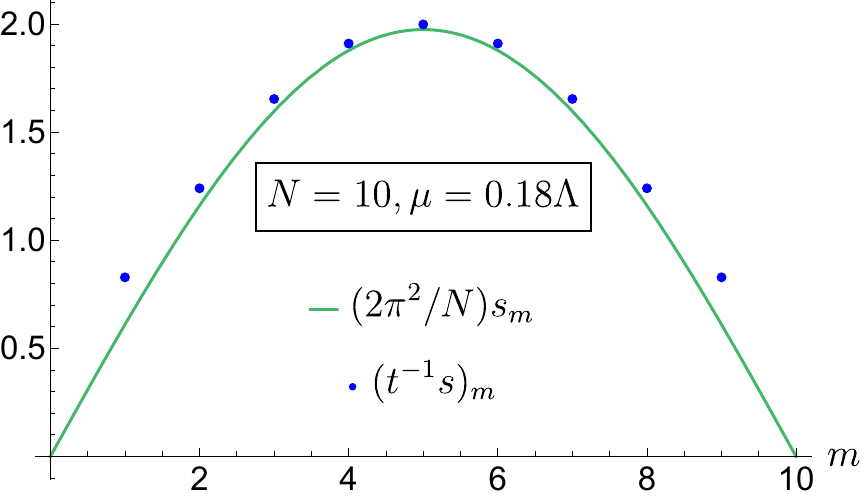}
    \bigskip
	\includegraphics[width=0.49\textwidth]{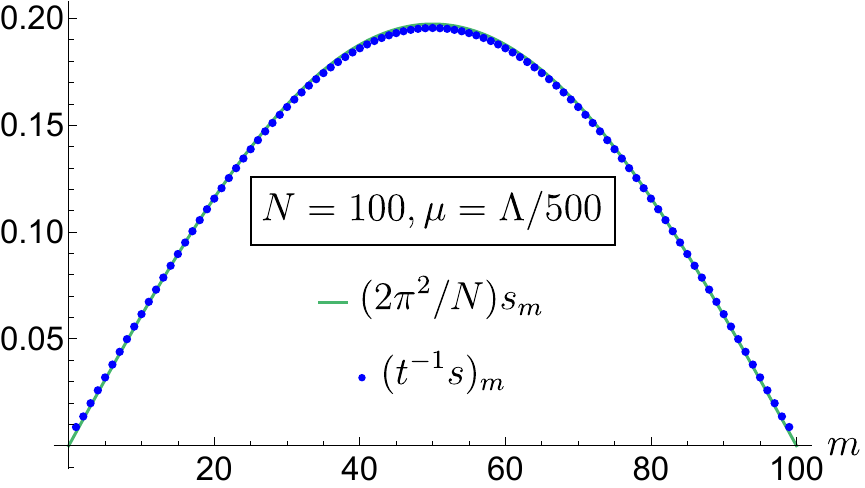}
	\caption{These plots demonstrate that the sum $\sum_{m=1}^{N-1} (t^{-1})_{mn} s_n \equiv (t^{-1} s)_m$ is well approximated by $(2\pi^2/N) \, s_m$, as claimed in \eqref{useful}. We depict the sums as a function of $m=1,\dots,N-1$ for both $N=10$ and $N=100$; we have chosen values of $\mu$ that are close to saturating the bound~\eqref{negbound}.  \label{fig:tinvs}}
	\end{figure}
	
Two comments are in order:
\begin{itemize}
\item The approximation~\eqref{useful}, is in fact excellent down to small values of~$N$. For $N=2,3$ it is exact; for~$N = 4$, the approximate eigenvector~$s_m$ is within a few percent of the true eigenvector, and the error decreases with increasing~$N$. This is shown in figure \ref{fig:tinvs}, which compares the values~$N = 10$ and~$N = 100$. 
\item As long as we only take the RG scale~$\mu$ in~\eqref{running} to scale at most like a power of~$N$, as in the bounds~\eqref{pdbound} and~\eqref{negbound}, the large-$N$ result in~\eqref{approxeigs} is robust, because the eigenvalues are only modified by subleading~$\CO(\log N)$ terms. This will enable us to estimate the leading large-$N$ scaling of various quantities in section~\ref{sec:ordparlargeN}, without having to precisely the specify the RG scale~$\mu$. 
\end{itemize}

\subsection{Sums over matrix elements of $t$ and their large-$N$ limit}

The sums evaluated in this subsection play an important role in section~\ref{sec:ordparlargeN}. We will compute the following two sums analytically -- first exactly, and then in the large-$N$ limit:
\begin{itemize}
\item[(i)] We begin by evaluating the sum over one index of~$t_{mn}$ defined by,\footnote{~Here we use the shorthand~$\cot_{\ell \over 2} = \cos (\ell / 2 \pi N) / \sin(\ell/2 \pi N)$.}
\bea
\label{eq:vn}
v_m = \sum_{n=1}^{N-1} t_{mn} &=  \frac{1}{(2\pi)^2}\left( 4  \sum_{\ell=1}^{m-1} \log \cot_{\frac{\ell}{2}}+ \log  \left({N s_m \cot_{\frac{m}{2}} }\right)  +  \log 16   \right) 
\eea	
which is exact in $N$, and derive its 	leading large-$N$ approximation as follows, 
\bea
\label{eq:vmlargeN}
v_m	 =   \frac{N}{\pi^2} f(\rho)  + \CO(\log N)\qquad  0 \leq f(\rho= \frac{m}{N}) \leq  \frac{2 G}{\pi}\simeq 0.58\quad  \text{as} \quad  N\to \infty
\eea
The function~$f(\rho = m/N)$ is a smooth~$\CO(1)$ function of~$\rho \in (0,1)$, which is~$C$-symmetric about its midpoint, $f(1-\rho) = f(\rho)$, and attains its maximum~$2G / \pi$ (where~$G$ is Catalan's constant) there. 

\item[(ii)]   We also evaluate the double sum over both indices of~$t_{mn}$,
	\begin{align}
	\label{eq:tmnexact}
	\sum_{m,n=1}^{N-1} t_{mn} 
	&= \frac{1}{(2\pi)^2}  \left(4 \sum_{m=1}^{N-1} (N-m) \log \cot_{\frac{m}{2}}+ N\left( \log N +  \log 2 \right) - \log 8 \right)  \\
	\label{eq:tmnlargeN}
	& = \frac{1}{(2\pi)^2} \left(   \frac{14 \zeta(3)}{\pi^2} N^2 + N \log N \right)+ \CO(N) \qquad \text{as}\qquad N\to \infty	
	\end{align}
Again, the first line is exact in $N$, and the second gives the behavior at large $N$. 
    
\end{itemize}
\noindent Throughout this subsection we take $\mu=\Lambda$, which generally violates the bounds on~$\mu$ in~\eqref{pdbound} and~\eqref{negbound}. Expressions at other values of~$\mu$ may be obtained using the RG running in~\eqref{running}. 

{\it Exact Results:} Let us begin by establishing~\eqref{eq:tmnexact}. 
The sum over the diagonal components yields,
	\begin{align}
	\label{dstart}
	\sum_m \log 16 N  s_m^3 
	= (N-1) \log 16  N + \log \prod_m s_m^3 
	= (N+2) \log N + (N-1) \log 2  
	\end{align}
while the off-diagonal sum yields,
	\begin{align}
	 \label{eq:start}
	\sum_{m\neq n} \log \Lambda_{mn} 
	= \sum_{m\neq n,n} \log \frac{(1-c_{m+n})^2}{(c_m-c_n)^2} 
	=2 \left[  \log \prod_{m,n \neq m} (1-c_{m+n}) - \log \prod_{m,n\neq m} (c_m-c_n) \right]
	\end{align}
By applying the following identities involving products of cosines,
	\begin{align}
	\label{cosid}
	\prod_{n \neq m} (c_m - c_n) = - \frac{(-1)^m N}{2^{N-1} s_m^2}\qquad\qquad \prod_{\ell = 1}^{N-1} (1-c_\ell) = \frac{N}{2^{N-1}}
	\end{align}
we manipulate the argument of the first log in \eqref{eq:start} as follows,
	\begin{align}
	\label{reduce}
	\begin{split}
	\prod_{n \neq m, n=1}^{N-1} (1-c_{m+n}) & = \frac{1}{(1-c_{2m})} \prod_{\ell =m+1}^{m+N-1} (1-c_\ell)  \\
	&= \frac{ N}{2^{N-1} (1-c_{2m})} \frac{ \prod_{\ell = N}^{m+N-1} (1 + c_{\ell  - N})    }{  \prod_{\ell = 1}^m (1-c_\ell)  } \\
	& = \frac{ N}{2^{N-1} s_m^2(1+c_{m})} \prod_{\ell = 1}^m \frac{(1+c_\ell)}{(1-c_\ell)} 
	\end{split}
	\end{align}
Substituting the last line of \eqref{reduce} along with the first equation of \eqref{cosid} into \eqref{eq:start}, and repeatedly using trigonometric identities, the expression simplifies as follows,
	\begin{align}
	\sum_{m \neq n} \log \Lambda_{mn} 
	&=2  \log  \prod_m \frac{1}{1+c_m} \prod_{\ell = 1}^m \frac{1+c_\ell}{1-c_\ell} \\
	&= 2 \log \prod_m \frac{1}{1+c_m}  + 4 \log \prod_{m=1}^{N-1} \prod_{\ell=1}^m \frac{s_\ell}{1-c_\ell}   \\
	&= 2 \log \prod_m \frac{1}{1+c_m}   + 4 \log \prod_{m=1}^{N-1} \left(\frac{s_m}{1-c_m}\right)^{N-m} \label{eq:lnnn}
	\end{align}
	The first term in \eqref{eq:lnnn} can be simplified using the following identity,
	\begin{align}
	\prod_{m=1}^{N-1} \frac{1}{1+c_m} 
	= \prod_{m=1}^{N-1} \frac{(1 - c_m)}{s_m^2} = \frac{1}{2N}
	\end{align}
Therefore, the entire expression \eqref{eq:start} simplifies as,
	\begin{align}
	\label{totaloffs}
	\sum_{m \neq n} \log \Lambda_{mn} 
	&= -2 \log 2N + 4 	\sum_{m=1}^{N-1} (N-m) \log \frac{s_m}{1-c_m}
	\end{align}
Combining \eqref{totaloffs} with \eqref{dstart}, and substituting $\cot(x/2) = \sin x / (1-\cos x)$, we finally obtain the entire sum $\sum_{m,n} \log \Lambda_{mn}$,
	\begin{align}
	\sum_m \log \Lambda_{mm} + \sum_{m\neq n} \log \Lambda_{mn} 
	= 4 	\sum_{m=1}^{N-1} (N-m) \log \cot_{\frac{m}{2}}+ N\left( \log N +  \log 2 \right) - \log 8
	\end{align}
This implies the desired result \eqref{eq:tmnexact}. 

Along the way, we have computed the single index sum
	\begin{align}
	v_m \equiv \sum_n t_{mn} 
	&= \frac{1}{(2\pi)^2} \left( \log \Lambda_{mm} + \sum_{n\neq m} \Lambda_{mn} \right) \\
	&= \frac{1}{(2\pi)^2}\left( \log 16 N  s_m^3  + \log \frac{1}{(1+c_m)^2} \prod_{\ell = 1}^m \frac{(1+c_\ell)^2}{(1 - c_\ell)^2} \right) \\
	&=   \frac{1}{(2\pi)^2}\left( \log 16 N  s_m^3  - 2 \log (1-c_m)  +4  \sum_{\ell=1}^{m-1} \log \cot_{\frac{\ell}{2}} \right) 
	\end{align}
establishing \eqref{eq:vn}. 

{\it Large~$N$ Approximations:} We will now expand the exact expressions in~\eqref{eq:vn} and~\eqref{eq:tmnexact} in the large-$N$ limit.\footnote{~The large-$N$ scaling of $v_m$ and $\sum_{m} v_m$, deduced analytically below, have also been verified numerically.} First, consider $v_m$ in \eqref{eq:vn}. The $N$-scaling of all of the terms except the sum over $\log \cot_{\ell/2}$ is straightforward. To determine the latter, we convert the sum to an integral, 
	\begin{align}
	\sum_{\ell = 1}^{m-1} \log \cot_{\frac{\ell}{2}} 
	&\to N \int_0^{\rho} \log \cot \frac{\pi x}{2}\  dx \equiv N f(\rho=m/N)
	\end{align}
The integral $f(\rho)$ is a finite, positive~$\CO(1)$ function of $\rho$, which is symmetric between the limits of $\rho$, $f(1-\rho) = f(\rho)$. The maximum value of $f(\rho)$ occurs at  $f(\rho = 1/2) = 2 G/ \pi \simeq 0.58$, where $G$ is Catalan's constant; its minima are at~$f(0) = f(1) = 0$. This establishes the large-$N$ limit in~\eqref{eq:vmlargeN}. 

Finally, we determine the large-$N$ scaling of the sum over both indices $\sum_{mn}t_{mn}=\sum_m v_m$ in~\eqref{eq:tmnexact}. As above, we convert the sum over $\log \cot_{m/2}$ to an integral, and then take the large-$N$ limit, leading to
	\begin{equation}\label{eq:largeNtmnlog}
    \begin{split}
        \sum_{m=1}^{N-1} (N-m) \log \cot_{\frac{m}{2}} \qquad \xrightarrow{N \to \infty} \qquad & N^2 \int_0^{1}  dx \, (1-x)  \log\cot\left(\frac{\pi x}{2}\right) \\
	& =\frac{7 \zeta(3)}{2\pi^2} N^2 + \CO(N)
        \end{split}
	\end{equation}
Substituting into \eqref{eq:tmnexact} the establishes the large-$N$ limit in~\eqref{eq:tmnlargeN}.

\subsection{Comparing~$t^{-1}s$ and~$T^{-1}s$ }
\label{sec:app.C.5}

	\begin{figure}[t!]
	\centering
	\includegraphics[width=0.95\textwidth]{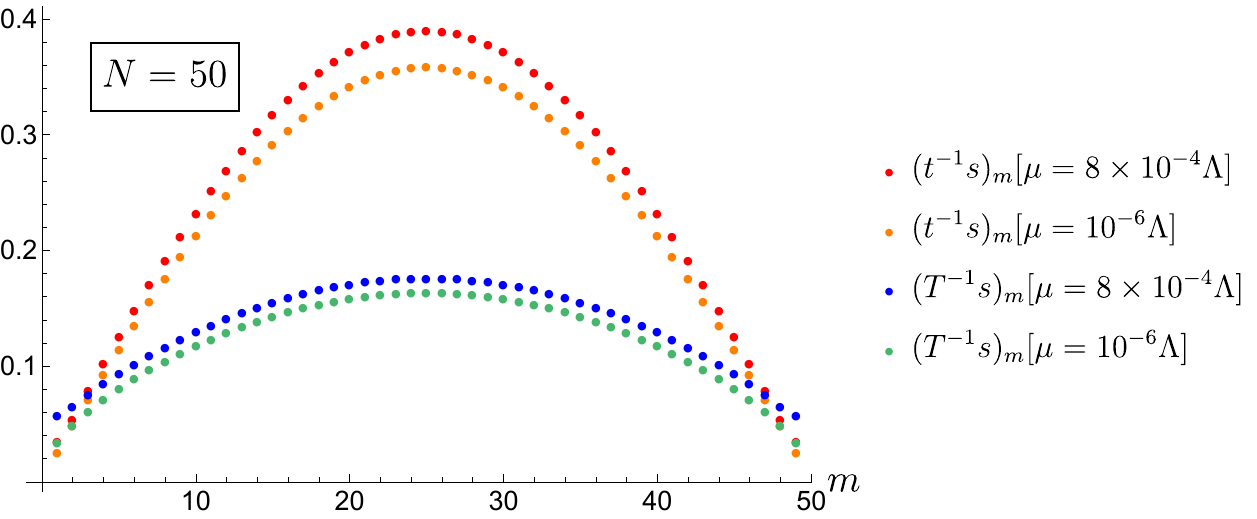}
	\caption{This plot depicts $(t^{-1}s)_m$ (the red and orange points) and $(T^{-1}s)_m$ (the blue and green points) as a function of $m=1,\dots,N-1$, for fixed $N=50$. We plot two values of the RG scale~$\mu$: one that saturates the bound~\eqref{negbound}, and the smaller value $\mu=10^{-6}\Lambda$. 
	\label{fig:tvTnequal50}}
	\end{figure}
	
	\begin{figure}[h]
	\centering
    	\includegraphics[width=0.48\textwidth]{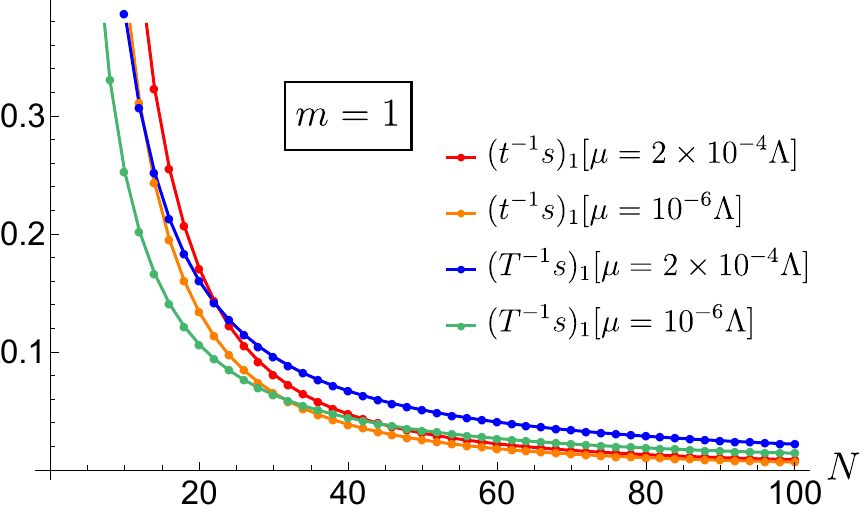}
	\includegraphics[width=0.48\textwidth]{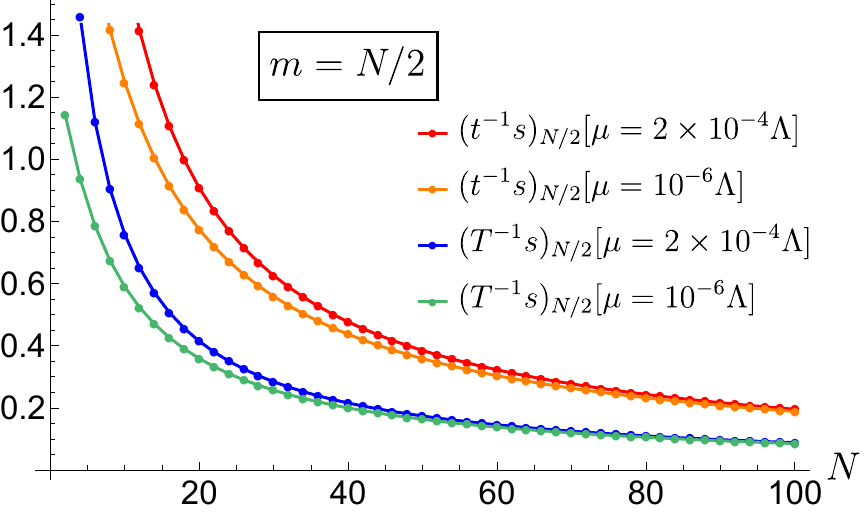}
	\caption{This plot depicts $(t^{-1}s)_m$ (the red and orange points) and $(T^{-1}s)_m$ (the blue and green points) as a function of even $N=2,\dots,100$, for~$m = 1$ (left panel) and~$m = N/2$ (right panel). We also indicate two values of~$\mu$: the first saturating the bound \eqref{negbound} for all~$N$ shown, and the second smaller value $\mu=10^{-6}\Lambda$.  
	\label{fig:tvTmiddlem}}
	\end{figure}

In this subsection, we compare the expressions~$(t^{-1} s)_m$ and~$(T^{-1} s)_m$, where the matrix~$T_{mn}$, which appears in the maximal Higgs branch solution~\eqref{6.TT}, is given by
\begin{equation}
    T_{mn} = t_{mn} + \delta_{mn} v_m~, \qquad v_m = \sum_{p=1}^{N-1} t_{mp}~.
\end{equation}
We will carry out this comparison numerically:
\begin{itemize}
    \item In figure \ref{fig:tvTnequal50} we plot $(t^{-1} s)_m$ and $(T^{-1} s)_m$ as a function of $m$ for $N=50$, and for different values of $\mu$. This figure shows that both $(t^{-1} s)_m$ and $(T^{-1} s)_m$ have similar shapes: both roughly look like a sine function~$s_m$ (see \eqref{useful}) with different amplitudes, with~$(T^{-1} s)_m \lesssim (t^{-1}s)$ by at most a factor~$\sim 3$. Moreover, these conclusions are not sensitive to~$\mu$. 

\item  In figure~\ref{fig:tvTmiddlem} we plot the two sums~$(t^{-1} s)_m$ and $(T^{-1} s)_m$ as a function of even $N$ and for different values of $\mu$. The left panel shows~$m =1$ (where the sums are smallest), and the right panel shows~$m \sim N/2$ (where the sums are largest). We find that the sums largely agree at~$m = 1$, while $(T^{-1} s)_m \lesssim (t^{-1}s)_m$ by  a factor ranging from~$\sim 1$ to ~$\sim 3$ for~$m = N/2$. Again there is no significant~$\mu$-dependence.
\end{itemize}

\newpage

	\section{Global stability of the maximal Higgs branch at $\kappa=0$}
	\label{sec:maxHglob}

In this appendix, we complement the analysis of the maximal Higgs branch HB with an investigation into its global stability. To this end, we compare the values of the potential, given in (\ref{redV}) for $\kappa=0$ and $x_k=0$, for different partitions $\AA |\BB$, and we obtain,
\bea
\label{Vu}
V_{\AA|\BB}= V_\text{CB} - {1 \over 8} \sum _{k, \ell \in \BB} u_{k \ell}
\eea 
where we have explicitly indicated the dependence of $V$ on the partition.  Let $\AA | \BB$ and $\AA' | \BB'$ be two partitions such that $\AA' \setminus \AA = \BB \setminus \BB' = \{p\}$. Equivalently the partition $\AA | \BB$ is such that the set $\BB$ contains the element $p$ and the partition $\AA' | \BB'$ is obtained by moving $p$ from $\BB$ to~$\AA'$. We shall now show that, for $\kappa=0$, we have,
\bea
\label{6.Vineq}
V_{\AA| \BB} < V_{ \AA' | \BB'}
\eea
This result states that, at $\kappa=0$, the branches of solutions are partially ordered by the cardinality $|\BB|$ of $\BB$: the larger $|\BB|$, the lower the minimum of the potential is. The absolute minimum is reached for the maximal Higgs branch where $\AA=\emptyset$. As a result, for $\kappa=0$, the maximal Higgs branch is locally and globally stable.

To prove (\ref{6.Vineq}) we consider the sub-matrices $u^{-1}$ and $(u')^{-1}$ of $t^{-1}$ of dimension $|\BB|$ and $|\BB'|$  corresponding to the partitions $\AA | \BB$ and $\AA' | \BB'$, respectively. The difference in the values of the potentials on these solutions is obtained from (\ref{Vu}),
\bea
V_ {\AA'| \BB'} - V_{\AA | \BB} 
= \frac{1}{8} \sum_{k,\ell \in \BB} u_{k\ell} - \frac{1}{8} \sum_{k',\ell' \in \BB'} u'_{k ' \ell '}
\eea
We may choose a basis, possibly by simultaneously permuting rows and columns,  in which $u^{-1}$ and $(u')^{-1} $ are related as follows,
\bea
u^{-1} = \left ( \bma (u')^{-1} & u_1 \cr u_1^t & u_2 \ema \right ) 
\hskip 1in
u = \left ( \bma v_0 & v_1 \cr v_1^t & v_2 \ema \right ) 
\eea
where $u_1$ and $v_1$ are column matrices of height $|\BB'|$ while $u_2,v_2 \in \RR$, $(u_1)_{k'} = (t^{-1})_{pk'}$ for all $k'\in \BB'$ and $u_2= (t^{-1})_{pp}$. Using the relations $(u')^{-1} v_0 + u_1 v_1^t=I$ and $(u')^{-1} v_1=-v_2 u_1$ implied by $u^{-1} u=I$, and eliminating $u_1$, we obtain $u' = v_0 - v_1 v_1^t/v_2$. 
Expressing the difference of the potentials in terms of these data, we find, 
\bea
V_{\AA'| \BB'} - V_{\AA | \BB} = \frac{1}{8 v_2} \bigg ( v_2+ \sum_{k'\in \BB'} v_{1k'} \bigg )^2 
\eea
Since $u$ is a positive definite matrix, we have $v_2>0$, which completes the proof of (\ref{6.Vineq}).

\newpage

\bibliographystyle{utphys}
\bibliography{main}


\end{document}